\documentclass[final,5p,times,twocolumn]{elsarticle}

\usepackage[T1]{fontenc}
\usepackage{ae,aecompl}
\usepackage{pslatex}
\usepackage[samesize]{cancel}
\usepackage{graphicx}
\usepackage{setspace}
\usepackage{amsmath}
\usepackage{amsfonts}
\usepackage{subcaption}
\usepackage{amssymb}
\usepackage{booktabs}
\usepackage{natbib}
\usepackage{float}
\usepackage{comment}
\usepackage{mathptmx} 
\usepackage{lineno}
\usepackage{url}
\usepackage{calc}
\usepackage{tabto}
\usepackage{lettrine}
\usepackage{multirow}
\usepackage{upgreek}

\journal{ArXiv}

\begin{document}

\begin{frontmatter}

\title{Investigation of Wavelength-induced Uncertainties in Full-Wave Radar Tomography of High Contrast Domain: An Application to Small Solar System Bodies}

\author[inst1,inst2]{Yusuf Oluwatoki Yusuf}
\ead{yusuf.yusuf@tuni.fi}
\affiliation[inst1]{organization={Computing Sciences, Tampere University (TAU)},
            city={PO Box 692},
            postcode={33101}, 
            state={Tampere},
            country={Finland}}
\author[inst2]{Astrid Dufaure}
\author[inst1]{Liisa-Ida Sorsa}
\author[inst2]{Christelle Eyraud}
\author[inst1]{Sampsa Pursiainen}
\affiliation[inst2]{organization={Aix-Marseille Univ, CNRS, Centrale Marseille, Institut Fresnel},
            state={Marseille},
            country={France}}
\begin{abstract}
This paper aims to reconstruct the internal structure of a two-dimensional test object via numerically simulated full-wave time domain radar tomography with the presence of wavelength-induced (WI) uncertainties, following from a complex domain structure, and domain diameters $21$ or $64$ times the wavelength of the signal propagating inside the target. In particular, we consider an application in planetary scientific studies of reconstructing the interior structure of an arbitrary high contrast small Solar System Body (SSSB), i.e., an asteroid, with a probing signal wavelength limited by the instrument and mission payload requirements. Our uncertainty reduction model finds the reconstruction via averaging multiple inverse solutions assuming that the WI deviations in the solutions correspond to random deviations, which we assume to be  independent and identically distributed (IID).  It incorporates error marginalisation via a randomised signal configuration, spatial-averaging of candidate solutions, frequency-based error marginalisation, and the truncated singular value decomposition (TSVD) filtering technique, based on our  assumptions of the phase discrepancy of the signal, domain parameters, and the full-wave forward model. The numerical experiments are performed for 20 and 60 MHz centre frequencies proposed for CubeSat-based radars, the latter being the centre frequency of the Juventas Radar which will be aboard Hera mission to investigate the interior structure of asteroid Dimorphos. A benchmark reconstruction of the target was obtained with the spatial averaging, sparse point density and frequency randomised configuration for both 20 and 60 MHz frequency systems. 
\end{abstract}

%\item Full-wave modelling is important to distinguish how  different parts of an object (e.g., void or surface) contribute to an observed scattered signal.
%\item Inverse modelling of scattered wavefield is an ill-posed problem  which can result into a less accurate reconstruction of the target domain as the signal frequency and, thereby, the measurement/modelling errors increase.
%\item Uncertainties following from full-wave modelling can be reduced using statistical approaches such as sparse configuration of measurement points, randomised signal configuration and averaging candidate solutions.
%\item This work aims to advance radar investigations targeting Solar System bodies, such as, the tomographic Juventas radar (JuRa) experiment which will perform measurements for the asteroid moon Dimorphos at 60 MHz centre frequency and 20 MHz signal bandwidth.
%\item We show that a full-wave forward simulation can be successfully coupled with inverse methods at such frequency to reconstruct the interior structure of a small solar body, i.e., asteroids/comets.


\end{frontmatter}

\section{Introduction}
\label{sec:Intro}

Radar tomography (RT) has emerged as a powerful technique for obtaining high resolution images of complex target domains in recent years. With the application of tomographic radar imaging spanning across several fields of research, such as biomedical imaging, geoscience, and engineering; sophisticated and computationally efficient inversion techniques to characterise the interior properties of such a target domain from the scattered field of the probing radar signal have also been developed extensively \cite{chew1990reconstruction,carlsten1995radar,semenov2005microwave,ernst2007application,pursiainen2016orbiter,wiskin2020full}. The inverse scattering problem of full-wave tomography, however, is an ill-posed inverse problem that can result into a less accurate reconstruction of the target domain as the signal frequency and, thereby, the measurement and modelling uncertainties increase. In particular, the inversion is very sensitive to the carrier, hence we propose a  method to help mitigate this sensitivity. This paper aims at reconstructing the interior structure of a complex domain, e.g., an arbitrary high contrast Small Solar System Body (SSSB), via full-wave time domain RT, and specifically, investigating the wavelength-induced uncertainties, which follow as a consequence of complex multipath signal propagation inside the target due to high contrast details in the permittivity distribution.

 The first  attempt to use radio waves transmission to infer the deep interior structure of an SSSB was the COmet Nucleus Sounding Experiment by Radio-wave Transmission (CONSERT) \cite{Kofman2007,Kofman2015}, which was part of the European Space Agency's (ESA) Rosetta mission to comet 67P/Churyumov-Gerasimenko. The emergence of small spacecraft technology as a part of deep space missions has improved the future possibilities to perform RT investigations of SSSBs with a sufficient signal coverage \cite{bambach2018discus,CDF2018}. Such a plan is included in ESA's coming Hera mission; the Juventas CubeSat carried by the Hera probe will perform tomographic radar measurements of Dimorphos, the asteroid moon of 65803 Didymos, with its Juventas Radar (JuRa)  \cite{herique2020jura,herique2020low}. 
 
 In this study, we concentrate on full-wave modelling, an important tool to distinguish how the different part of the domain (e.g., the voids or surface) contribute to the observed  scattered signals \citep{Eyraud2020analog}. This requires that the differences between the modelled and measured field are small enough compared to the measurement inaccuracies. Our focus is suppressing the effects of these inaccuracies in the reconstruction process, which is of utmost importance, e.g., based on the recent numerical study by \citep {deng2021ei+}, where the envelope inversion method was used in suppressing local minima points in the observed signal with a misfit function. The asteroid model in this study has an average real relative permittivity of $\varepsilon_r' \approx 4$ which is higher than that of some comets  at $\varepsilon_r ' \approx$ 1.4, i.e., comet 67P and Wild 2 \cite{brownlee2012overview,Kofman2015, Herique2018}. In JuRa, the centre frequency is 60 MHz, hence, from $\lambda={\mathtt{c}_0\varepsilon_r'^{-1/2}/f}$, the wavelength is 5 m outside and approximately 2.5 m inside the target, where $\mathtt{c}_0$ is the speed of light in vacuum.   The  diameter of Dimorphos is about 160 m, which is  approximately 64 times larger than the wavelength of the signal propagating inside it. We define the wavelength-induced uncertainties as modelling errors in form of phase inaccuracies brought about by the nonuniformity of the permittivity distribution, change in wave velocity and irregular propagation path in the target domain.

 In full-wave modelling with the finite element time domain (FETD) method \cite{pursiainen2016orbiter,takala2018multigrid}, the WI uncertainties in the modulated signal can result in modelling or measurement errors produced by (1) wavelength-induced uncertainties' factors i.e., nonuniform permittivity distribution, change in wave velocity and wave path length in the target domain, (2)  discretization, (3) wave propagation accuracy, and (4) numerical inaccuracy due to noise in demodulation of the carrier signal. This study concentrates on the errors due to (1), wavelength-induced uncertainties, which are caused by the short wavelength of the signal as compared to the target object diameter, and propagation through a high contrast domain. To marginalise these errors, we propose a Gaussian prior model for the uncertainty of the phase in form of a discrepancy function. This discrepancy gives the maximum limit for the applicable baseband frequency considering the real part of the permittivity that can be reconstructed. The numerical experiments performed on a two-dimensional domain show how the discrepancy-based  prior model, combined with total variation regularised-inversion approach, allows for the marginalisation of the wavelength-induced uncertainties via independence sampling. The experiments also show how data obtained for a dense spatial point distribution can be effectively utilised in the process of marginalising wavelength-induced uncertainties, taking into account the Nyquist sampling condition which depends on both the modulated and demodulated wave \cite{khare2002direct,chaparro2018signals}.
 
 This article is organised into Sections \ref{sec:materials}-- \ref{sec:discuss}. Section \ref{sec:materials} focuses on the model description, and statistical formulations of spatial and frequency-based error marginalisation process. Furthermore, the total variation inversion technique and the truncated singular value decomposition filtering are discussed in Section \ref{sec:inversion}. In Section \ref{sec:exp}, the experimental setup and signal configurations are highlighted, and the corresponding parameters related to the numerical implementation are documented.  Section \ref{sec:results} includes the numerical results for the two-dimensional analogue model and the discussion of these results is  presented in Section \ref{sec:discuss}.
 
\section{Materials and methods}
\label{sec:materials}

\subsection{Inverse wave propagation problem}

We consider transverse electric (TE) wave propagation in which the total electric field $u$ propagating in the horizontal plane is oriented along the vertical direction. The TE-field satisfies the wave equation  
\begin{equation}
\label{wave}
\varepsilon_r' \frac{\partial^2 u}{\partial t^2} + \sigma \frac{\partial u}{\partial t} - \Delta u = \frac{\partial \mathfrak{V}}{\partial t} 
\end{equation}
in a given spatio-temporal domain $[0,T] \times \Omega$, in which the spatial part includes the scattering target $\Omega_1$ and its near surroundings $\Omega_2$.  Here, $\varepsilon_r'$ is the real part of the relative permittivity ($\varepsilon_r' = \varepsilon_r - j \varepsilon_r''$), $\varepsilon_r''$  the imaginary part of the relative permittivity, $\varepsilon_r$  the relative permittivity, $\sigma$ the conductivity $\sigma = 2 \pi f  \varepsilon_r''$, $f$ denotes the signal frequency, and $\partial \mathfrak{V}/ \partial t$ is a point source term for which $\mathfrak{V} = \mathfrak{V}(t, \vec{p}_0)$ represents the current density of a vertical antenna set at point $\vec{p}_0$. The spatial scaling is assumed to be such that the velocity of the wave in vacuum is one ($\mathtt{c}_0= 1$). The time $t$, position ${\vec r}$,  permittivity $\varepsilon'_r$, conductivity $\sigma$, and velocity ${\mathtt c}= \varepsilon_r'^{-1/2}$   can all be scaled to SI-units through the expressions $s \mathtt{c}_0^{-1} t$, $s {\vec r}$, $\varepsilon_0 \varepsilon'_r$, $s^{-1} \varepsilon_0  \mathtt{c}_0 \sigma$, and  $\mathtt{c}_0 {\mathsf c}$,  respectively. Here, $s$  is  a spatial scaling factor (meters), $\mathtt{c}_0 = (\varepsilon_0  \mu_0)^{-1/2}$ is the speed of the electromagnetic wave in vacuum,   $\varepsilon_0 = 8.85 \cdot 10^{-12}$ F/m is the electric permittivity of vacuum, and $\mu_0 = 4 \pi \cdot 10^{-7}$ H/m is the magnetic permeability which is assumed to be constant in $\Omega$.

The domain is assumed to be decomposed by a triangular mesh $\mathcal{T} = \{ T_1, T_2, \ldots, T_M \}$, whose j-th triangle $\mathcal{T}_j$  corresponds to a set indicator function $\chi_j$, with  $\chi_j(\vec{r}) = 1$ if $\vec{r} \in T_j$ and $\chi(\vec{r}) = 0$ otherwise. The mesh $\mathcal{T}$ discretises the real relative permittivity as 
\begin{equation}
\label{perturbation}
\varepsilon_r' =  \tilde{\varepsilon}_r' + \sum_{j = 1}^M s_j \chi_j,
\end{equation}
where $\tilde{\varepsilon}_r'$ is a constant background permittivity, $s_j$ is the corresponding coefficient of $\chi_j$, and $\tilde{{\bf x}} = \sum_{j = 1}^M s_j \chi_j$ is the perturbation of the discretised real relative permittivity. 
{Replacing the total electric field $u$ in the wave equation \eqref{wave} which is a smooth function, by its partial derivative with respect to $s_j$ results in}
\begin{equation}
\label{totalfield}
\varepsilon_r' \frac{\partial^2}{\partial t^2} \left( \frac{\partial u}{\partial s_j} \right) + \sigma \frac{\partial }{\partial t} \left( \frac{\partial u}{\partial s_j} \right)  - \Delta \frac{\partial u}{\partial s_j} =  \frac{\partial}{\partial t} \left( \frac{\partial h}{\partial s_j}\right),  
\end{equation}
where $h$ on the right-hand side is referred to as the scattering source which follows from Ampere's law, and  relates directly to $\mathfrak{V}$, as $\mathfrak{V}=\frac{\partial h}{\partial s_j}$. Here,  $h$ is denoted by
\begin{equation}
\label{source}
    h =  \varepsilon_r' \frac{\partial u}{\partial t}\vert_{s=0} =  \sum_{j = 1}^M s_j \chi_j \frac{\partial u}{\partial t}\vert_{s=0}  + \tilde{\varepsilon}_r' \frac{\partial u}{\partial t }\vert_{s=0} ,
\end{equation}
hence, its partial derivative with respect to $s_j$ gives $\partial h/\partial s_j = \chi_j \frac{\partial u}{\partial t}\vert_{s=0} $. This scattering source can be interpreted as the source of the differentiated full-wave as, e.g., in \cite{liu2018full}. We obtained the measurement data ${\bf y}$ as a difference between a full-wave detailed model and a full-wave constant background model $(u(t, \vec{p}_i) -  \tilde{u}(t, \vec{p}_i))$. Here, the detailed model can be interpreted as the superposition of the background model and the perturbation $\tilde{{\bf x}}$.  Furthermore, a differential operator $J_\ell$ is used to obtain a linearised approximation of the detailed model taking into account the coupling of the respective scatterers and multi-path effects with the background. However, it omits the coupling between the scatterers (scatterer-to-scatterer interaction). This approximation is similar to the Extended-Born Approximation (EBA) \cite{abubakar2005green,gao2006high}, i.e., it is a model which tries to estimate multiple scattering effects in addition to the direct ones.  Unlike the classical Born approximation in which the total field $u$ has been replaced with the incident field \cite{van1954correlations,chew1990reconstruction,sorsa2020time}, this linearisation incorporates higher-order or multi-path effects between the scattering obstacle and the target body. It is one of the two common ways to approach full-wave inversion, the second method being to solve a local nonlinear least-squares optimisation problem \cite{virieux2009overview}. 

The wavefield is assumed to consist of two complex quadrature amplitude modulated (QAM) components \cite{sorsa2021analysis}. The components of QAM consist of the in-phase and quadrature component, of which, the latter has a $\pi/2$ phase difference compared to the in-phase signal. Consequently, the amplitude of the signal can be obtained accurately anywhere in the spatio-temporal domain. The reconstruction process utilises  baseband frequency data ${\bf y}$ obtained via QAM demodulation as a full  wave recorded for time steps $t_1, t_2, \ldots, t_N$ at spatial points $\vec{p}_1, \vec{p}_2, \ldots, \vec{p}_K$. The data corresponds to the  difference between the wavefields  $u(t, \vec{p}_i)$ and  $\tilde{u}(t, \vec{p}_i)$ associated with the actual real relative permittivity distribution $\varepsilon_r'$  and its background estimate $\tilde{\varepsilon}_r'$.  Using the differentiated signal, this difference can be estimated as follows:
\begin{equation}
\label{linear}
   y_j(t) = \mathfrak{DM} [u(t, \vec{p}_i) -  \tilde{u}(t, \vec{p}_i)]  =   \mathfrak{DM}  \left[ \sum_{j = 1}^M s_j \frac{\partial }{\partial s_j} u(t, \vec{p}_i) \right], 
    \end{equation}
where $\mathfrak{DM}$ denotes a QAM demodulation operator. Defining Jacobian matrices ${\bf J}_\ell$ with 
\begin{equation} (J_\ell)_{i,j}  = \mathfrak{DM} \left[ \frac{\partial }{\partial s_j} u(t_\ell, \vec{p}_i) \right], \end{equation}  
    where $\frac{\partial }{\partial s_j} u(t_\ell, \vec{p}_i)$ denotes a linearised approximation of the full wavefield with respect to the permittivity  perturbation of a single point scatterer (triangle), one obtains a linearized forward model
    \begin{equation} 
    \label{linear_inverse_problem}
    {\bf y}_\ell =  {\bf J}_\ell \tilde{{\bf x}} + {\bf n}_\ell \quad \ell = 1, 2, \ldots, N,  \quad \hbox{i.e.,} \quad {\bf y} =  {\bf L} \tilde{{\bf x}} + {\bf n} 
    \end{equation}
for the inverse problem of reconstructing ${\bf x}$  given the data ${\bf y}$. Here,  \begin{equation} 
 {\bf y} =   \begin{pmatrix} {\bf y}_1 \\ {\bf y}_2 \\ \vdots \\ {\bf y}_N \end{pmatrix}, \quad 
{\bf L} =   \begin{pmatrix} {\bf J}_1 \\ {\bf J}_2 \\ \vdots \\ {\bf J}_N \end{pmatrix}, \quad \hbox{and} \quad {\bf n} =   \begin{pmatrix} {\bf n}_1 \\ {\bf n}_2 \\ \vdots \\ {\bf n}_N \end{pmatrix}.  \end{equation} The matrix ${\bf L}$ determines the linearized effect of the scatterers on the data ${\bf y}$ as a function of the perturbations of the exact discretised real relative permittivity $\tilde{{\bf x}}$, and the vector ${\bf n}$ is the noise term. We assume here that the noise term includes both measurement and modelling inaccuracies. Of these, the latter are associated with the difference $u - u_s$, where $u$ is the actual wave and $u_s$ is its simulated counterpart which can include both discretization and general modelling errors.

\subsection{Wave propagation in high and low contrast medium}
\label{sec:prop}

We focus on inverting equation \eqref{linear_inverse_problem} under the assumption that   ${\bf L}$ contains inevitable modelling errors following from wavelength-induced uncertainties in the wave simulations. The carrier wave, in which the inputs are imposed, has 4 and 6 times the frequency of the baseband signal for the two systems of this study. In QAM, the transmitter uses the input signal to vary the carrier's amplitude, hence creating a modulated signal.   
\begin{figure}[!ht]
\centering
\begin{scriptsize}
 \begin{minipage}{8.0cm} \centering
 \begin{minipage}{3.8cm} (A) \\ \vskip0.1cm 
   \begin{minipage}{3.8cm} 
    \includegraphics[width=3.3cm]{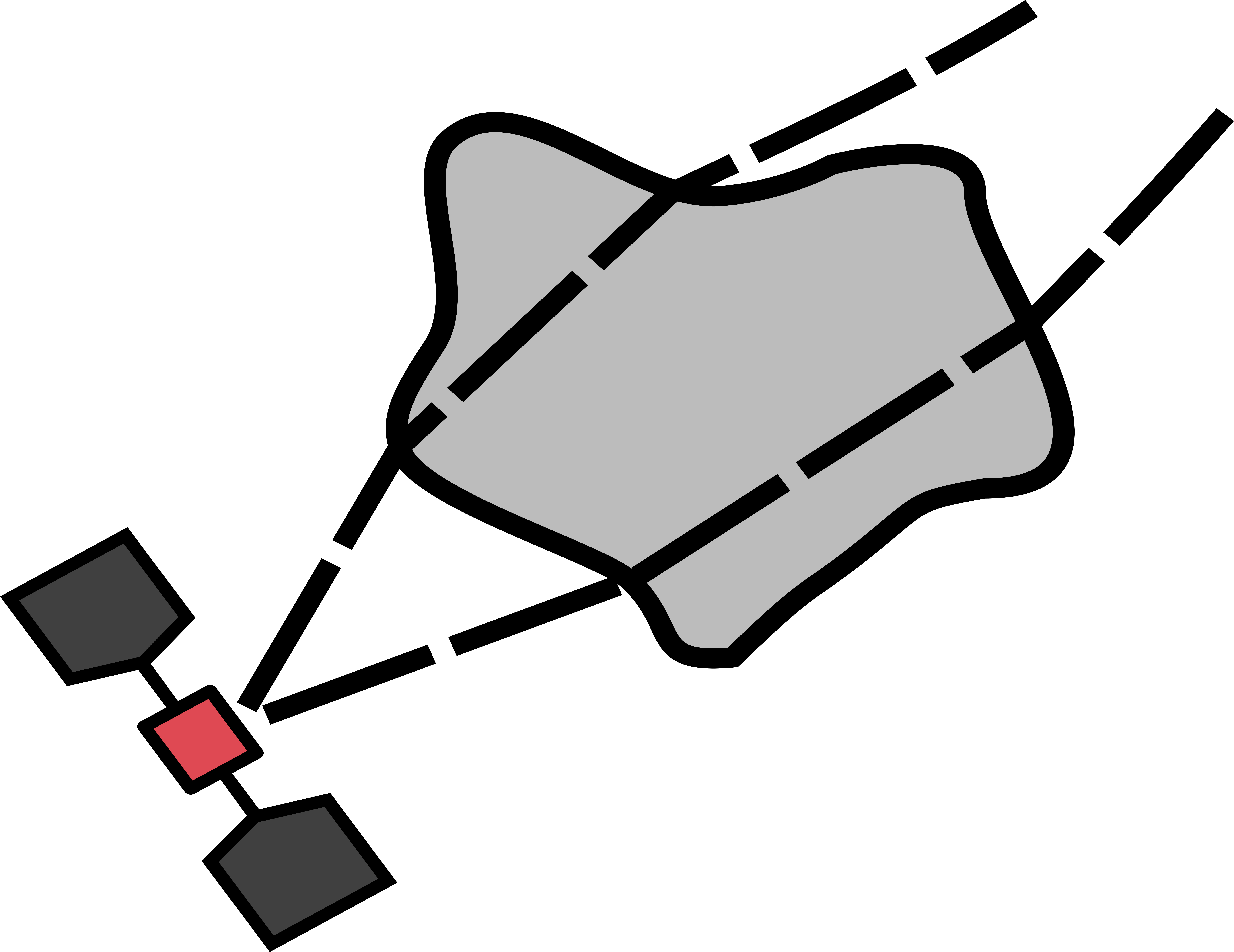}
    \end{minipage}
    \end{minipage}
    \begin{minipage}{3.9cm} (B) \\ \vskip0.1cm 
            \begin{minipage}{3.4cm} 
    \includegraphics[width=3.4cm]{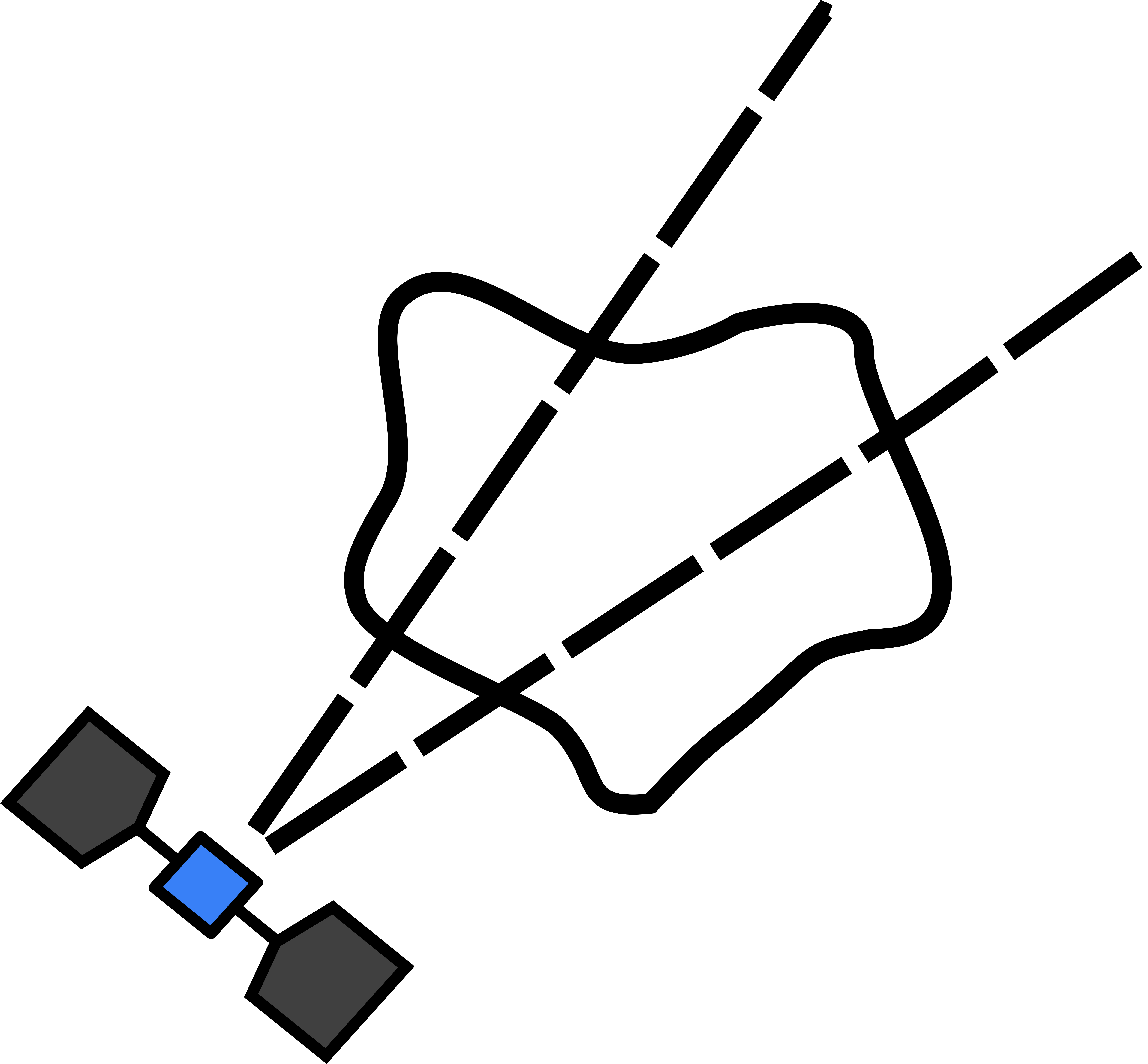}
    \end{minipage}
    \end{minipage}
    \begin{minipage}{3.9cm} (C) \\ \vskip0.5cm 
    \includegraphics[width=3.8cm]{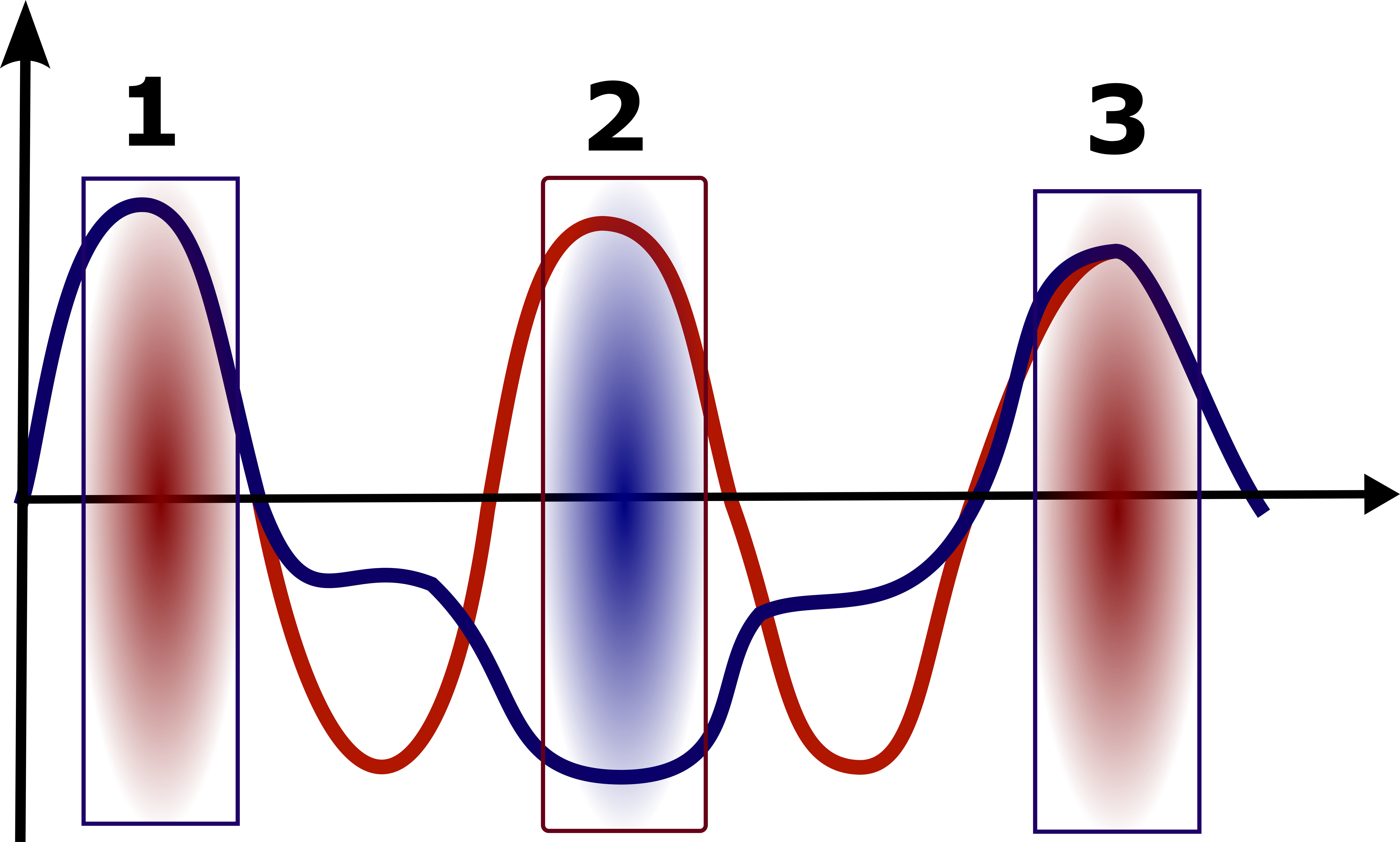}
    \end{minipage} 
        \begin{minipage}{3.8cm}  (D) \\ \vskip0.1cm 
    \includegraphics[width=3.6cm]{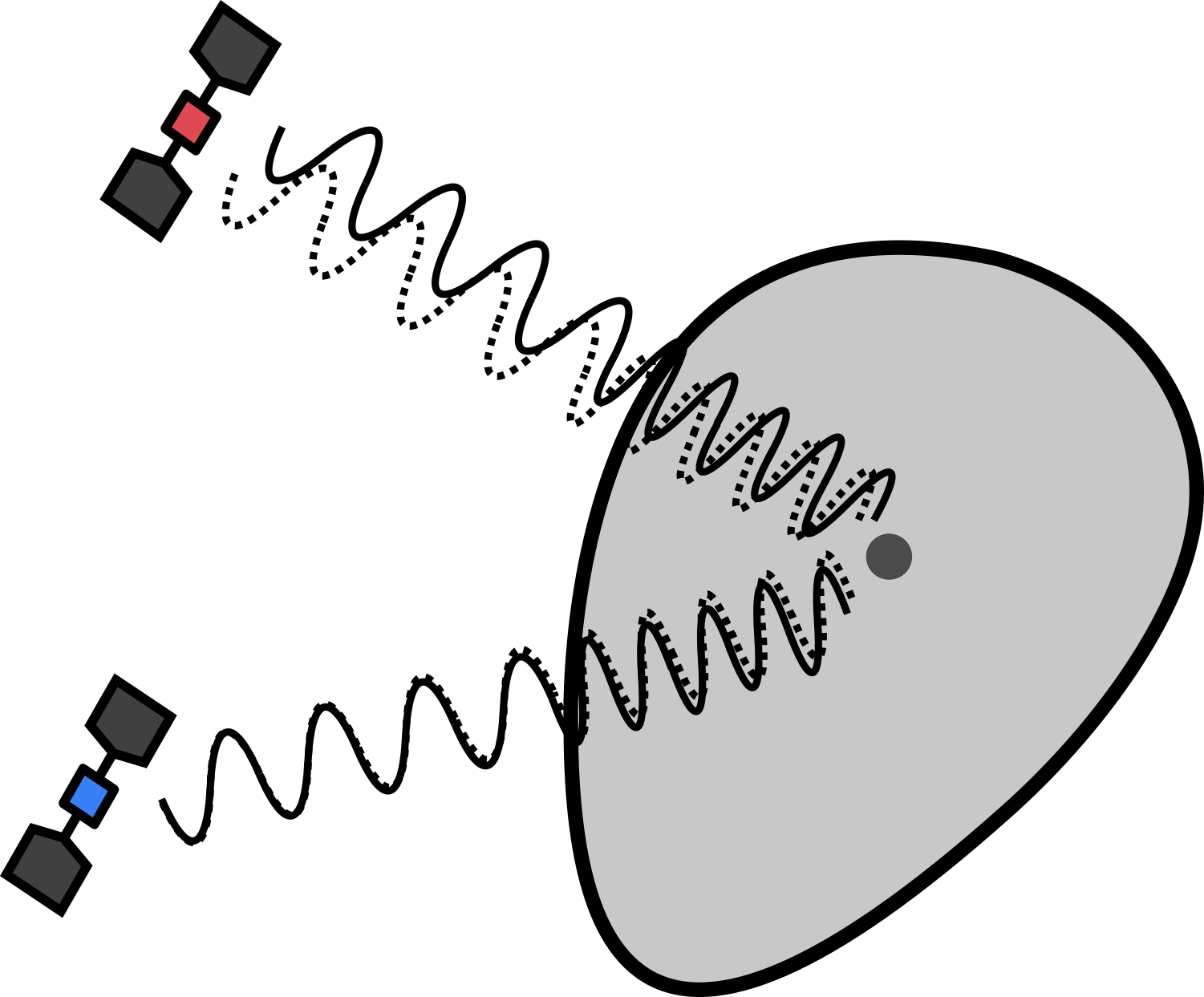}  
     \end{minipage}\\ \vskip0.2cm 
    \end{minipage}
    \end{scriptsize}
    \caption{Forward modelling error sources,  their effect and discrepancy.  Schematic illustration of signal propagated through a high contrast (A)  and low contrast (B) domain. Since the velocity of the wave, and thereby wavelength decreases when the contrast increases, there is a phase shift of $\pi$ as the signal propagates from a medium of low to a medium of high refractive index  \cite{daniels2004ground}. {Moreover, the increase in the probing signal energy concentrated at the target surface is likely to increase the amplitude errors, resulting into larger refraction in the high contrast medium due to increased deviations from the signal energy}.(C) Wavelength-induced uncertainties as a result of phase misfit can lead to large random fluctuations in the demodulated baseband data;  three (1,2 and 3) of the local maxima of the demodulated red wave coincide with those of the blue wave. The second peak has a flipped sign due to an opposite phase of the carrier (areas with matching peaks shaded in red and flipped peaks shaded in blue). More generally, phase misfit can cause any phase error in the complex plane while the difference between the sign corresponds to the phase angle of $\pi$ in the complex plane. (D) Two signals (solid and dotted wave) are considered to be in the same phase if the phase angle difference is less than or equal to a quarter of pi, i.e., $\phi \leq \pi/4$. We assume that this discrepancy condition is satisfied for a wavefield propagating the length $\ell$ inside the target $\Omega$ with real relative permittivity $\epsilon_r'$, if the maximum frequency of the wavefield is less than or equal to the baseband frequency (pulse bandwidth) $f_B$, and if the uncertainty of these parameters satisfies the discrepancy condition in equation \eqref{discrepancy_inequality}.
    \label{fig:discrepancy}}
\end{figure}

Specifically, a wavefield propagated through a complex domain undergoes changes as a result of permittivity variations within the spatial domain leading to  effects such as refraction and reflection, on the material boundaries. The boundary effects give rise to changes in the wave amplitude, phase, and direction of propagation depending on the contrast of the medium described by the refractive index \cite{Bousquet2017}. The greater the refractive index is, the stronger the effects are. The refractive index $\mathtt{n}={\mathtt{c}_0}/{\mathtt{c}}=\sqrt{\varepsilon'_r}$ is obtained as the ratio of the free space velocity ${\mathtt{c}_0}$ and the velocity $\mathtt{c}$ in the medium,  via  the real relative permittivity $\varepsilon_r'$. Since the frequency is maintained everywhere in the medium,  there is a smaller velocity and a shorter wavelength in the high-contrast (high permittivity) target than in free space or low-contrast target. Hence, the complexity of the wave propagation, and thereby the occurrence of wavelength-induced uncertainties increases along with the value of the refractive index causing an increase in the point-wise difference between the actual and simulated wave; See Figure \ref{fig:discrepancy}.

\subsection{Structure detectability in inverse reconstruction }

The inverse wave modelling in this study is characterised by a limited spectrum as described in the wave propagation phenomena in Section \ref{sec:prop}. It can be expected that there are intense scattering peaks from different parts of the domain that cannot be predicted by our forward model, meaning that the deviations are replicated or amplified in the inverse solution. Therefore, it is difficult to distinguish the surface and void back scattering data with the full-wave modelling approach in the absence of prior information on the domain structure. While the existence of a scattering obstacle, for example, a void beneath a surface layer might be predicted based on the difference between the simulated and background data, obtaining the magnitude of the scattering perturbation for the full domain is difficult under limited spectral coverage of the data \cite{dogan2017detection}. Hence, our focus in this study is on the overlap obtained between the exact structure and reconstruction. 

\subsection{Prior model for wavelength-induced uncertainties}
\label{sec:uc}

\begin{figure}[!ht]
\centering
\begin{scriptsize}
  \begin{minipage}{4.7cm}  ({\bf A}) Modulated signal pulse  \centering \\ \vskip0.1cm 
    \includegraphics[width=4.7cm]{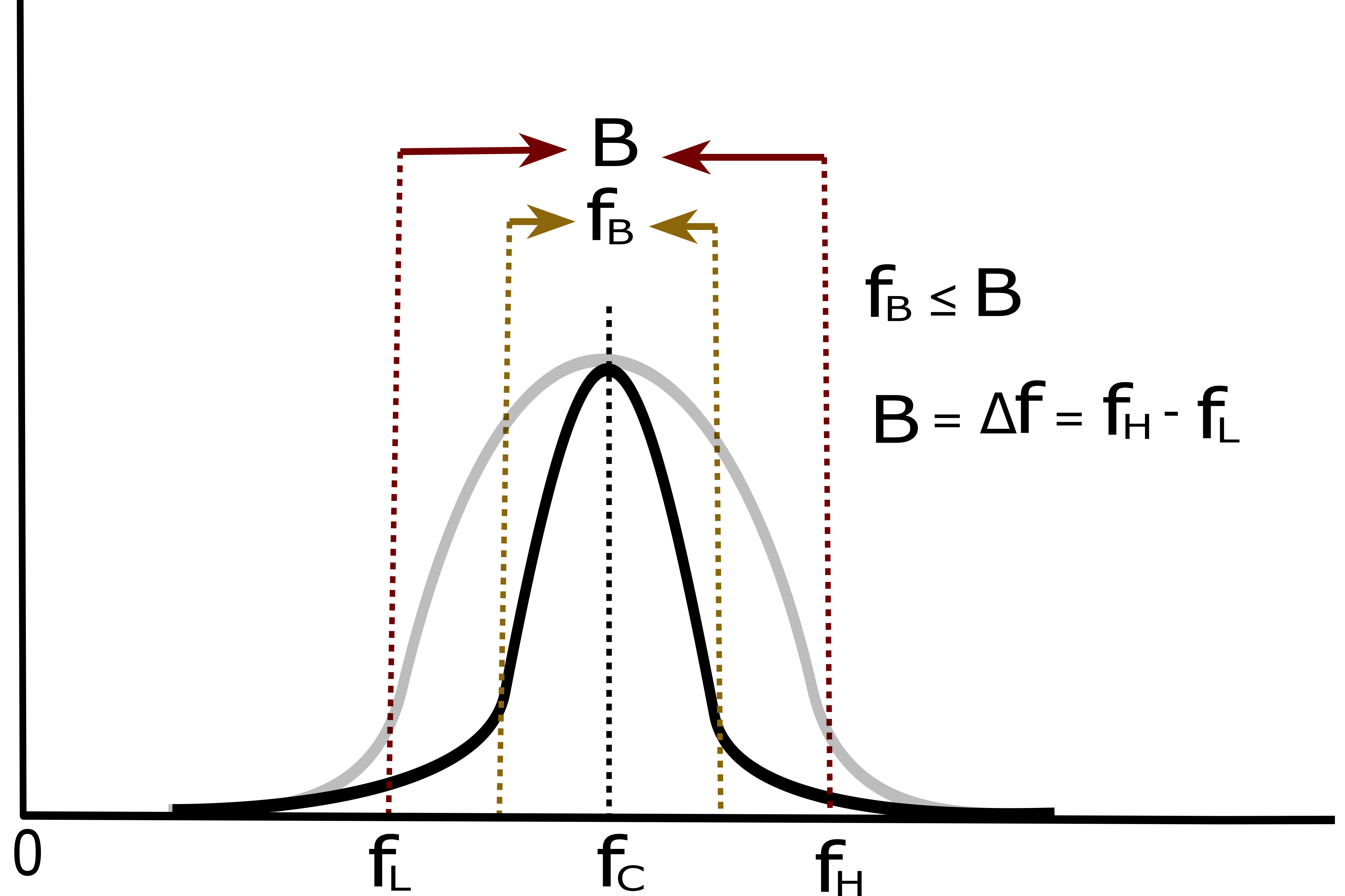}
    \end{minipage} 
    \begin{minipage}{3.3cm}  ({\bf B}) Demodulated signal pulse  \centering \\ \vskip0.1cm 
    \includegraphics[width=3.3cm]{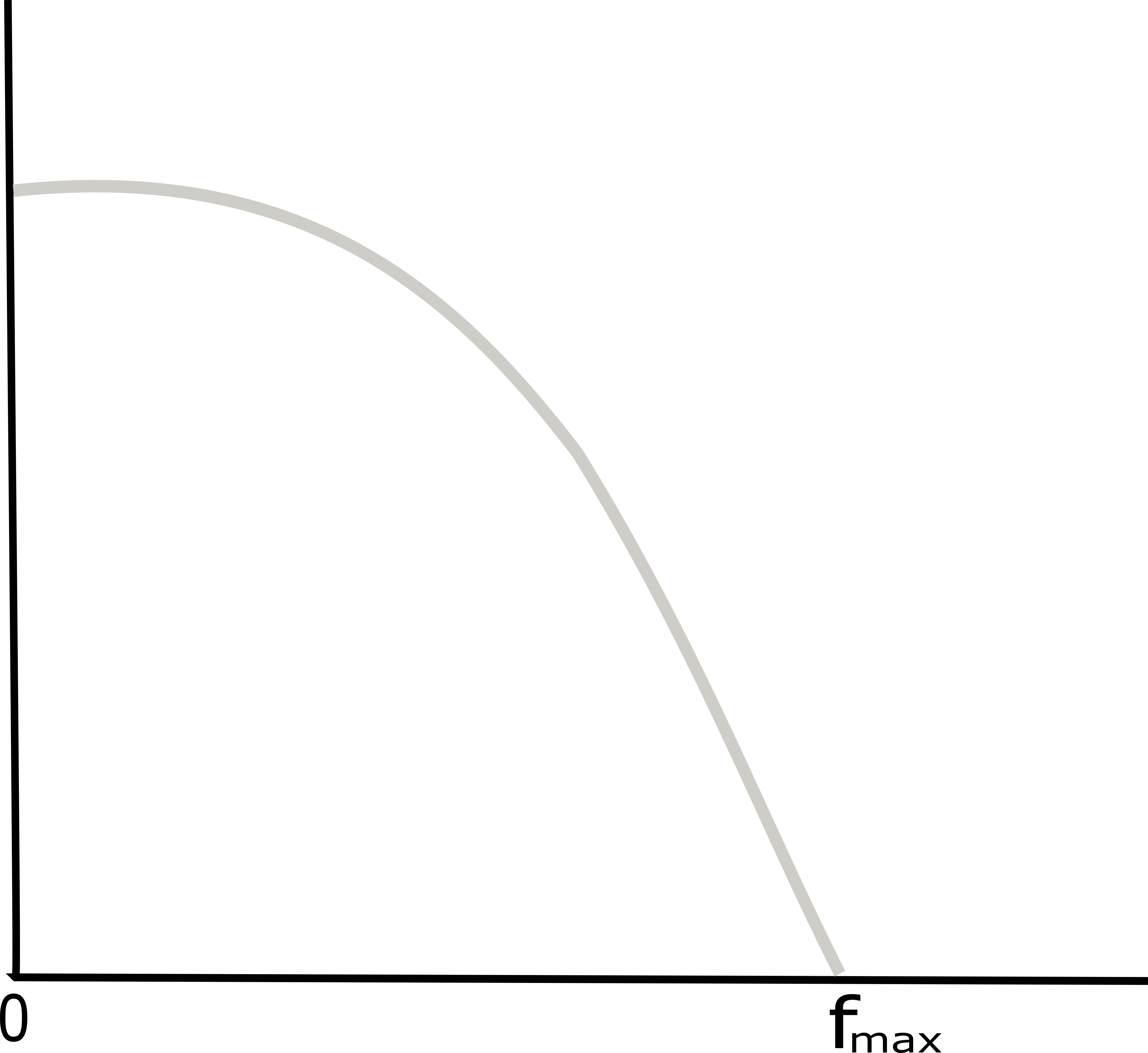}
    \end{minipage}
    \end{scriptsize}
    \caption{Schematic diagram of (A) modulated and (B) demodulated signal pulses. The radar bandwidth $B$ bounds the pulse bandwidth $f_B$ from above, hence the propagated pulse is always limited by the instrument capacity. Here, $f_H$ and $f_L$ are the upper and lower cut-off frequencies respectively, $f_C$ is the centre frequency of the modulated signal pulse, while  $f_{\max}$ is the maximum frequency of the demodulated signal pulse.}
    \label{fig:frequency} 
\end{figure}

Wavelength-induced uncertainty in a propagated wave signal can be viewed from the spatial and spectral perspectives. The Nyquist sampling criterion plays an important role in determining the number of points, sampled from a wave signal. We consider a pulse of length $T$ and pulse bandwidth $1/T = f_B \leq  B$ which matches the twice maximal frequency of the demodulated signal and is bounded by the radar bandwidth $B$ from above; See Figure \ref{fig:frequency}. 

The pulse is modulated via the QAM approach which incorporates an in-phase and quadrature components into the signal resulting in the centre frequency $f_{\hbox{\scriptsize centre}}$, corresponding to the centre of the spectrum with a pulse bandwidth $f_B$. The Nyquist criterion is defined by the spatial and temporal  sampling rate  $\mathrm{S}_\mathrm{s}$ and $\mathrm{S}_\mathrm{t}$, stating $\mathrm{S}_\mathrm{s} \geq 2f_{\max}/ \mathtt{c}$ and $\mathrm{S}_\mathrm{t} \geq 2f_{\max}$. This implies that, the QAM demodulated baseband signal  can be captured fully if the sampling frequency is greater than double the highest frequency  $f_{\max}$  contained in the baseband signal \cite{orfanidis1995introduction}. In case of QAM, $f_{\max} = f_B/2$ i.e., the maximum pulse frequency corresponds to half the pulse bandwidth for a smooth pulse. We use the first derivative Blackman-Harris (BH) pulse, which is employed often in geophysical FETD modelling applications as the probing pulse of a radar signal. The modulated signal is transmitted and propagated through the tomographic target $\Omega_1$, after which it is demodulated to obtain the final simulated  measurement data ${\bf y}$ containing frequencies on the baseband frequency interval $[0,f_{\max}]$. 

\subsubsection{Model for grid reduction}
\label{sec:grid}

The principles of compressed sensing, see, e.g., \cite{egiazarian2007compressed}, provide a viable means to reduce the wavelength-induced uncertainties. That is, a sparse and a stable reconstruction of the unknown permittivity perturbation can be obtained utilising a reduced data coverage in the reconstruction process. To avoid phase shifts in the eventual signal, i.e., the demodulated data wave, a statistically motivated approach is needed to reduce spatial uncertainty based on the assumption that the wavelength-induced uncertainties are random.  Among other possible approaches are  the classical regularised inversion methods or nonlinear data fitting in the form of least squares optimisation \cite{liu2018full,tejero2018appraisal,dong2020convolution}. We assume that the pulse bandwidth has been set so that any data fluctuation ${\bf z}$ related to  frequencies higher than that of the demodulated wave, i.e., the pulse bandwidth $f_B$, are noisy due to modelling inaccuracies, and therefore, constitute a nuisance to be marginalised out of the final reconstruction. Because the higher frequency fluctuations are regarded as noise, we set the spatial sparse grid density to half the wavelength corresponding to the bandwidth frequency. Given a simulated measurement data ${\bf y}$, and the corresponding data fluctuations ${\bf z}$, the permittivity perturbation distribution of a target domain can be approximated by a reconstruction {$({\bf x}  = \tilde{{\bf x}} + \eta)$} assuming the distribution of the permittivity perturbation is unknown for a target domain with a low-to-moderate complexity in terms of its geometry, where $\eta$ is the deviation from the exact permittivity perturbation. Associating the unknown permittivity perturbation distribution ${\bf x}$ with a conditional posterior probability density of ${\bf x}$ given ${\bf y}$ and ${\bf z}$, $\pi({\bf x} \mid {\bf y}, {\bf z})$, the effect of ${\bf z}$ on ${\bf x}$ in which both are dependent, can be marginalised as \cite{finetti1972probability}:  \begin{equation} \label{margin}
\pi(\bf {\bf x} \mid {\bf y}) = \int_{\mathbb{R}^m} \pi ({\bf z)} \pi ({\bf x} \mid {\bf y}, {\bf z}) \, \hbox{d} {\bf z},
    \end{equation}
where $\pi ({\bf z})$ is a probability density for the modelling errors, especially wavelength-induced uncertainties; { $\pi(\bf {\bf x} \mid {\bf y})$  is the marginalised density, and ${\mathbb{R}^m}$ is the space of ${\bf z}$}. Furthermore, the expectation of ${\bf x}$ given the data ${\bf y}$ can be obtained as: 
\begin{equation}
\label{posterior}
\mathbb{E}(\bf {\bf x} \mid {\bf y}) = {\bf x}^\dag = \int_{\mathbb{R}^n} {\bf x} \,  \pi ({\bf x} \mid {\bf y}) \,  \hbox{d} {\bf x},
\end{equation}
where ${\mathbb{R}^n}$ is the space of ${\bf x}$. In our current approach, ${\bf x}$ given ${\bf y}$ and ${\bf z}$ is determined by a classical regularised inversion scheme, that is, the conditional distribution of ${\bf x}$ given ${\bf y}$ and ${\bf z}$ corresponds to Dirac's delta function.
Substituting equation \eqref{margin} in equation \eqref{posterior} where  $ \pi ({\bf x} \mid {\bf y}, {\bf z})$ is replaced with the Dirac's delta function $\delta(F({\bf y},{\bf z}) - {\bf x})$. According to the  delta function's shifting property, this gives
\begin{equation}
    \label{substitute}
    {\bf x}^\dag = \int_{\mathbb{R}^m} \int_{\mathbb{R}^n} {\bf x} \,  \delta(F({\bf y},{\bf z}) - {\bf x} )  \,  \hbox{d} {\bf x} \,  \pi ({\bf z}) \, \hbox{d} {\bf z}.
\end{equation}

The simplest model is to assume that the wavelength-induced  uncertainties are IID or nearly IID and that the posterior is set by a  deterministic function ${\bf x} = F({\bf y}, {\bf z})$. Thus, the expectation is obtained as 
\begin{equation}
\label{expectdagger}{\bf x}^\dag =\int_{\mathbb{R}^m} F({\bf y},{\bf z}) \pi({\bf z}) \, \hbox{d} {\bf z}.  
\end{equation}
Equation \eqref{expectdagger} can be resolved  by the Monte Carlo sampling to save computing resources \cite{liu2008monte}. However, to ensure convergence of the sampling, the realisations ${\bf z}_k$ should be sampled from the distribution $\pi({\bf z})$ or an arbitrary distribution very close to $\pi({\bf z})$ and assumed to be uniform in this case. The law of large numbers and the generalised central limit theorem \cite{liu2008monte} motivate the convergence if the sample in question is weakly enough correlated  and the convergence rate $\mathcal{O}(K^{-1/2})$ of the sample-based mean  
\begin{equation}
\label{sample_based}
 \frac{1}{K} \sum_{k = 1}^K F({\bf y}, {\bf z}_k) \to {\bf x}^\dag, \quad \hbox{when} \quad K \to \infty. 
\end{equation} 

To formulate this approach for spatial point sets, we define sets $\mathcal{P}_k$ and $\mathcal{Q}$ in which the mutual distance between any two points is less than or equal to $\lambda_{f_{\max}}/2 = \mathtt{c}_0/(2f_{\max})$  and $\lambda_{\hbox{\scriptsize centre}}/2 = \mathtt{c}_0/(2 f_{\hbox{\scriptsize centre}})$, respectively. That is, $\mathcal{P}_k$ is dense enough to satisfy the spatio-temporal Nyquist criterion w.r.t.\ the information content of the scattered field \cite{Bucci1987}. It is further assumed that $\mathcal{Q}$ contains the full set of measurement points and $\mathcal{P}_k$, $k = 0,1,2,\ldots, k$ with an equal number of points in each is its subset, $\mathcal{P}_k \subset \mathcal{Q}$. As the measurements are inverted after the QAM demodulation, if modelling errors are absent, each dataset to be inverted has a full coverage when the criterion w.r.t.\ $f_{\max}$ is satisfied.  However, due to the existence of the modelling errors, each dataset satisfying this criterion needs to be considered rather as a subset of the full data fulfilling the criterion w.r.t.\ $f_{\hbox{\scriptsize centre}}$.  As we consider the modelling error for a given set to be a random variable,  the sample-based mean in equation \eqref{sample_based} can be realised by first finding a suitable point set sample $\mathcal{P}_k$ which can be obtained either by a regular or randomised point selection process. 

\subsubsection{Model for phase error formation}
\label{sec:phase}

A complementary approach to reduce uncertainty in the spectral sense is to apply a discrepancy condition. Assuming that the phase angle shift between the modelled and measured demodulated signal is maximally $\pi/4$, the signal is measured for a two-way path whose length inside the target is $\ell$. If the signal penetrates to  the centre of the target from each direction, providing a full signal coverage for the interior, then $\ell$ can be associated with the largest diameter of the target. In order to maintain the phase angle difference within the interval $[0,\pi/4]$, the path length per wavelength inside the target, defined as
\begin{equation}
N  = \ell/\lambda_{f} = \frac{\sqrt{\varepsilon_r'} f \ell}{\mathtt{c}_0},
\label{eq:N}
\end{equation}
for the modelled and measured wavefield, must not differ more than $\pi/4$ in phase which is equivalent to one-eighth of the pulse cycle. This gives rise for the following definition of the phase discrepancy, $\mathtt{d}$, which is based on the maximum relative difference of $N$:
\begin{equation}
\label{discrepancy}
    \mathtt{d}= \frac{N+\frac{1}{8}}{N} -1 = \frac{8\sqrt{\varepsilon_r'} f \ell + \mathtt{c}_0}{8\sqrt{\varepsilon_r'} f \ell } - 1,  
\end{equation}
where $\varepsilon_r'$ represents the average real relative permittivity of the target. The maximum deviation between $N_{sim}$ and $N$ resulting from the simulated and actual wave propagation is, thereby, 
\begin{equation}
\label{discrp_N}
    1- \mathtt{d} \leq \frac{N_{\hbox{\scriptsize sim}}}{N}  \leq 1 + \mathtt{d} \quad \hbox{or} \quad   \left|  \frac{\Delta N}{N}\right| \leq \mathtt{d},
\end{equation}
where $\Delta N = N_{sim} - N$.
By combining the equations \eqref{eq:N} and \eqref{discrp_N}, this can be expressed in terms of absolute total derivative as
\begin{equation}
\label{discrepancy_inequality}
    \frac{|\Delta \varepsilon_r'|}{2 \sqrt{\varepsilon_r'}}   +  \frac{|\Delta f|}{f} + \frac{|\Delta \ell|}{\ell} \leq \mathtt{d}.
\end{equation}
Firstly, this inequality allows for estimating the discrepancy, given the uncertainty of the {\em a priori} real relative permittivity estimate $\varepsilon_r'$ and that of the path length $\ell$, i.e., the more complex the geometry and the higher the contrast of the domain the more uncertain $\ell$. Secondly, together with the definition of the discrepancy in equation \eqref{discrepancy}, it also enables estimating the maximal bandwidth applicable in the measurement; that is, the discrepancy implied by the permittivity, path length and frequency deviation should coincide with the {\em a priori} uncertainty estimates obtained for these quantities. Thirdly, the uncertainty given by equation \eqref{discrepancy_inequality} can be incorporated into the forward modelling process through a surrogate approach, assuming that all uncertainty follows from frequency deviation $\Delta f$, which refers to the possible frequency fluctuations of the modulated signal which are reflected to the baseband data through the demodulation process. Since the effect of such deviations can be evaluated via signal demodulation, one can use  surrogate approach which allows avoiding repetitive and computationally expensive wave simulations. We define a surrogate frequency $f_s$ to be a Gaussian random variable with mean $f_m$ and standard deviation $\mathtt{d}$ matching with the discrepancy, i.e., 
\begin{equation}
\label{freq}    
f_s = f_m  + \mathtt{d} \mathtt{r},
\end{equation}
where $\mathtt{r}$ is a zero-mean Gaussian random variable with standard deviation equal to one and $f_m$ is the actual centre frequency $f_{\hbox{\scriptsize centre}}$. The formulation in equation \eqref{freq} models the uncertainty as a surrogate frequency $f_s$ for other uncertainty types explained
in equation \eqref{discrepancy_inequality}. Equation \eqref{discrepancy_inequality} shows how frequency, velocity, and path-length uncertainties contribute to the wavelength-induced uncertainties. Thus, we can use frequency perturbations as a surrogate for the velocity and path-length uncertainties. Namely, it is difficult to perturb the permittivity distribution for each case and similarly the domain structure to account for the uncertainties relating to velocity and path-length. This would require performing an extensive set of computationally costly full-wave simulations, increasing the already expensive calculations by a large factor. Hence, perturbing the frequency is used to create similar effect as what would be obtained by perturbing the permittivity and domain structure. This is further described in section \ref{sec:sig-spec} and implemented in the case of randomised averaging approach experiments in this study.

\subsection{Inversion techniques}
\label{sec:inversion}

Full-wave tomography is challenging, partly because there is a need to select an inversion method that appropriately solves the problem of reconstructing scattered waves from the target domain. Expectedly, there is a trade-off between the performance and complexity of several inversion methods available.

\subsubsection{Total variation}
\label{sec:tv}
Total variation (TV) is an inversion method that has evolved from image denoising application to more robust applications of inverse problems \cite{Luisierr2013}. 
It has been found to be stable in tomographic reconstruction of  distributional information based on sparse data and to even enable finding a robust reconstruction assuming that the distribution to be reconstructed is also sufficiently sparse \cite{sorsa2020time,candes2006stable,candes2006robust}. We obtain a  TV regularised solution of the linearized forward model, i.e., ${\bf x} = \min_{{\bf x}} F(x)$ with  
\begin{equation}
\label{tv_solution}
F(\mathbf{x})=\left\|\mathbf{L} \mathbf{x}-\mathbf{y}\right\|_{2}^{2}+2 \sqrt{\alpha} \|\mathbf{D} \mathbf{x}\|_{1},
\end{equation}
following the approach of  \cite{pursiainen2016orbiter,takala2018multigrid}, i.e., 
{\setlength\arraycolsep{2 pt} \begin{eqnarray}
    \label{rweTV}
    {\bf x}^\ddag & = & {\bf L}^\ast {\bf y}, \label{backpropagation} \\
    {\bf x}_{\kappa + 1} & = & ({\bf L}^\ast {\bf L} +  {\bf D}^\mathrm{T}\Gamma_{\kappa} {\bf D})^{-1} {\bf x}^\ddag. 
\end{eqnarray}}
Here, ${\bf x}^\ddag$ is a backpropagated estimate for ${\bf x}$.  $\Gamma_{\kappa} = \mathrm{diag}(|{\bf D x}_{\kappa}|)^{-1}$ for all $ \kappa \geq 0$, where $\Gamma_{0} = {\bf I}$, is the weighting matrix limiting the magnitude of ${\bf x}$ to avoid numerical instability due to division by a very small value. $D$ is a normalised positive definite and invertible derivative operator computed over the edges of $\mathcal{T}$,  defined as
\begin{equation}
\label{tvparam}
D_{i,j}= \frac{\| {\bf L } \|_{\hbox{\scriptsize fro}} }{\sqrt{N} } \left(  \beta  I_{i,j} + \alpha \frac{(2 I_{i,j}-1) \int_{\mathcal{T}_i\cap \mathcal{T}_j}  ds}{\max_{i,j} \int_{\mathcal{T}_i\cap \mathcal{T}_j} ds} \right), 
\end{equation} where $I_{i,j} = 1$ if $i = j$, $I_{i,j} = 0$ if $i \neq j$, and $\alpha$ and $\beta$ are the regularisation and smoothing parameters of the solution, and  the average Frobenius or column norm $\| {\bf L } \|_{\hbox{\scriptsize fro}}/{\sqrt{N}} $ is regarded as a normalising constant. The final reconstruction is found as a vector containing the absolute value of the entries in the final iterate.

According to the general theory of regularisation \cite{engl1996regularization,kaipio2006statistical}  the noise effects in the reconstruction should be suppressed by ensuring that the regularisation level is great enough to prevent  noise corruption of  the objective function. Here, the TV regularisation is considered to be balanced when the two terms in equation \eqref{tv_solution} are roughly equal, i.e.,  if  $\sqrt{\alpha}$ approximately matches the expected relative noise fluctuations in the data. For example, if the total effect of known and unknown noise is $\approx$ 20 dB, then $\alpha \approx 0.01$.  The smoothing parameter $\beta$  defines a diagonal weight for TV, ensuring that ${\bf D}$ is invertible. We choose the magnitude of this parameter  to be slightly less than $\alpha^2$ to ensure the appropriate invertibility of ${\bf D}$ of the TV regularisation while keeping the contribution of the diagonal weighting below the expected noise effects.

\subsubsection{Filtering via truncated singular value decomposition}

Singular value decomposition (SVD) \cite{golub2013matrix} is a traditional means of data compression.  We combine the statistical approach described in Section \ref{sec:uc} with the truncated SVD (TSVD) to filter spatio-temporal data \cite{ludeno2020comparison}. This is motivated by the assumption that scattering consists of separate scattering patterns  $n = 1, 2, 3, \ldots, m$, organised in descending order by their amplitude. The spatial pattern and time-dependence are described by two orthogonal sets of vectors ${\bf u}_1, {\bf u}_2, \ldots, {\bf u}_m$ and ${\bf v}_1, {\bf v}_2, \ldots, {\bf v}_m$. The vector entries in ${\bf u}_n$ and ${\bf v}_n$ correspond to the spatial points and time steps of the measurements, respectively. Consequently, the matrix ${\bf Y} = ({\bf y}_1, {\bf y}_2, \ldots, {\bf y}_N)$ contains the full data  and the TSVD filter is defined as follows: 
\begin{equation}
\label{tsvd}
{\bf x}^\ddag = {\bf L}^\ast {\bf y}  \approx  \sum_{n = 1}^p \sigma_n    {\bf K}_n^\ast  {\bf u}_n, \quad \! \! \hbox{where} \! \!  \quad {\bf K}_n = \sum_{\ell = 1}^{N}  \overline{ v}_{\ell,n} {\bf J}_\ell, 
\end{equation}
 ${\sigma}_n$ denotes the amplitude of the $n$-th most intense pattern, and $p \leq m$ is selected so that the energy corresponding to the remaining sum is approximately that of the noisy signal, i.e., $\sum_{n = {p+1}}^m \sigma_n^2 = \| {\bf n} \|_2^2$. In essence, we choose $p$ such that $\sum_{i = 1}^p\sigma_i^2/\sum_{n = 1}^m \sigma_n^2 \geq 1/(1+10^{-\frac{\mu}{20}})^2$, where $i= 1, \ldots, m$ and $\mu$ is the noise level. The right-hand side provides a lower bound by the assumption that the amplitude of the signal is a sum of the noiseless and noisy signals.

In this study, the TSVD approach is applied in the backpropagation part in equation \eqref{backpropagation} of the TV regularisation process. We compare its performance to the case of full data.  Here, ${\bf K}_n^\ast {\bf u}_n$ can be interpreted as a backpropagated pattern ${\bf u}_n$ with intensity $\sigma_n$ and time dependence ${\bf v}_n$. This formulation of TSVD is used since the scattering patterns are assumed to be well localised in the time domain. 

\subsection{Numerical Experiments}
\label{sec:exp}

\subsubsection{Two-dimensional domain}
\label{sec:two-dim}

\begin{figure*}[!ht]
\begin{scriptsize}
\begin{minipage}{16cm} 
    \centering  \begin{minipage}{5.1cm} ({\bf A})  \centering \\ \vskip0.1cm 
     \includegraphics[width=5.1cm]{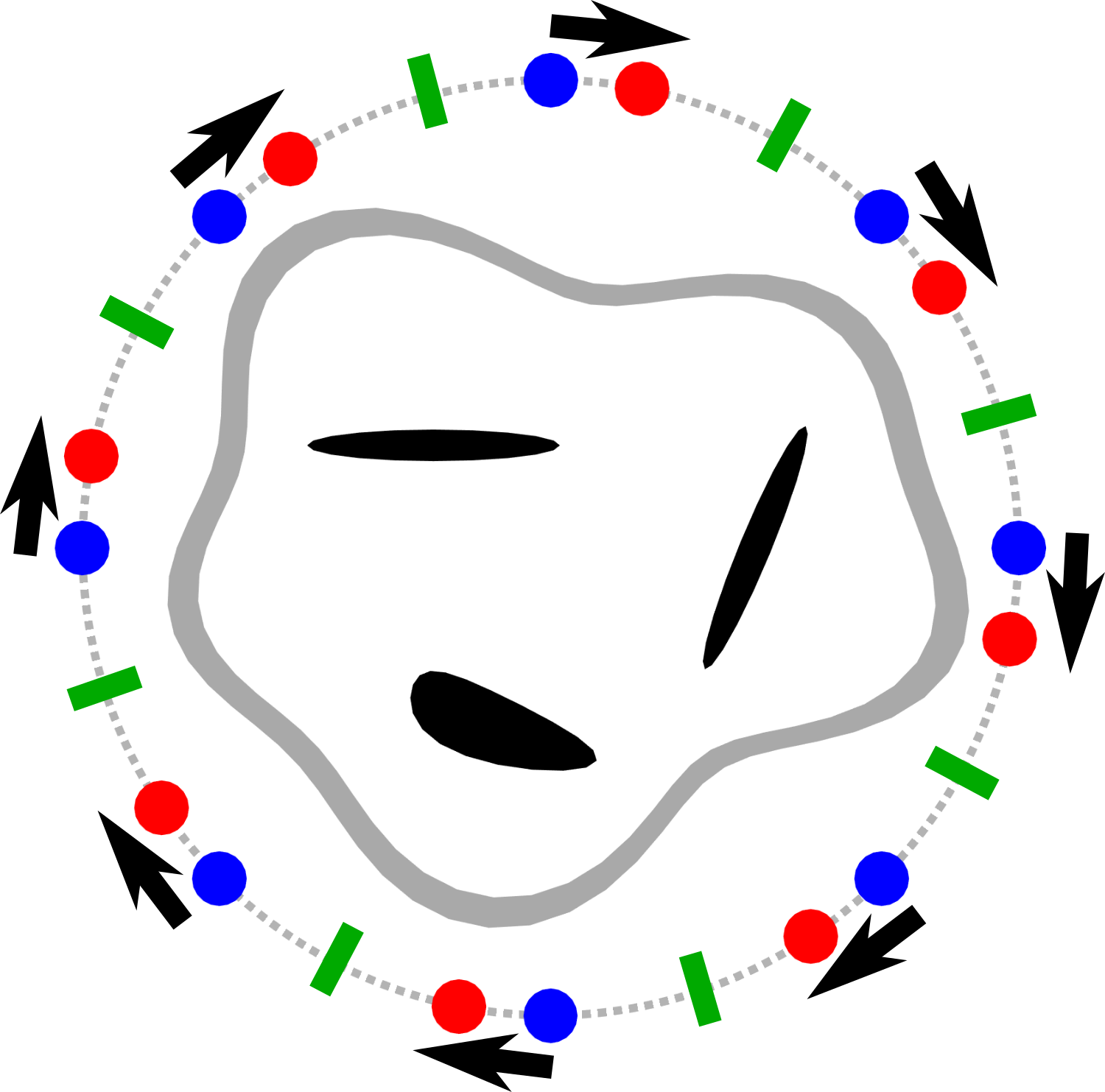} 
    \end{minipage} 
    \begin{minipage}{5.1cm} ({\bf B})  \centering \\ \vskip0.1cm 
     \includegraphics[width=5.1cm]{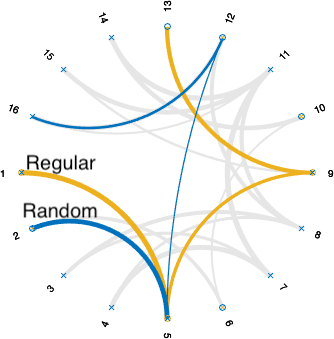} 
    \end{minipage} 
   \begin{minipage}{5.1cm} ({\bf C})  \centering \\ \vskip0.1cm 
     \includegraphics[width=5.1cm]{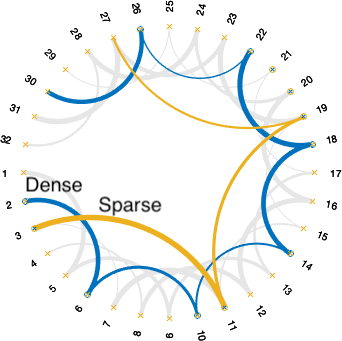}  
    \end{minipage} 
    \end{minipage} 
    \end{scriptsize}
    \caption{The target $\Omega_1$ applied in the two-dimensional numerical simulations.  ({\bf A}) The domain consists of an interior part, three voids and a surface layer (mantle) with complex relative  permittivity values $\varepsilon_r = 4.0 + j 0.03$, $\varepsilon_r = 1.0$, and $\varepsilon_r = 3.0 + j 0.02$, respectively. The synthetic orbit used in the numerical experiments is visualised by the circle surrounding the target. The blue and red dots depict two monostatic signal configurations with uniform distance between the points on the synthetic orbit. The sectors are depicted by the intervals between the green lines. The complete measurement point set including all the sectors satisfies the Nyquist criterion w.r.t.\ the centre frequency of the signal. ({\bf B}) In the regular turn, such monostatic point sets are obtained by rotating an initial set of points from different sectors as shown by the arrows and, thus, covering each point in the full set of measurements. In the randomised approach, a uniformly  distributed  random point is picked  independently from each sector, whose length satisfies the Nyquist criterion w.r.t.\ the pulse bandwidth as described in Section \ref{sec:grid}.  ({\bf C}) In the sparse density configuration, points are selected from each sector with intervals large enough to cover the total receiver points on the synthetic orbits. The dense point configuration has twice as many points as the sparse case i.e., half the interval for point selection as compared to the sparse configuration.}
    \label{fig:p_exact}
\end{figure*}

In the numerical experiments,  we investigate inverting full-waveform data in two dimensions using the target  $\Omega_1$ of \cite{sorsa2020time} with unitless diameter 0.28 scaled to the estimated diameter 160 m ($s = 571$) of Dimorphos and a synthetic orbit of diameter 0.32; see Figure \ref{fig:p_exact}. The far-field formulation is omitted due to the difference in attenuation between 2- and 3-Dimensional domain \cite{takala2018far}.  We considered a  background model with a constant permittivity  $\tilde{\varepsilon}_r' = 4$ and a detailed (exact) model  $\Omega_1$ as shown in Figure \ref{fig:p_exact}.  The interior part of the detailed model is given a real relative permittivity $\varepsilon'_r = 4$, excluding the three voids having a vacuum permittivity ($\varepsilon'_r = 1$) and the surface layer (mantle) with real relative permittivity value of $\varepsilon'_r = 3$.  The existence of a surface layer for an asteroid target is predicted in the impact studies by \cite{jutzi2017formation} and the overall structure is considered to have a relatively low density \cite{carry2012density}. The permittivity of asteroid minerals such as kaolinite and dunite match roughly these values regarding solid and powder composition (with grain size distribution of approximately 1-10 $\mu$m) for the interior and surface layer, respectively \cite{herique2002dielectric,Herique2018}. The loss tangent $\updelta$ which determines the imaginary part of the complex relative permittivity $\varepsilon_r = (1 + i \updelta ) \varepsilon'_r$  is considered as a nuisance parameter, as the internal absorption distribution is known to be challenging to be reconstructed \cite{barriot1999two}. Hence, we are not inverting for this parameter, but it causes an error source in the simulation. We set  $\updelta$  to be in the range from 0.0005 to 0.04 proposed for asteroid minerals in \cite{kofman2012}. Here, we consider the internal absorption simply as a source of error; however, it should be noted that for an average loss tangent $\tan(\updelta) = 0.008$ and $\varepsilon'_r = 4$, an attenuation of 30 dB/km is fully realistic when using a frequency of 20 MHz and 78 dB/km when using 60 MHz given by $-8.68\alpha$, where $\alpha = \frac{2\pi f}{\mathtt{c}_0} \sqrt{ \frac{\varepsilon'_r}{2} \sqrt{1+\left(\frac{\varepsilon''_r}{\varepsilon'_r}\right)^2}-1}$ \cite{ulaby2014microwave}.

\subsubsection{Signal specifications}
\label{sec:sig-spec}

 The data ${\bf y}$ was obtained as the (noisy) difference between the two simulated full wavefields corresponding to the background and detailed domain described above. To avoid an inverse crime, i.e., an overly accurate data fit, these domains were discretised using two different finite element meshes.

\begin{figure}[ht!]
\centering
\begin{scriptsize}
  \begin{minipage}{3.9cm} ({\bf A}) 20  MHz  \centering \\ \vskip0.1cm
    \includegraphics[width=3.9cm]{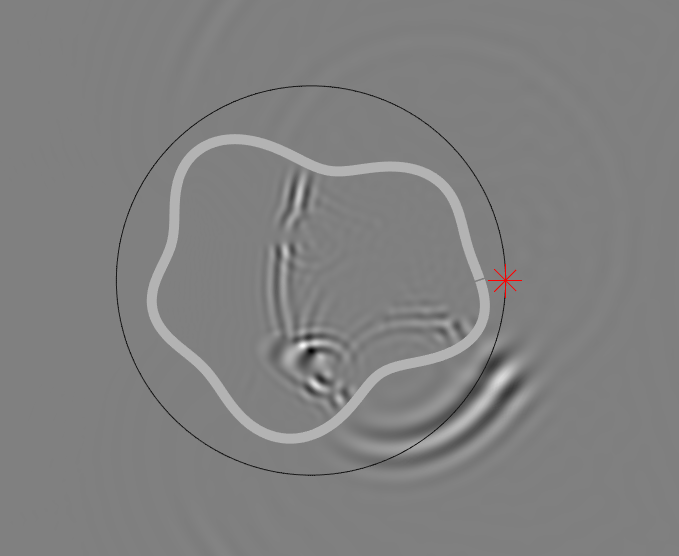}  
    \end{minipage} 
    \begin{minipage}{3.8cm} ({\bf B}) 60 MHz \centering \\ \vskip0.1cm 
    \includegraphics[width=3.8cm]{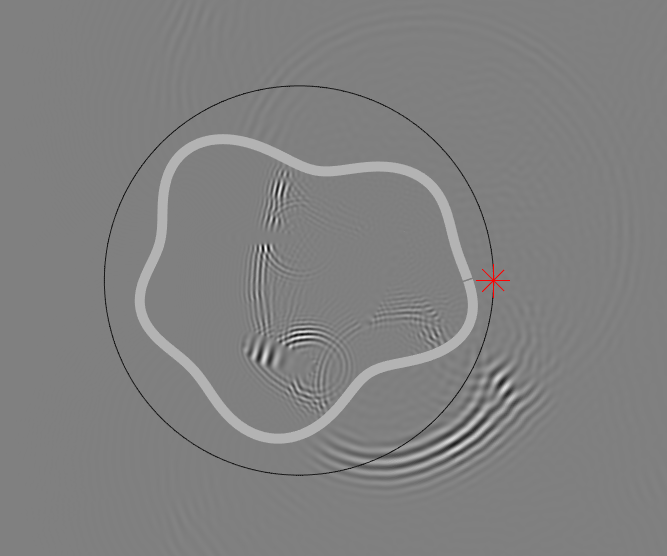}
    \end{minipage}
    \end{scriptsize}
    \caption{Monostatic signal propagation path in the two-dimensional domain for the signal pulse ({\bf A}) 20 and ({\bf B}) 60 MHz at $0.22\mu$s measurement time ($s \mathtt{c}_0^{-1} t$)  after propagation from the transmission point marked with red asterisk. The 20 MHz signal has a single strong wavefront propagating in the domain while the 60 MHz signal has multiple weak wavefronts propagating through the domain, which is a potential source of wavelength-induced uncertainty.}
    \label{fig:wavefront} 
\end{figure}

We apply QAM modulated Blackman-Harris window as a signal pulse centred at ({\bf A}) 20 and ({\bf B}) 60 MHz frequencies covering the bandwidths of 10 and  20 MHz, respectively; See Table \ref{tab:signal_frequency} and Figure \ref{fig:wavefront}. The centre frequency of 20 MHz has been proposed as a potentially feasible low end for a CubeSat-based penetrating radar exploration in a recent concept study \cite{bambach2018discus}, while 60 MHz will be applied in the JuRa investigation  \cite{herique2019juventas, herique2020jura}. The general knowledge of asteroid composition and radar signal penetration suggests that the centre frequency should be between 10 and 100 MHz \cite{binzel2005internal,kofman2012} to enable full coverage of the wavefield inside the target.

\begin{table*}[h!]
    \centering 
        \caption{Blackman-Harris signal pulses and their corresponding frequencies, bandwidths, pulse durations, cycles per pulse, wavelengths, signal wavelength-diameter ratios, and sampling rates ($\mathrm{S_{s}}$)}
    \begin{tabular}{@{}llllllll@{}}
         &   &  &  &  \\
       Signal   & Centre   & Bandwidth  & Pulse & Cycles & $\lambda$ & Diameter/$\lambda$  & S$_{\mathrm{s}}$\\ 
      pulse   &  freq. (MHz) & (MHz) & duration ($\mu$s) &  per pulse & (m)& &\\
       \toprule
        ({\bf A}) & 20 & 10  & 0.190 & 4 & 7.5 & 21 &0.067 \\ 
          ({\bf B}) & 60 & 20  & 0.095 & 5.5 & 2.5 & 64 & 0.134\\ 
    \end{tabular}%
    \label{tab:signal_frequency}
\end{table*}

We consider the following four monostatic signal configurations ({\bf I})--({\bf IV})  (Figure \ref{fig:p_exact}, Table \ref{tab:signal_configurations}) for obtaining the full-wave data. These configurations differ by the averaging and filtering strategy applied in processing the data. The averaging process utilises the simulated full-wavefield which has been obtained around the tomographic target, with spatial and temporal resolution satisfying the Nyquist criterion with respect to the centre frequency.

\begin{table*}[!t]
        \caption{Spatial resolution of the full dataset before and after the averaging process with their corresponding averaging techniques. The approach to the spatial averaging in configurations ({\bf I})--({\bf IV}) of the numerical experiments is described in Figure \ref{fig:p_exact}, and the signal pulses ({\bf A}) 20 and ({\bf B}) 60 MHz described in Table \ref{tab:signal_frequency}. The 3rd and 4th columns give the total and averaged number of spatial points in the full dataset and after the averaging process, respectively.  The 5th column includes the number of turns in the regular averaging process for each averaging resolution, and the 6th gives the number of spatial point sets and frequencies applied in the randomised averaging. The last column indicates the type of filtering applied to each case in the experiment. }
            \centering 
    \begin{tabular}{@{}lllllll@{}}
        Signal   & Signal& Total  & Spatial & Regular  & Randomised  \\
       pulse   & configuration & points & averaging  &  turns & turns/frequency & Filter \\
       \toprule
        ({\bf A}) & ({\bf I})& 128 & dense/ 64  & 2 & - & Unfiltered\\ 
        & ({\bf II}) & 128 & dense/ 64  & 2 & - & TSVD\\
        & ({\bf III})& 128 & dense/ 64  & - & 8/5 & Unfiltered\\ 
        & ({\bf IV}) & 128 & dense/ 64  & - & 8/5 & TSVD\\
          & \\
        & ({\bf I})& 128 & sparse/ 32  & 4 & - & Unfiltered \\ 
        & ({\bf II})& 128 & sparse/ 32  & 4 & - & TSVD\\ 
        & ({\bf III}) & 128 & sparse/ 32  & - & 8/5 & Unfiltered \\ 
        & ({\bf IV})  & 128 & sparse/ 32  & - & 8/5 & TSVD\\ 
          & \\
          \toprule
          ({\bf B}) &({\bf I}) & 384 & dense/ 128  & 3 & -  & Unfiltered\\ 
         & ({\bf II}) & 384 & dense/ 128  & 3 & -  & TSVD\\ 
         & ({\bf III})  & 384 & dense/ 128  & - & 8/5 & Unfiltered\\
          & ({\bf IV})& 384 & dense/ 128  & - & 8/5 & TSVD\\
            & \\
          & ({\bf I}) & 384 & sparse/ 64  & 6 & - & Unfiltered\\ 
          & ({\bf II}) & 384 & sparse/ 64  & 6 & - & TSVD\\ 
          & ({\bf III}) & 384 & sparse/ 64  & - & 8/5 & Unfiltered\\ 
           & ({\bf IV})  & 384 & sparse/ 64  & - & 8/5 & TSVD\\ 
    \end{tabular}%
    \label{tab:signal_configurations}
\end{table*}

\subsubsection{Averaging approaches}
The regular averaging approach refers to a systematic method of selecting measurement points located at uniform angular distances from each other and averaging the (reconstructed) candidate solutions corresponding to such signal configurations. The points are independently selected at intervals corresponding to the spatial sampling rate $\mathrm{S}_\mathrm{s}$ in Table \ref{tab:signal_configurations}. Two different spatial resolutions, sparse and dense, are considered in the reconstruction averaging process in this study. The sparse cases have an average point density corresponding to the Nyquist criterion for the pulse bandwidth, and the dense cases have twice as much of the point density of the sparse cases. Thus,   32 (sparse) and 64 (dense) points are sampled for system ({\bf A}), while 64 (sparse) and 128 (dense) points are sampled for system ({\bf B}), where systems ({\bf A}) and ({\bf B}) have a total of 128 and 384 points, respectively. To ensure that the resulting point sets cover the full set of measurement data, multiple constellations of measurement points are sampled. In essence, the regular averaging approach essentially has $4 \times 32$ (4 constellations of sparse) and $2 \times 64$ (2 constellations of dense) points for system ({\bf A}), and $6 \times 64$ (6 constellations of sparse) and $3 \times 128$ (3 constellations of dense) points for system ({\bf B}). The number of constellations (4, 2, 6, and 3) in the regular averaging approach are referred to as the regular turns in Table \ref{tab:signal_configurations} and illustrated in Figure \ref{fig:p_exact}. 

In the randomised averaging approach, the spatial points are selected in a uniformly distributed random manner (no equal distance between points) from the sectors constituted by the regular turn.  In addition, the centre frequency is slightly perturbed by varying the surrogate centre frequency in the demodulation process according to the discrepancy condition in equation \eqref{freq}. However, the number of constellations of the measurement points was fixed to 8 for both sparse and dense points in the two systems.  In essence, we demodulate the wavefield data for 5 different surrogate centre frequency realisations, using the discrepancy $\mathtt{d}$ as the standard deviation and actual centre frequency $f_{\hbox{\scriptsize centre}}$ as the mean, $f_m$. Hence, the estimated inverse solution for each point is averaged over 5 different frequencies and then all 8 spatial constellations. The sampling rate of the regular averaging approach was retained, resulting to a system with similar number of measurement points i.e., 32 (sparse) and 64 (dense) points for system ({\bf A}), and 64 (sparse) and 128 (dense) points for system ({\bf B}). This guarantees that every point has the chance to be selected and the full coverage of the data is ensured since a uniformly distributed random point is selected from each sector illustrated by the green lines in Figure \ref{fig:p_exact}. 

In configurations ({\bf I}) and ({\bf II}),  regular spatial averaging  is applied to unfiltered (full) data in ({\bf I}) and TSVD filtering ({\bf II}). Configurations ({\bf III}) and ({\bf IV}) utilise a randomised point set  and frequency perturbed data  with full data ({\bf III}) and TSVD filtering ({\bf IV}). 

\subsubsection{Noise}

The measurement noise was simulated by adding a constant noise level relative to the maximum signal amplitude (Peak SNR). The inversion results are obtained for two different magnitudes of the  noise; with the lower noise level, the signal-to-noise ratio (SNR) of the simulated data was 20 dB and with the higher one, it was 12 dB. The higher SNR is motivated by the observations of the CONSERT team suggesting that the main signal peak of the measurement was found to have at least 20 dB SNR with respect to noise peaks \cite{Kofman2015}. The lower SNR has been set based on recent numerical and experimental modelling studies \cite{Sorsa2019,Eyraud2020analog,sorsa2021analysis} suggesting that a SNR above 10 dB might be sufficient for reconstructing the internal permittivity of a complex-structured asteroid analogue and that such an accuracy between a numerically modelled field and a laboratory measurement can be achieved experimentally. The experimental setup in Table \ref{tab:signal_configurations} was implemented for both 20- and 12-dB noise cases.

\subsubsection{Similarity and Error Measures}

To analyse the accuracy of the reconstructions in different parts of the target $\Omega_1$, we define sub-domains $\mathcal{S}_1$ (voids) and $\mathcal{S}_2$ (surface layer) restricted by the contours of the permittivity distribution in a descending order. The structural similarity (SSIM) of the reconstruction to the exact distribution is computed to measure the quality of the reconstructed image relative to the background \cite{wang2004image}.  We also evaluate the Root Mean Squared Error (RMSE) between the exact distribution and reconstruction for $\Omega_1$, $\mathcal{S}_1$ and $\mathcal{S}_2$. The Overlap Error (OE) between the detail $\mathcal{S}_i$ and a reconstructed permittivity perturbation is defined as the ratio \begin{equation} \hbox{area}(\mathcal{S}_i \cup \mathcal{O}_{i})/\hbox{area}(\mathcal{S}_{i}), \end{equation} where $\mathcal{O}_i = \bigcup_{i \in I_\zeta}  T_i$ is a set composed by triangles $T_i$ in the index set $I_\zeta$ which includes the most intense part of the reconstructed distribution with surface area equals to that of the given detail excluding the sets $\mathcal{O}_j$ for $j=1,2,\ldots, i-1$, that is,  $I_\nu =  \{ i \, | \, |x_i| \geq \nu, T_i \in \Omega_1 \setminus \bigcup_{j = 1}^{i-1}  \mathcal{O}_j \} $ and $\zeta  =  \arg \max_{\nu} \{  \sum_{i \in I_\nu} \hbox{area} (T_i) \leq \hbox{area} (\mathcal{S}_i) \} $.

The Wilcoxon rank sum (WRS) test, a nonparametric alternative to the two-sample t-test for two populations $X$ and $Y$ of  assumably independent samples, was used to analyse the significance of the differences observed. The WRS tests the null hypothesis that data in $X$ and $Y$ are samples from continuous distributions of equal medians, against the alternative that they are not, with the assumption that the samples are independent \cite{gibbons2014nonparametric}.

\subsubsection{Numerical implementation}

The computations are implemented using the openly available GPU-Torre package\footnote{https://github.com/sampsapursiainen/GPU-Torre} \cite{sorsa2020time}, which utilises the Matlab platform (Mathworks, Inc.). GPU-Torre combines a GPU-accelerated FETD forward routine and a multigrid-based inversion approach, where the triangular mesh $\mathcal{T}$ applied to discretise the permittivity distribution in the inversion stage is coarser than the one applied in the forward wave propagation. Each matrix ${\bf L}$ was  found by a GPU-accelerated deconvolution process, obtained in less than 10 seconds. Using Dell 5820 Workstation equipped with 256 GB RAM and 8GB NVIDIA Quadro RTX 8000 GPU RAM, the complete FETD simulation of the signals took approximately 30 and 134 hours, while it took 14  and 21 minutes for a single point of the ({\bf A}) 20 and ({\bf B}) 60 MHz frequencies, respectively. The inversion mesh $\mathcal{T}$ consisted of 896 triangles. The total variation regularisation parameter $\alpha$ in equation \eqref{backpropagation} was chosen experimentally to be 0.02 and 0.005 for the 20 and 60 MHz systems respectively, while $\beta$  in equation \eqref{tvparam} was chosen to be $0.5 \alpha^2$ for both cases. In all, a total of 32 numerical experiments are considered as described in Table \ref{tab:signal_configurations} given that two noise level cases are implemented in this study. 

\section{Results}
\label{sec:results}

\begin{figure}[!ht]
    \centering \begin{scriptsize} SNR 20 dB \\ \vskip0.2cm \begin{minipage}{7.7cm} \centering
 \rotatebox{90}{\hskip-0.5cm dense}    \begin{minipage}{1.8cm} \centering
 \includegraphics[width=1.8cm]{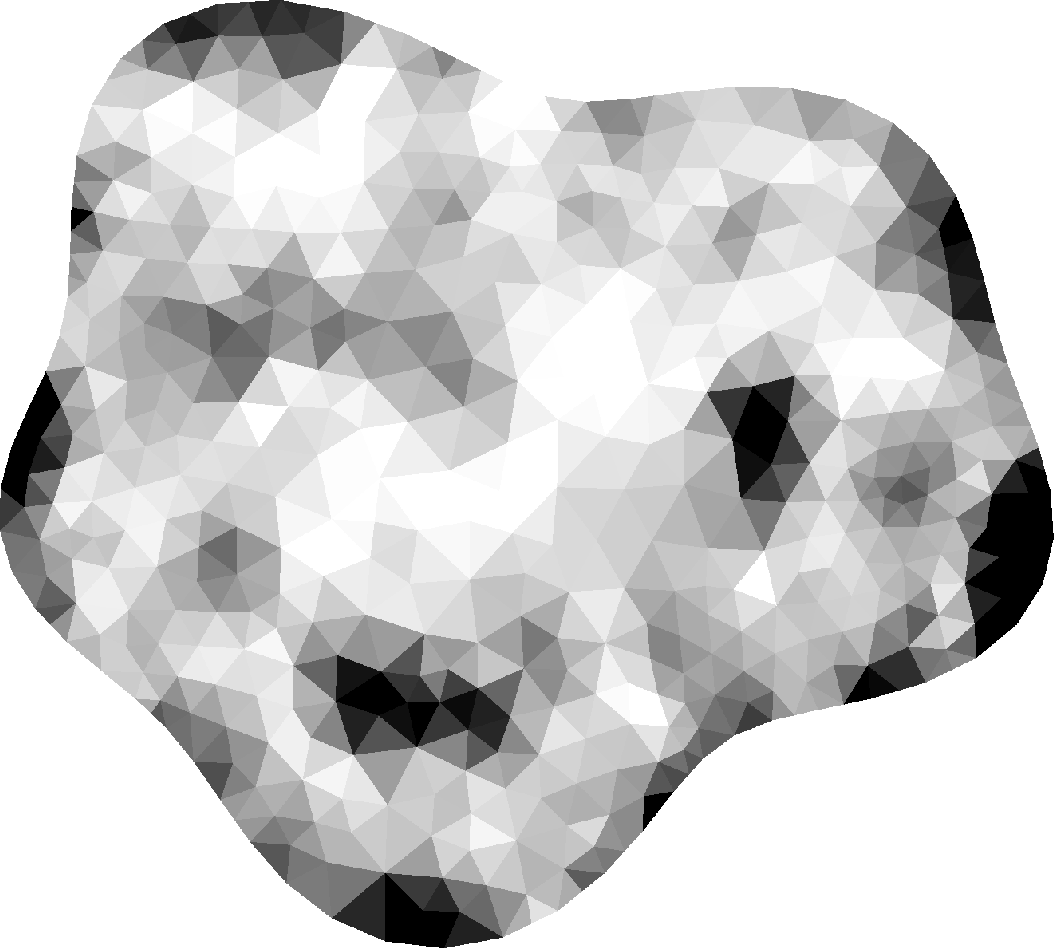} \\
  \includegraphics[width=1.8cm]{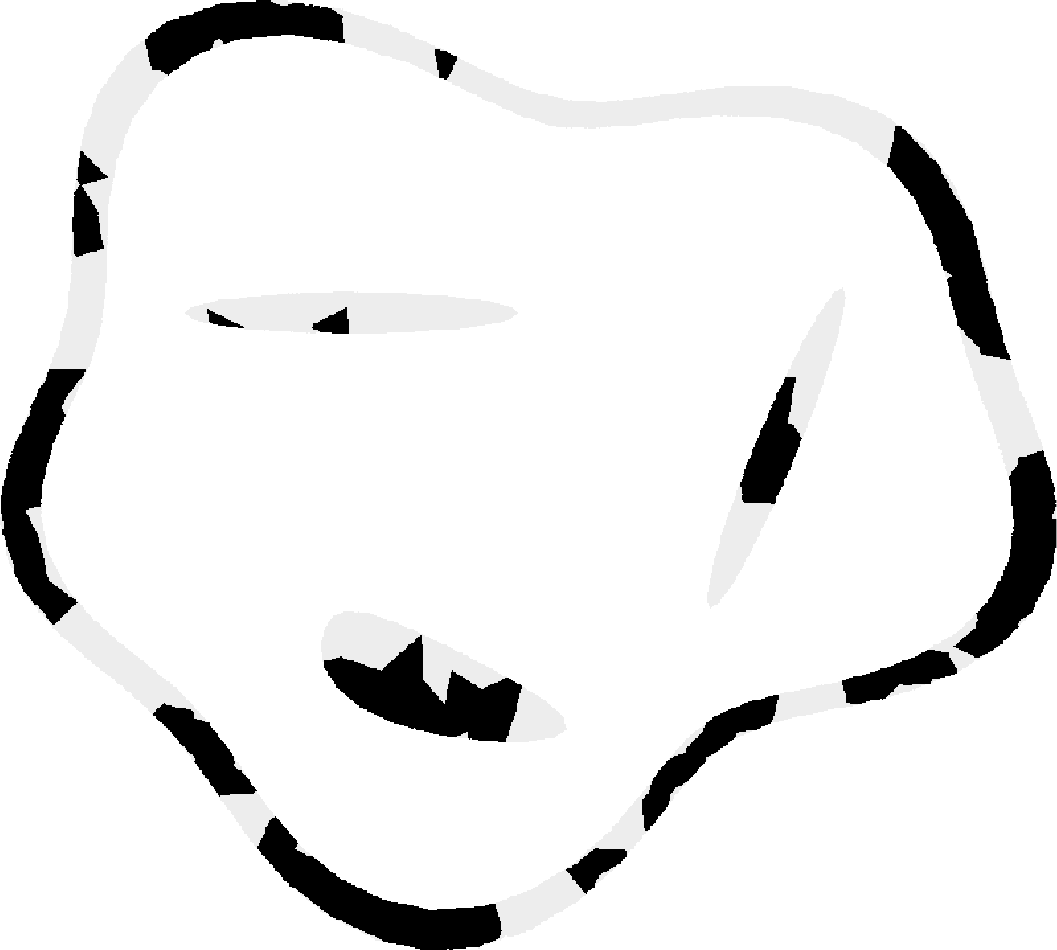} 
    \end{minipage}
    \begin{minipage}{1.8cm} \centering   
   \includegraphics[width=1.8cm]{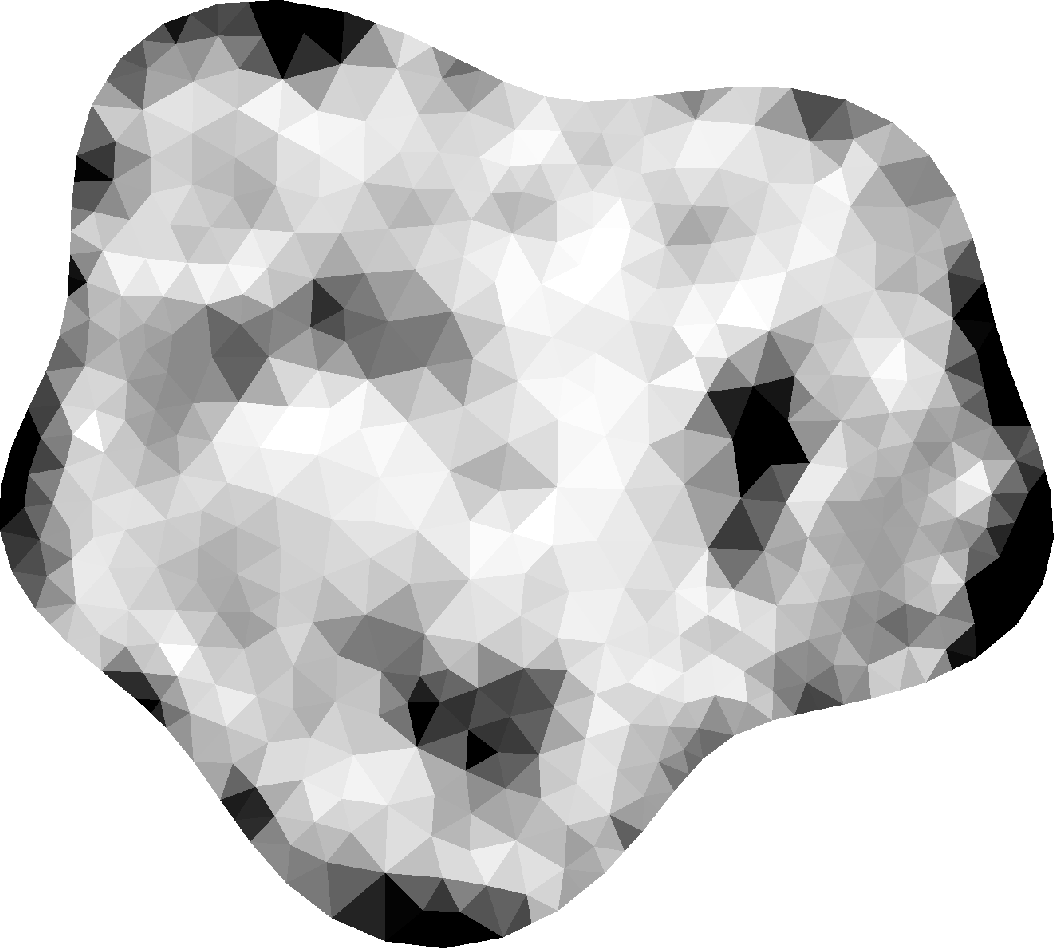} \\
   \includegraphics[width=1.8cm]{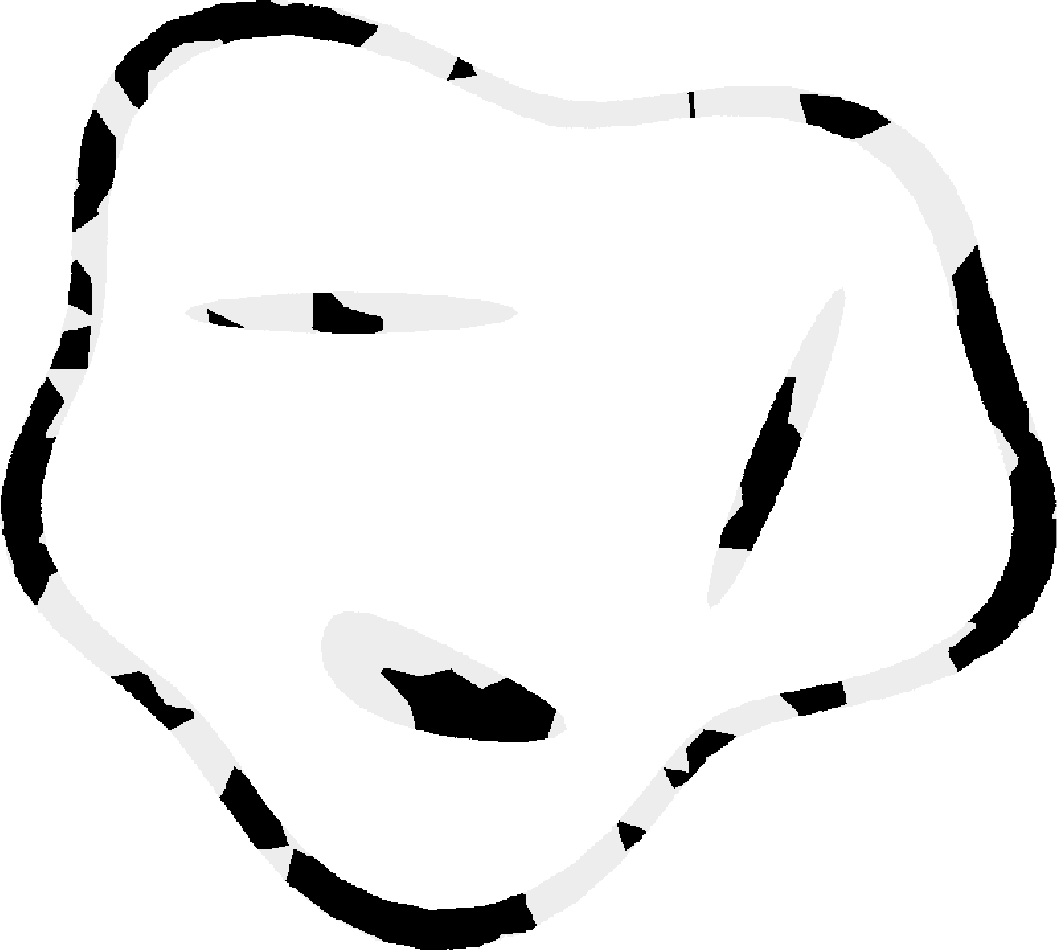} 
          \end{minipage}
    \begin{minipage}{1.8cm} \centering 
           \includegraphics[width=1.8cm]{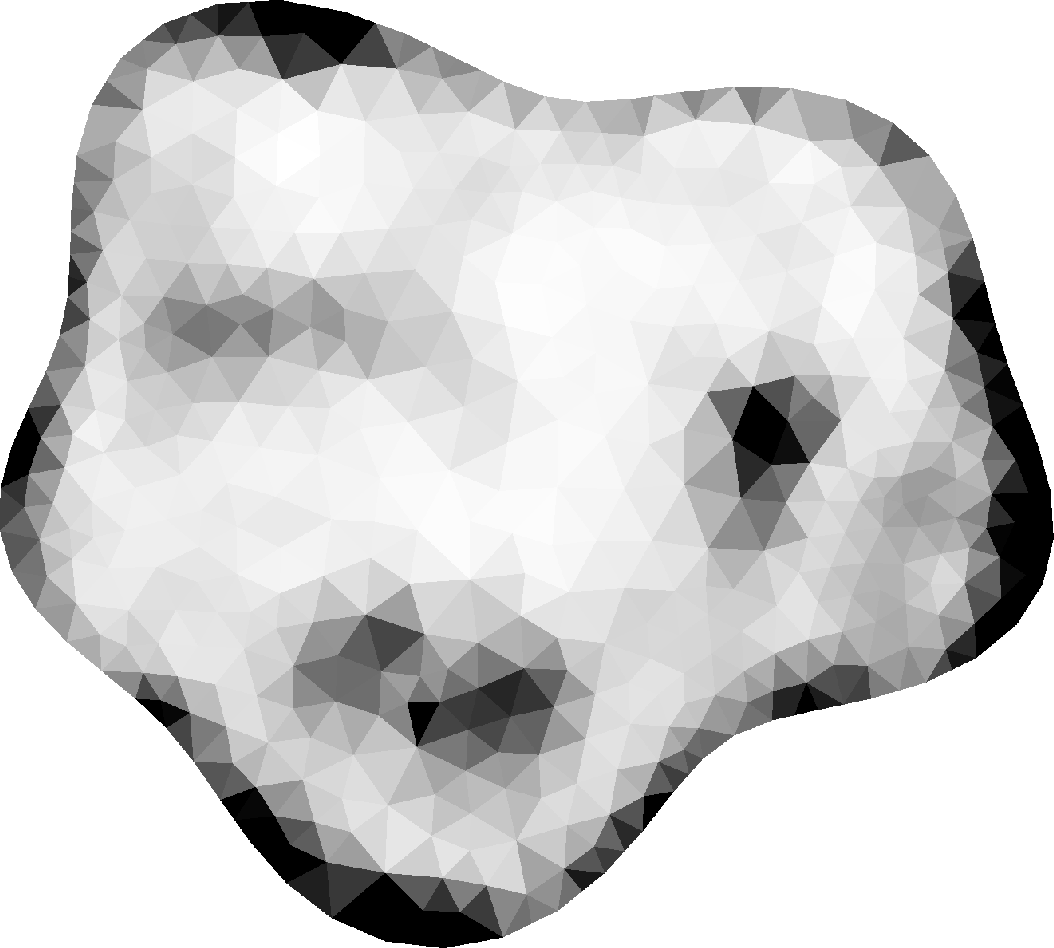} \\ 
           \includegraphics[width=1.8cm]{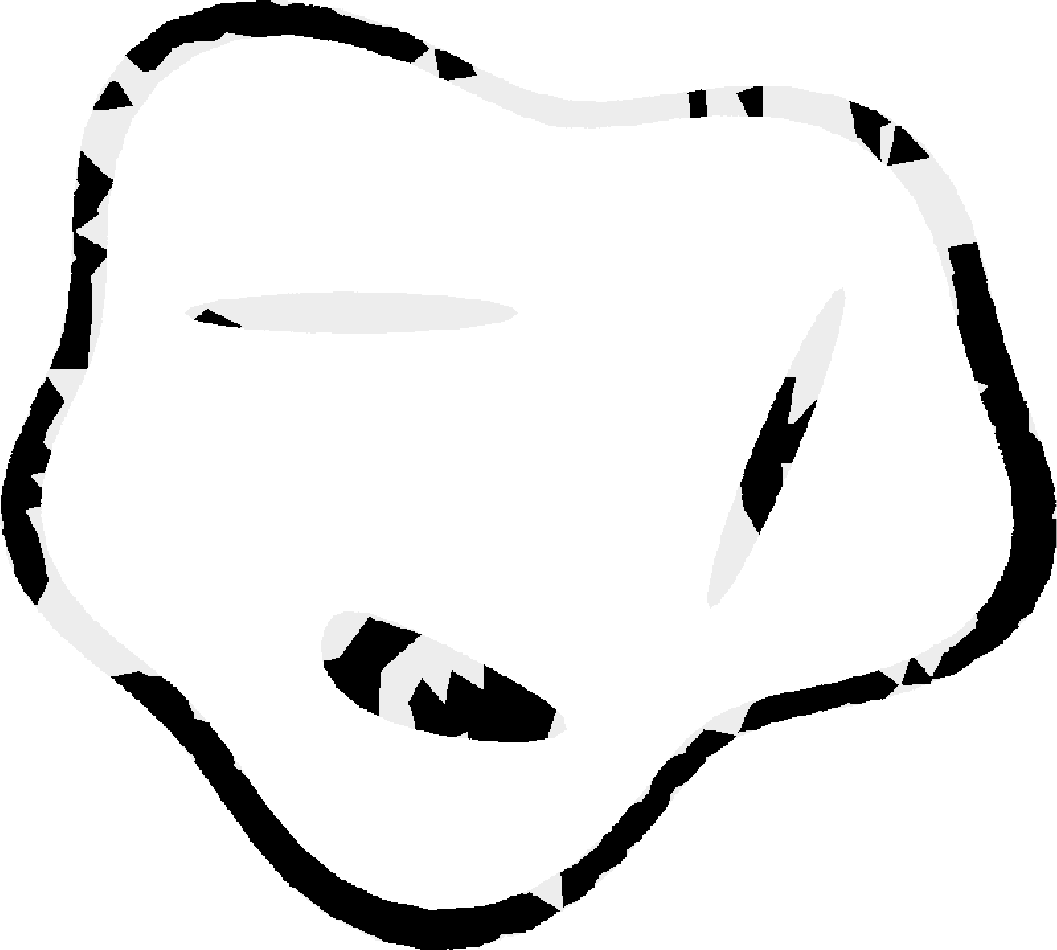} 
               \end{minipage}
    \begin{minipage}{1.8cm} \centering 
         \includegraphics[width=1.8cm]{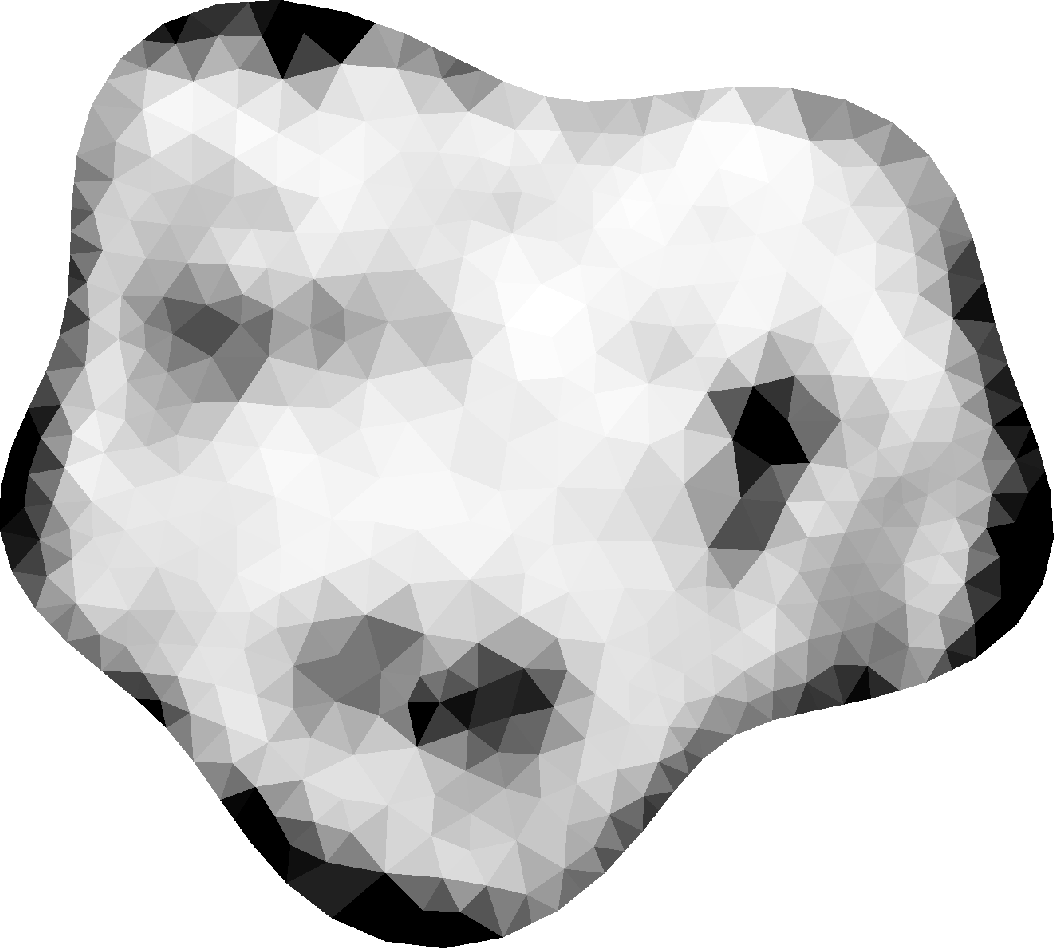}  \\ 
         \includegraphics[width=1.8cm]{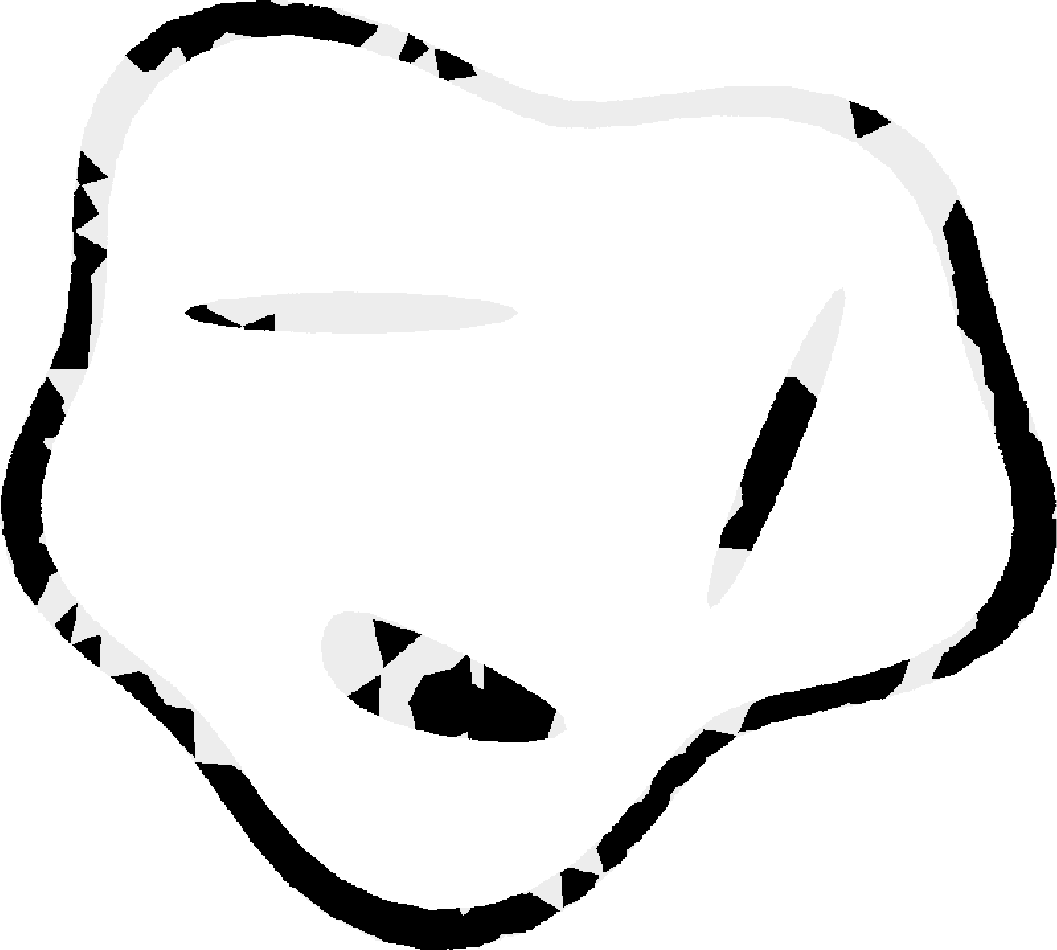} 
         \end{minipage} \\  
\rotatebox{90}{\hskip-0.5cm sparse}     \begin{minipage}{1.8cm} \centering        
         \includegraphics[width=1.8cm]{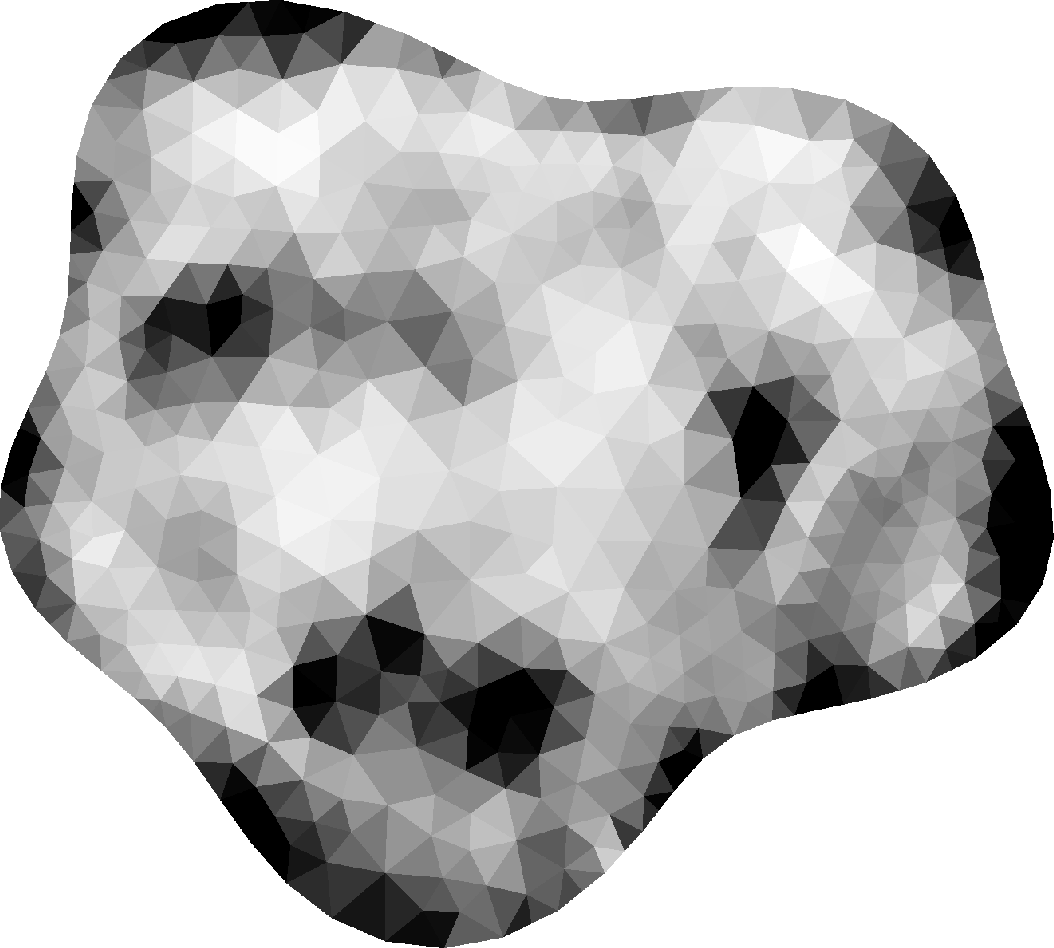} \\ 
         \includegraphics[width=1.8cm]{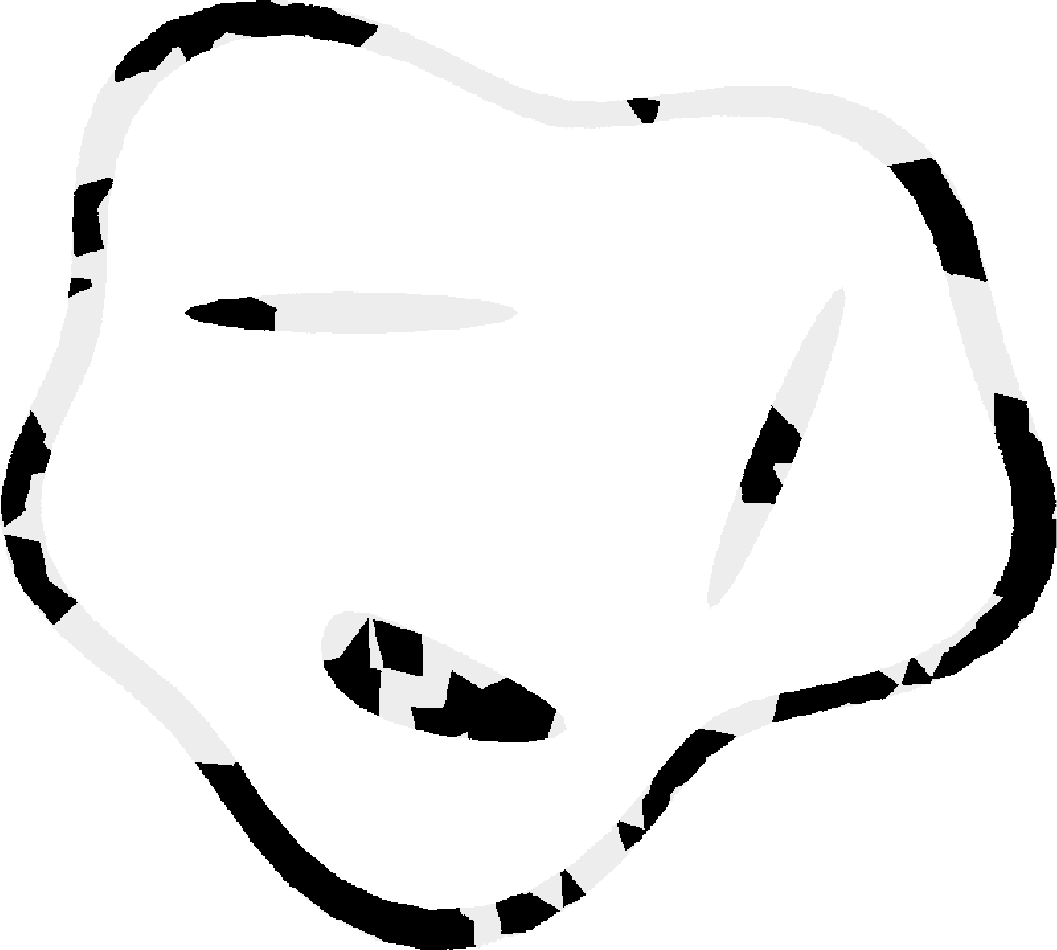}  
             \end{minipage}
    \begin{minipage}{1.8cm} \centering 
        \includegraphics[width=1.8cm]{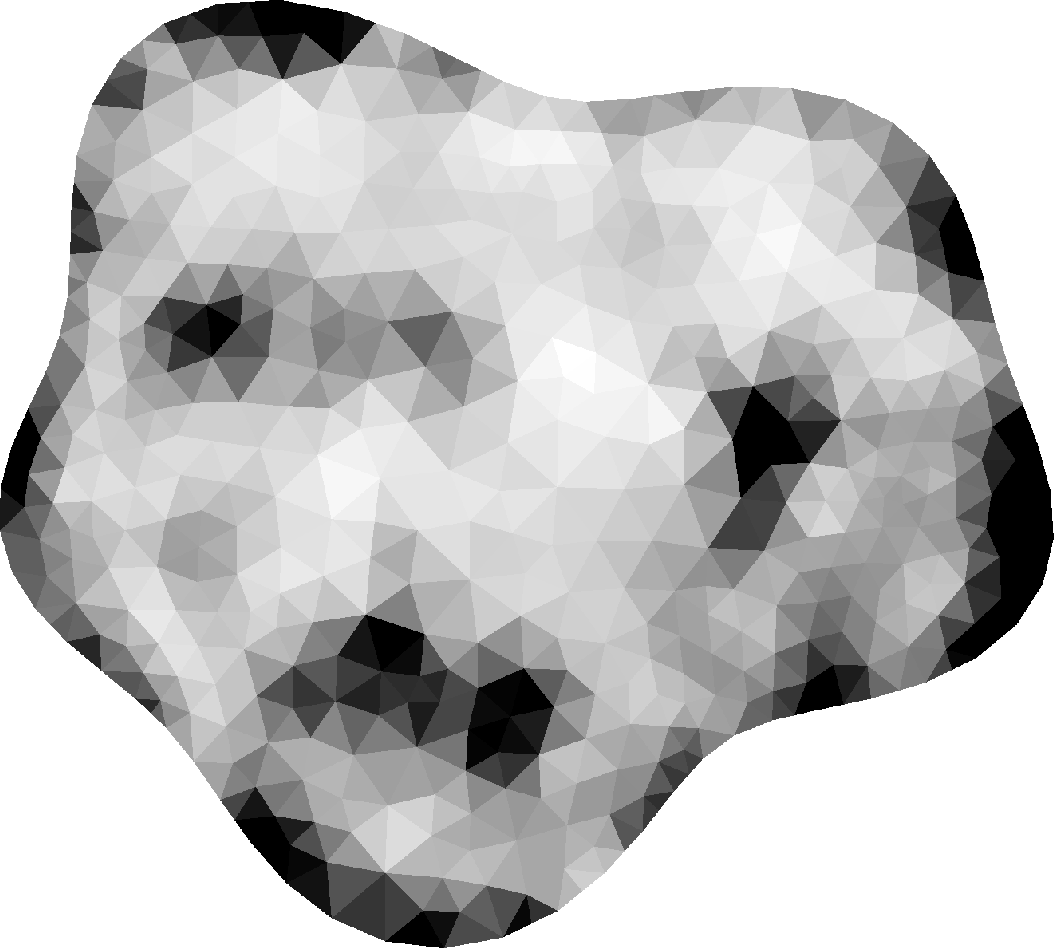} \\
         \includegraphics[width=1.8cm]{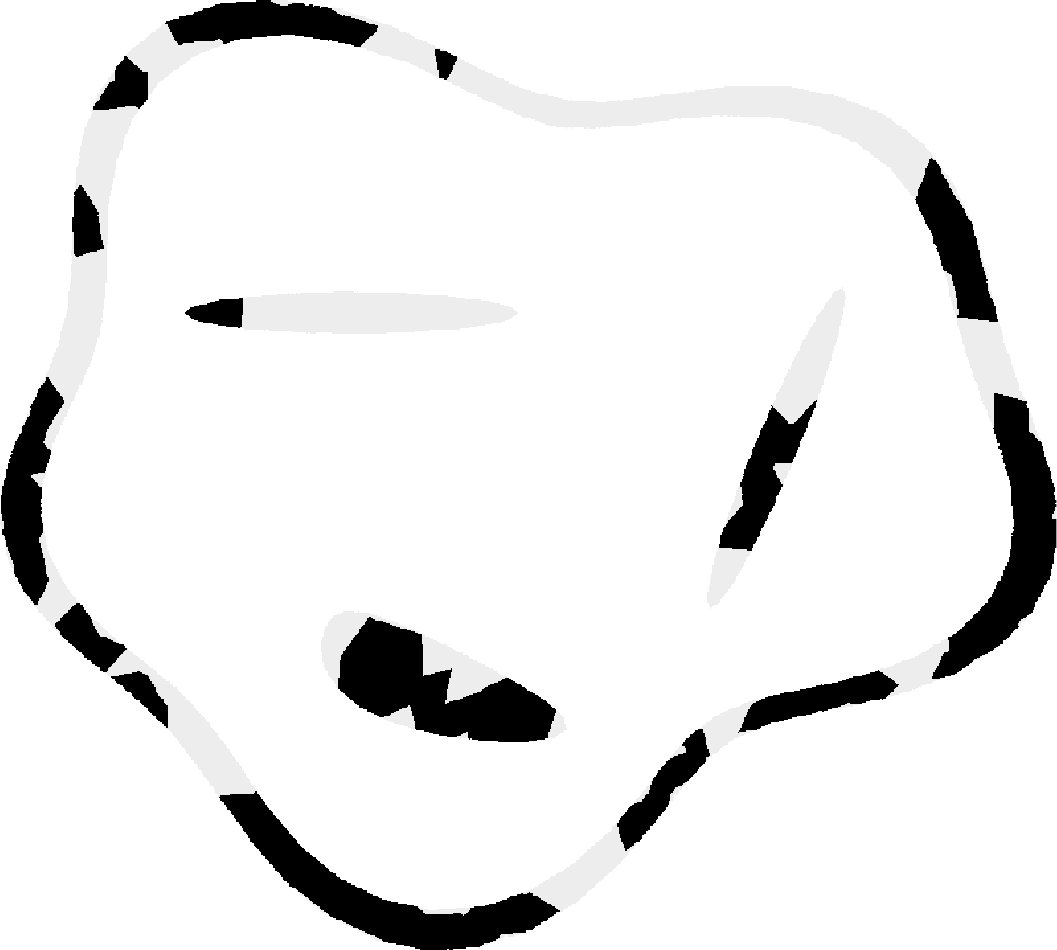} 
            \end{minipage}
    \begin{minipage}{1.8cm} \centering 
           \includegraphics[width=1.8cm]{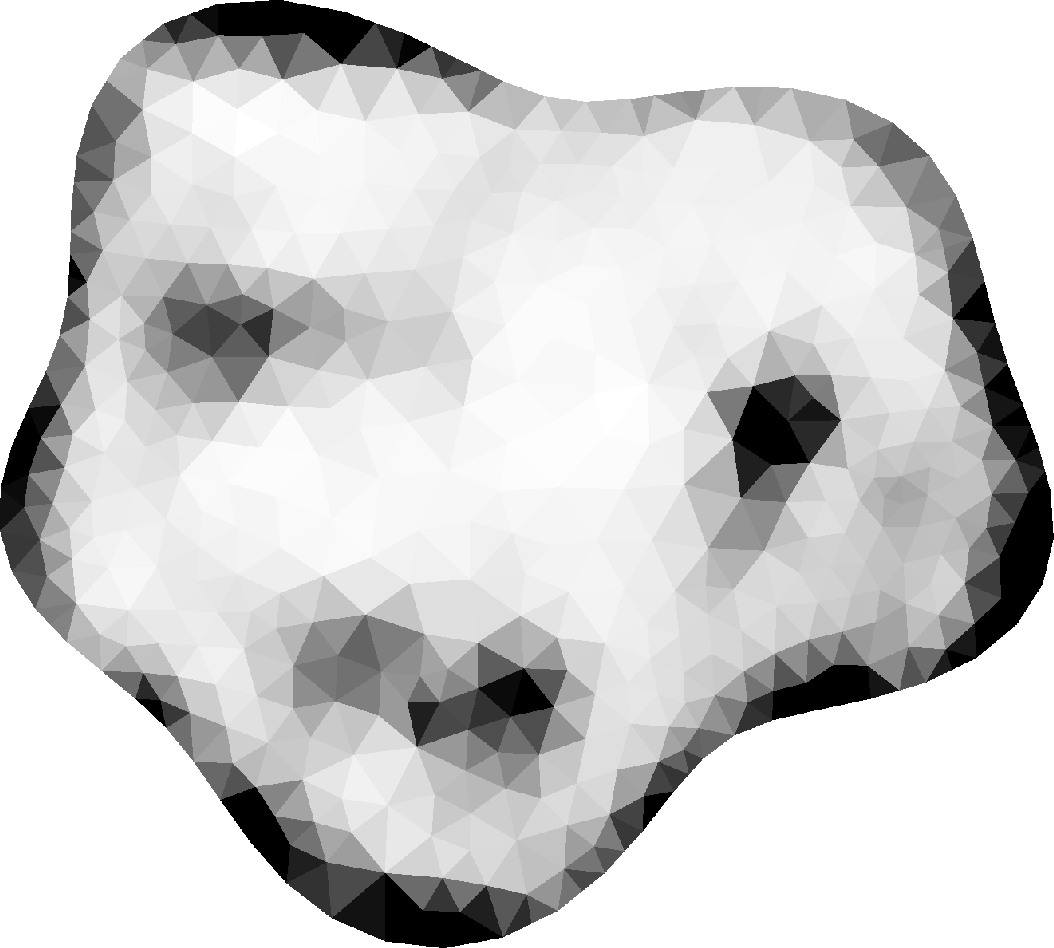} \\ 
            \includegraphics[width=1.8cm]{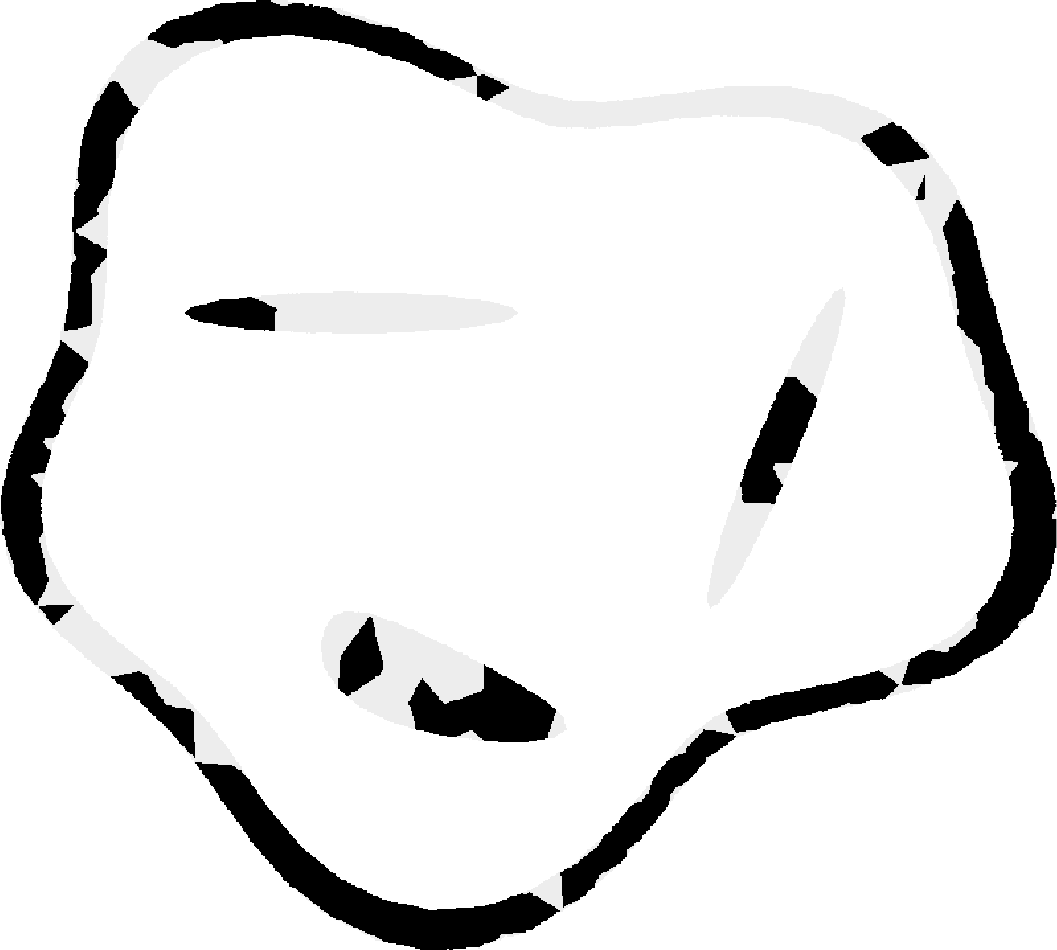} 
               \end{minipage}
    \begin{minipage}{1.8cm} \centering 
         \includegraphics[width=1.8cm]{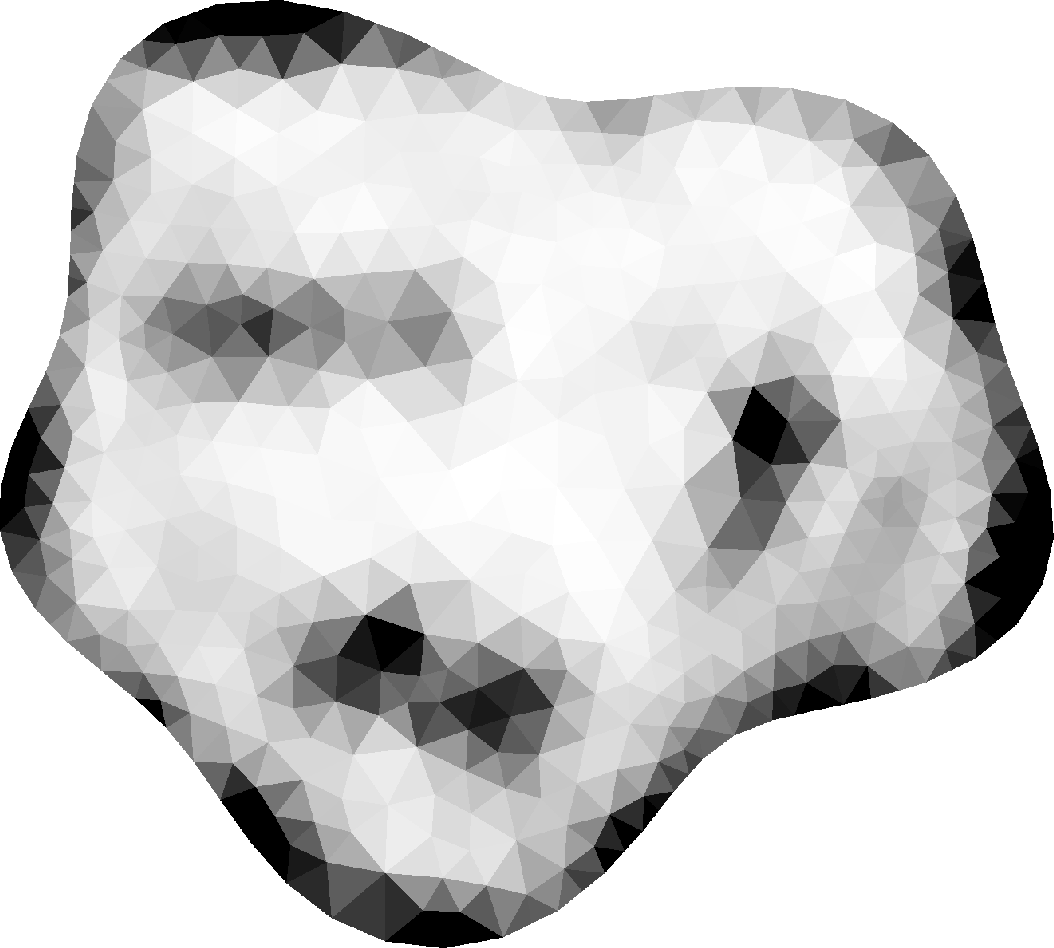} \\ 
          \includegraphics[width=1.8cm]{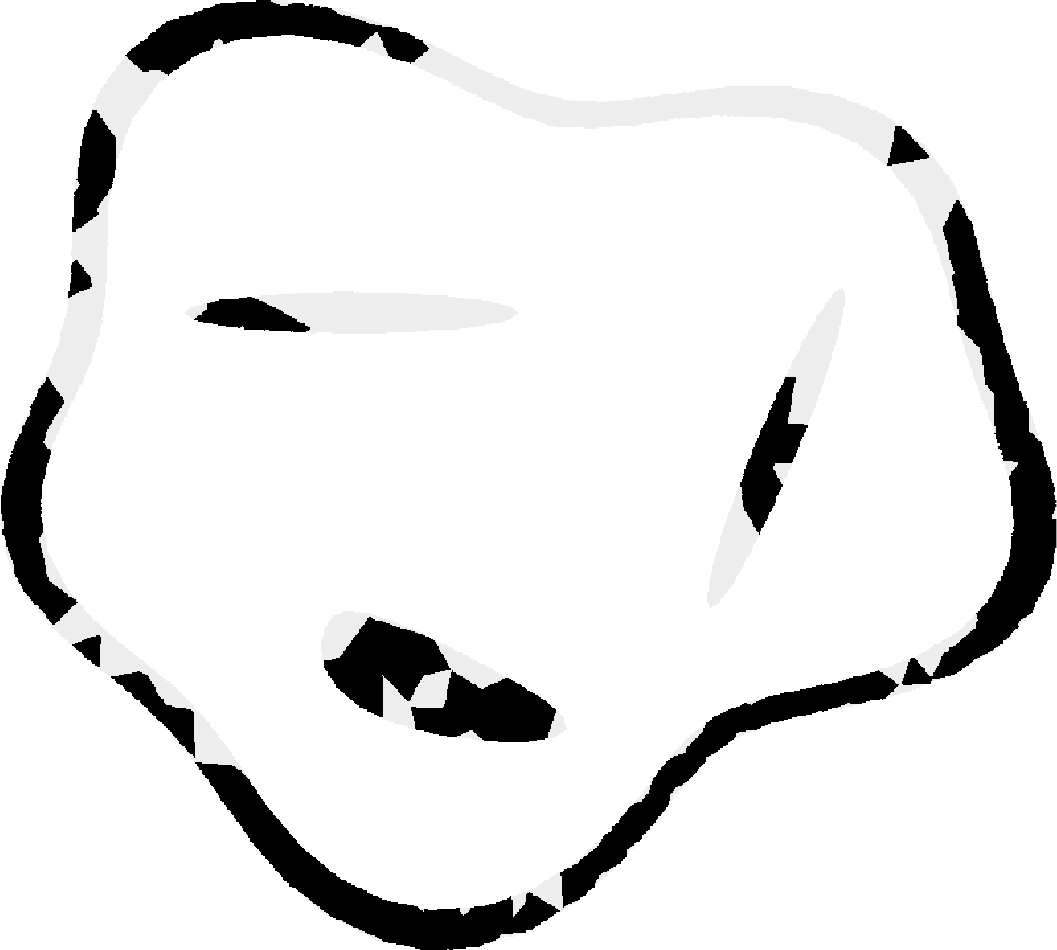} 
             \end{minipage} \\ \vskip0.2cm
              \begin{minipage}{1.8cm} \centering        
        ({\bf I}): Unfiltered 
             \end{minipage}
    \begin{minipage}{1.8cm} \centering 
   ({\bf II}): TSVD 
            \end{minipage}
    \begin{minipage}{1.8cm} \centering 
   ({\bf III}): Unfiltered \&  randomised
               \end{minipage}
    \begin{minipage}{1.8cm} \centering 
   ({\bf IV}): TSVD \&  randomised 
             \end{minipage} \end{minipage} 
                         \begin{minipage}{0.6cm} \centering
             \includegraphics[height= 5cm]{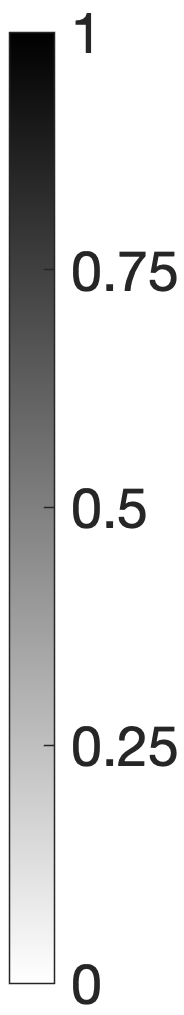}
             \end{minipage} \\ \vskip0.4cm SNR 12 dB \\ \vskip0.2cm 
    \begin{minipage}{7.7cm} \centering
 \rotatebox{90}{\hskip-0.5cm dense}    \begin{minipage}{1.8cm} \centering
 \includegraphics[width=1.8cm]{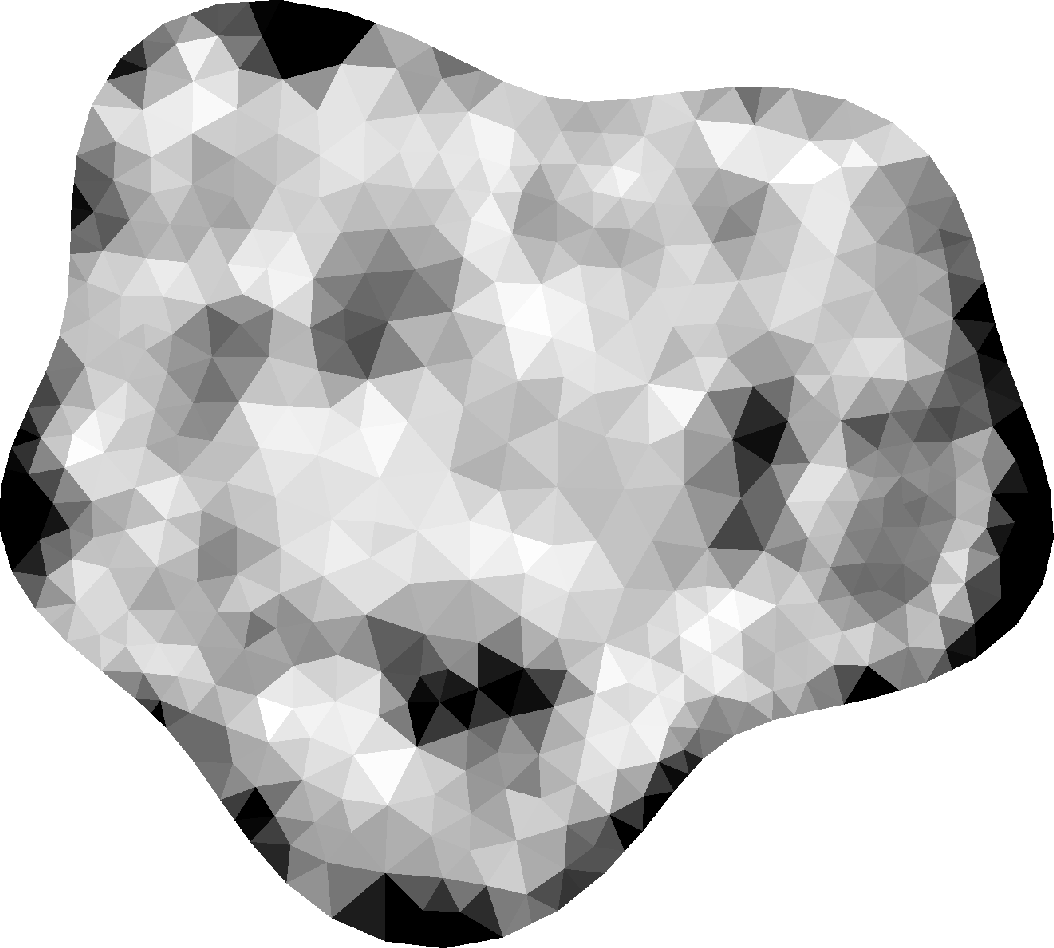}  \\
  \includegraphics[width=1.8cm]{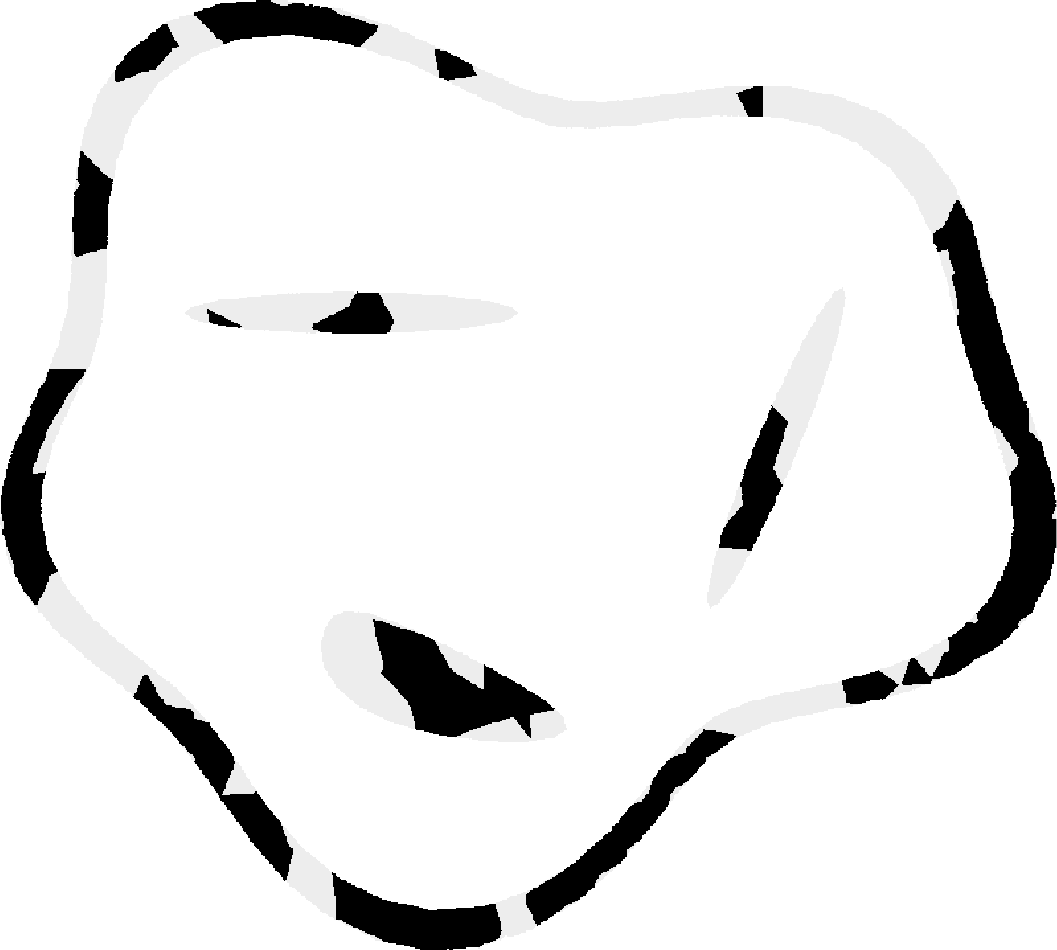} 
    \end{minipage}
    \begin{minipage}{1.8cm} \centering   
   \includegraphics[width=1.8cm]{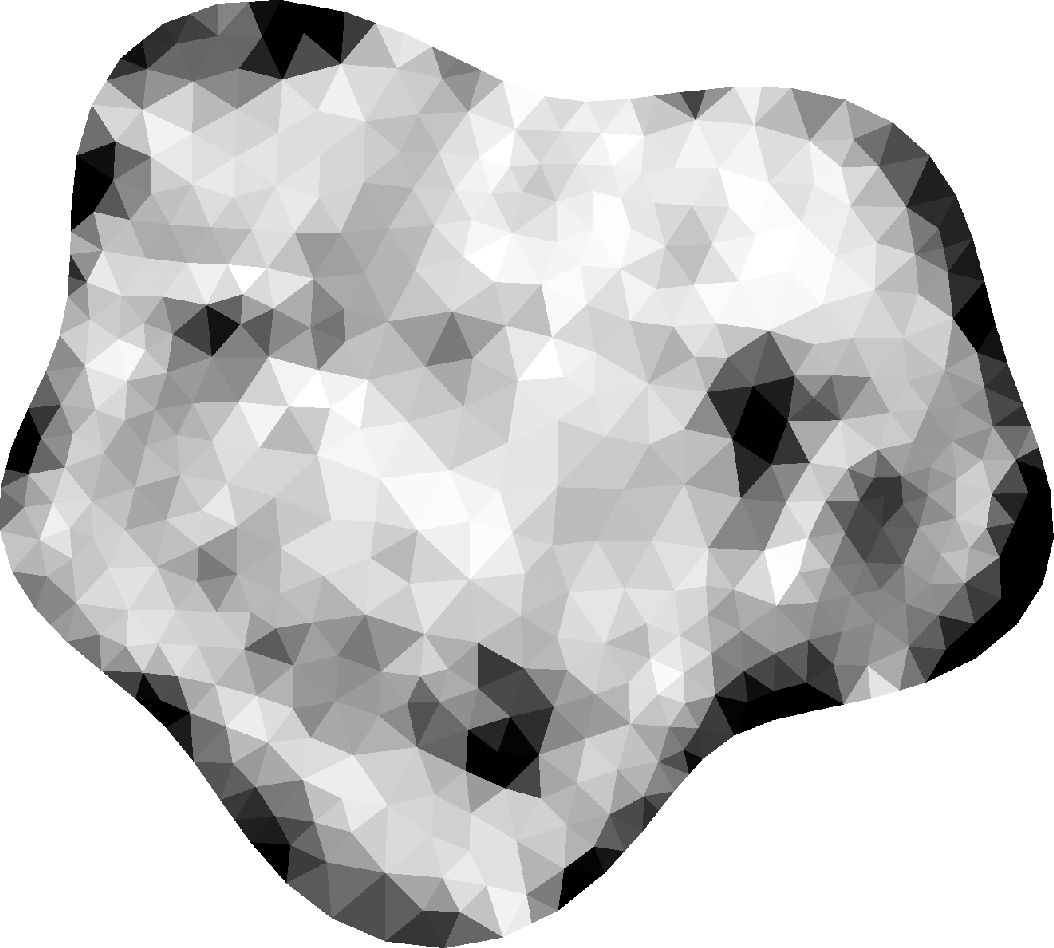} \\
  \includegraphics[width=1.8cm]{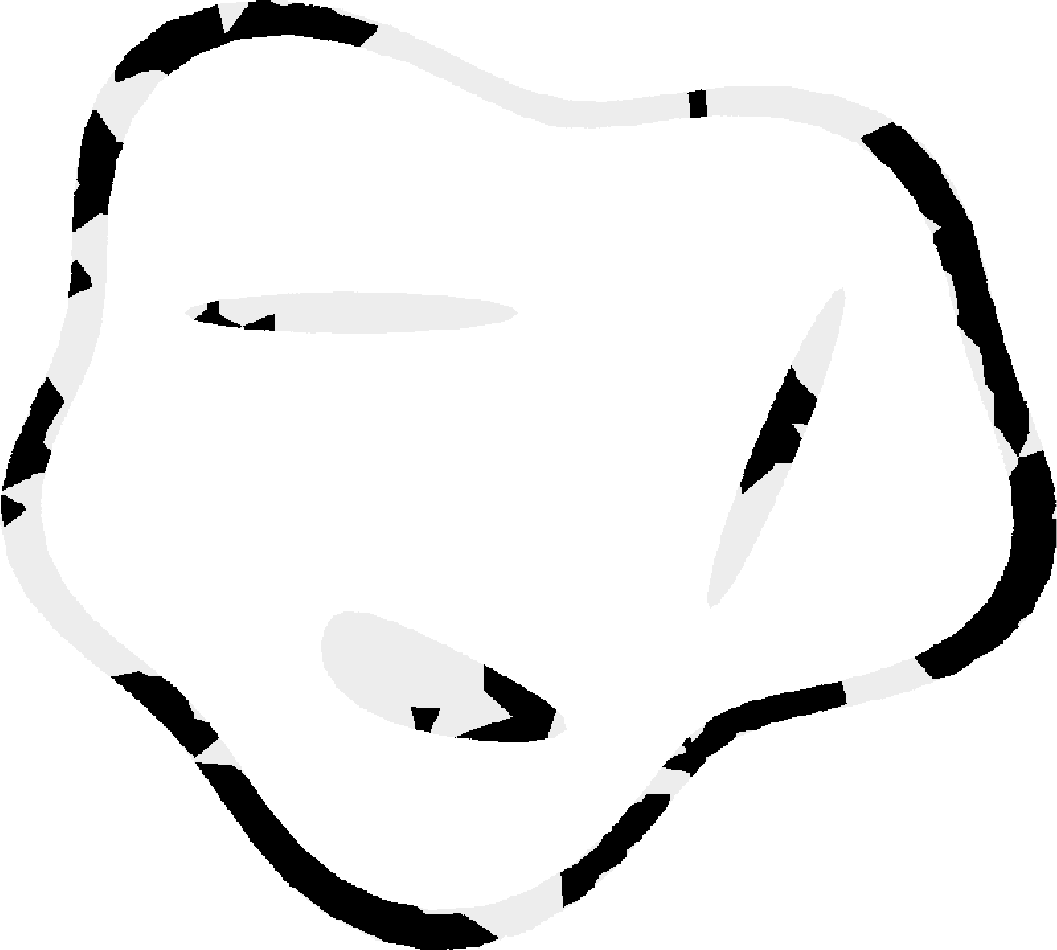} 
          \end{minipage}
    \begin{minipage}{1.8cm} \centering 
           \includegraphics[width=1.8cm]{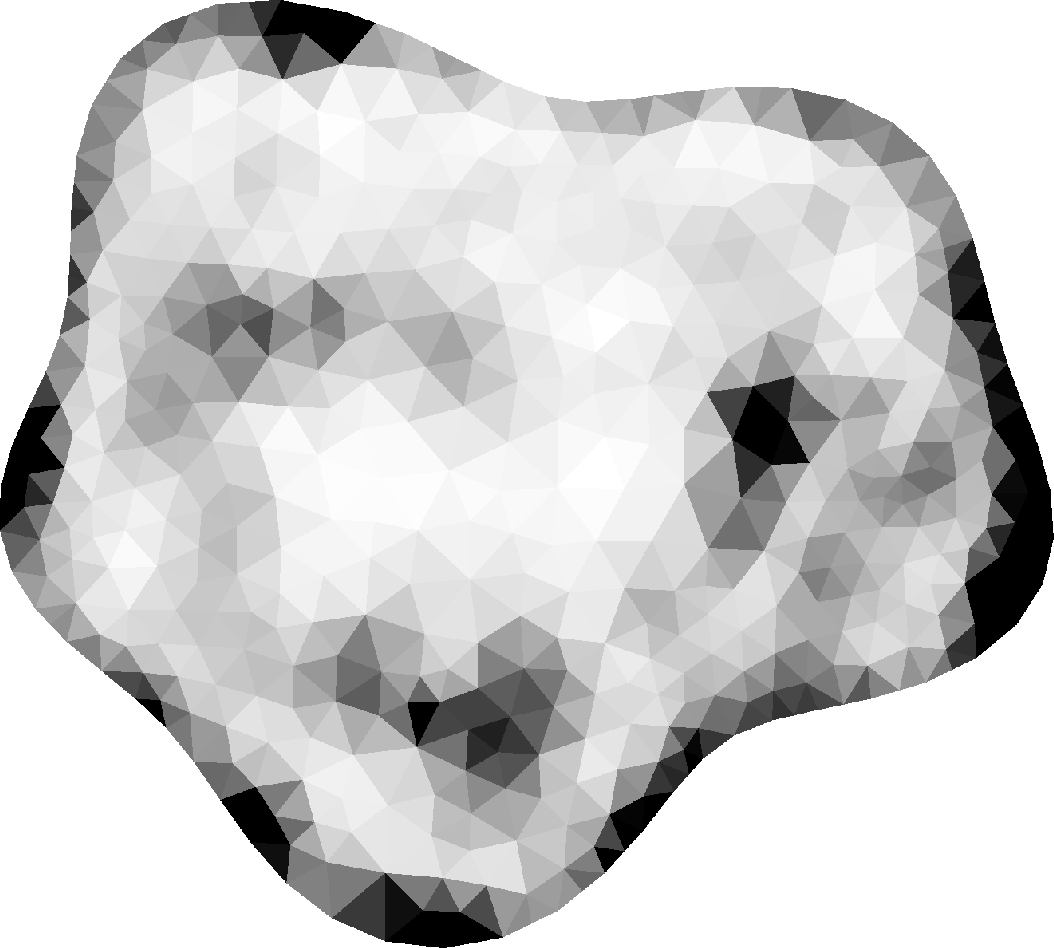}  \\
  \includegraphics[width=1.8cm]{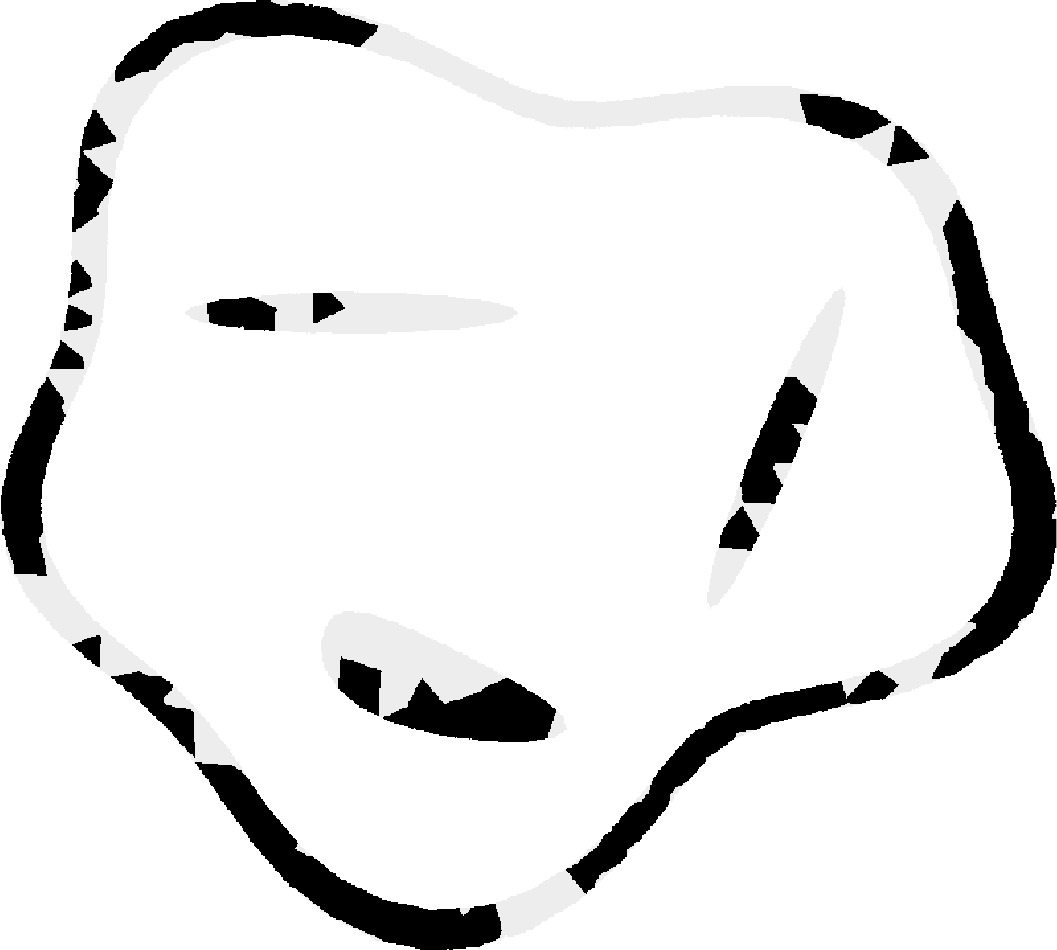} 
               \end{minipage}
    \begin{minipage}{1.8cm} \centering 
         \includegraphics[width=1.8cm]{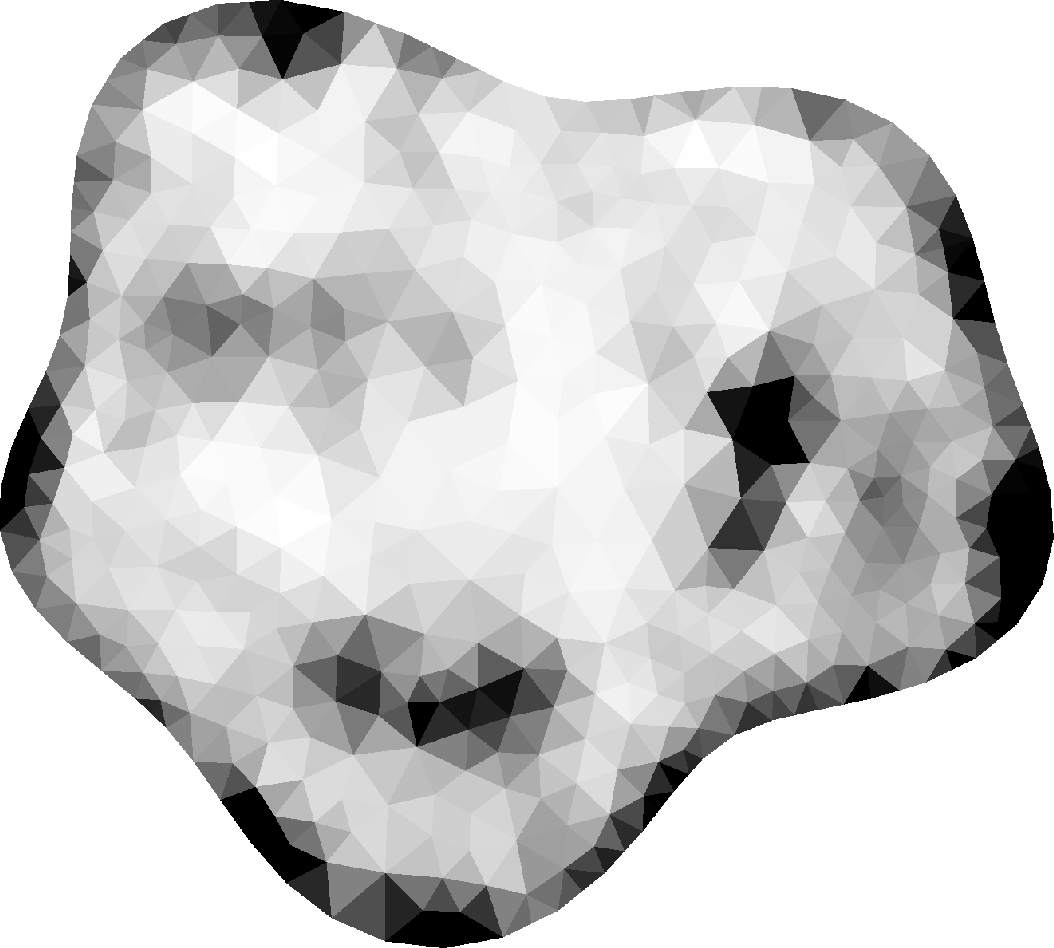} \\
  \includegraphics[width=1.8cm]{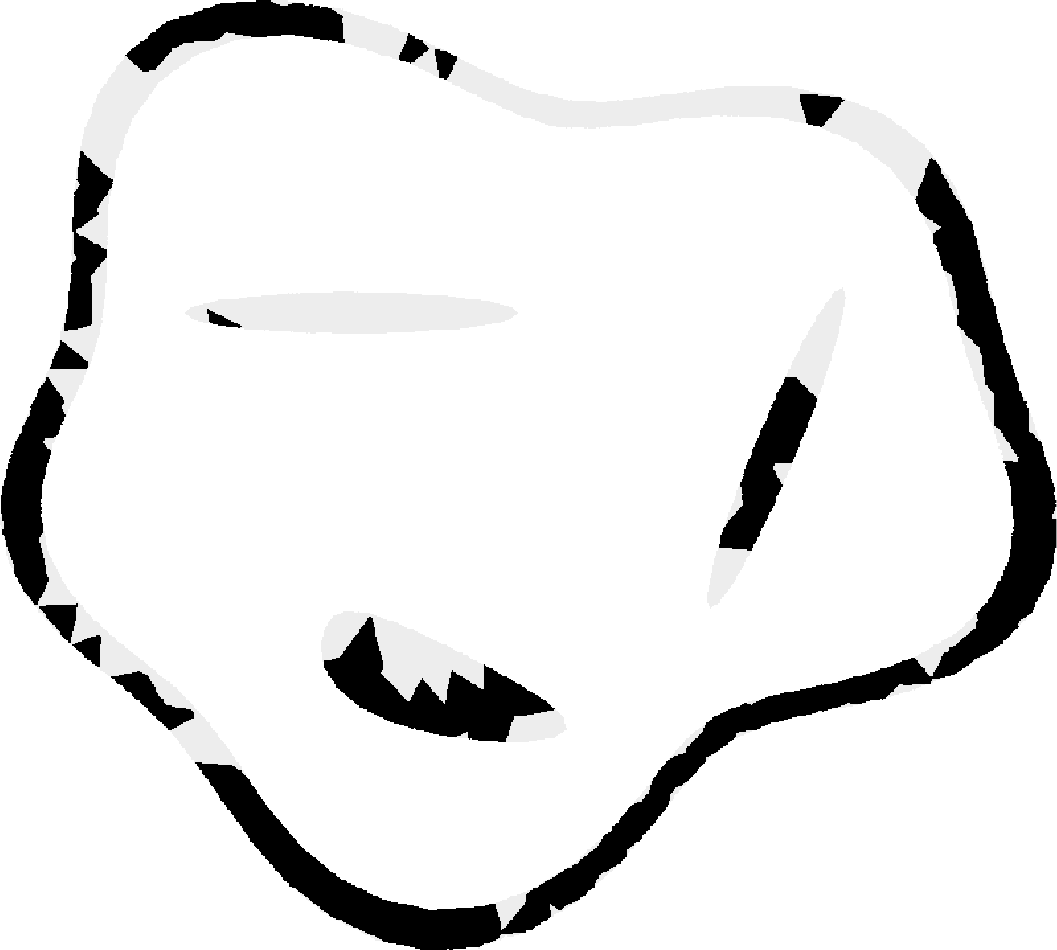}    \end{minipage} \\  
\rotatebox{90}{\hskip-0.5cm sparse}     \begin{minipage}{1.8cm} \centering        
         \includegraphics[width=1.8cm]{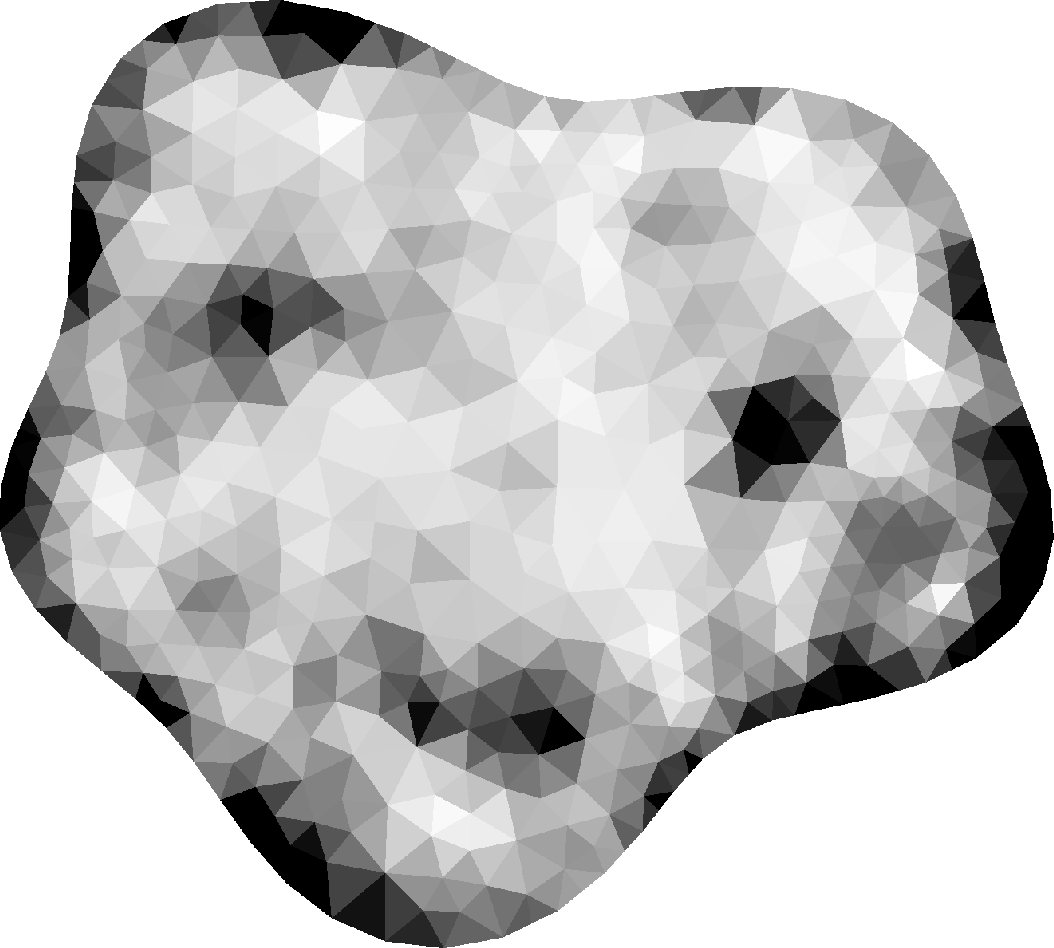} \\
  \includegraphics[width=1.8cm]{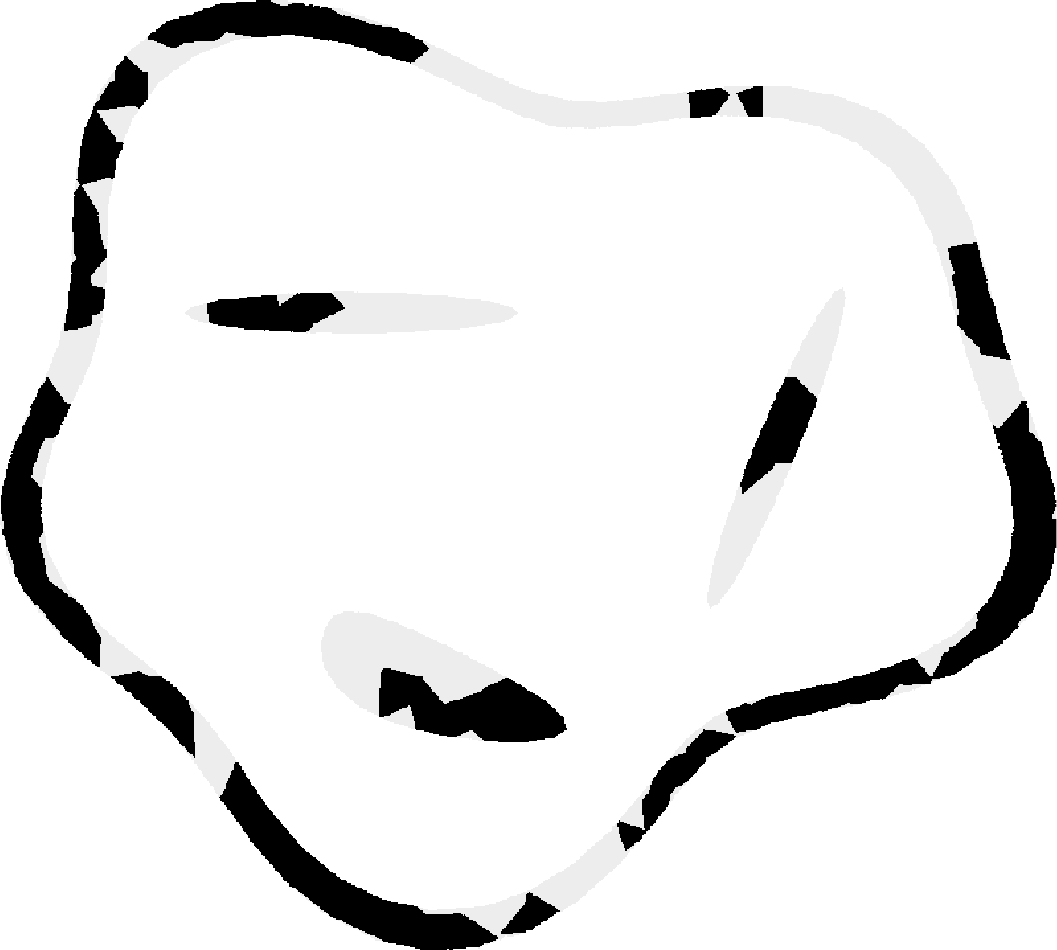} 
             \end{minipage}
    \begin{minipage}{1.8cm} \centering 
        \includegraphics[width=1.8cm]{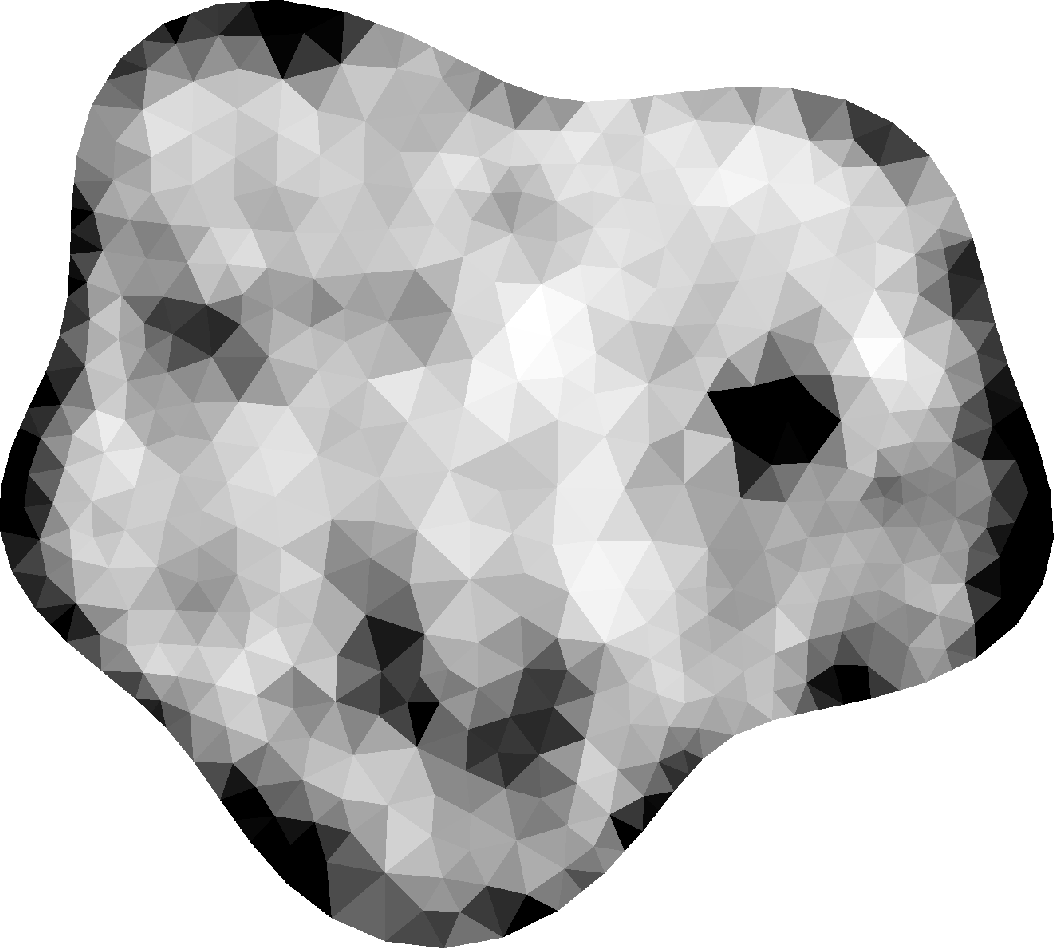} \\
  \includegraphics[width=1.8cm]{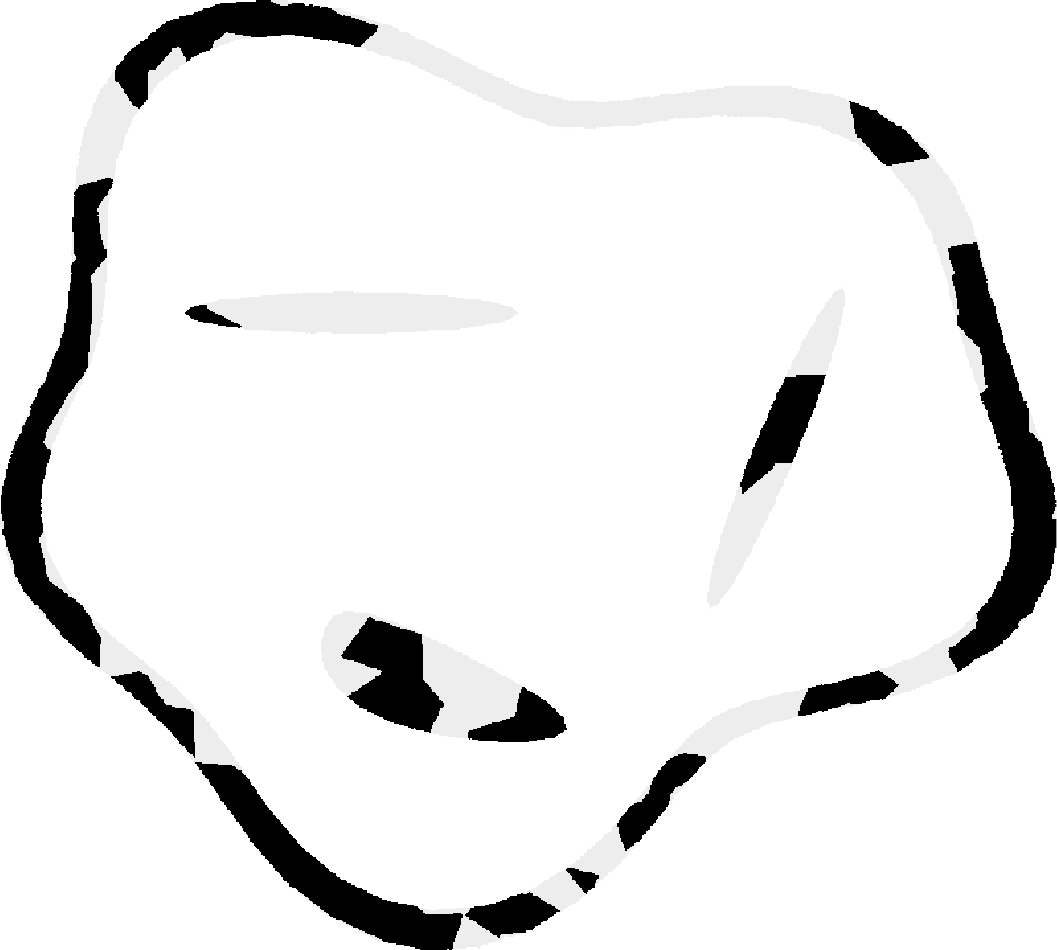} 
            \end{minipage}
    \begin{minipage}{1.8cm} \centering 
           \includegraphics[width=1.8cm]{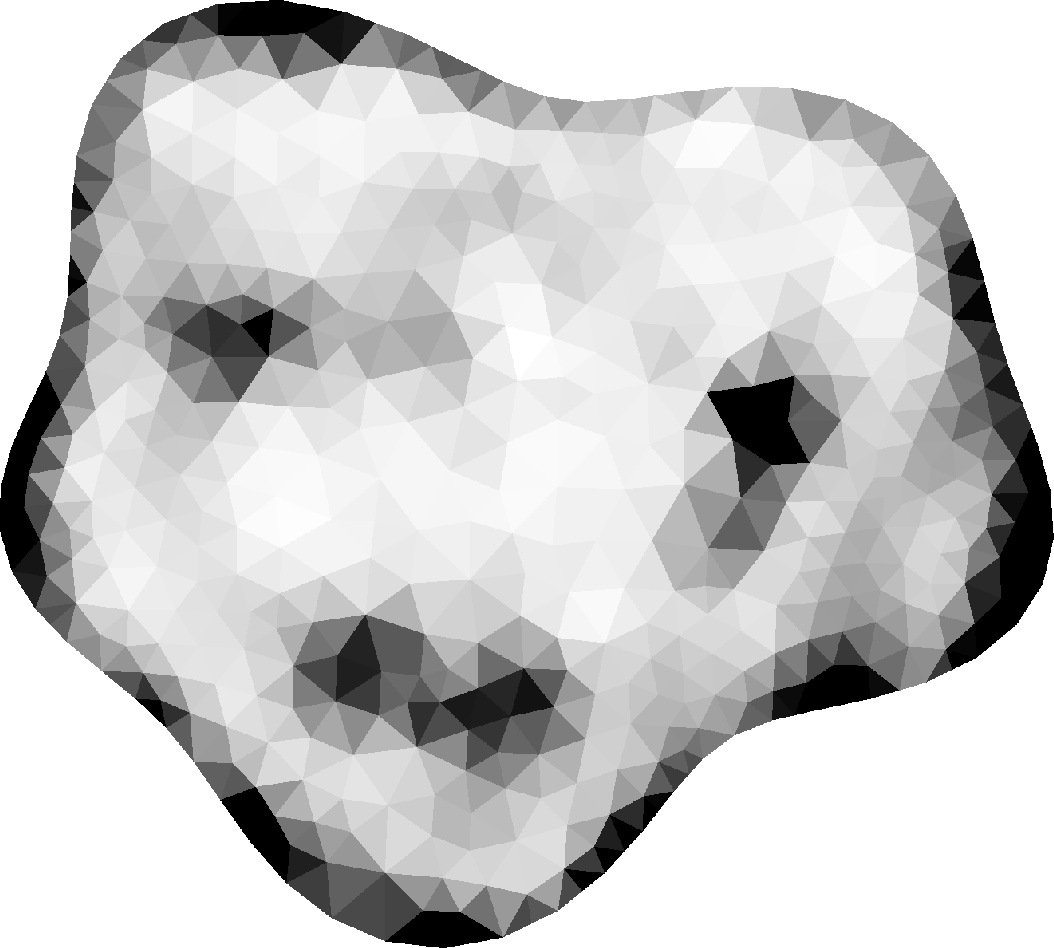}\\
  \includegraphics[width=1.8cm]{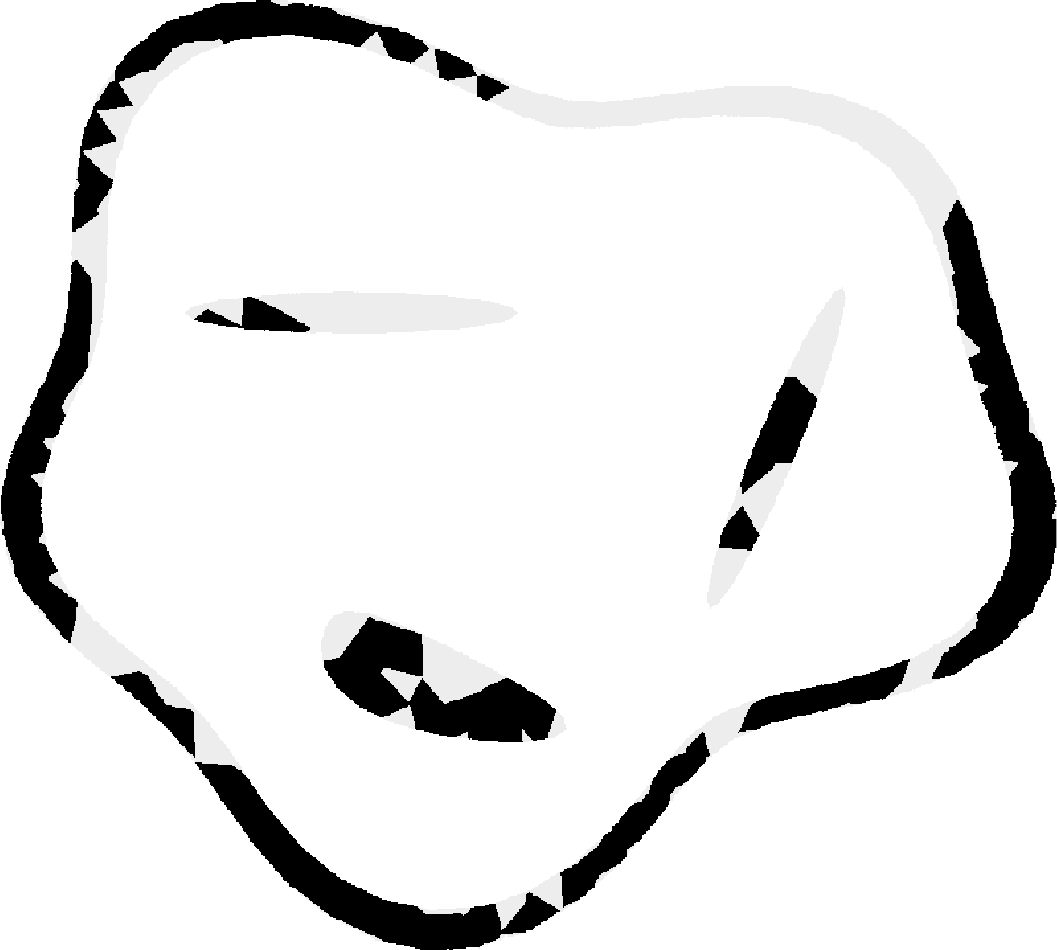} 
               \end{minipage}
    \begin{minipage}{1.8cm} \centering 
         \includegraphics[width=1.8cm]{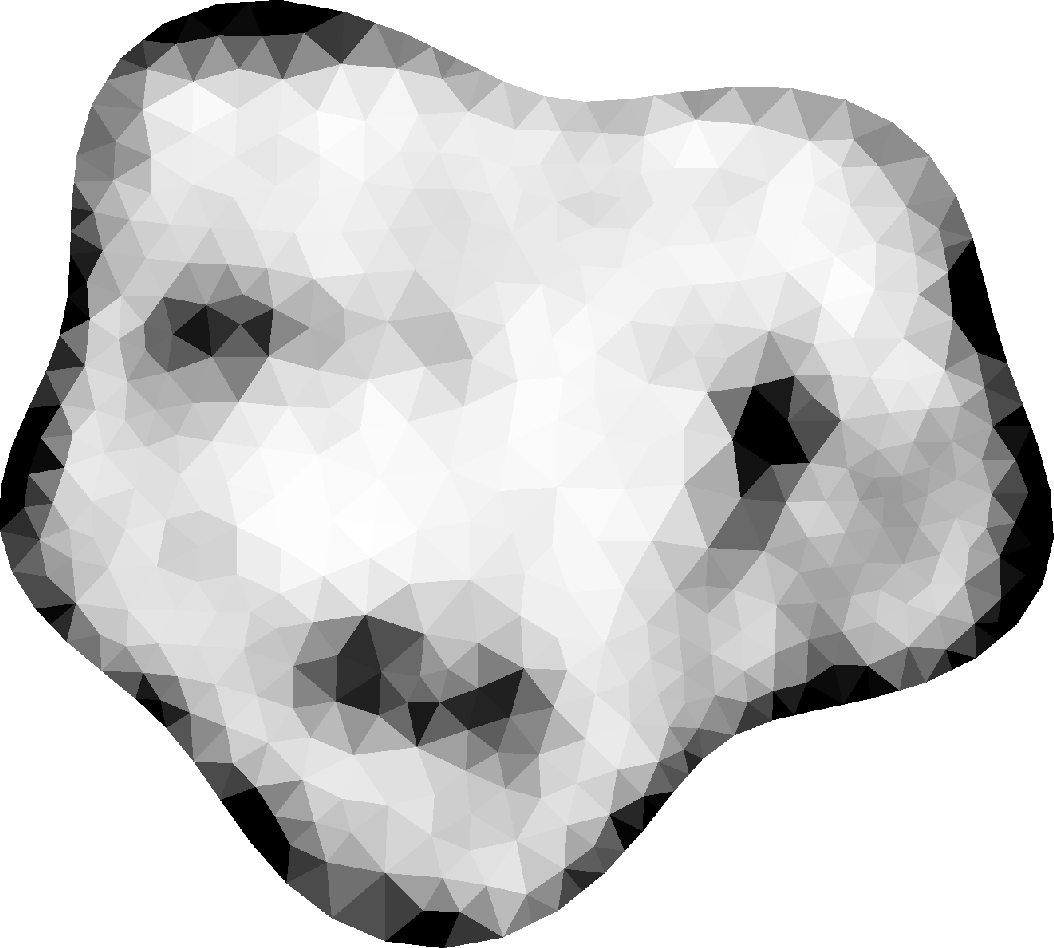}\\
  \includegraphics[width=1.8cm]{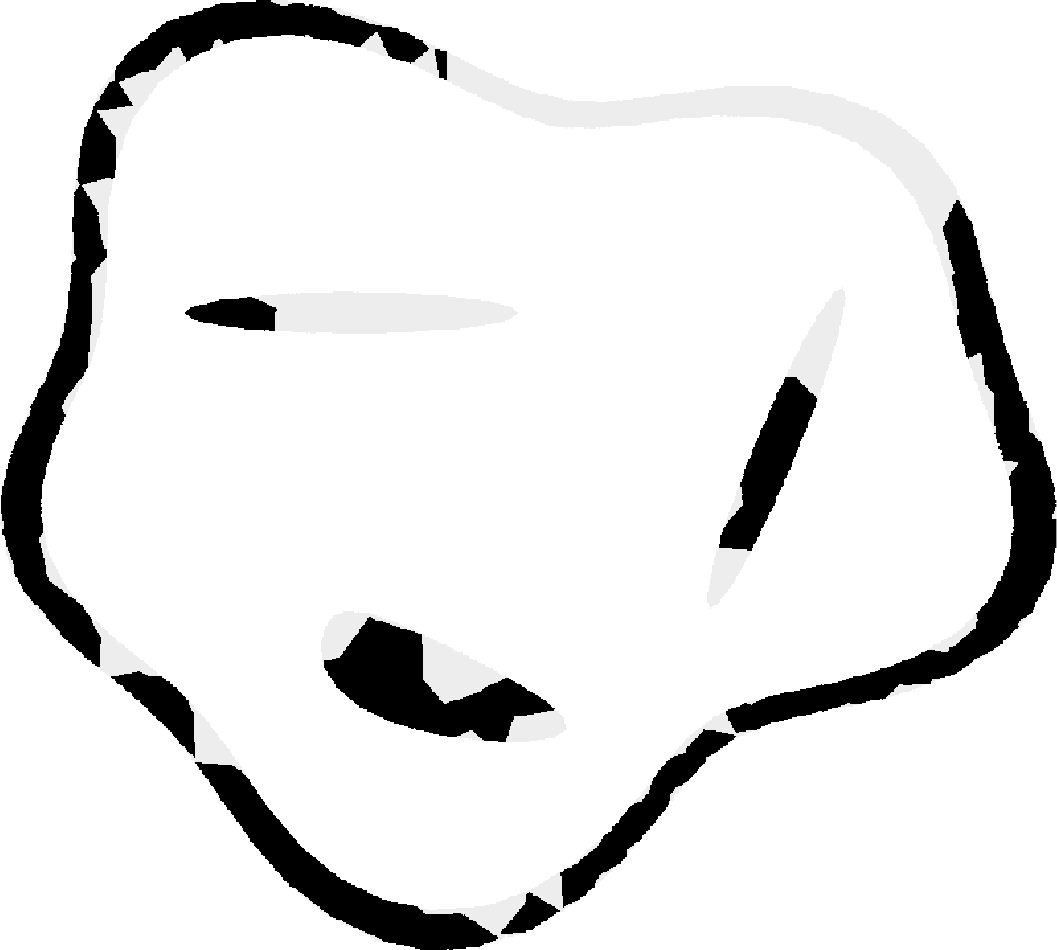} 
             \end{minipage} \\
             \vskip0.2cm
              \begin{minipage}{1.8cm} \centering        
        ({\bf I}): Unfiltered
             \end{minipage}
    \begin{minipage}{1.8cm} \centering 
   ({\bf II}): TSVD 
            \end{minipage}
    \begin{minipage}{1.8cm} \centering 
   ({\bf III}): Unfiltered \&  randomised 
               \end{minipage}
    \begin{minipage}{1.8cm} \centering 
   ({\bf IV}): TSVD  \&  randomised
             \end{minipage}  \end{minipage} 
                         \begin{minipage}{0.6cm} \centering
             \includegraphics[height= 5cm]{images/color_bar_lin_1.png}
             \end{minipage}
             \\ \vskip0.2cm 
             \end{scriptsize}
    \caption{Domain reconstructions obtained using the 20 MHz signal pulse ({\bf A}) with SNR of 20 dB (1st--4th row) and 12 dB (7th--10th row). The reconstructions are presented in the odd rows while their corresponding overlap reconstructions (even rows) depict how much of the surface and void details are reconstructed in black (grey parts are those which do not overlap). The top row shows the results for the dense configuration with 64 points  and the bottom row for the sparse one with 32 points. The results for  configurations ({\bf I})--({\bf IV}) are shown in columns from left to right, respectively.  }
    \label{fig:B_low_freq}
\end{figure}

\begin{figure}[!ht]
    \centering  \begin{scriptsize} SNR 20 dB \\ \vskip0.2cm 
    \begin{minipage}{7.7cm} 
 \rotatebox{90}{\hskip-0.5cm dense}    \begin{minipage}{1.8cm} \centering
 \includegraphics[width=1.8cm]{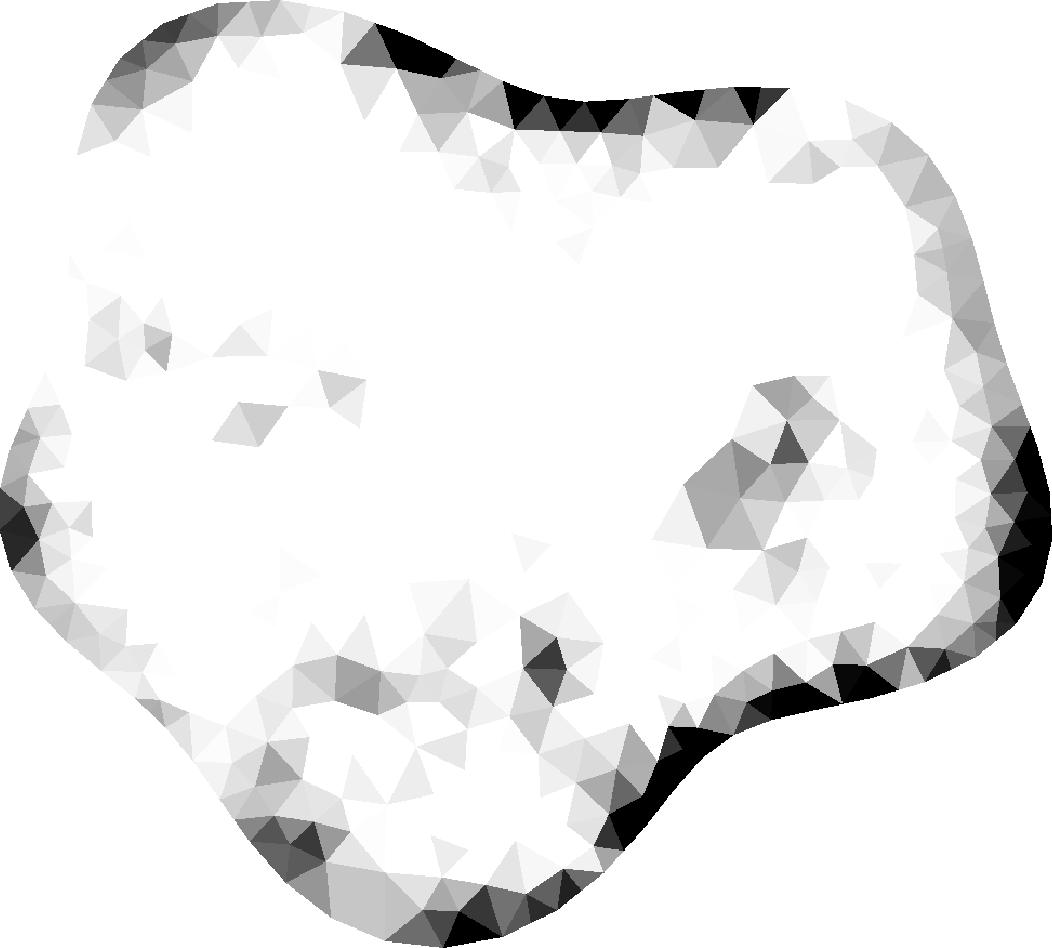} \\
  \includegraphics[width=1.8cm]{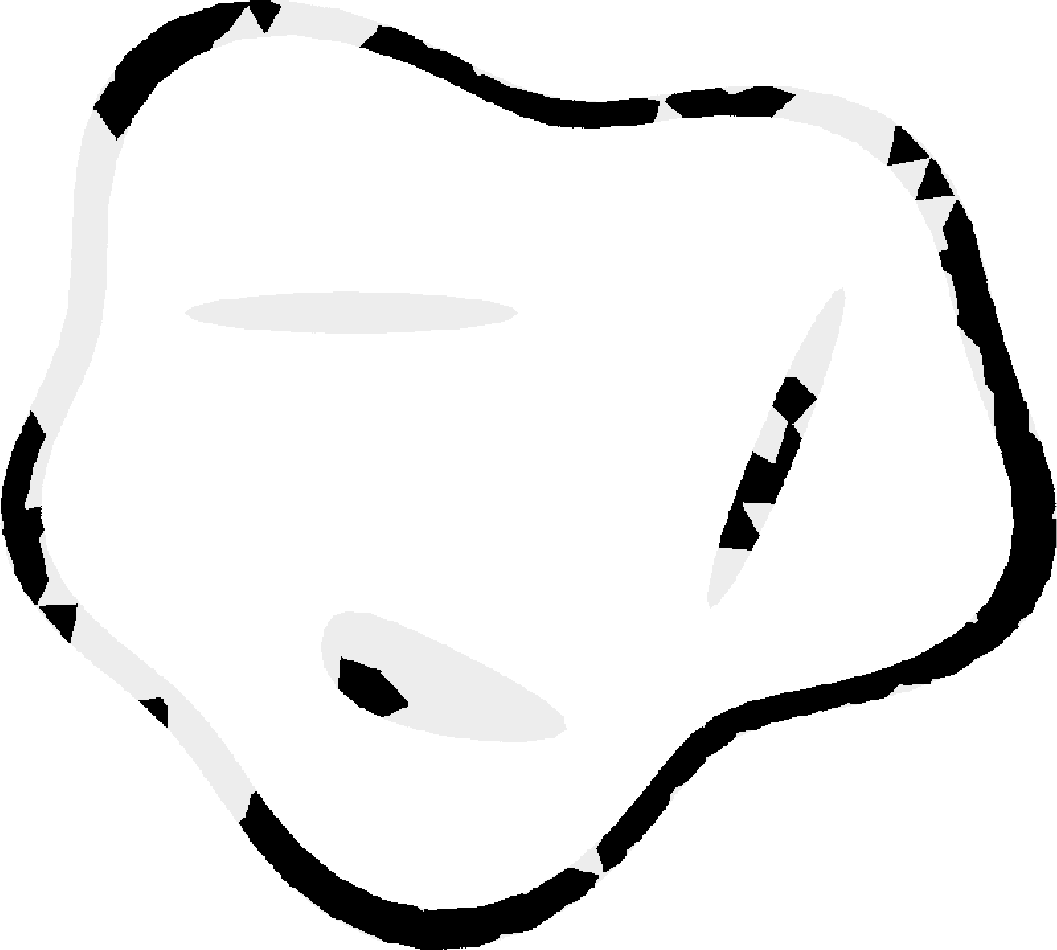} 
    \end{minipage}
    \begin{minipage}{1.8cm} \centering   
   \includegraphics[width=1.8cm]{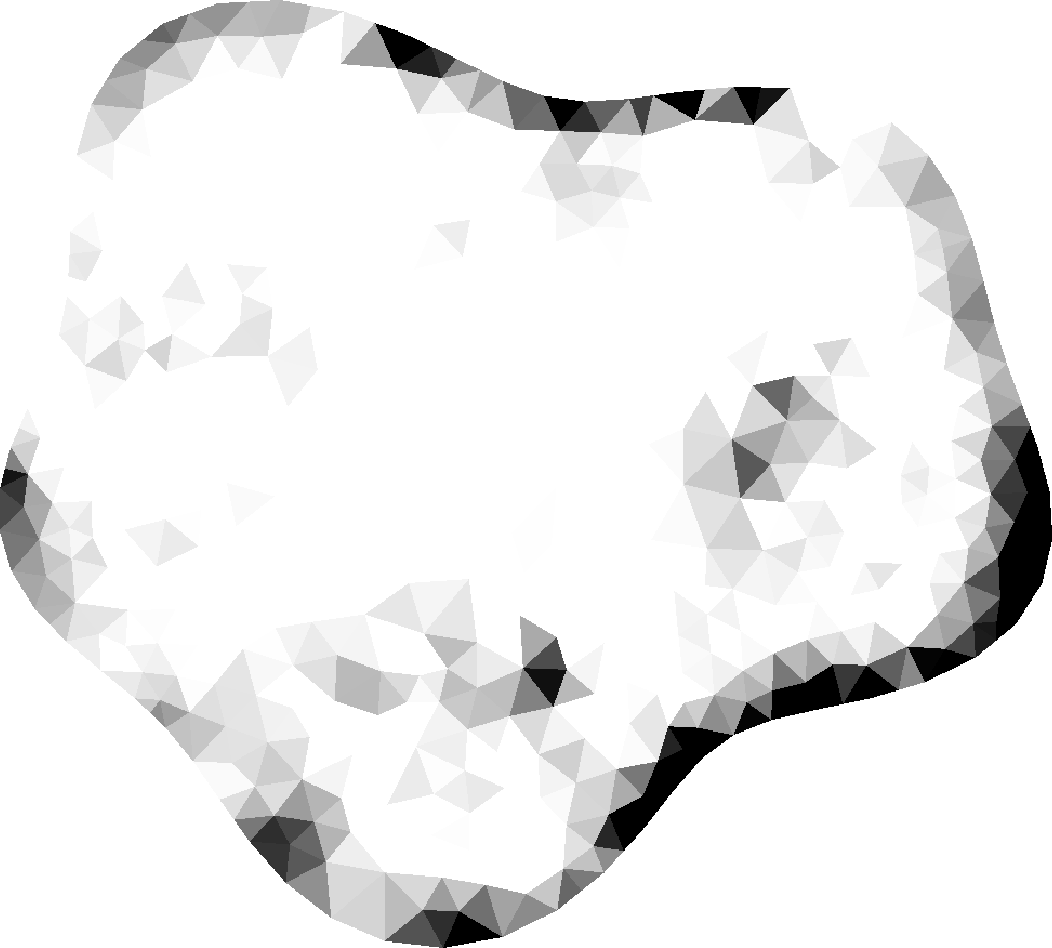} \\
   \includegraphics[width=1.8cm]{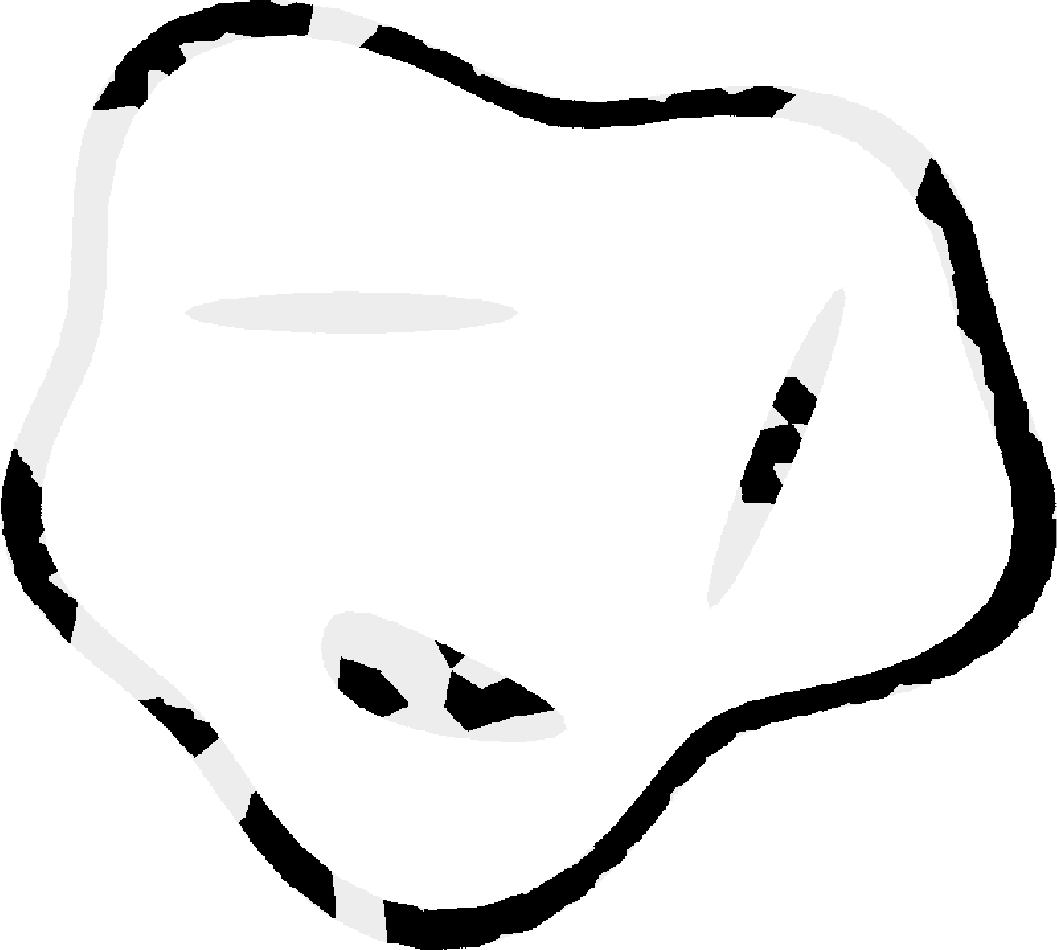} 
          \end{minipage}
    \begin{minipage}{1.8cm} \centering 
           \includegraphics[width=1.8cm]{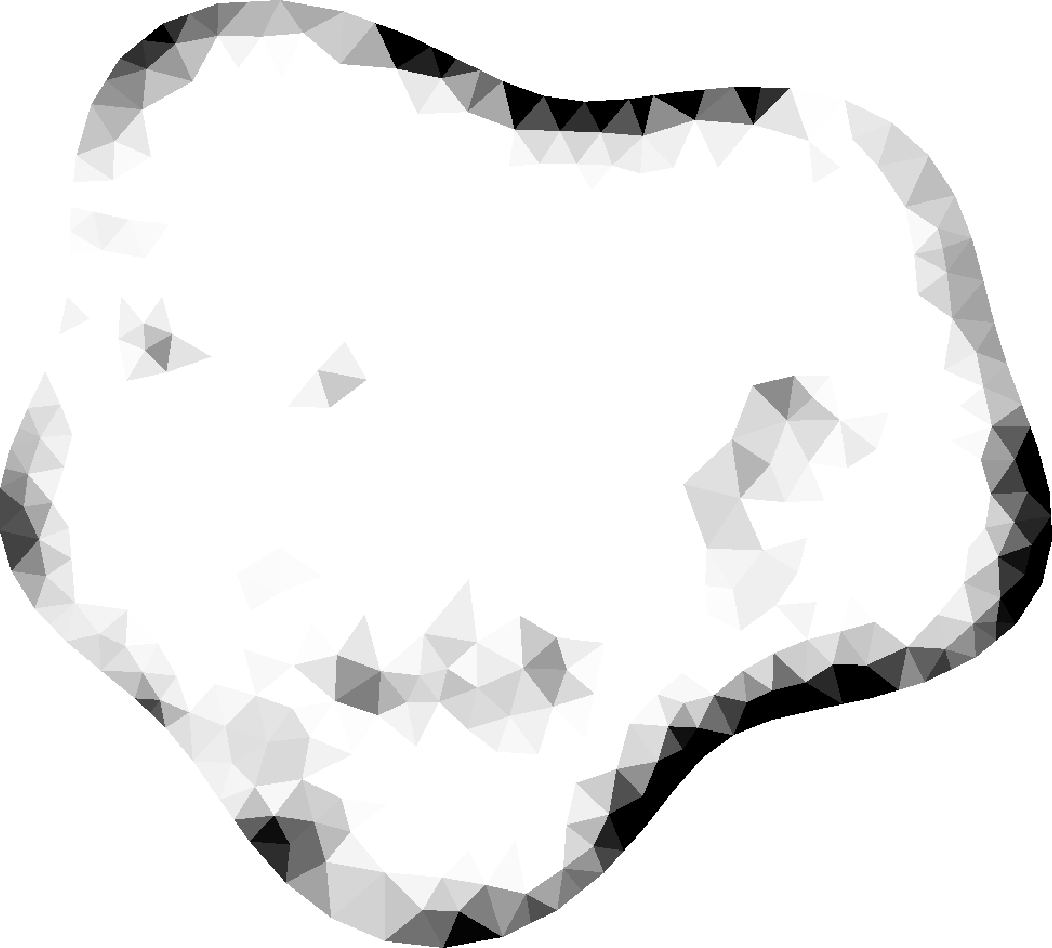} \\ 
           \includegraphics[width=1.8cm]{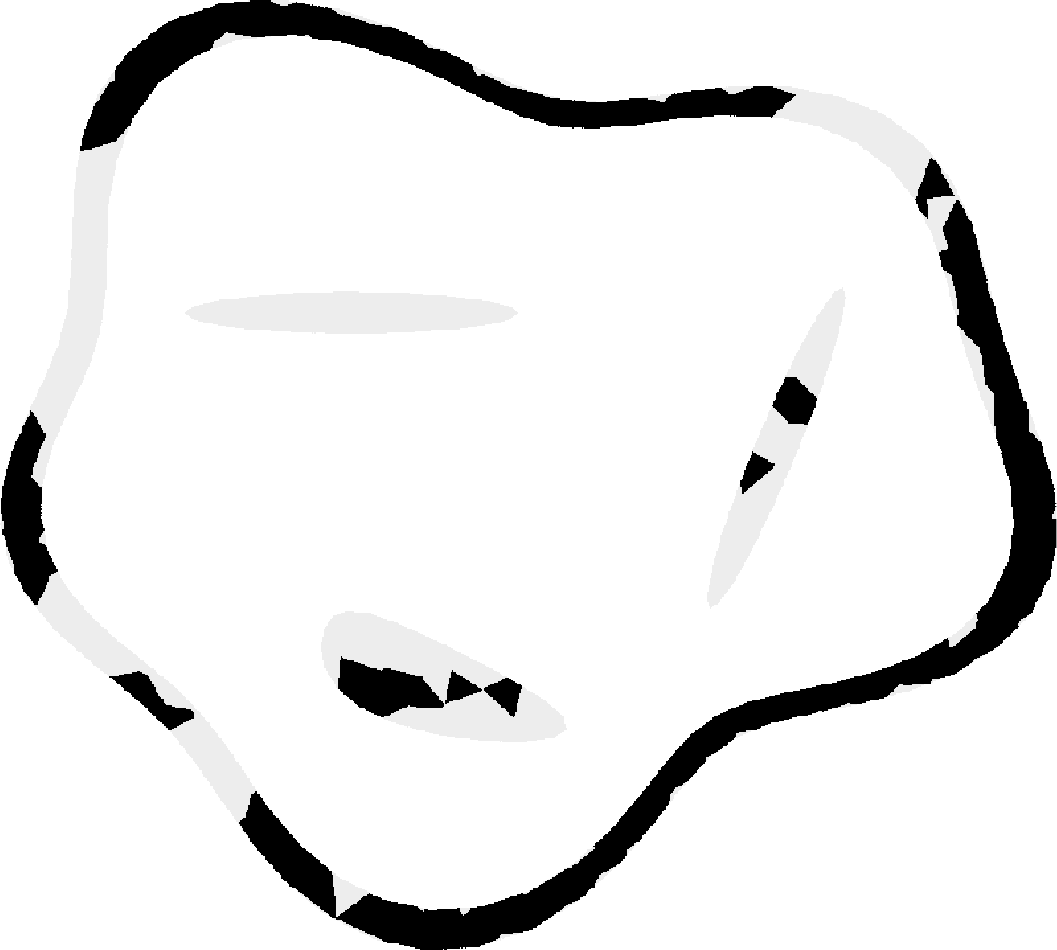} 
               \end{minipage}
    \begin{minipage}{1.8cm} \centering 
         \includegraphics[width=1.8cm]{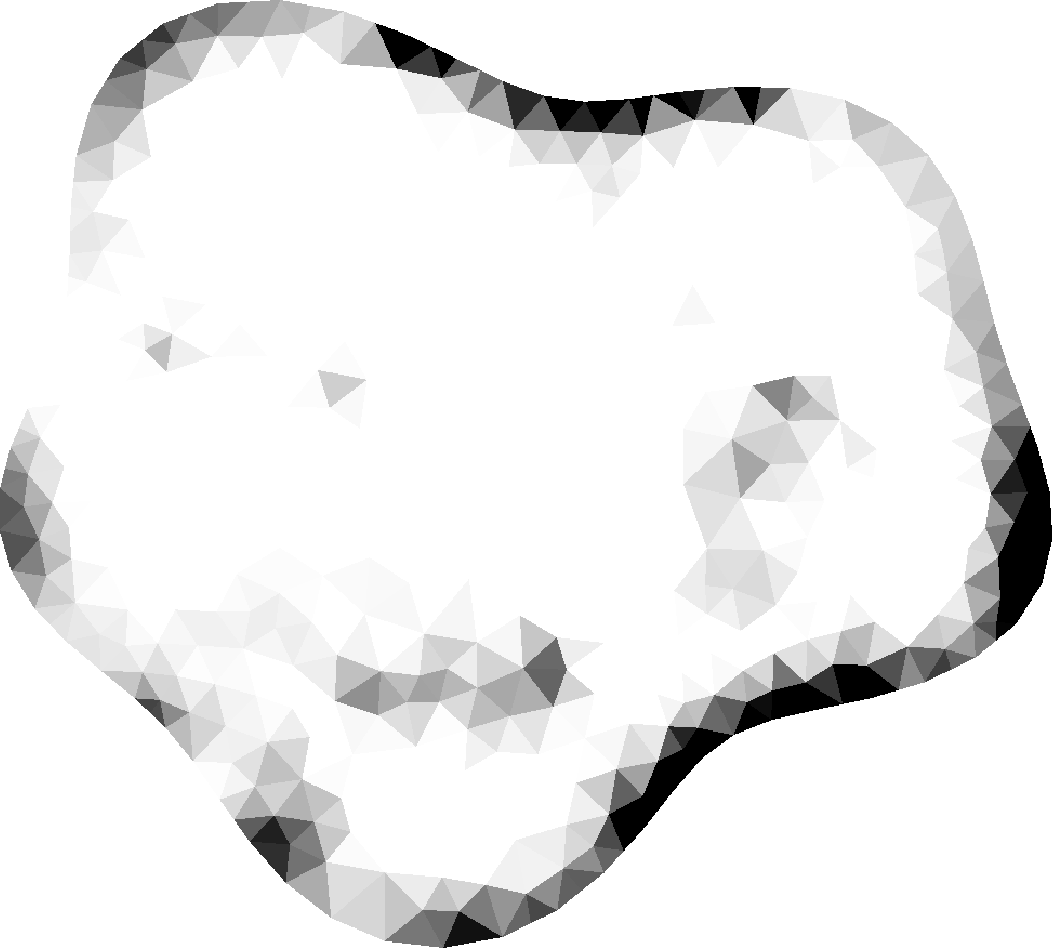}  \\ 
         \includegraphics[width=1.8cm]{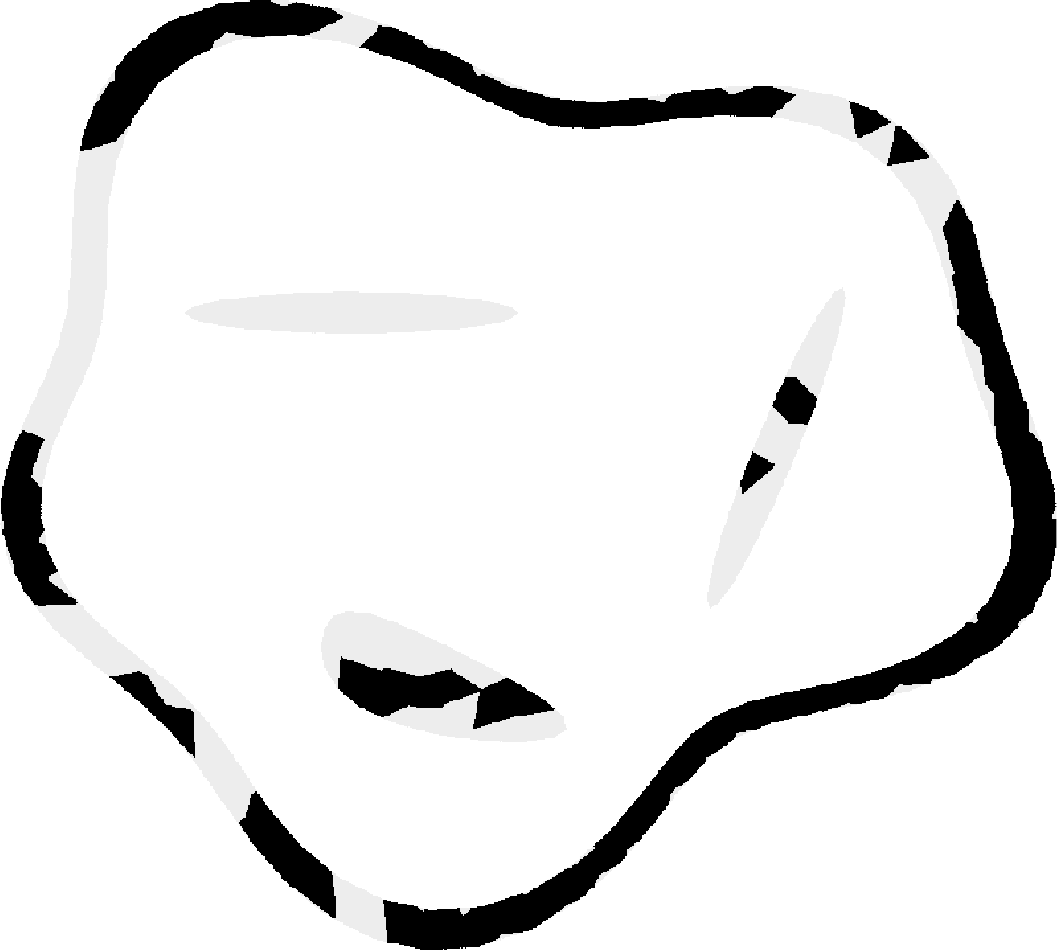} 
         \end{minipage} \\  
\rotatebox{90}{\hskip-0.5cm sparse}     \begin{minipage}{1.8cm} \centering        
         \includegraphics[width=1.8cm]{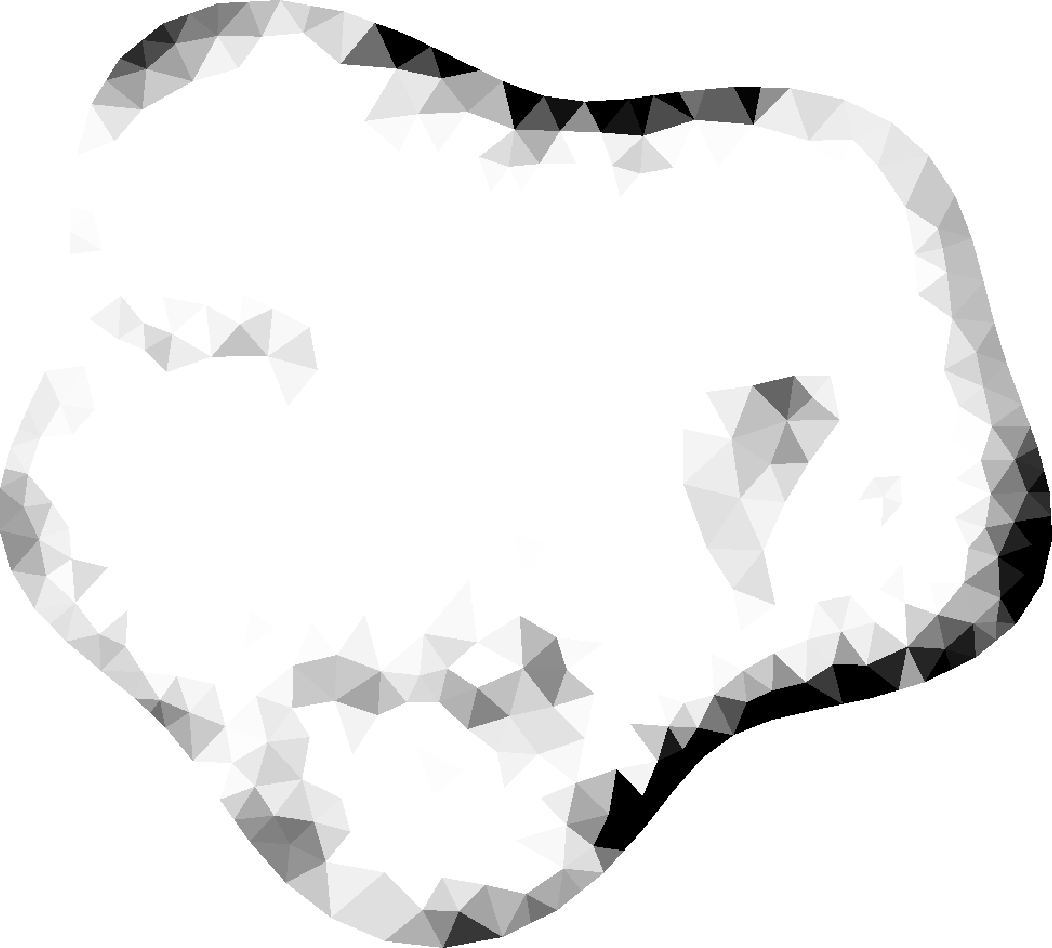} \\ 
         \includegraphics[width=1.8cm]{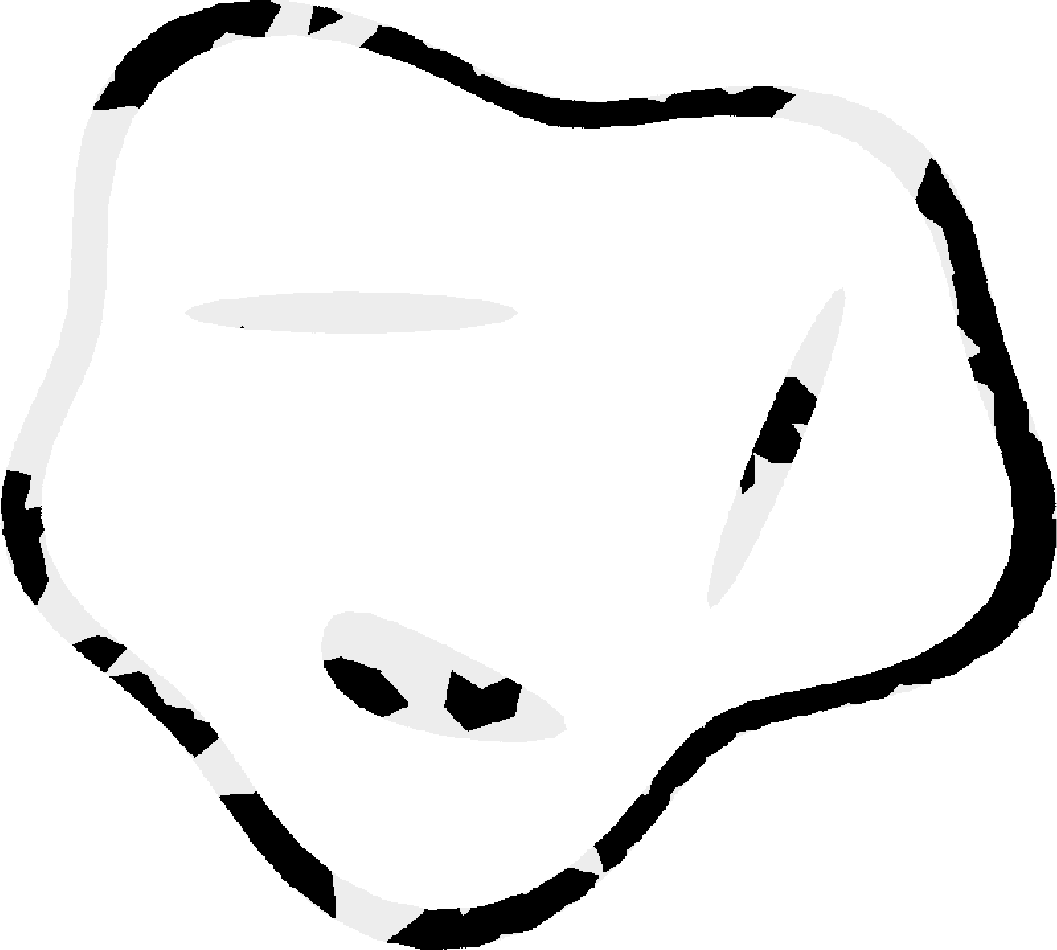}  
             \end{minipage}
    \begin{minipage}{1.8cm} \centering 
        \includegraphics[width=1.8cm]{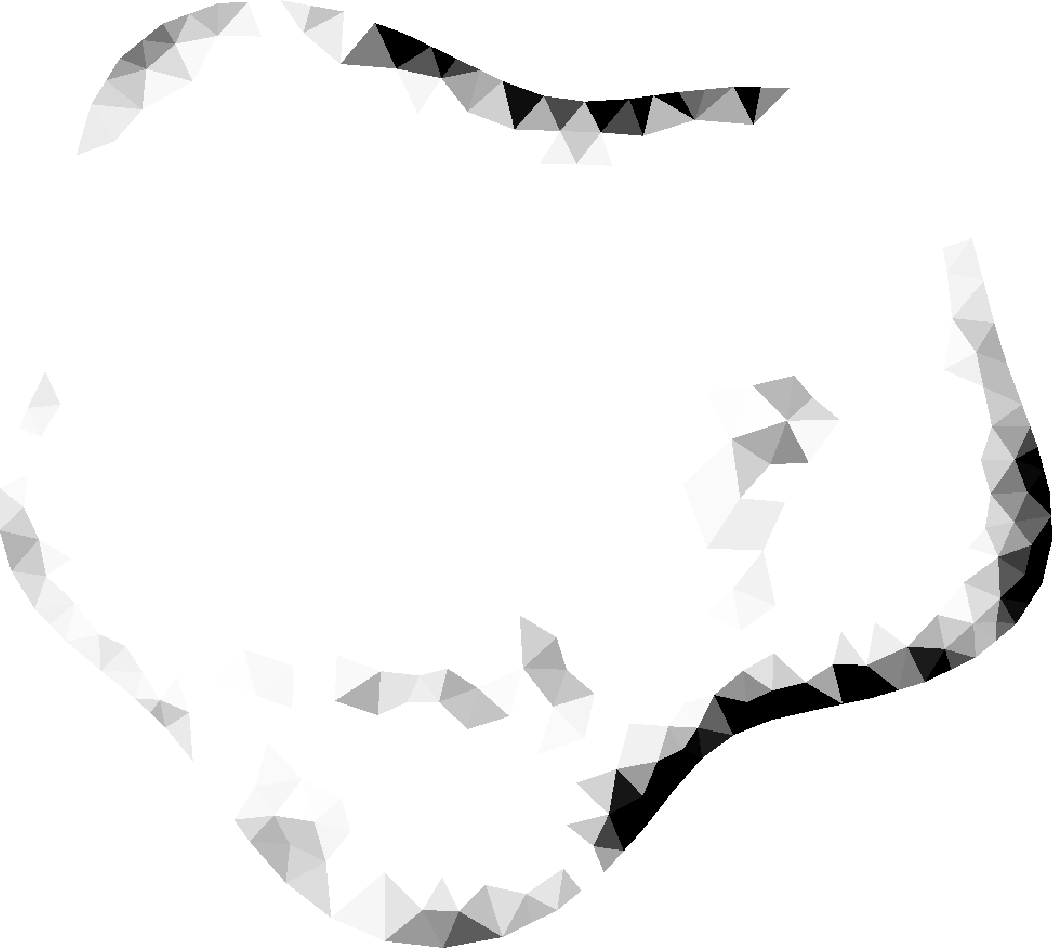} \\
         \includegraphics[width=1.8cm]{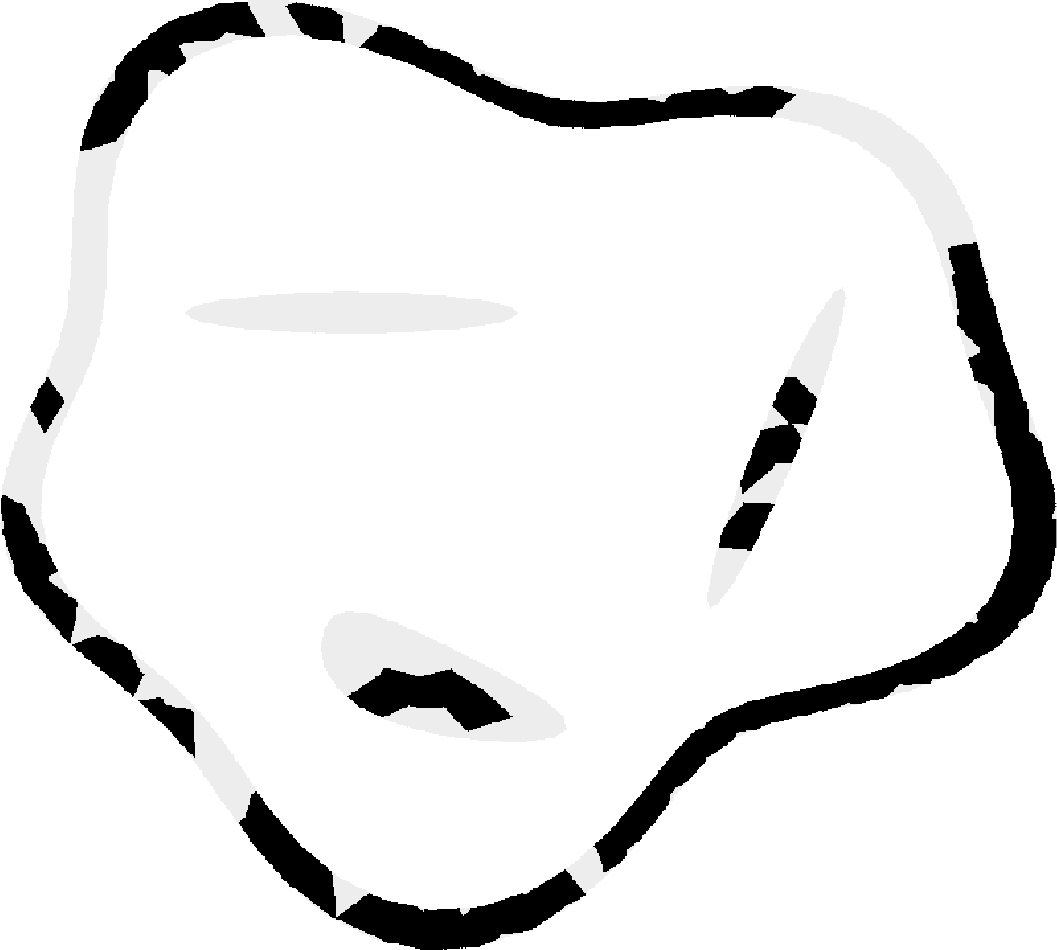} 
            \end{minipage}
    \begin{minipage}{1.8cm} \centering 
           \includegraphics[width=1.8cm]{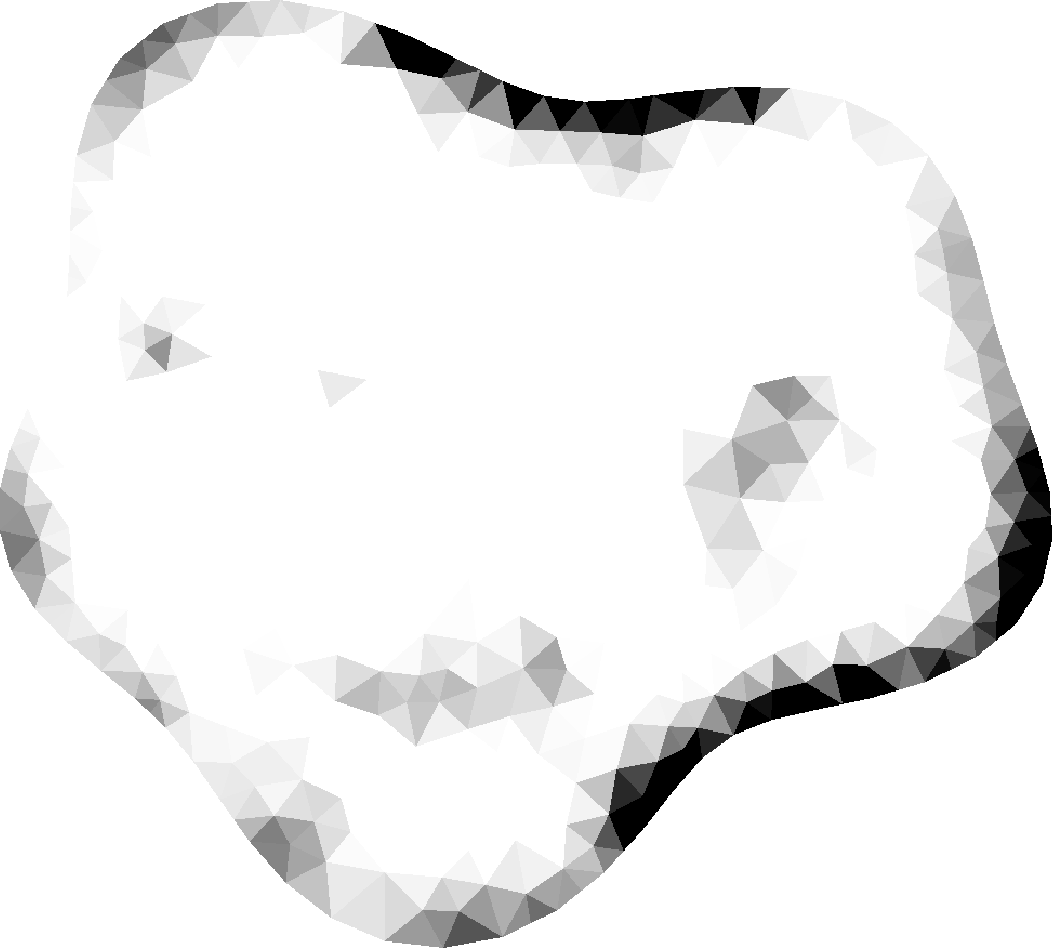} \\ 
            \includegraphics[width=1.8cm]{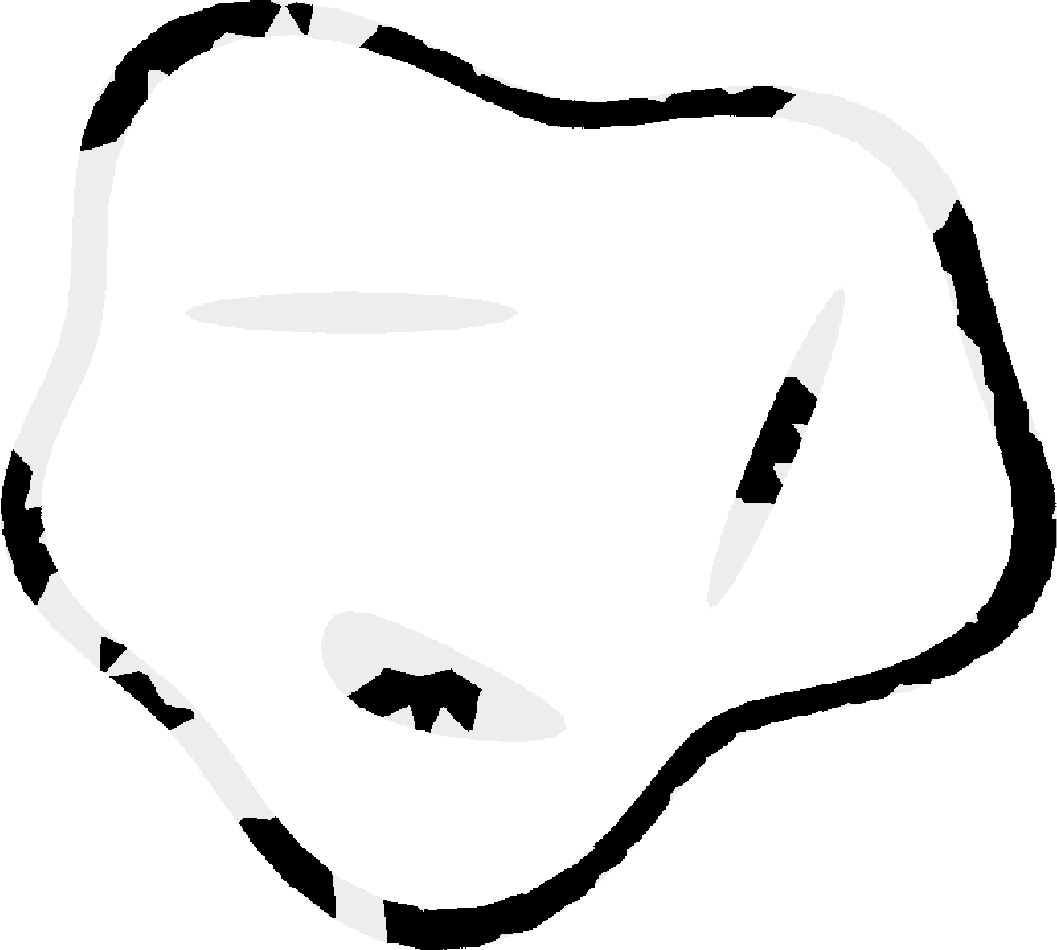} 
               \end{minipage}
    \begin{minipage}{1.8cm} \centering 
         \includegraphics[width=1.8cm]{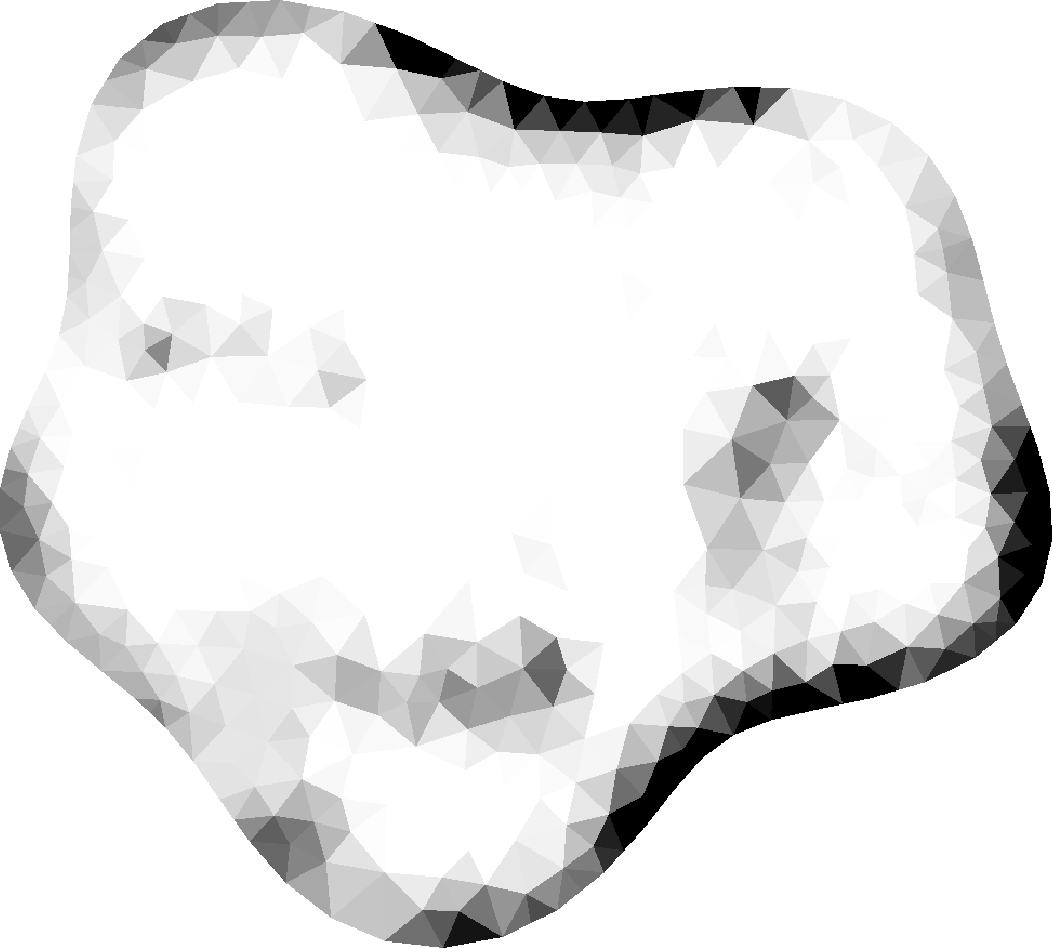} \\ 
          \includegraphics[width=1.8cm]{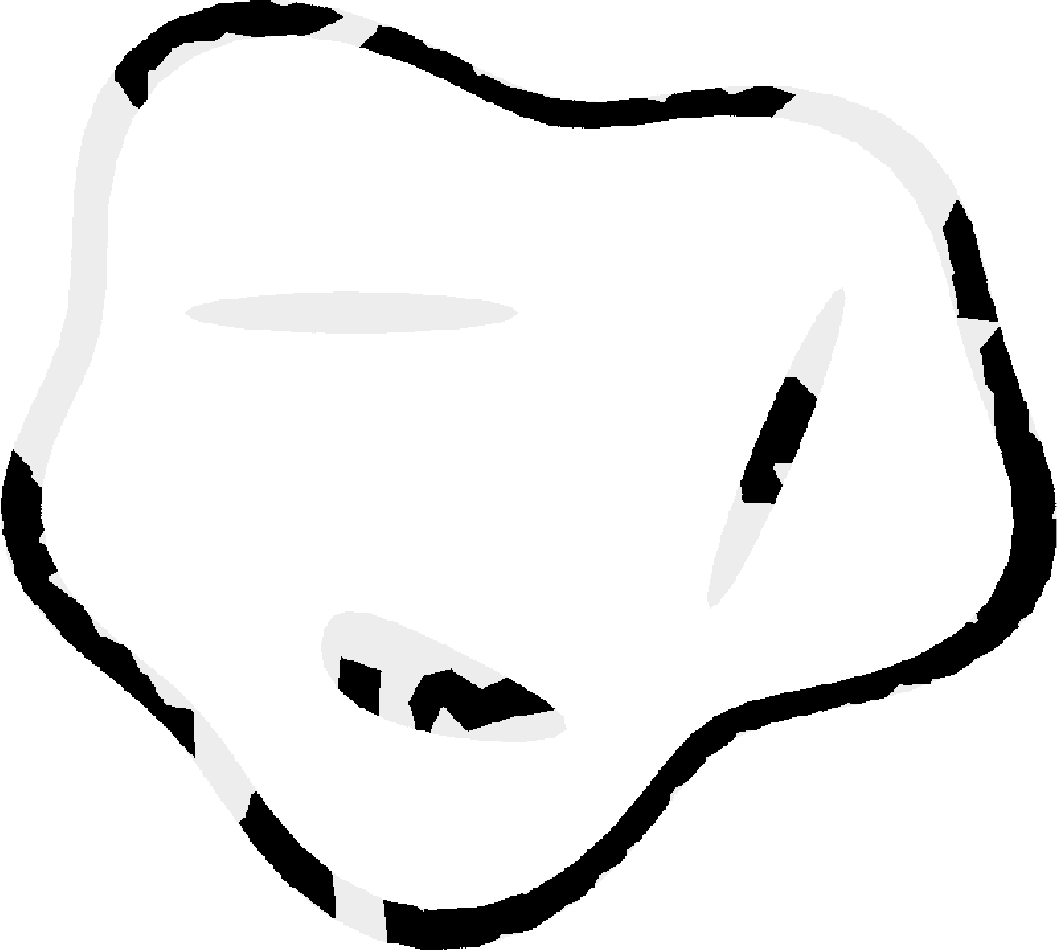} 
             \end{minipage} \\ \vskip0.2cm
              \begin{minipage}{1.8cm} \centering        
        ({\bf I}): Unfiltered 
             \end{minipage}
    \begin{minipage}{1.8cm} \centering 
   ({\bf II}): TSVD 
            \end{minipage}
    \begin{minipage}{1.8cm} \centering 
   ({\bf III}): Unfiltered \&  randomised
               \end{minipage}
    \begin{minipage}{1.8cm} \centering 
   ({\bf IV}): TSVD \&  randomised
             \end{minipage} \end{minipage}
                         \begin{minipage}{0.6cm} \centering
             \includegraphics[height= 5cm]{images/color_bar_lin_1.png}
             \end{minipage}
             \\ \vskip0.4cm SNR 12 dB \\ \vskip0.2cm
    \begin{minipage}{7.7cm}
 \rotatebox{90}{\hskip-0.5cm dense}    \begin{minipage}{1.8cm} \centering
 \includegraphics[width=1.8cm]{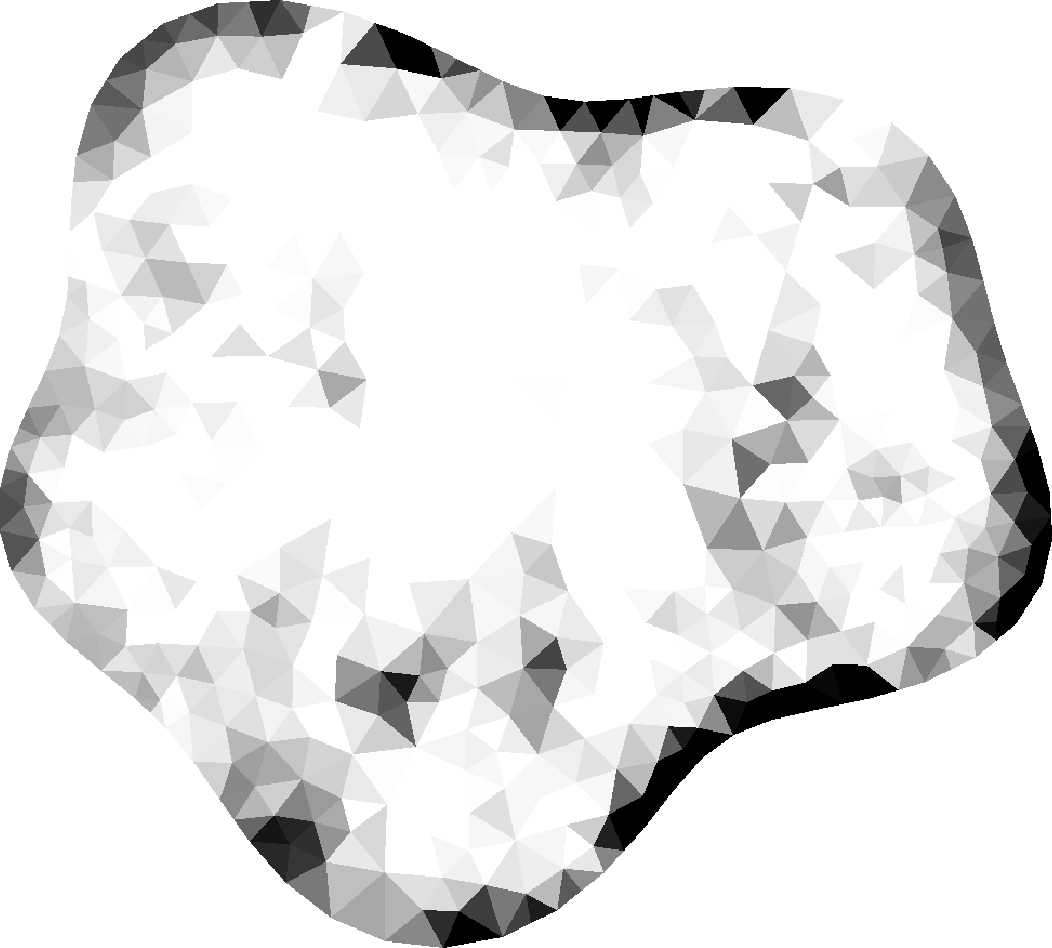}  \\
  \includegraphics[width=1.8cm]{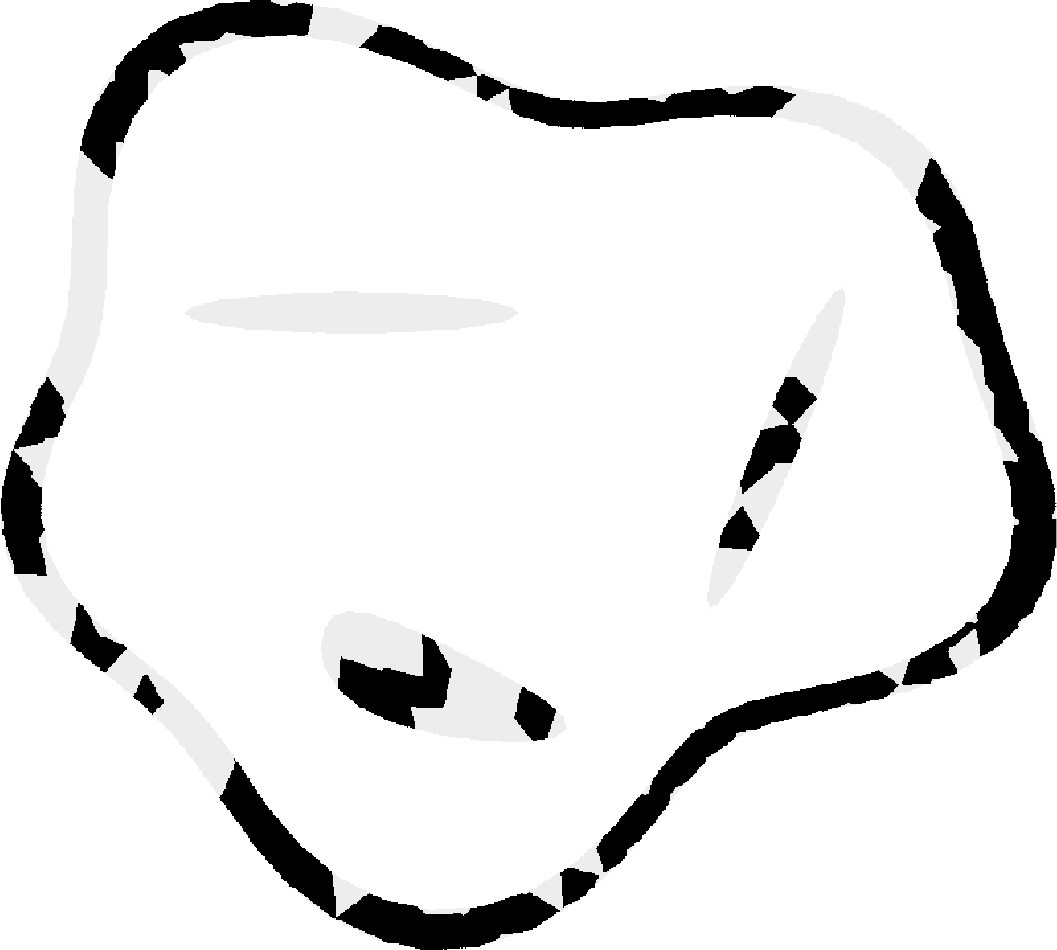} 
    \end{minipage}
    \begin{minipage}{1.8cm} \centering   
   \includegraphics[width=1.8cm]{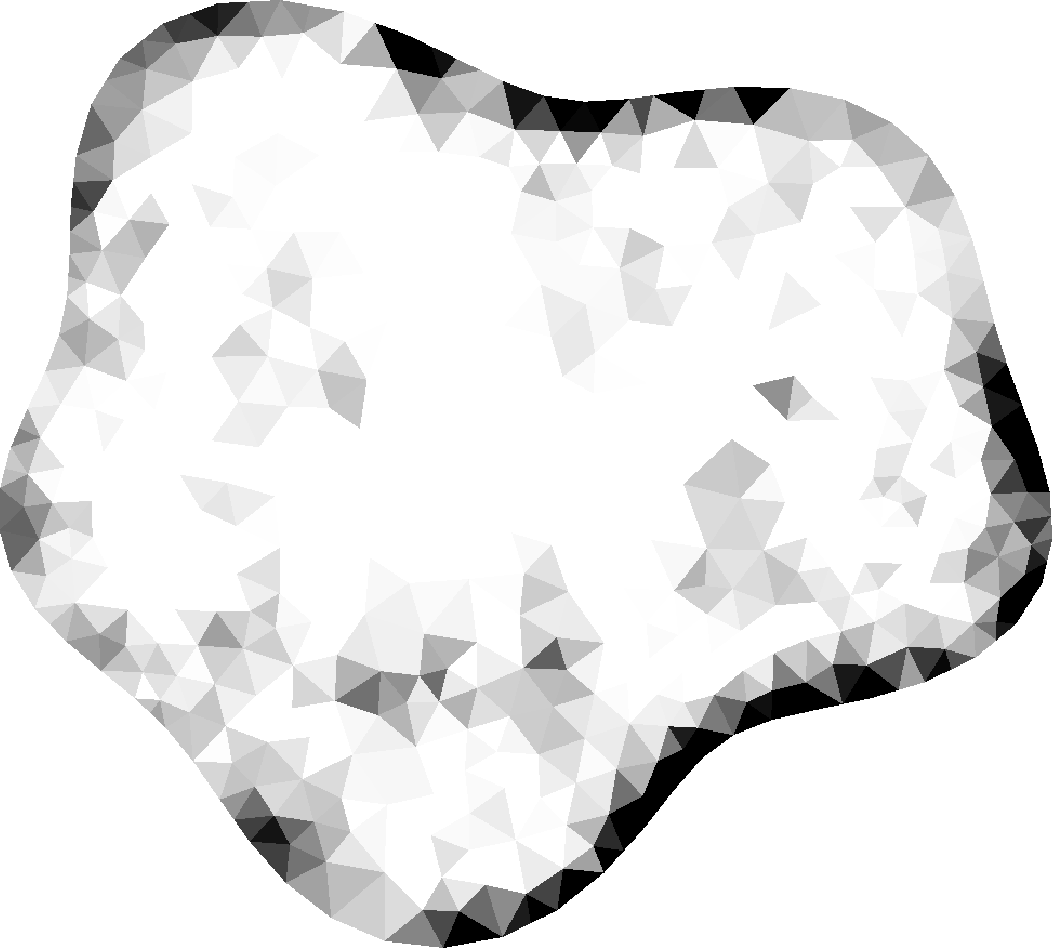} \\
  \includegraphics[width=1.8cm]{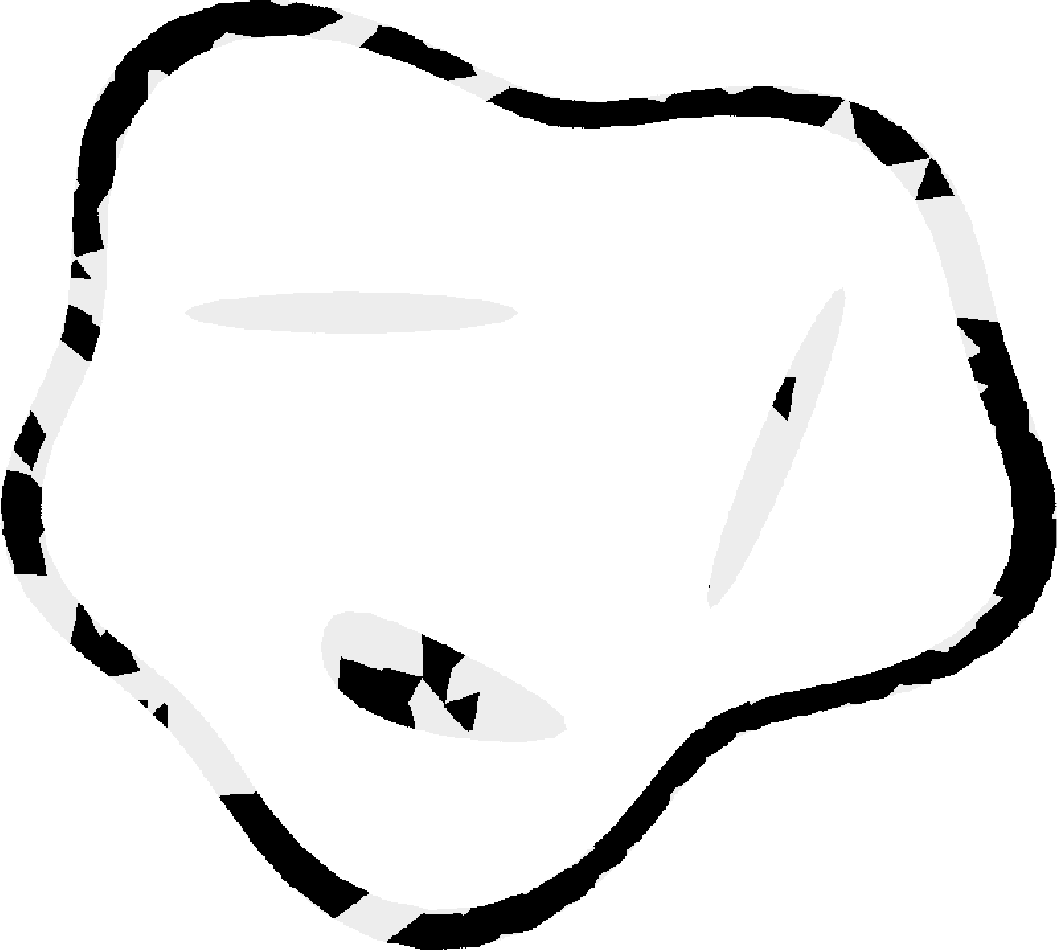} 
          \end{minipage}
    \begin{minipage}{1.8cm} \centering 
           \includegraphics[width=1.8cm]{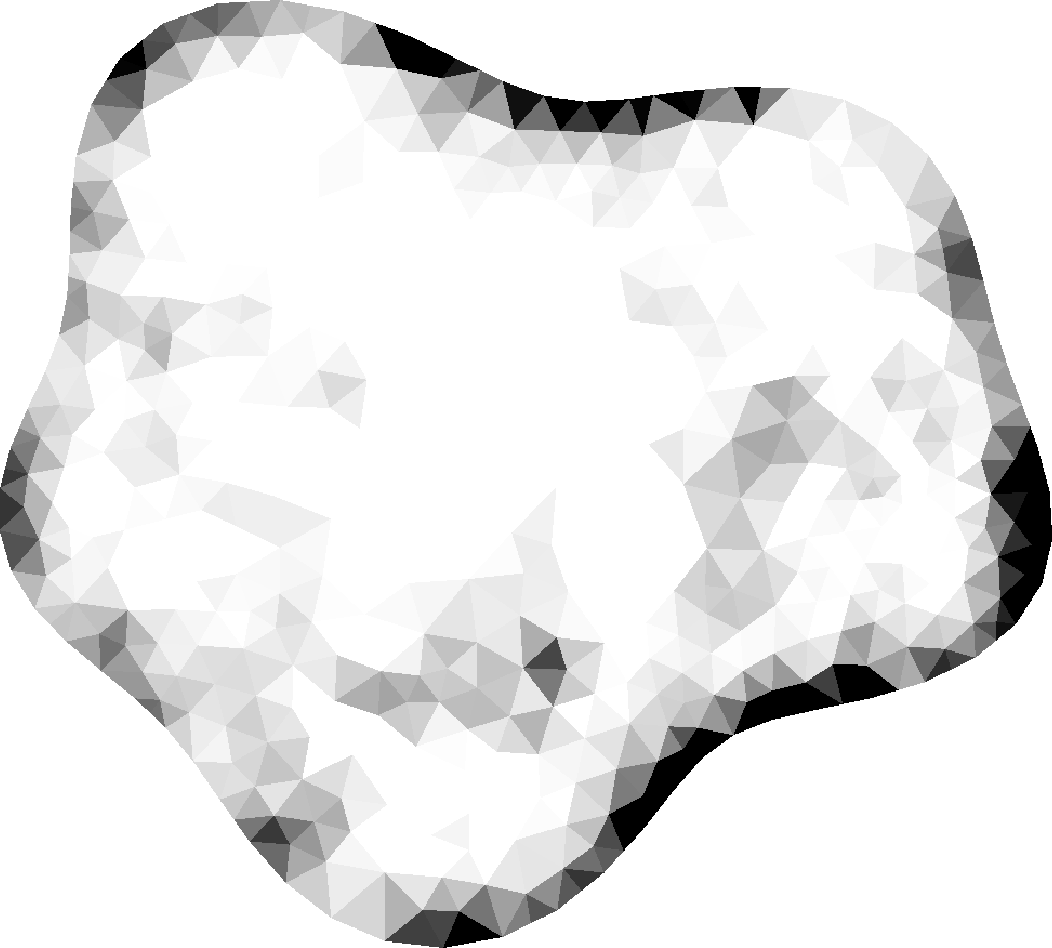}  \\
  \includegraphics[width=1.8cm]{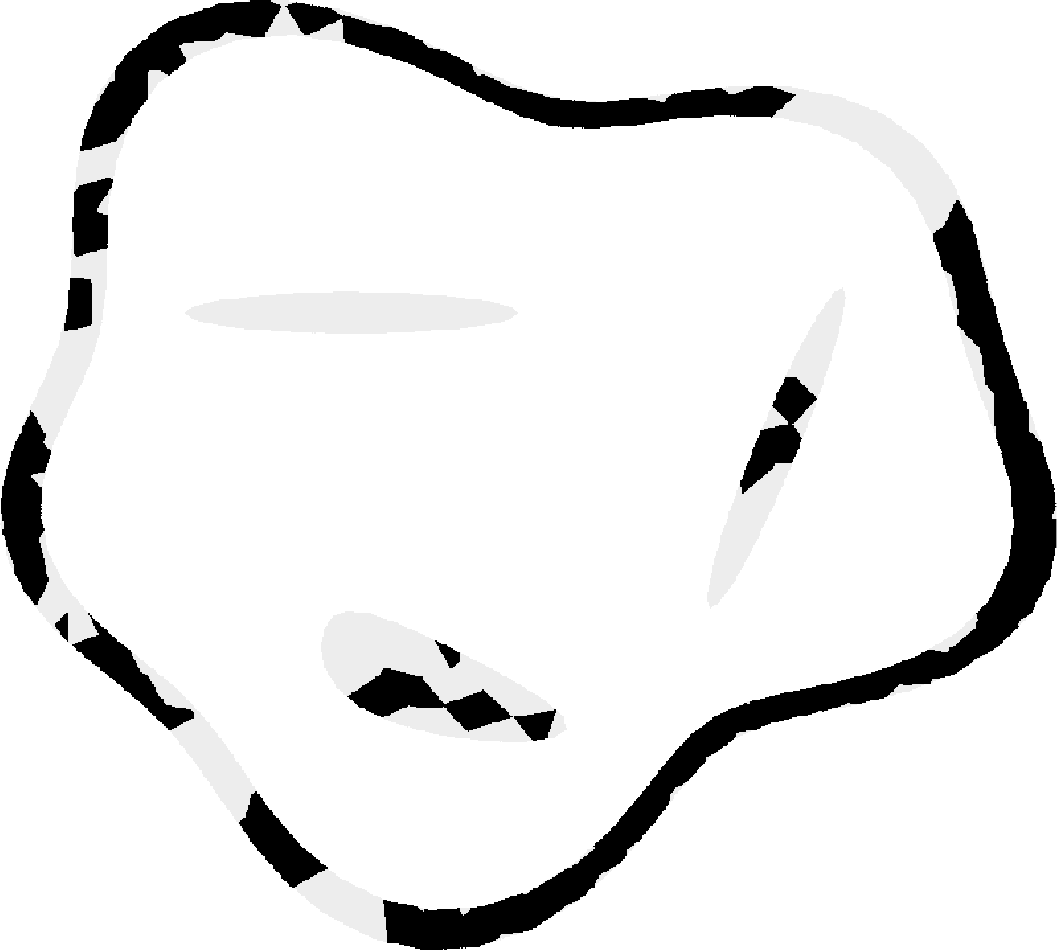} 
               \end{minipage}
    \begin{minipage}{1.8cm} \centering 
         \includegraphics[width=1.8cm]{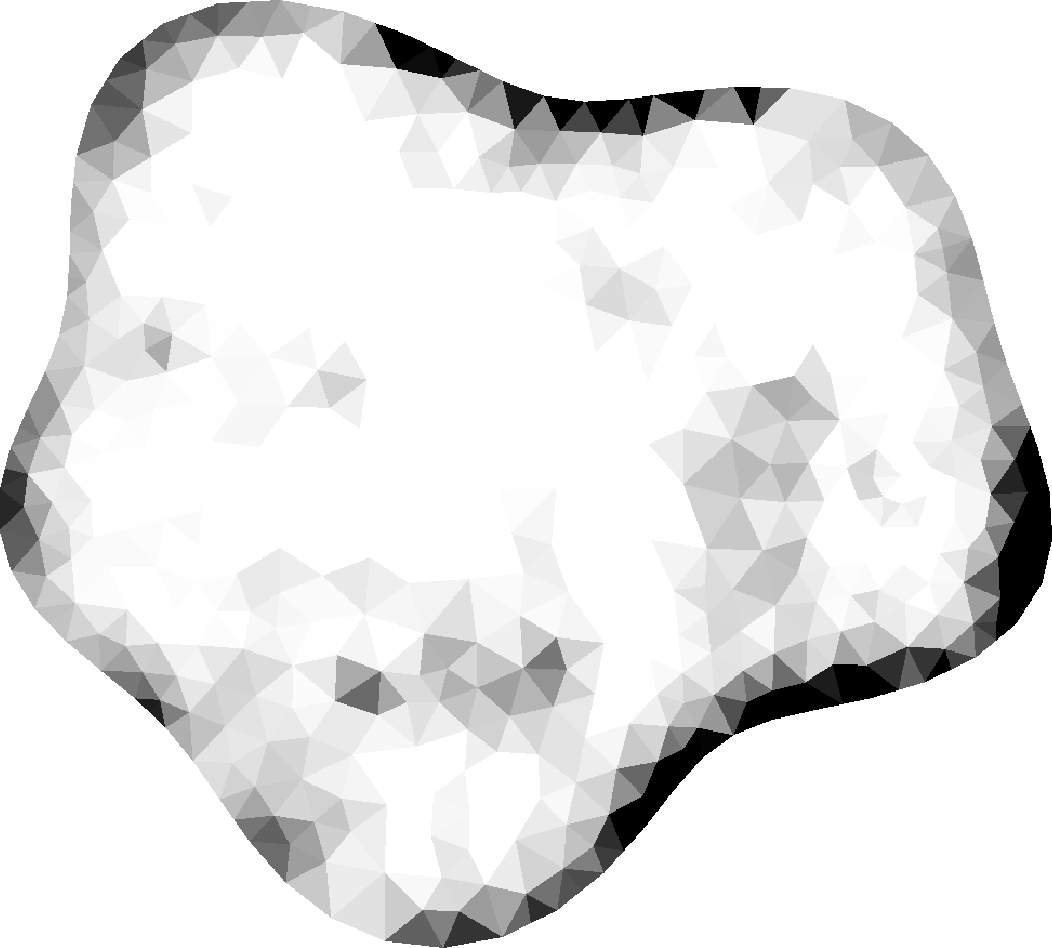} \\
  \includegraphics[width=1.8cm]{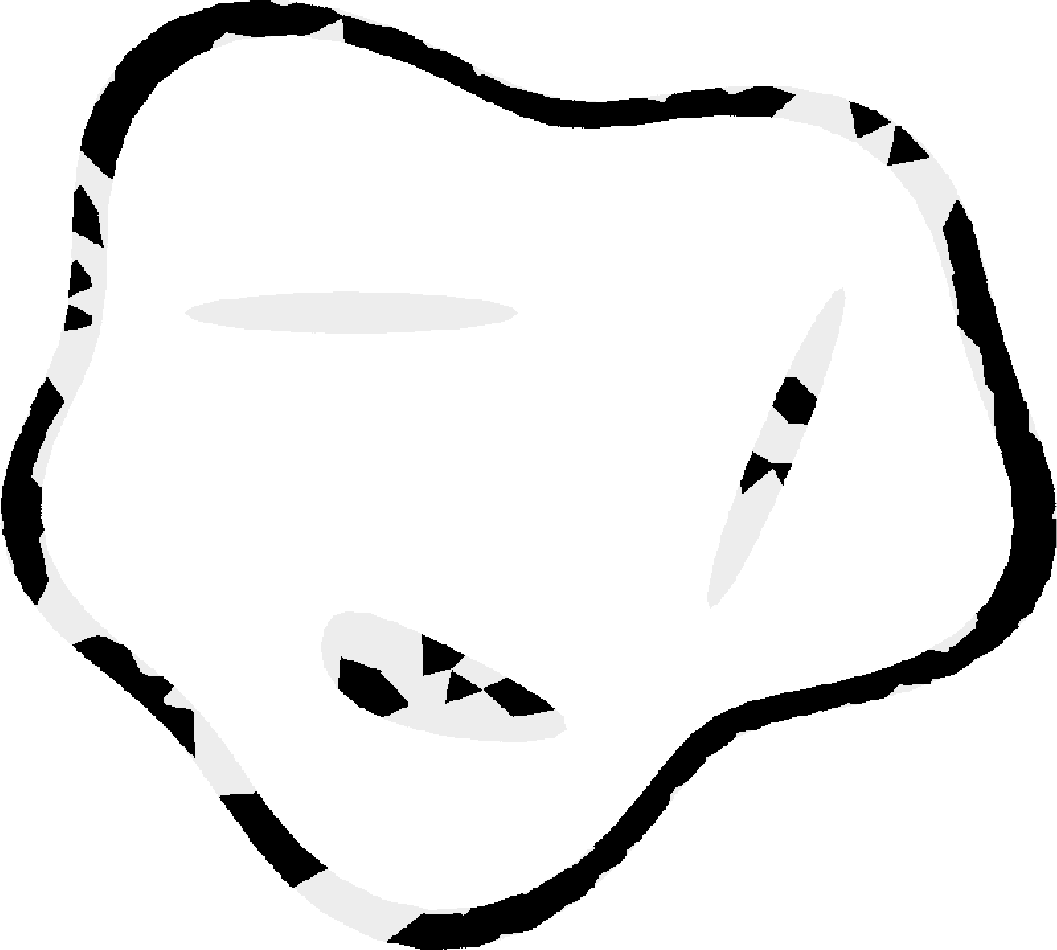}    \end{minipage} \\  
\rotatebox{90}{\hskip-0.5cm sparse}     \begin{minipage}{1.8cm} \centering        
         \includegraphics[width=1.8cm]{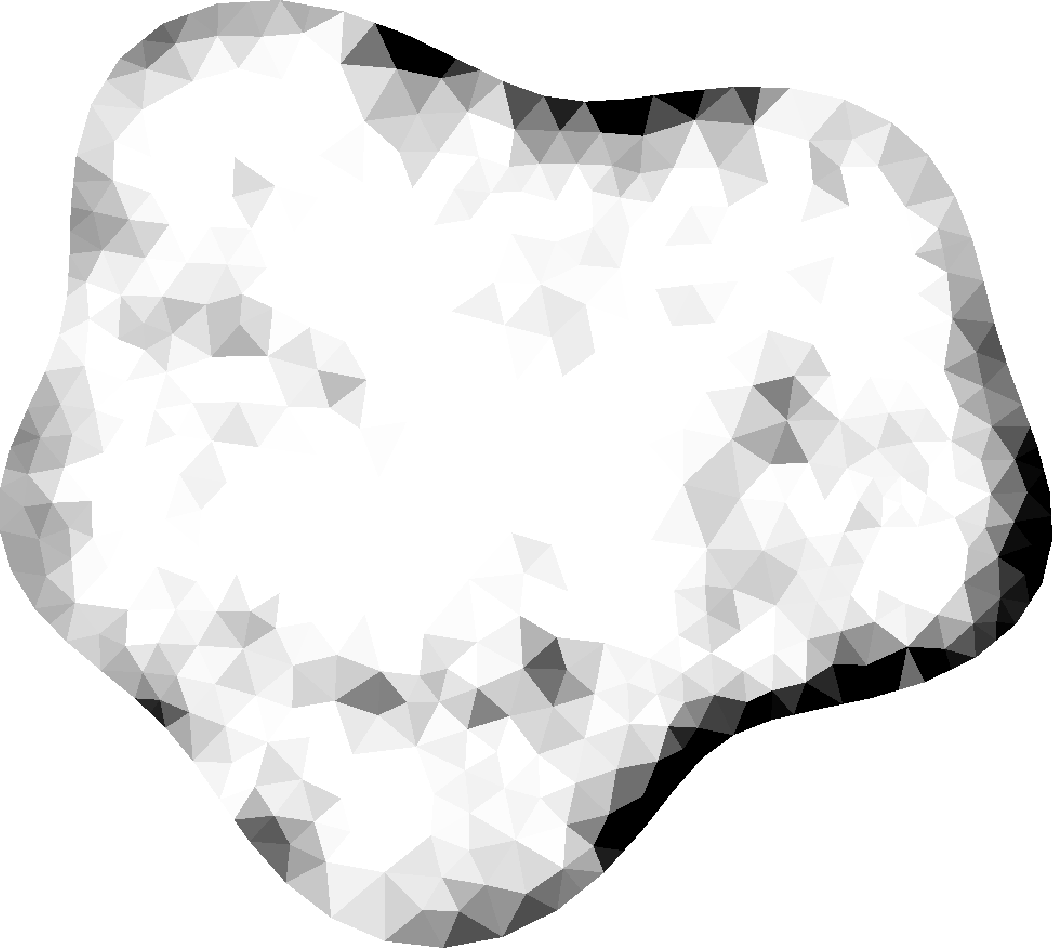} \\
  \includegraphics[width=1.8cm]{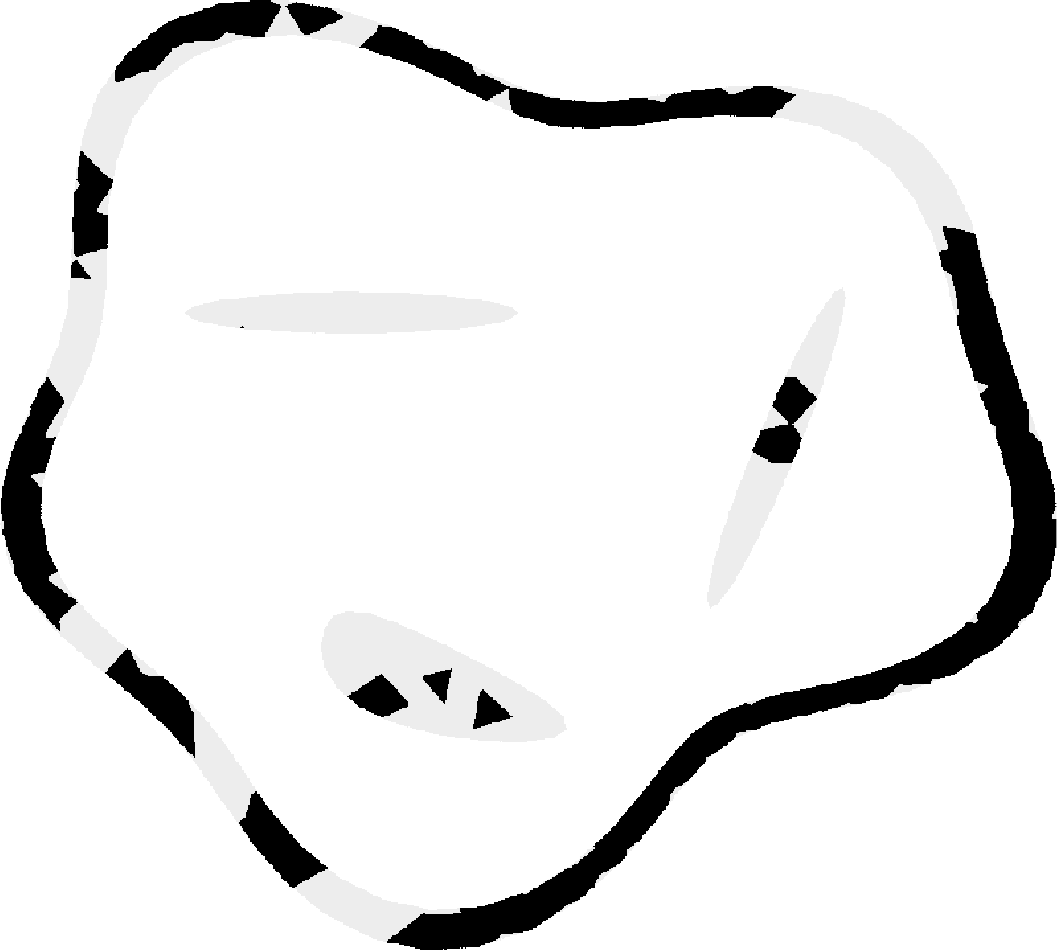} 
             \end{minipage}
    \begin{minipage}{1.8cm} \centering 
        \includegraphics[width=1.8cm]{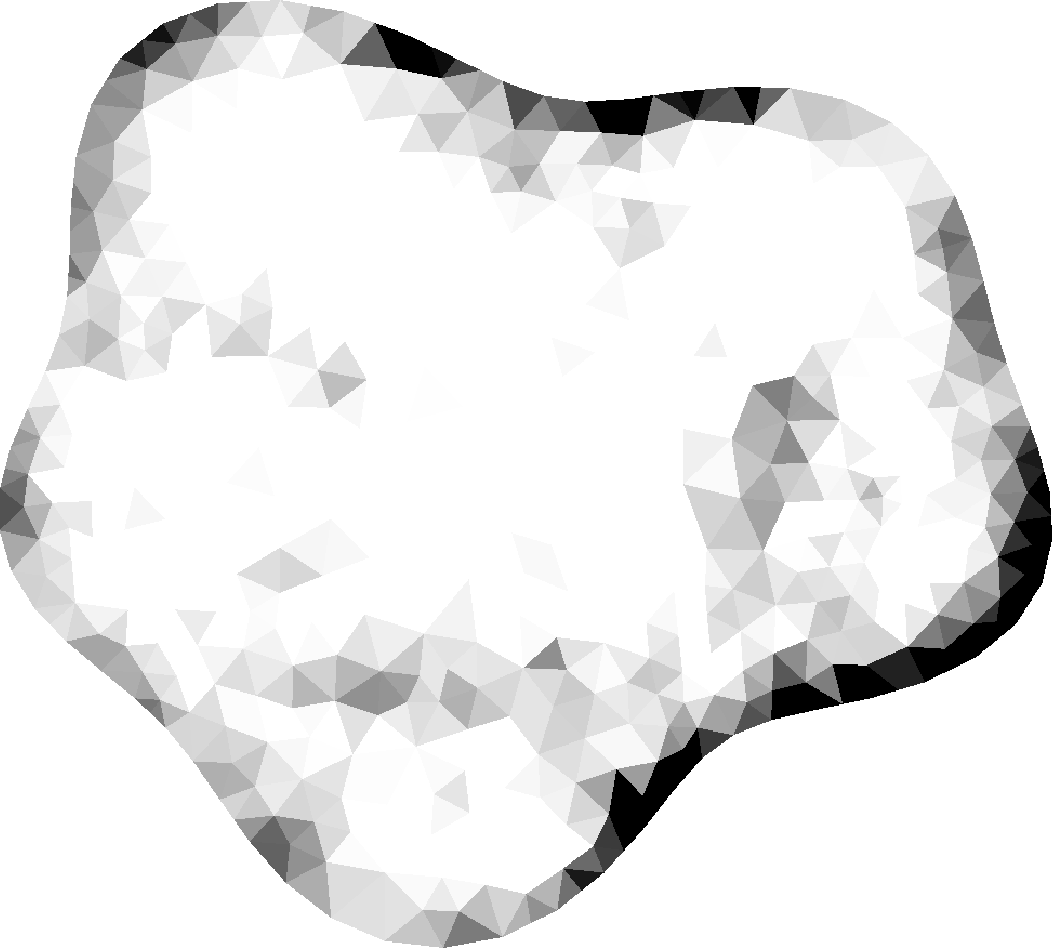} \\
  \includegraphics[width=1.8cm]{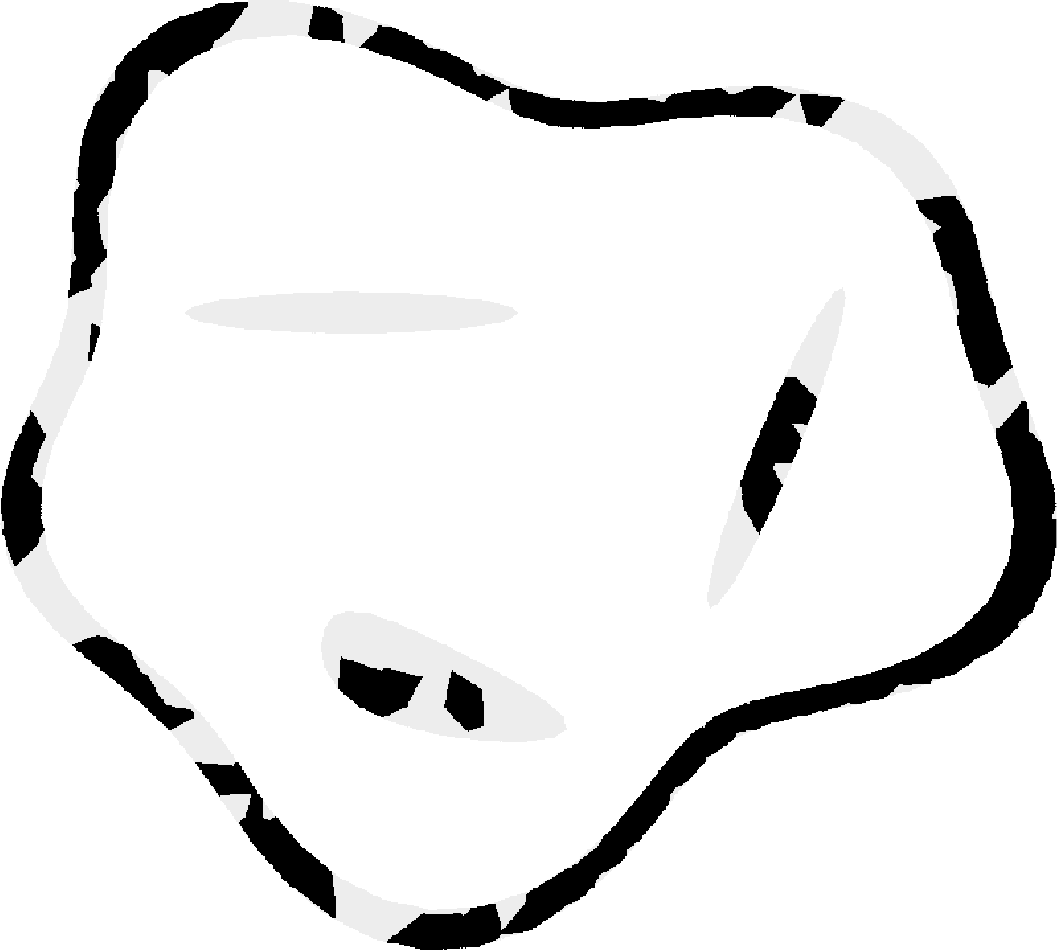} 
            \end{minipage}
    \begin{minipage}{1.8cm} \centering 
           \includegraphics[width=1.8cm]{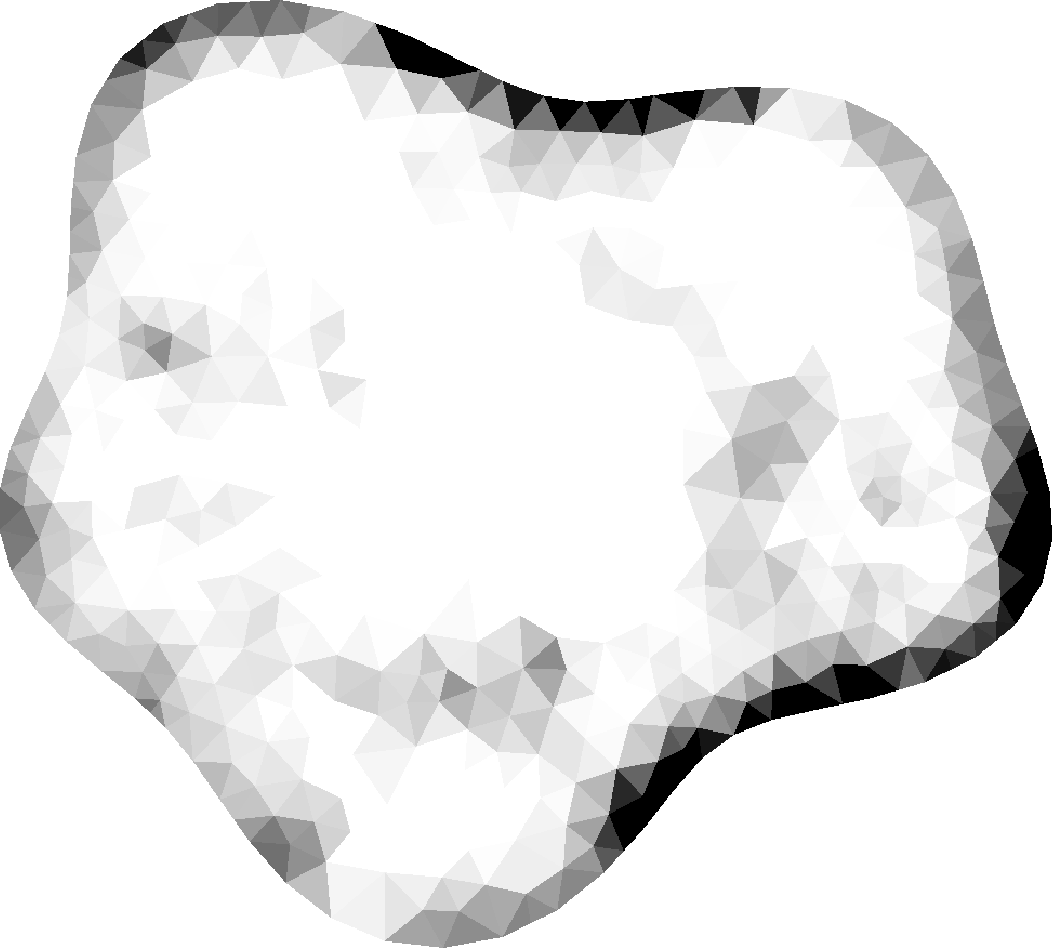}\\
  \includegraphics[width=1.8cm]{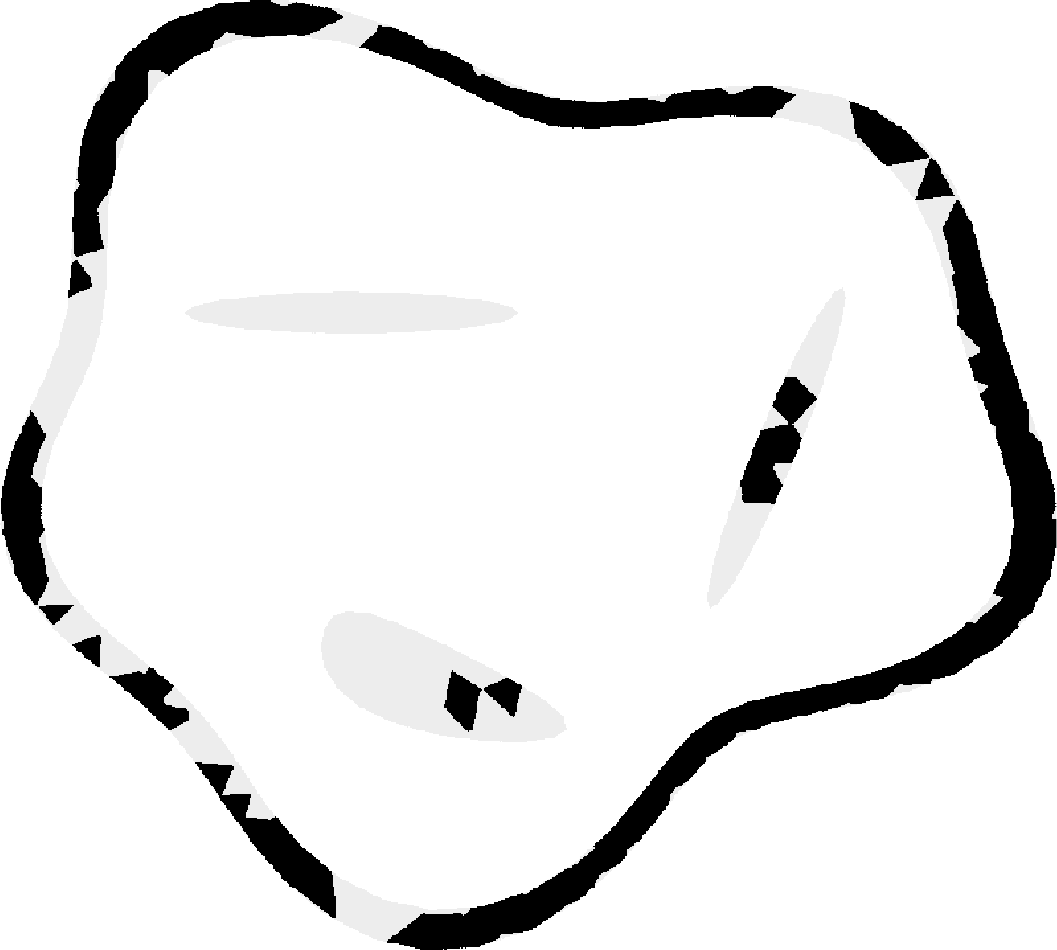} 
               \end{minipage}
    \begin{minipage}{1.8cm} \centering 
         \includegraphics[width=1.8cm]{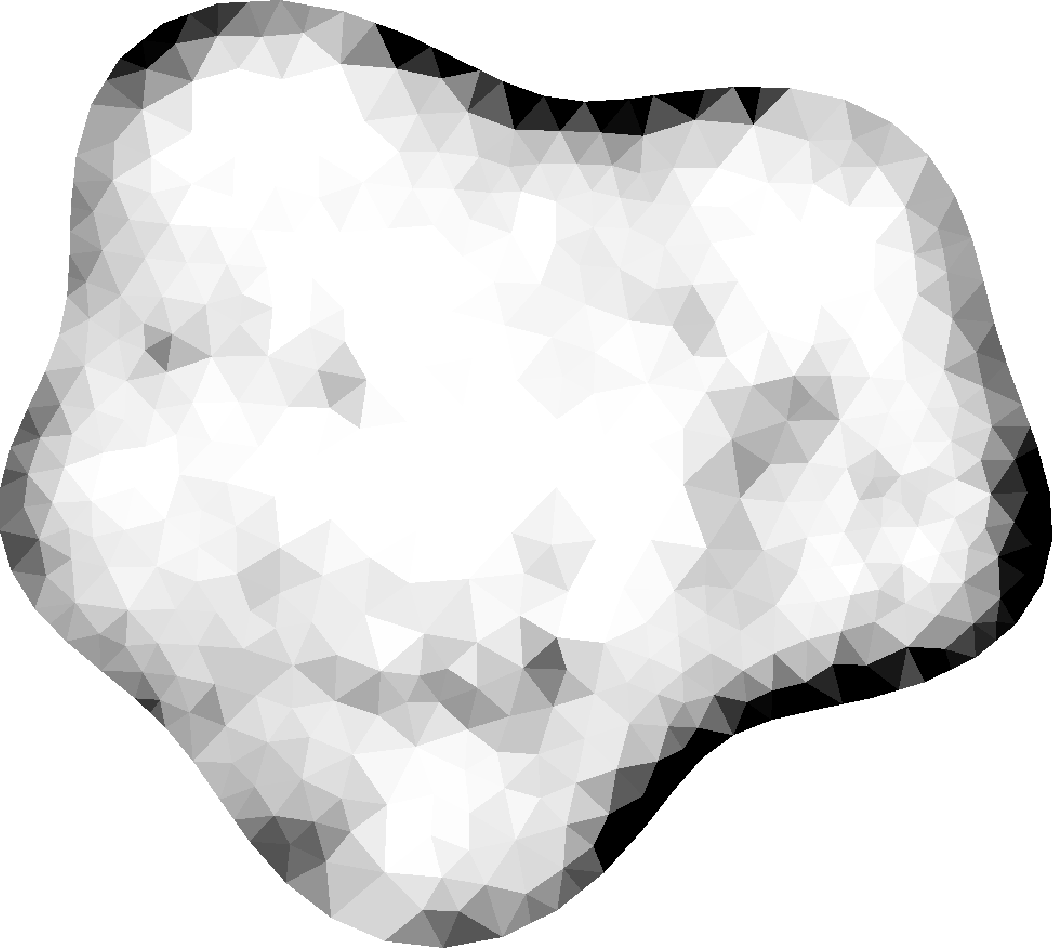}\\
  \includegraphics[width=1.8cm]{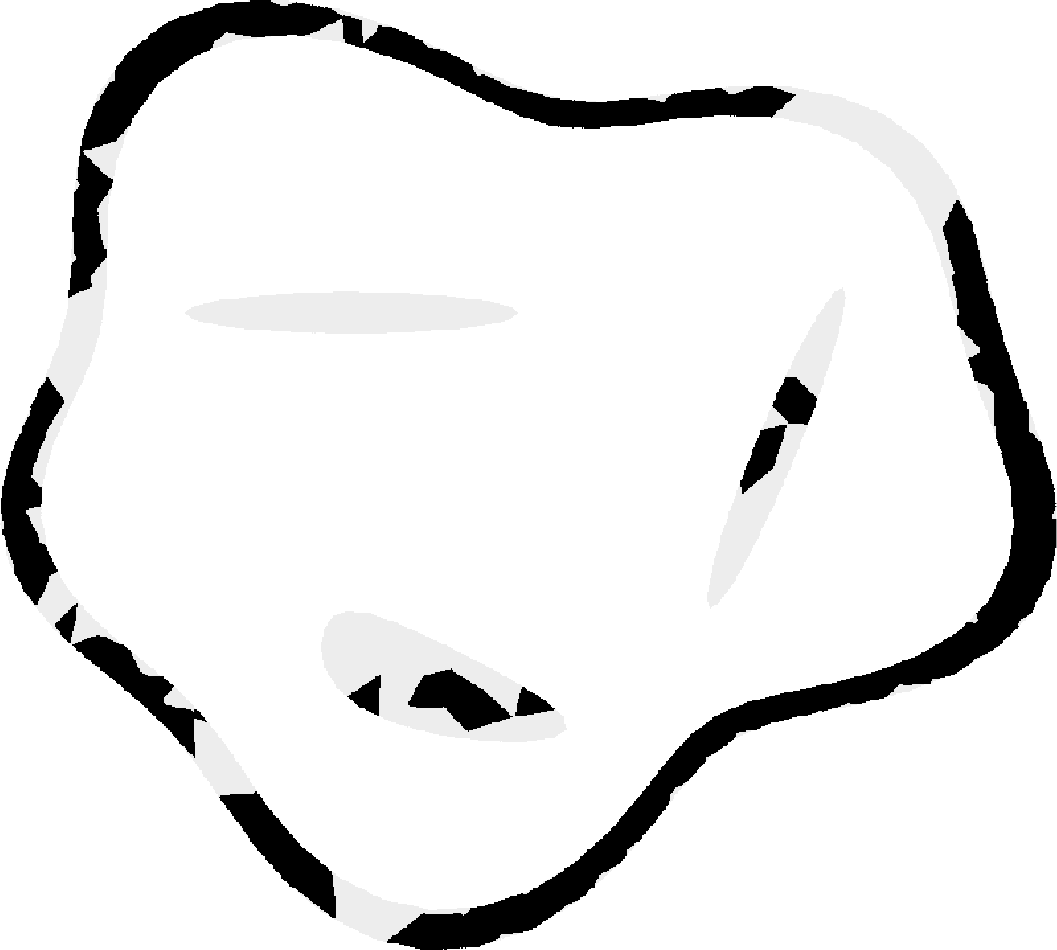} 
             \end{minipage} \\ \vskip0.2cm
              \begin{minipage}{1.8cm} \centering        
        ({\bf I}): Unfiltered 
             \end{minipage}
    \begin{minipage}{1.8cm} \centering 
   ({\bf II}): TSVD 
            \end{minipage}
    \begin{minipage}{1.8cm} \centering 
   ({\bf III}): Unfiltered \&  randomised
               \end{minipage}
    \begin{minipage}{1.8cm} \centering 
   ({\bf IV}): TSVD   \&  randomised
             \end{minipage} \\
             \end{minipage}
             \begin{minipage}{0.6cm} \centering
             \includegraphics[height= 5cm]{images/color_bar_lin_1.png}
             \end{minipage}
              \\ \vskip0.2cm 
             \end{scriptsize}
    \caption{As in Figure \ref{fig:B_low_freq} obtained using the 60 MHz signal pulse ({\bf B}). The top row shows the results for the dense configuration with 128 points  and the bottom row for the sparse one with 64 points. The results for  configurations ({\bf I})--({\bf IV}) are shown in columns from left to right, respectively.}
    \label{fig:B_high_freq}
\end{figure}

\begin{figure}[!ht]
    \centering  
    \begin{scriptsize}  SSIM \\\vskip0.1cm 
    \begin{minipage}{2.05cm} \centering
    \includegraphics[width=2.05cm]{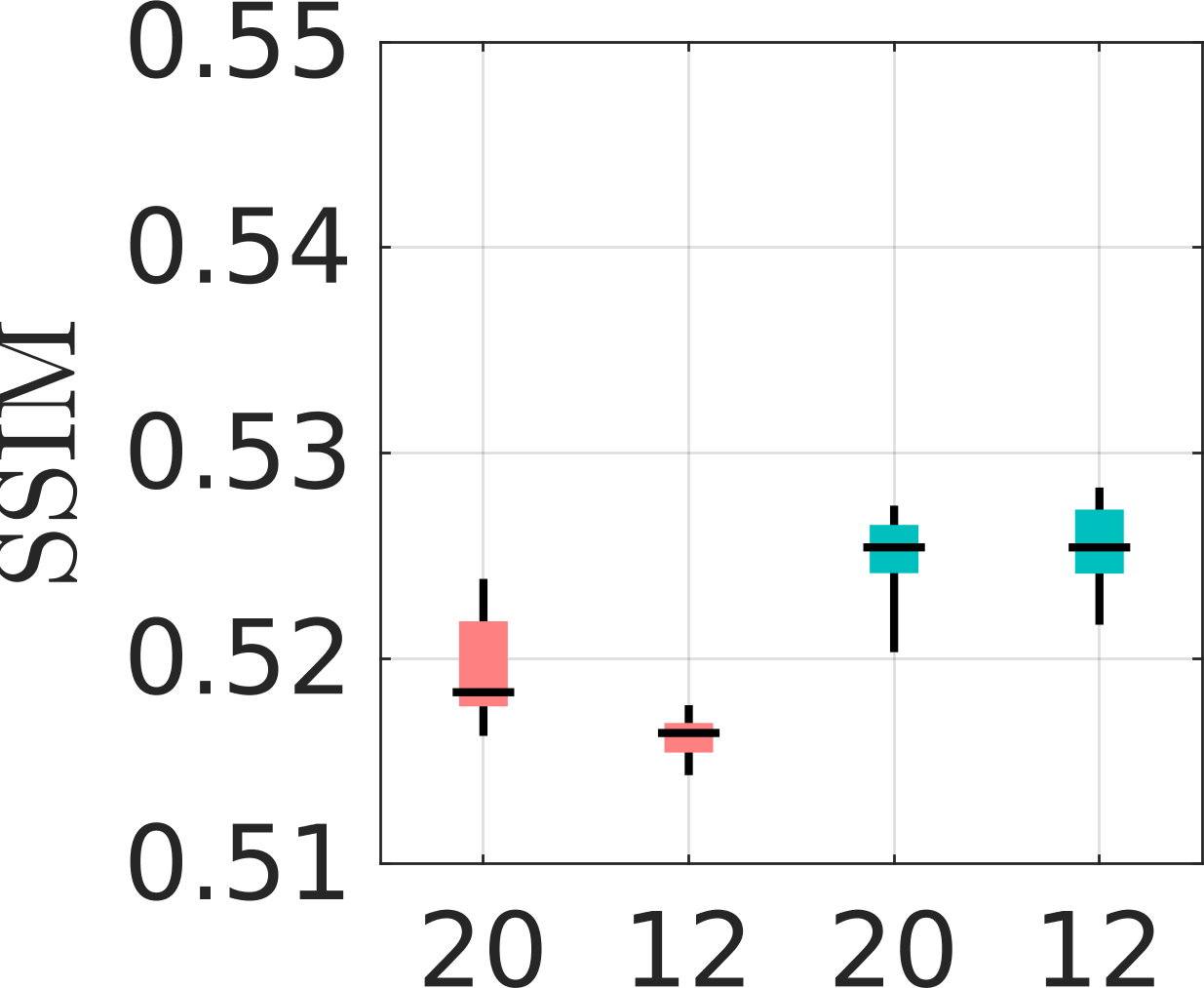} \end{minipage} \begin{minipage}{1.9cm}  \centering
        \includegraphics[width=1.8cm]{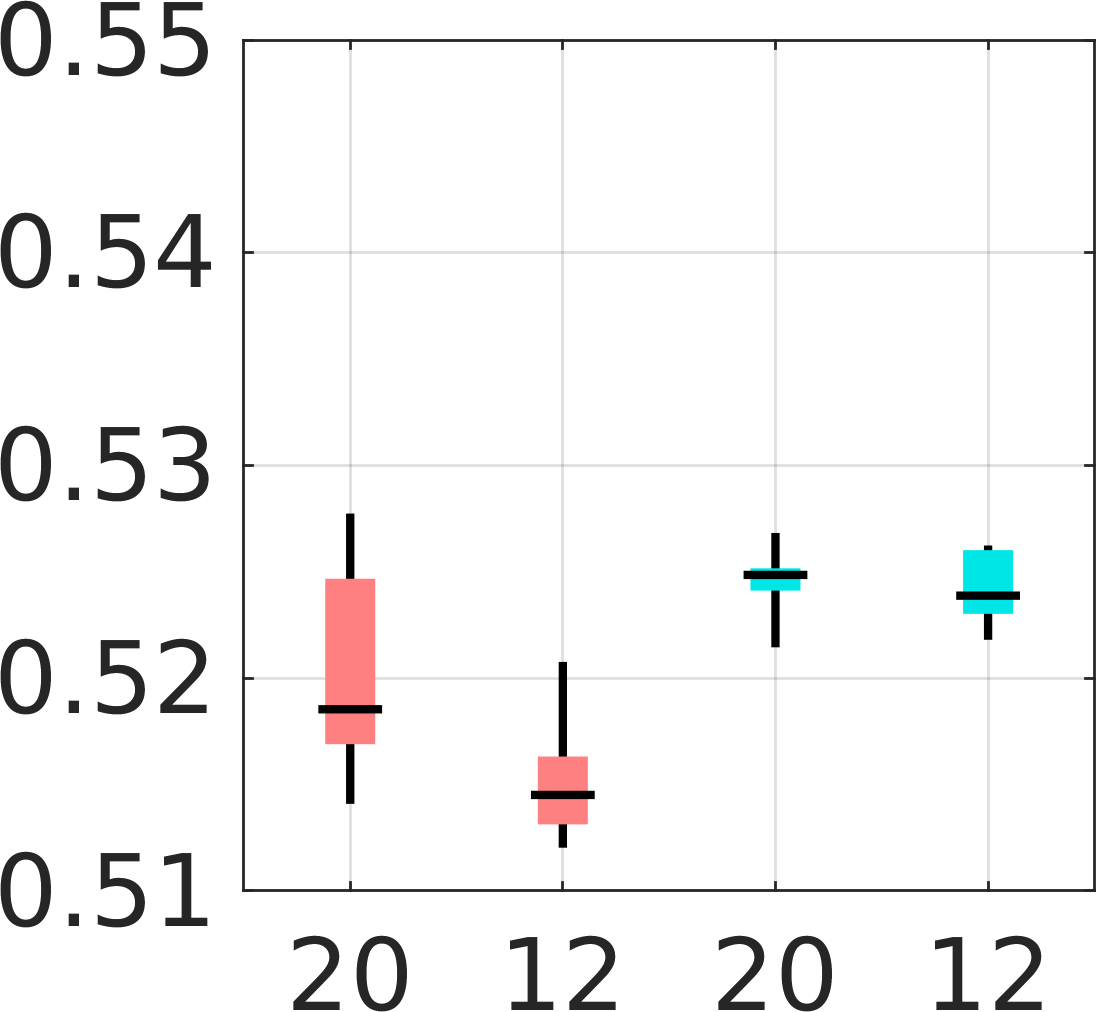} \end{minipage} \begin{minipage}{1.9cm}  \centering
           \includegraphics[width=1.8cm]{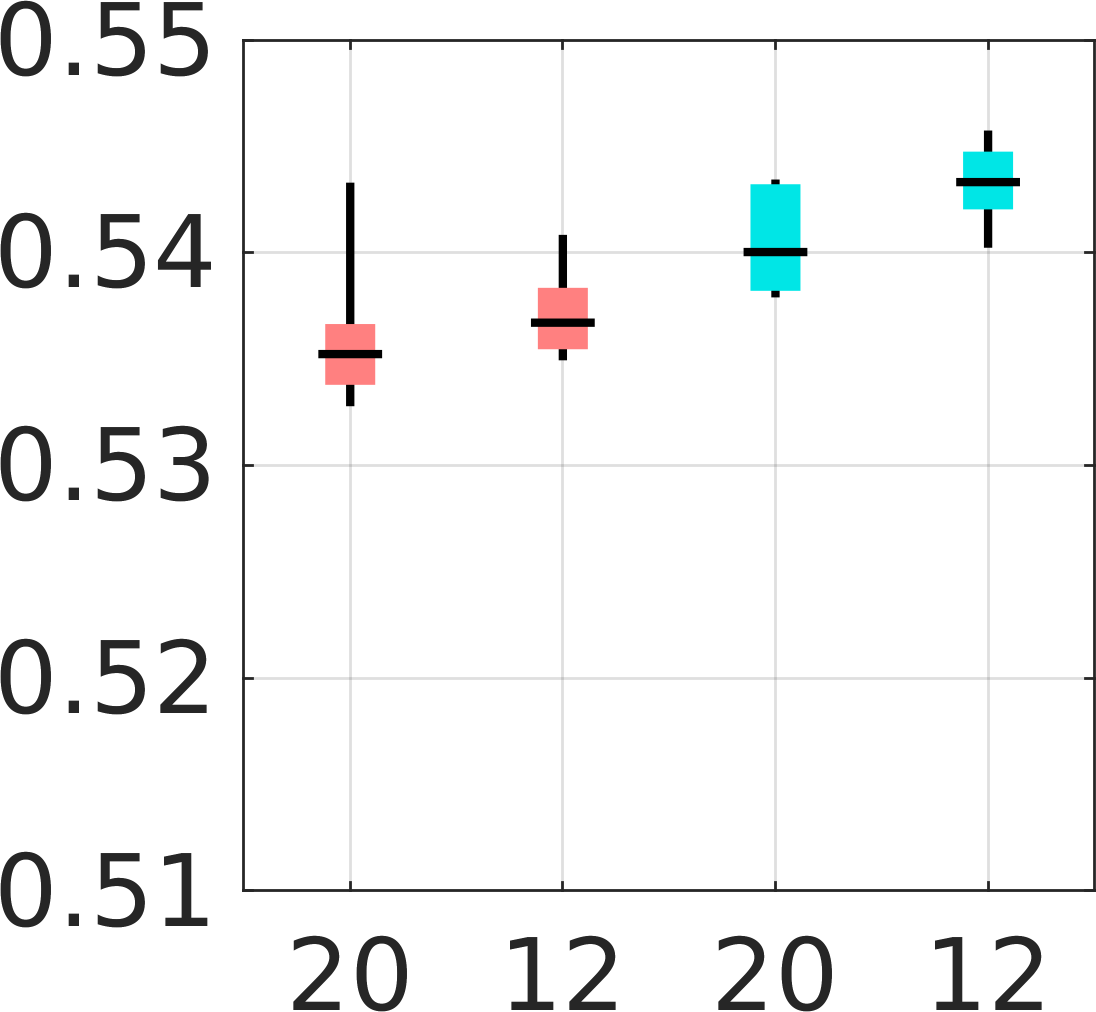} \end{minipage} \begin{minipage}{1.9cm} \centering
         \includegraphics[width=1.8cm]{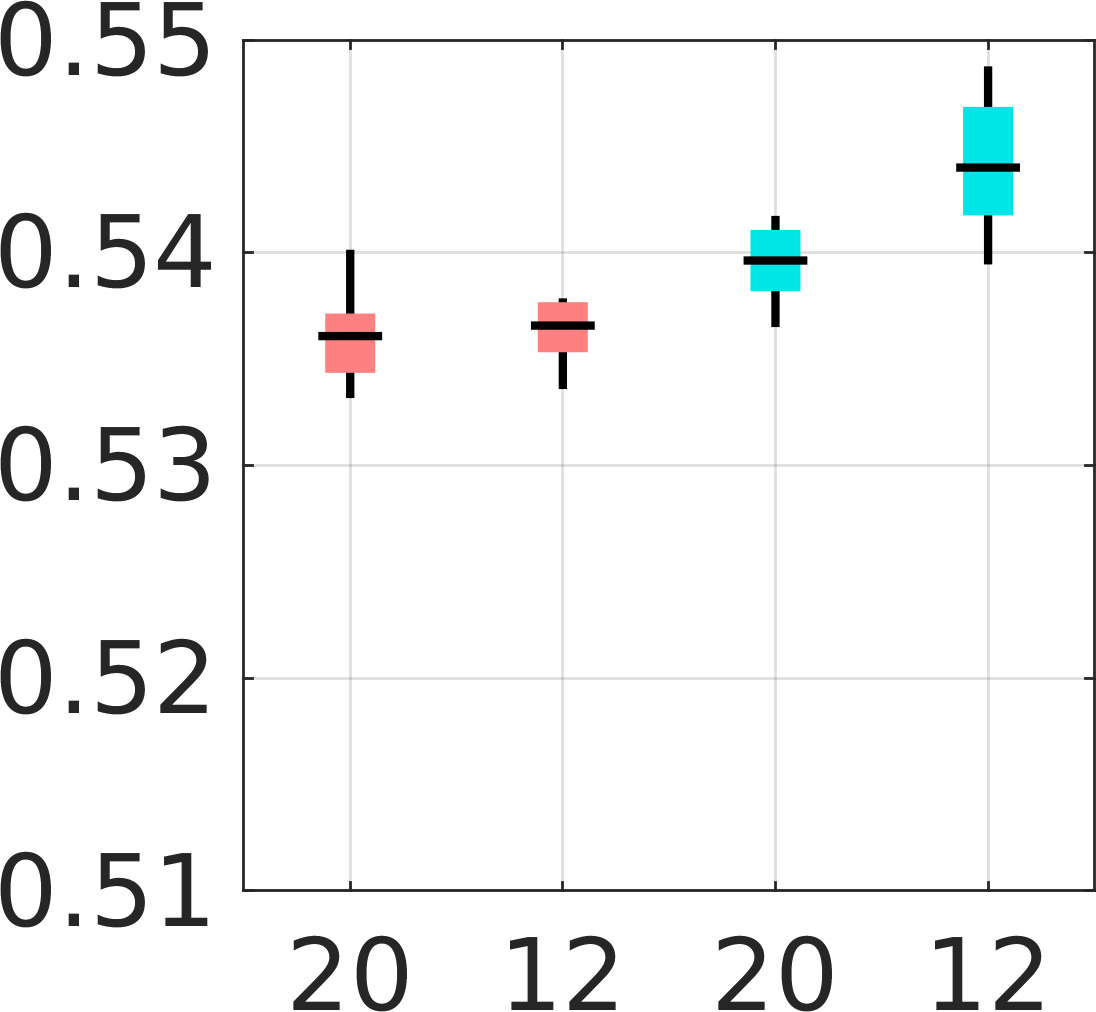} \end{minipage}  \\ \vskip0.2cm Surface RMSE \\ \vskip0.1cm 
       \begin{minipage}{2.0cm} \centering
           \includegraphics[width=2.0cm]{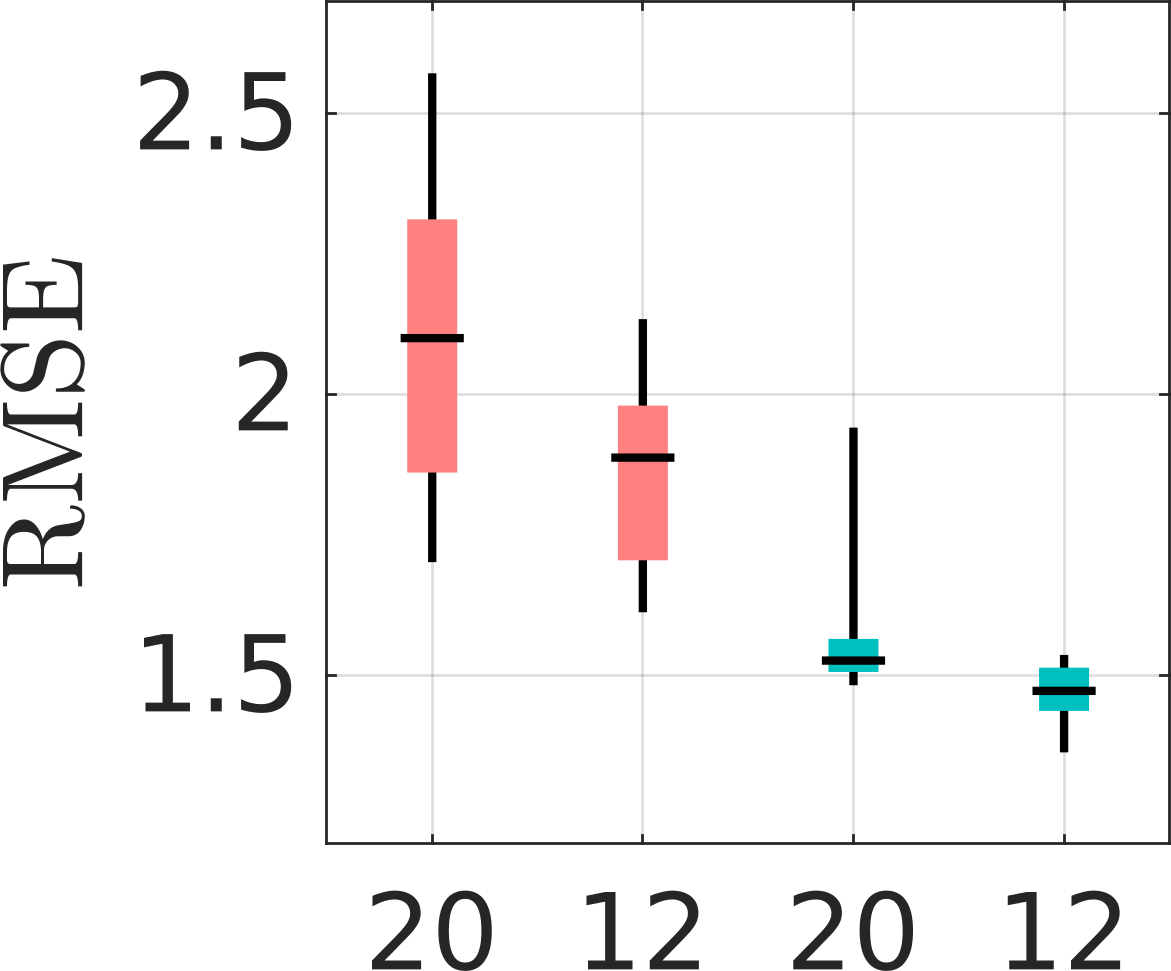} \end{minipage} \begin{minipage}{1.9cm} \centering
        \includegraphics[width=1.8cm]{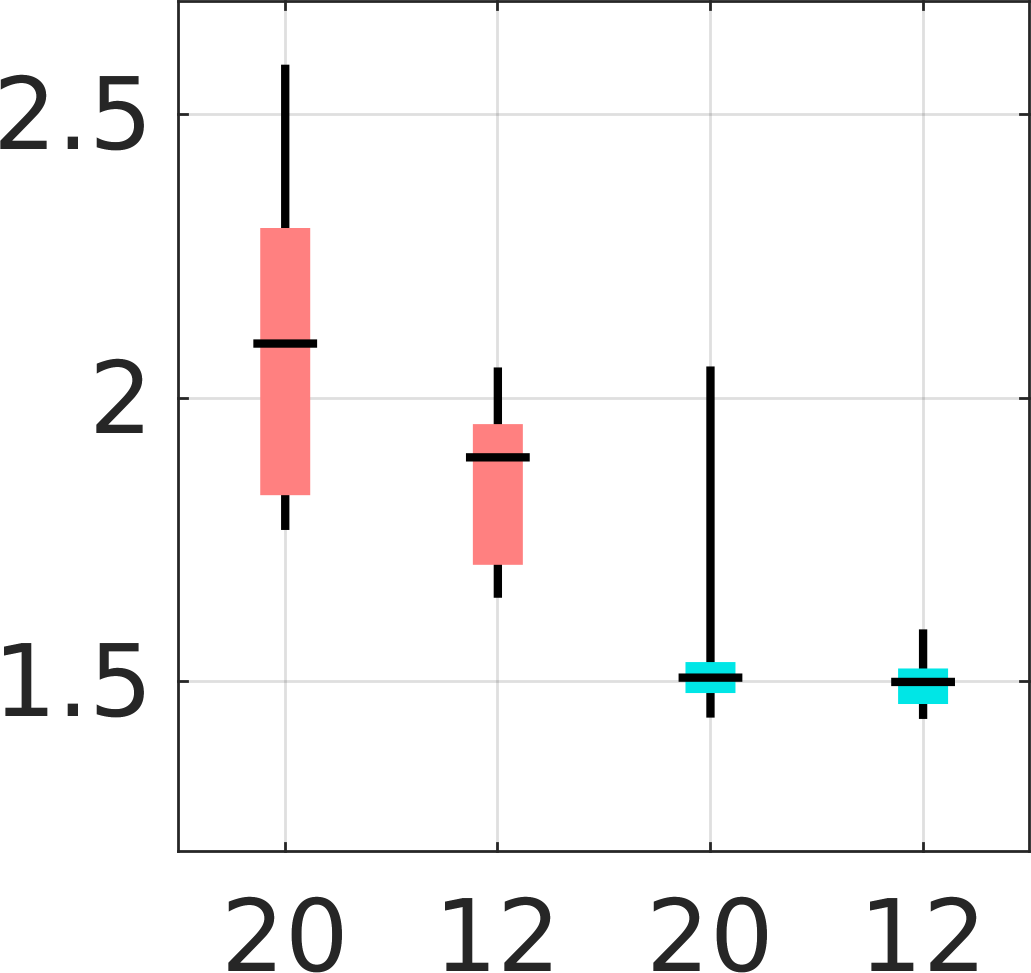} \end{minipage} \begin{minipage}{1.9cm}  \centering
           \includegraphics[width=1.8cm]{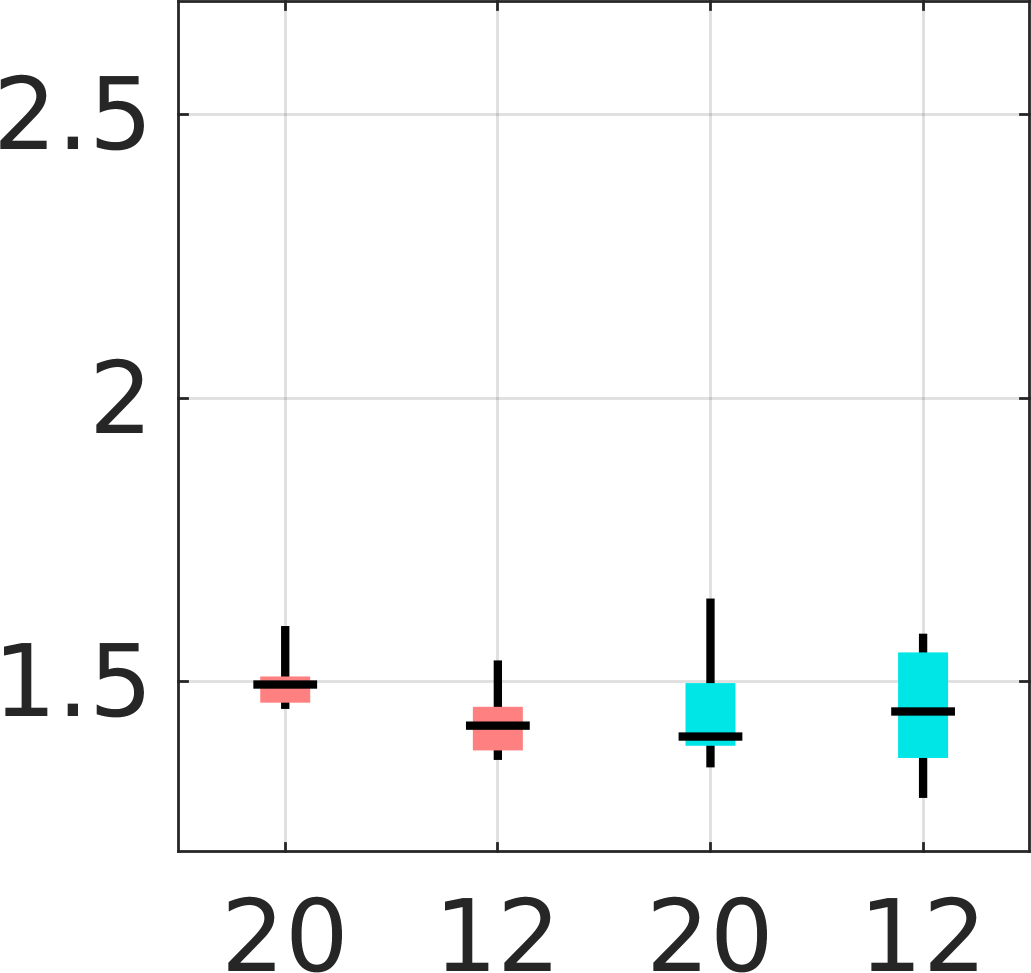} \end{minipage} \begin{minipage}{1.9cm}  \centering
         \includegraphics[width=1.8cm]{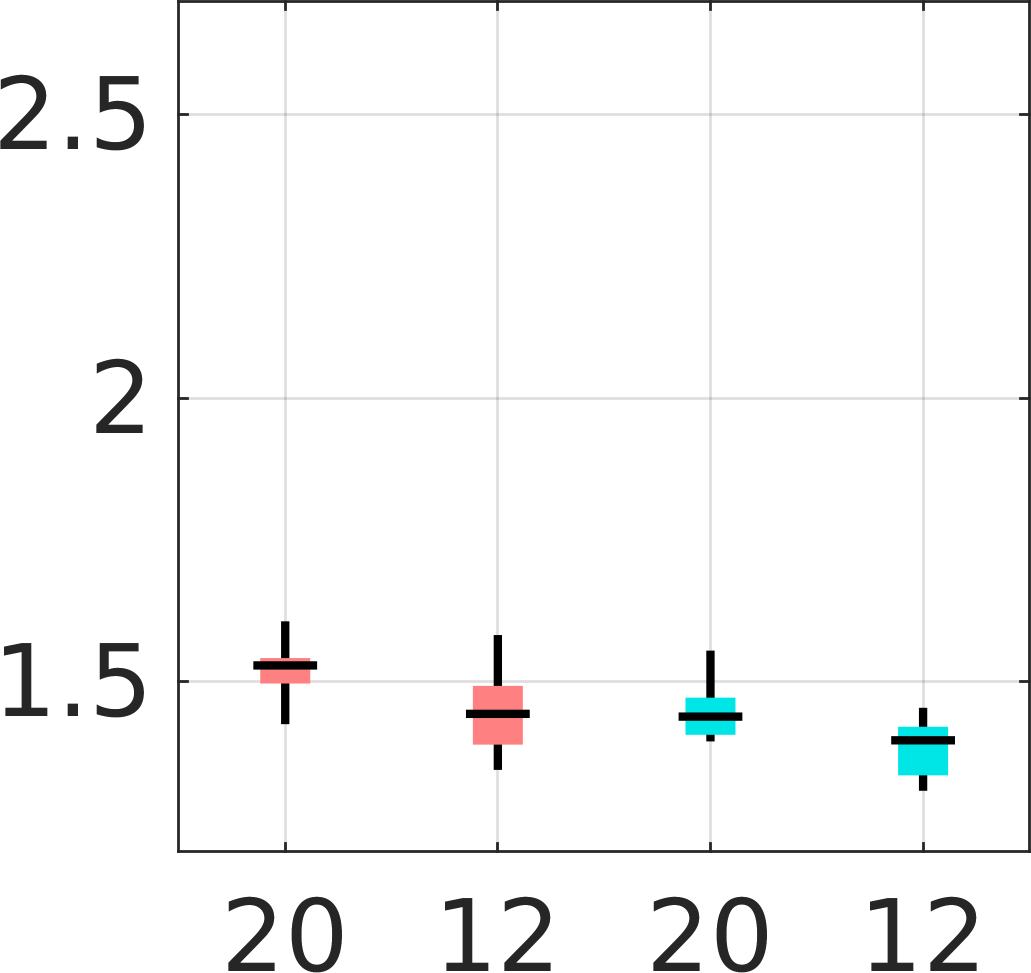} \end{minipage}  \\ \vskip0.2cm   Void RMSE  \\ \vskip0.1cm 
        \begin{minipage}{2.0cm}  \centering
           \includegraphics[width=2.0cm]{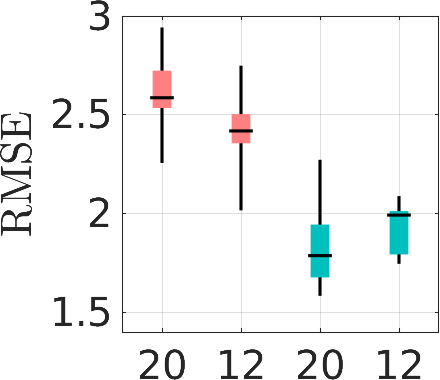} \end{minipage} \begin{minipage}{1.9cm}  \centering
        \includegraphics[width=1.8cm]{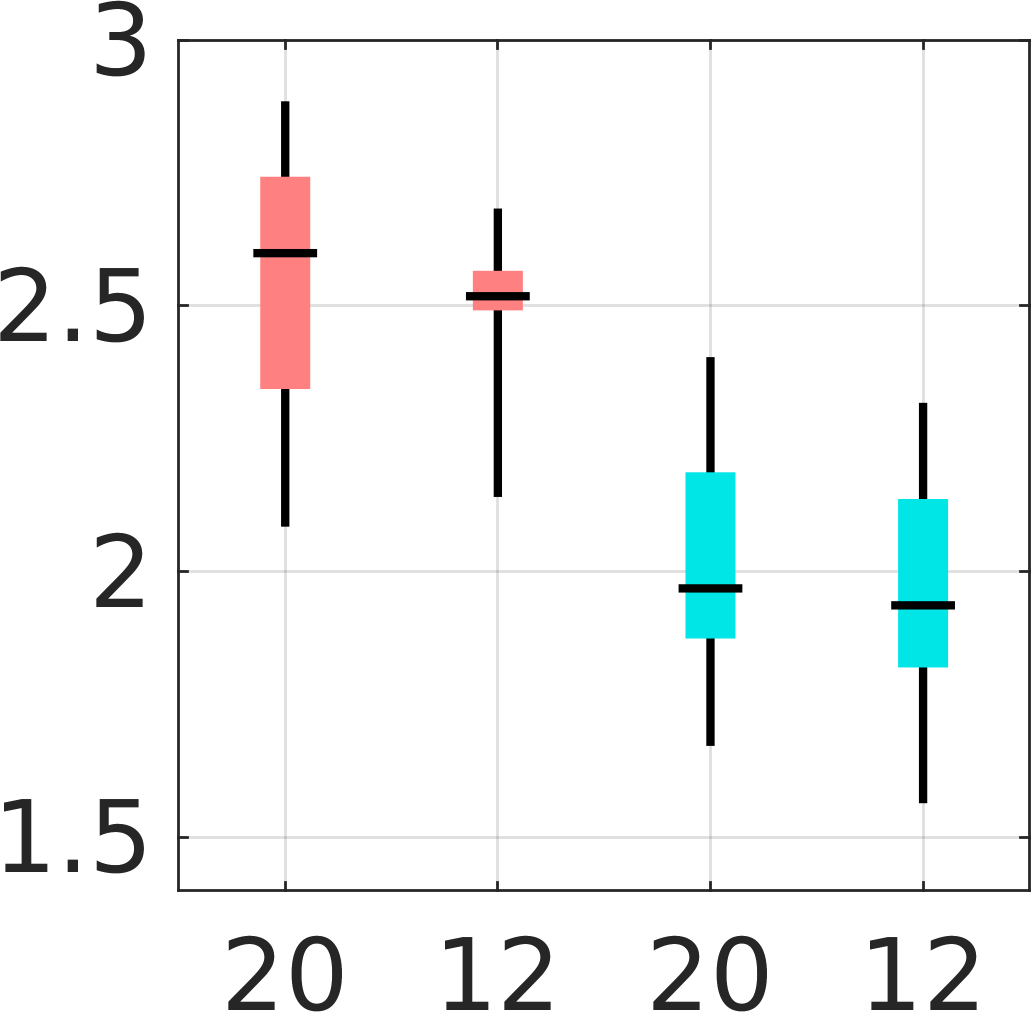} \end{minipage} \begin{minipage}{1.9cm}  \centering
           \includegraphics[width=1.8cm]{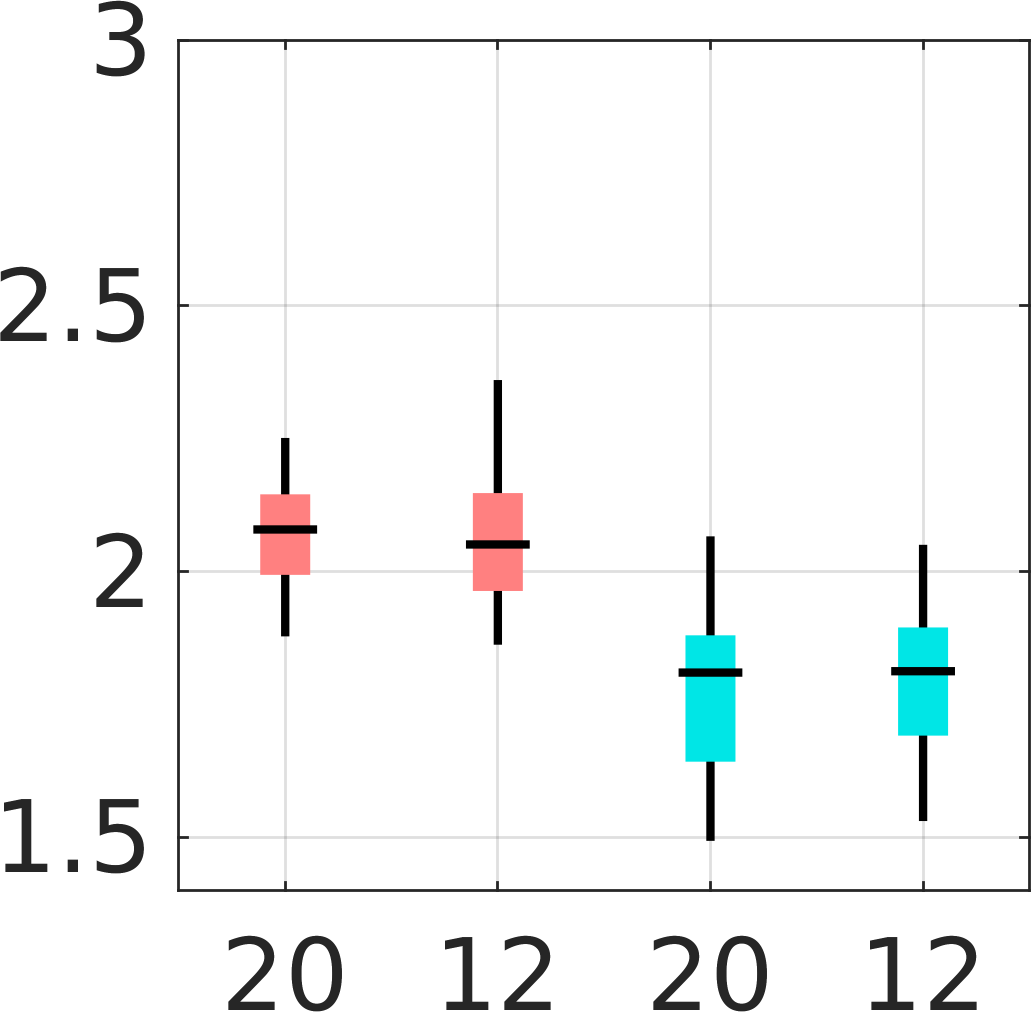} \end{minipage} \begin{minipage}{1.9cm}  \centering
         \includegraphics[width=1.8cm]{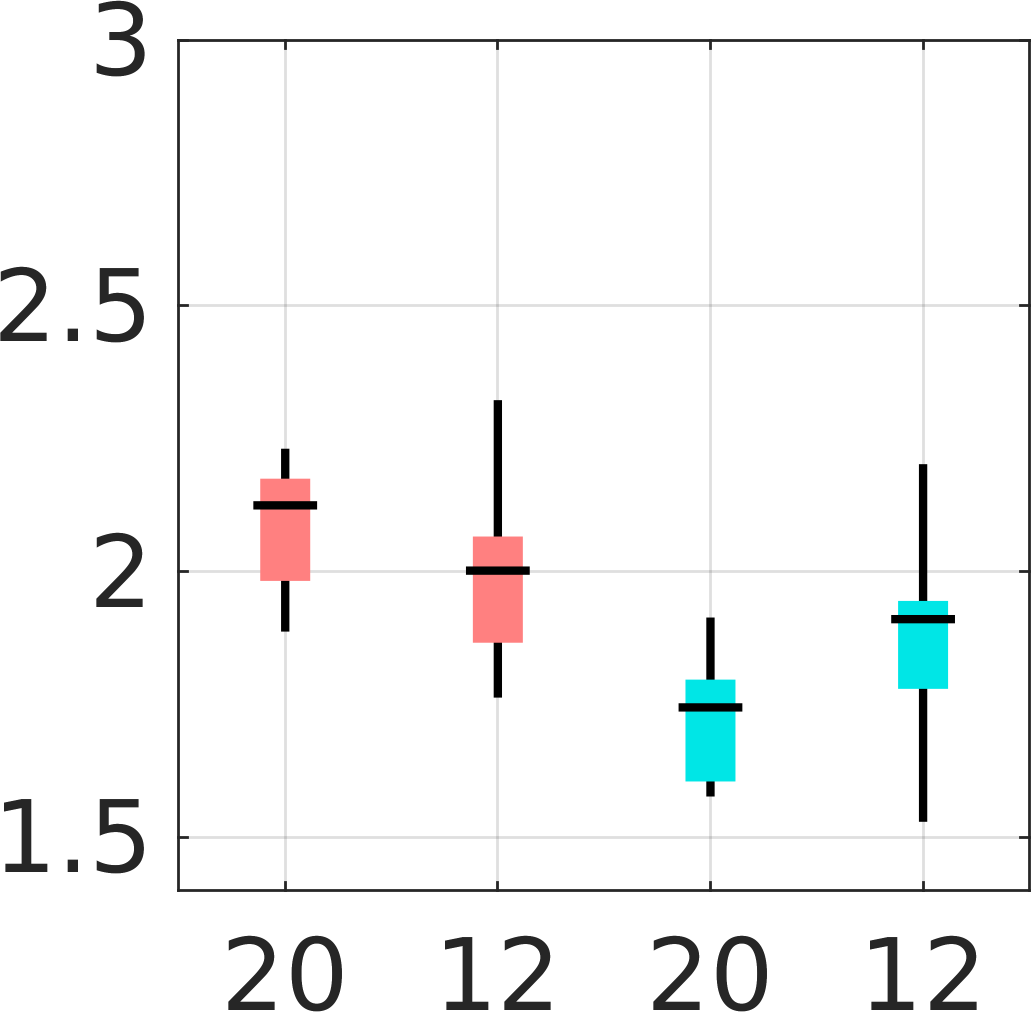}  \end{minipage} 
         \\ \vskip0.2cm Surface OE  \\  \vskip0.1cm  
        \begin{minipage}{2.0cm}  \centering
    \includegraphics[width=2.0cm]{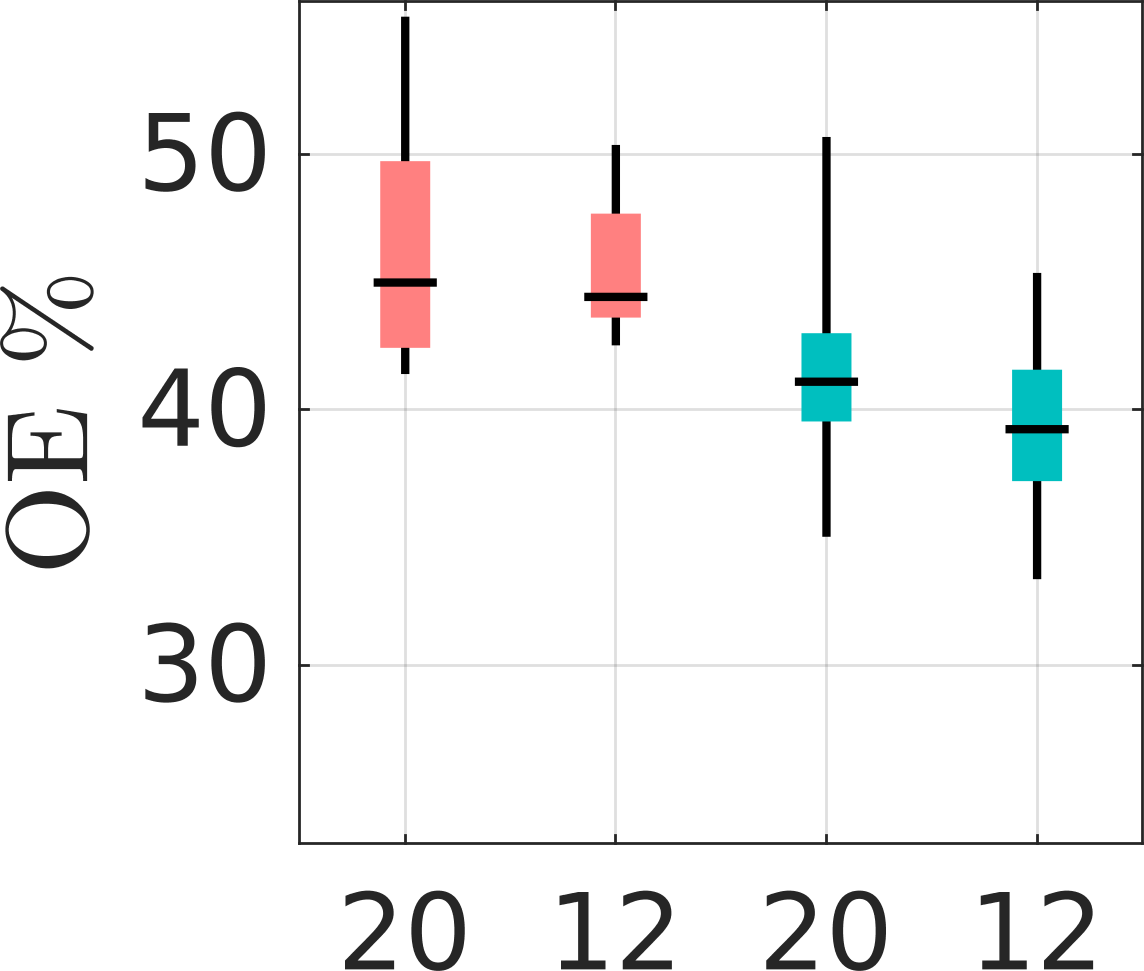} \end{minipage} \begin{minipage}{1.9cm}  \centering
        \includegraphics[width=1.8cm]{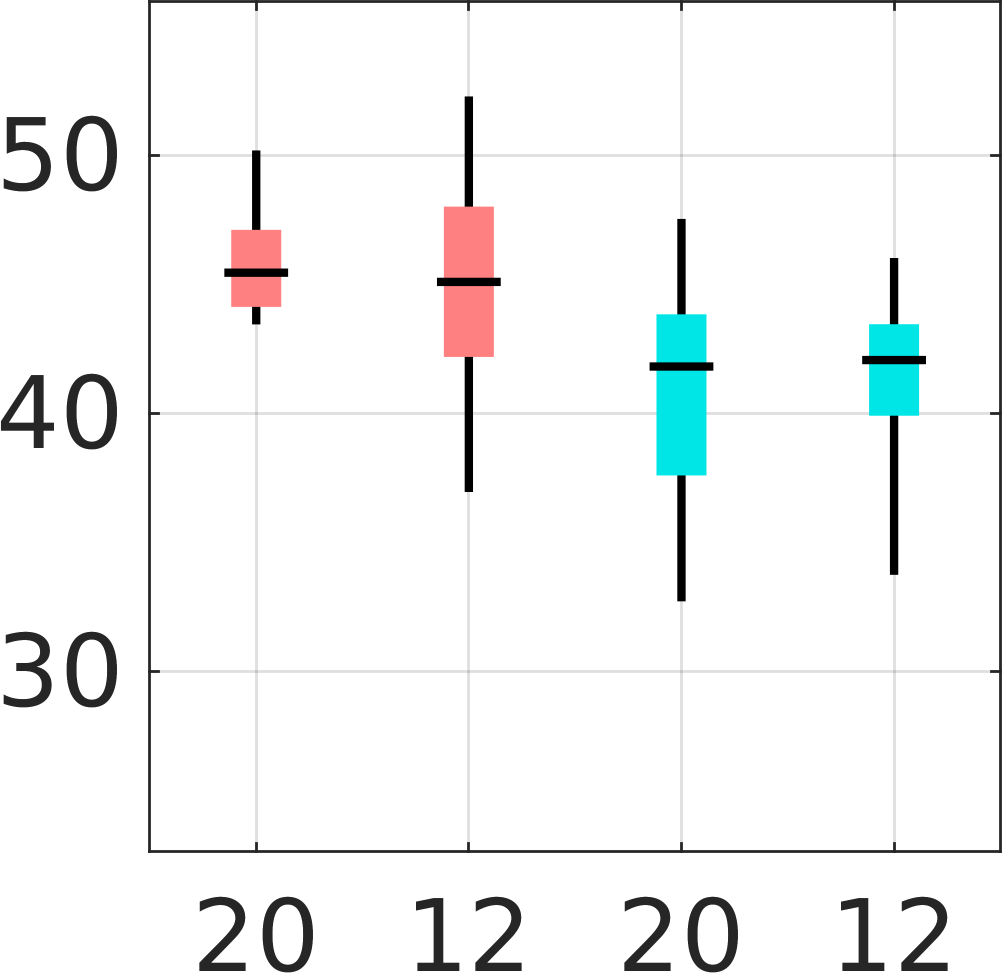} \end{minipage} \begin{minipage}{1.9cm}  \centering
           \includegraphics[width=1.8cm]{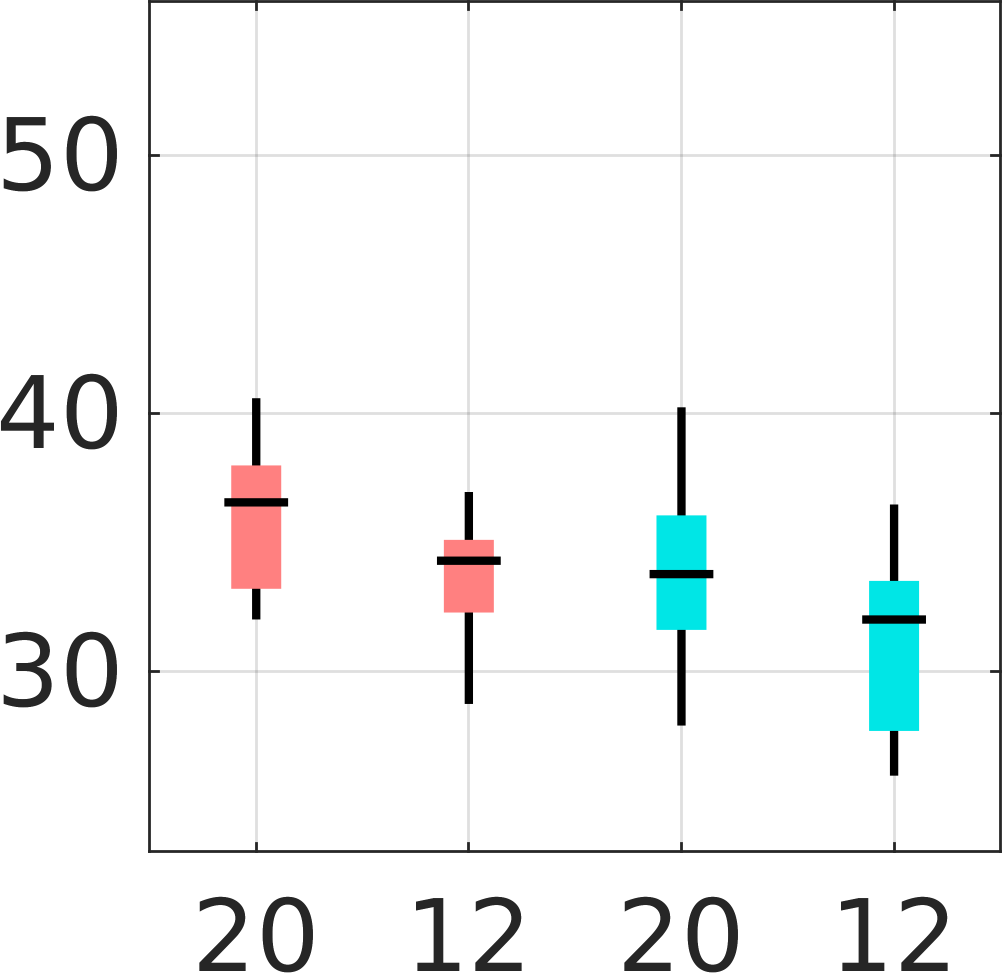} \end{minipage} \begin{minipage}{1.9cm}  \centering
         \includegraphics[width=1.8cm]{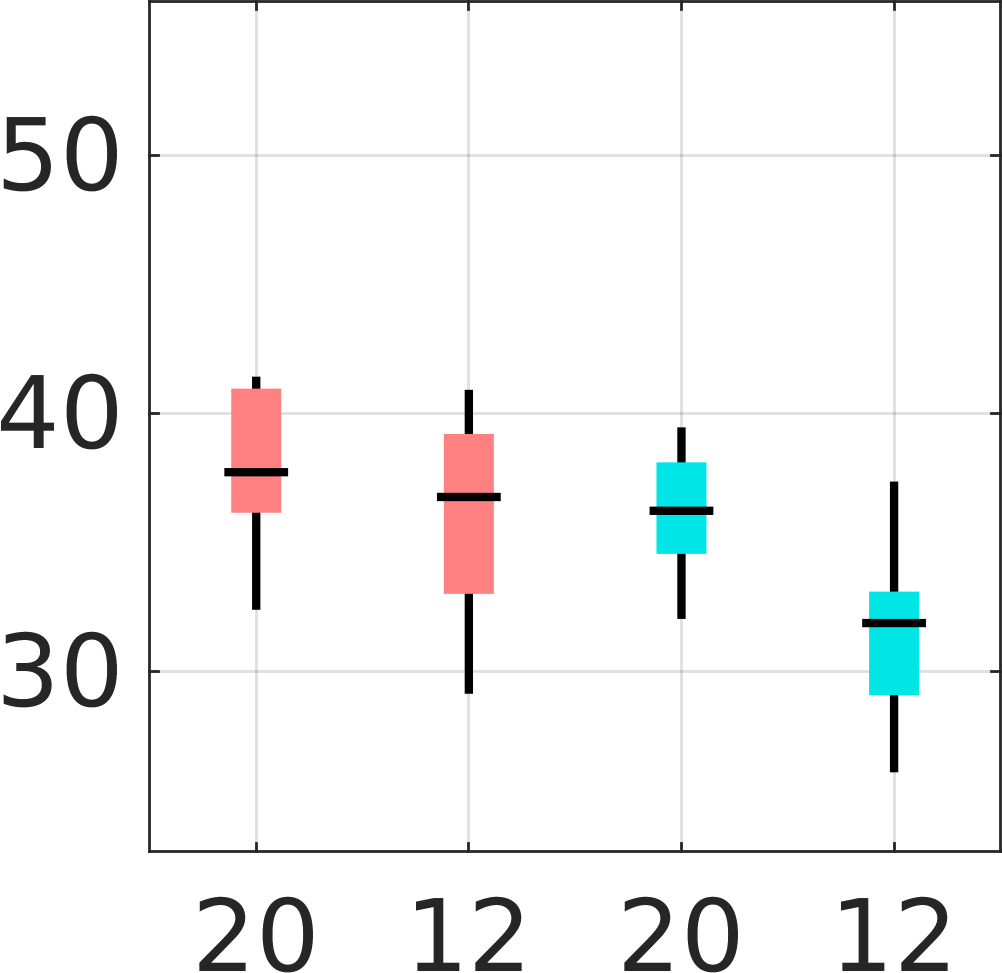} \end{minipage} \\ \vskip0.2cm Void OE \\ \vskip0.1cm 
         \begin{minipage}{2.0cm}  \centering
           \includegraphics[width=2.0cm]{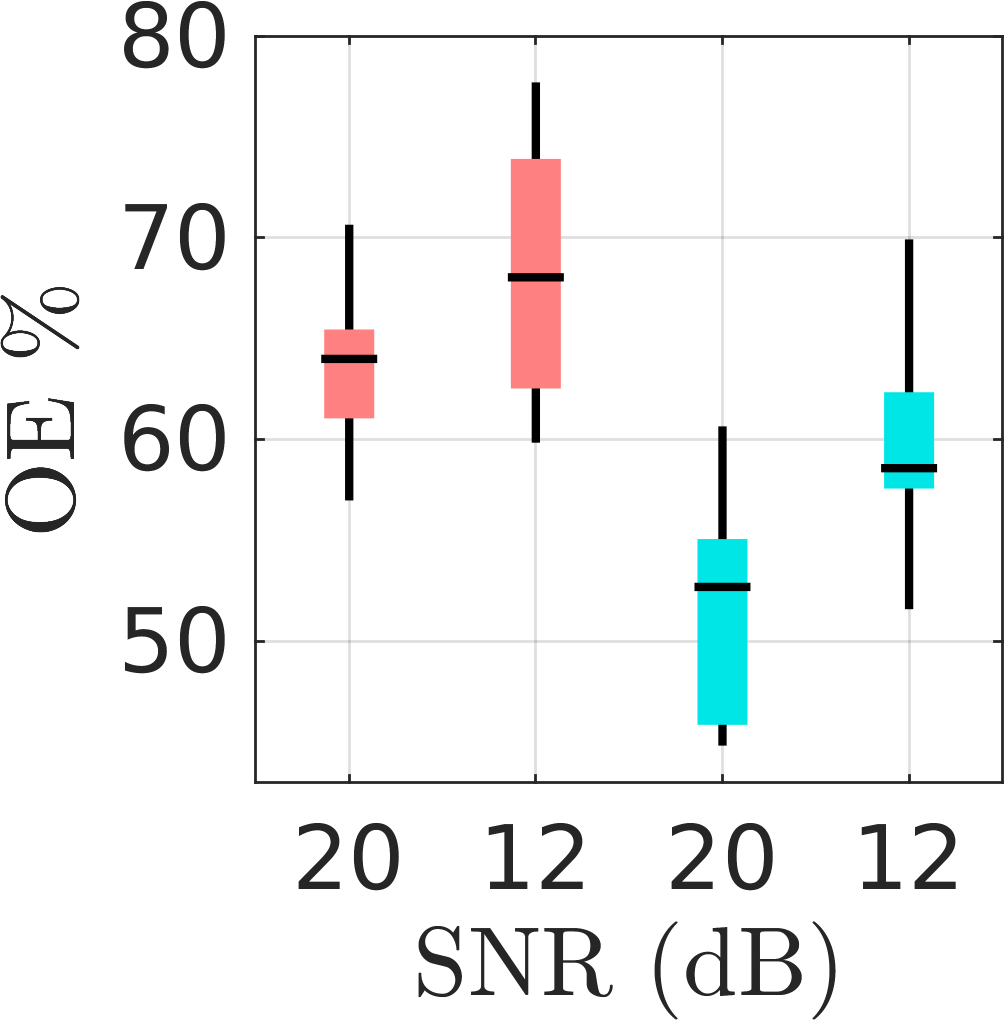} \end{minipage} \begin{minipage}{1.9cm}  \centering
        \includegraphics[width=1.8cm]{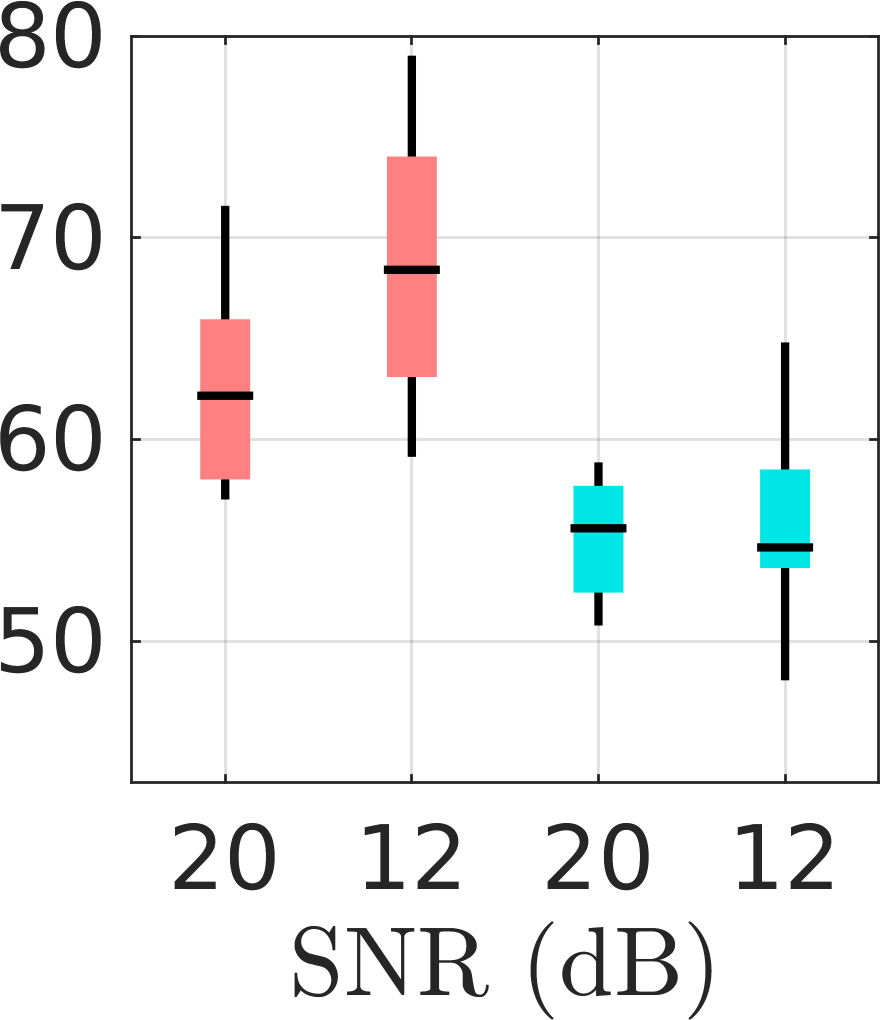} \end{minipage} \begin{minipage}{1.9cm} \centering
           \includegraphics[width=1.8cm]{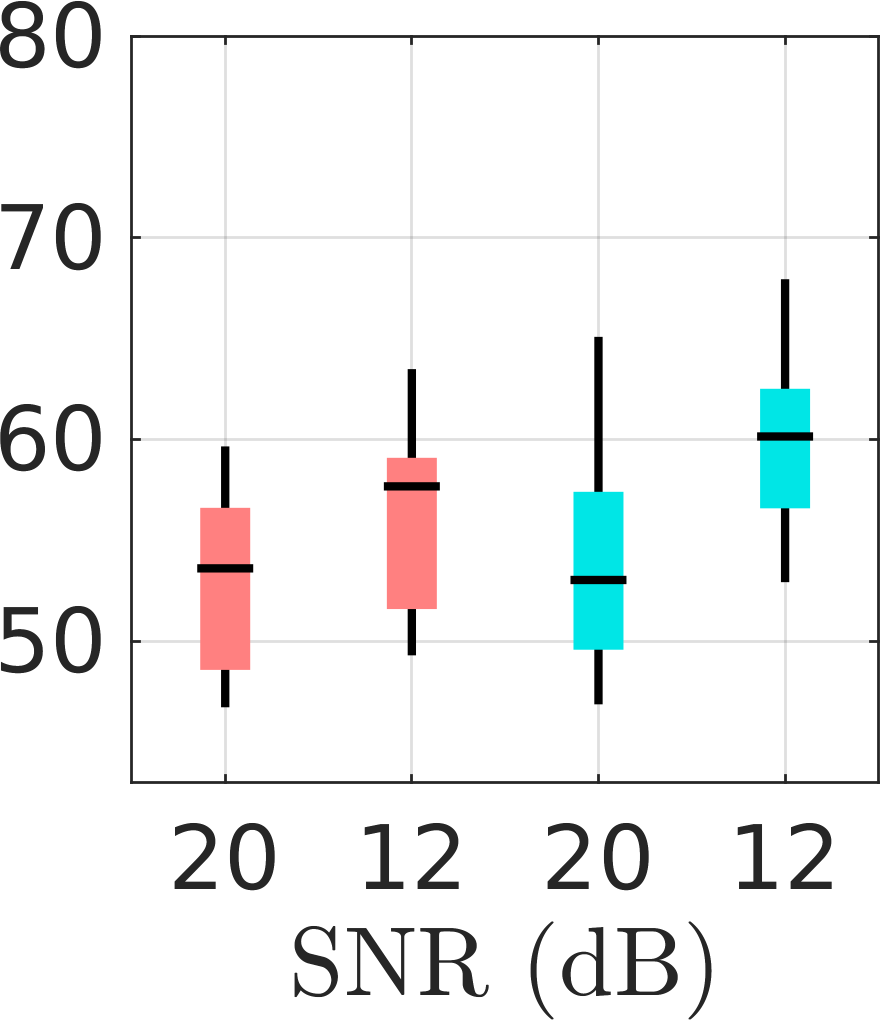} \end{minipage} \begin{minipage}{1.9cm}  \centering
         \includegraphics[width=1.8cm]{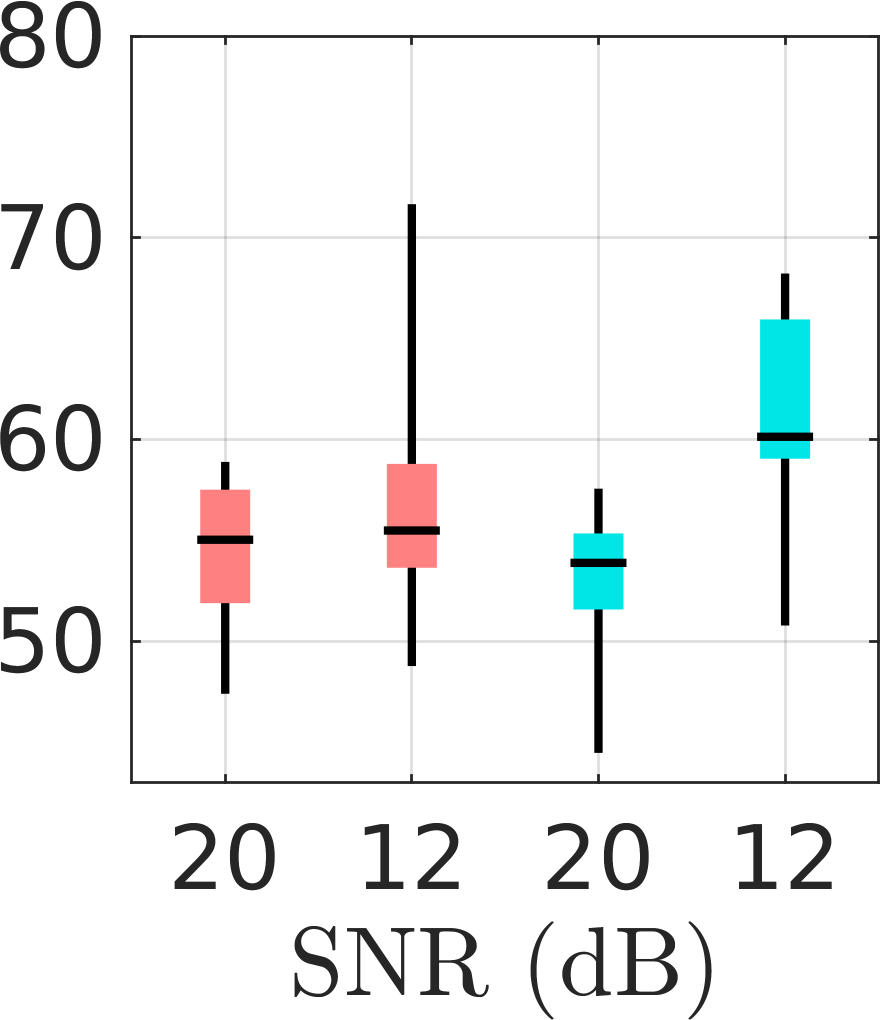} \end{minipage}   \\ \vskip0.2cm 
          \begin{minipage}{1.9cm} \centering        
        ({\bf I}): Unfiltered 
             \end{minipage}
    \begin{minipage}{1.9cm} \centering 
   ({\bf II}): TSVD 
            \end{minipage}
    \begin{minipage}{1.9cm} \centering 
   ({\bf III}): Unfiltered \&  randomised 
               \end{minipage}
    \begin{minipage}{1.9cm} \centering 
   ({\bf IV}): TSVD  \&  randomised
             \end{minipage} 
         \end{scriptsize}
    \caption{Structural Similarity (SSIM, scale of 0 - 1 and 1 shows the best result), Root Mean Squared Error (RMSE, lower limit of 0, which shows the best result) and Overlap Error (OE \%, scale of 0 - 100 and 0 shows the best result) for the 20 MHz signal pulse ({\bf A}) with SNR of 20 and 12 dB. The red boxes correspond to the dense configuration and the cyan boxes correspond to the sparse configuration, both having their respective noise levels indicated on the horizontal axis. This figure indicates that the RMSE \& OE  of the surface and void are in most cases ({\bf I})--({\bf IV}) lower for the sparse as compared to the dense configuration. Similarly, the SSIM is higher for the sparse case across all the levels of configurations ({\bf I})--({\bf IV}) shown in columns from left to right, respectively.  Here the most ideal reconstructions are those obtained from the sparse randomised configurations. } 
    \label{fig:A_boxplot}
\end{figure}

\begin{figure}[!ht]
    \centering  \begin{scriptsize}  SSIM \\\vskip0.1cm 
      \begin{minipage}{2.05cm}  \centering
    \includegraphics[width=2.05cm]{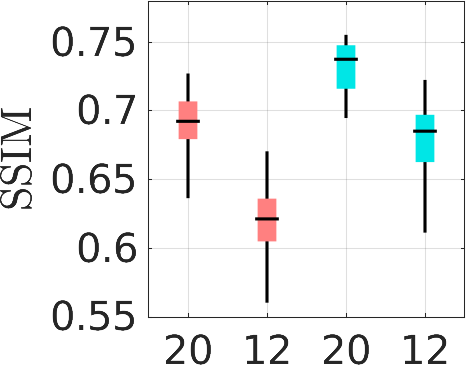}  \end{minipage} \begin{minipage}{1.9cm}  \centering
        \includegraphics[width=1.8cm]{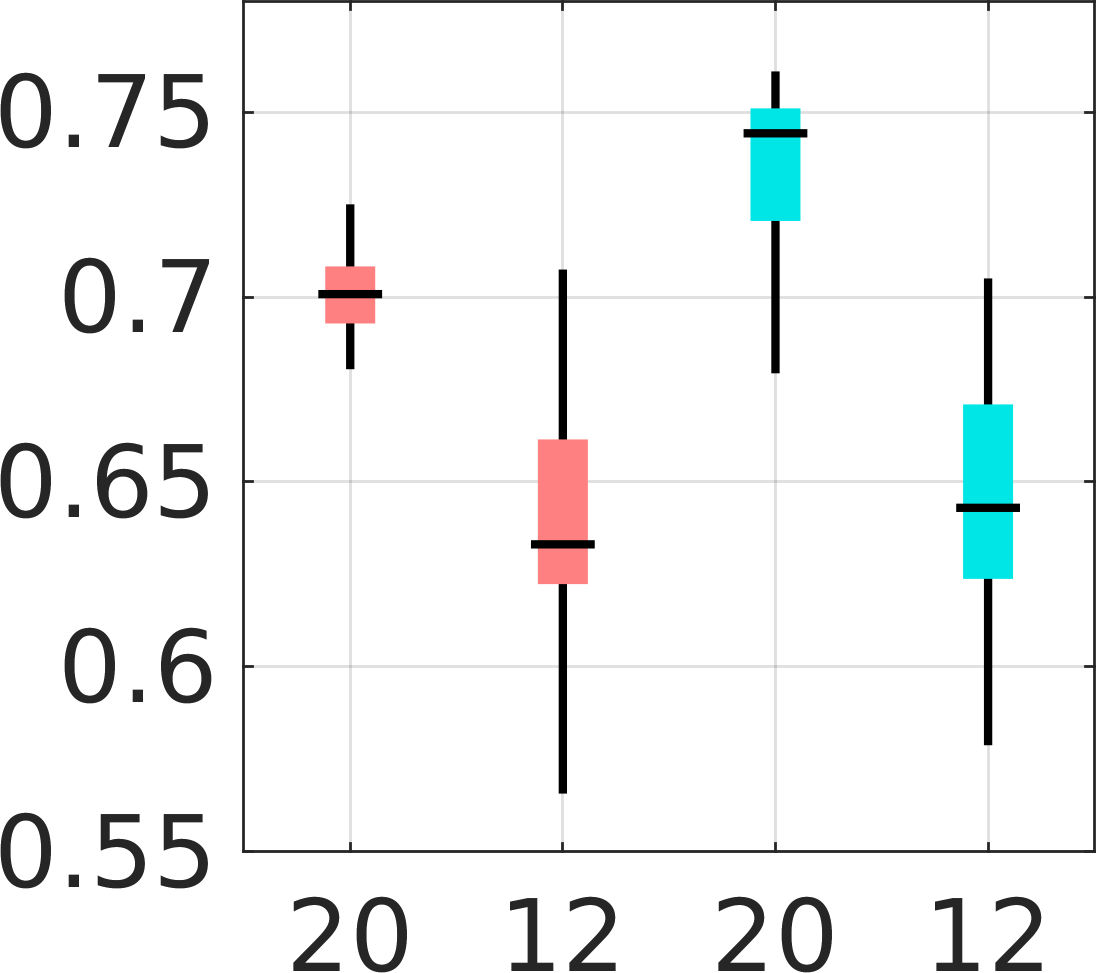}  \end{minipage} \begin{minipage}{1.9cm}  \centering
           \includegraphics[width=1.8cm]{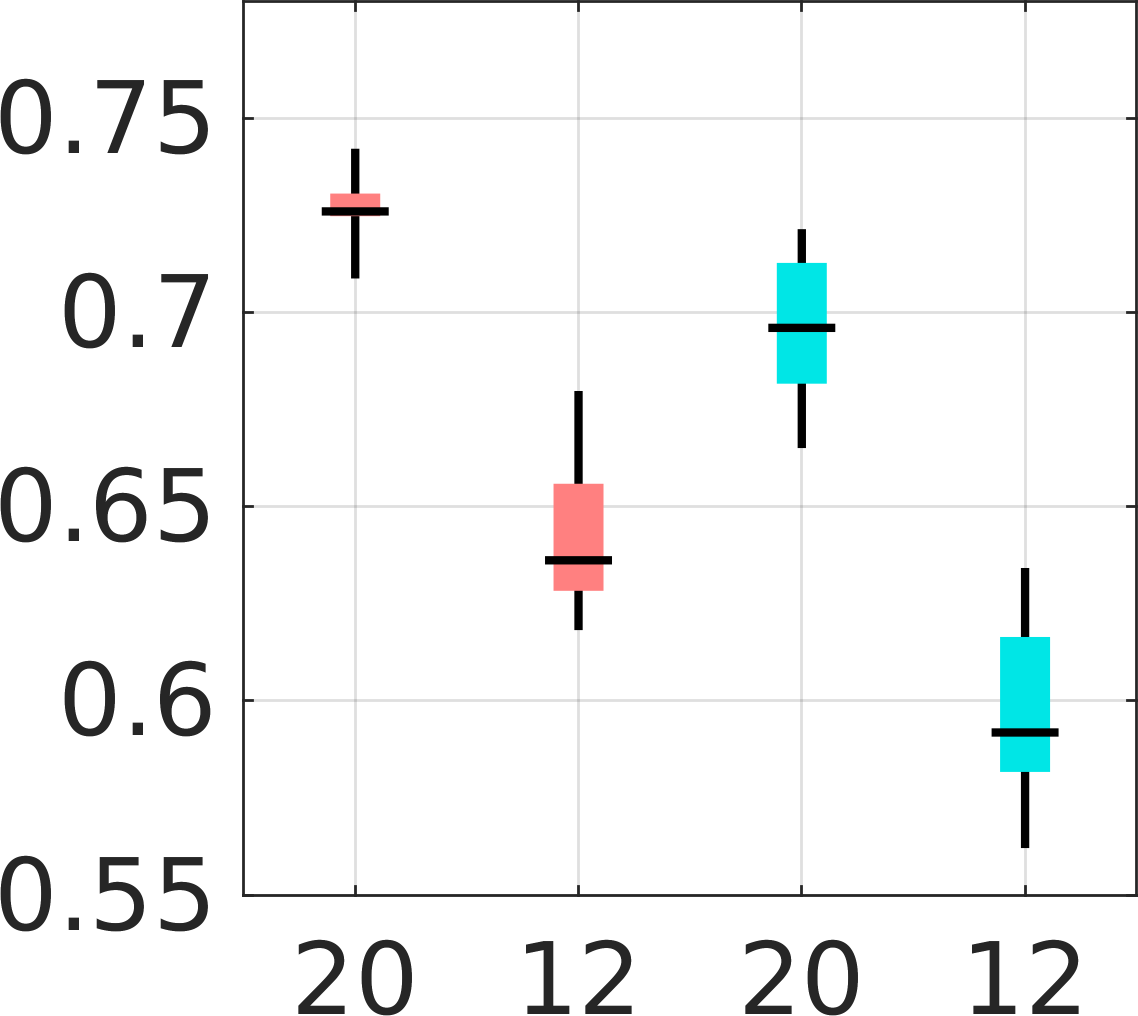}  \end{minipage} \begin{minipage}{1.9cm}  \centering
         \includegraphics[width=1.8cm]{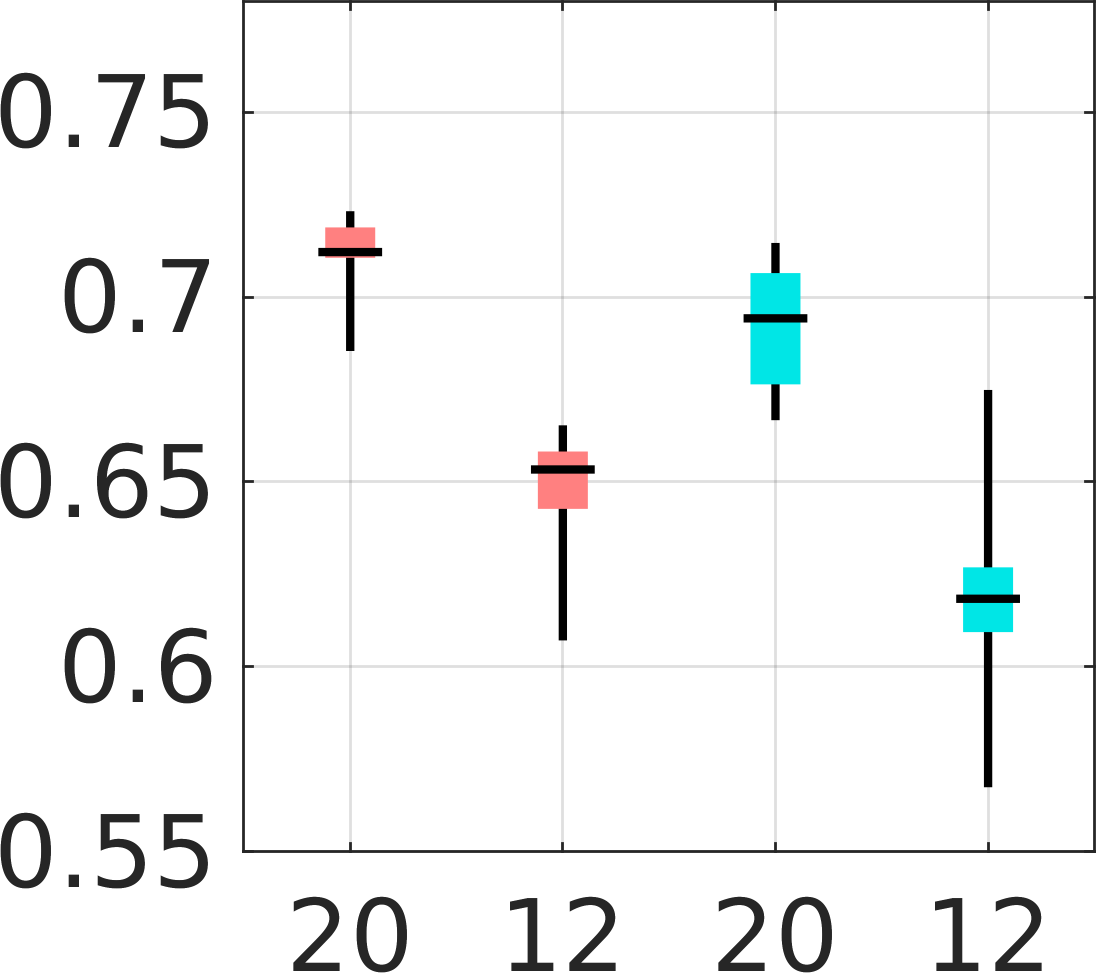}  \end{minipage}   \\ \vskip0.2cm Surface RMSE  \\ \vskip0.1cm 
           \begin{minipage}{2.0cm}  \centering
           \includegraphics[width=2.0cm]{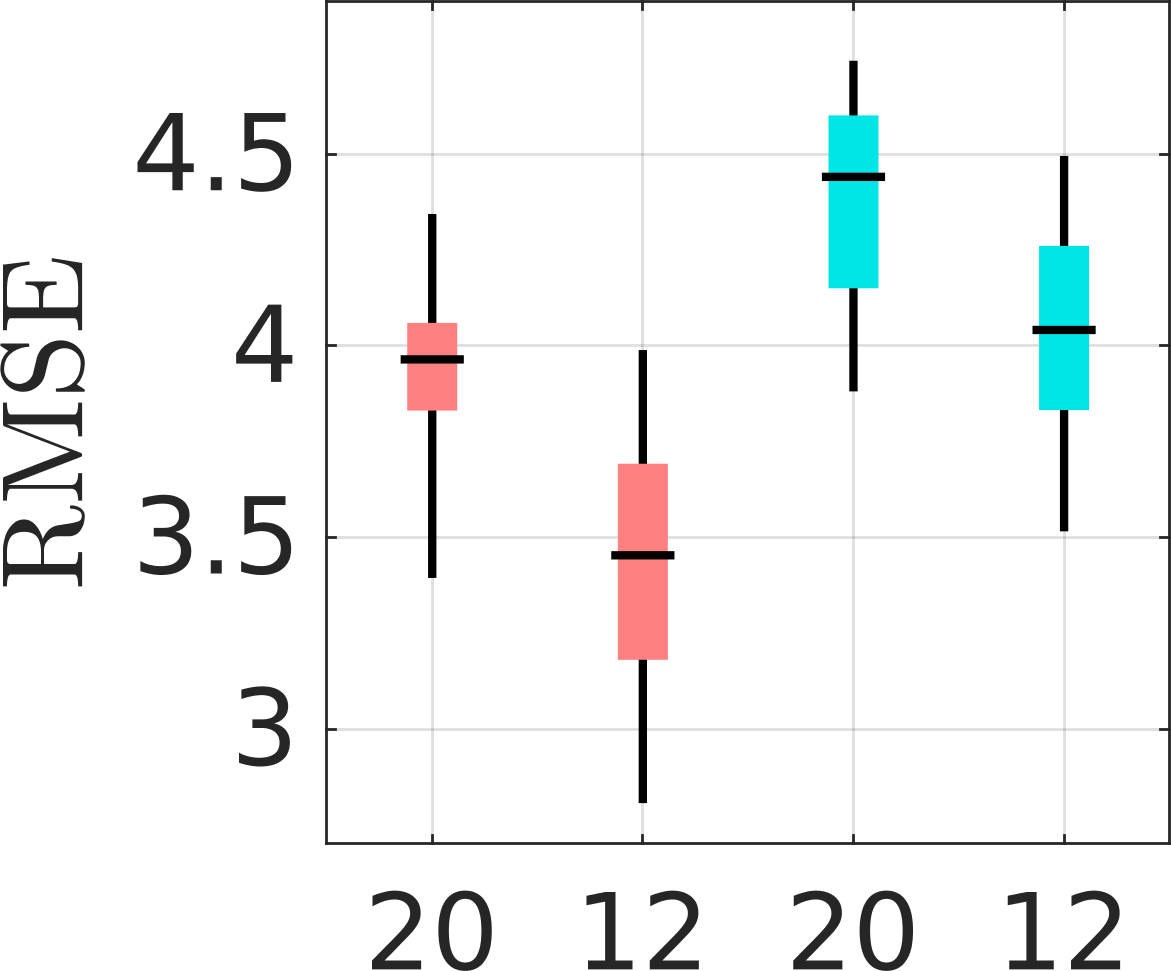}  \end{minipage} \begin{minipage}{1.9cm}  \centering
        \includegraphics[width=1.8cm]{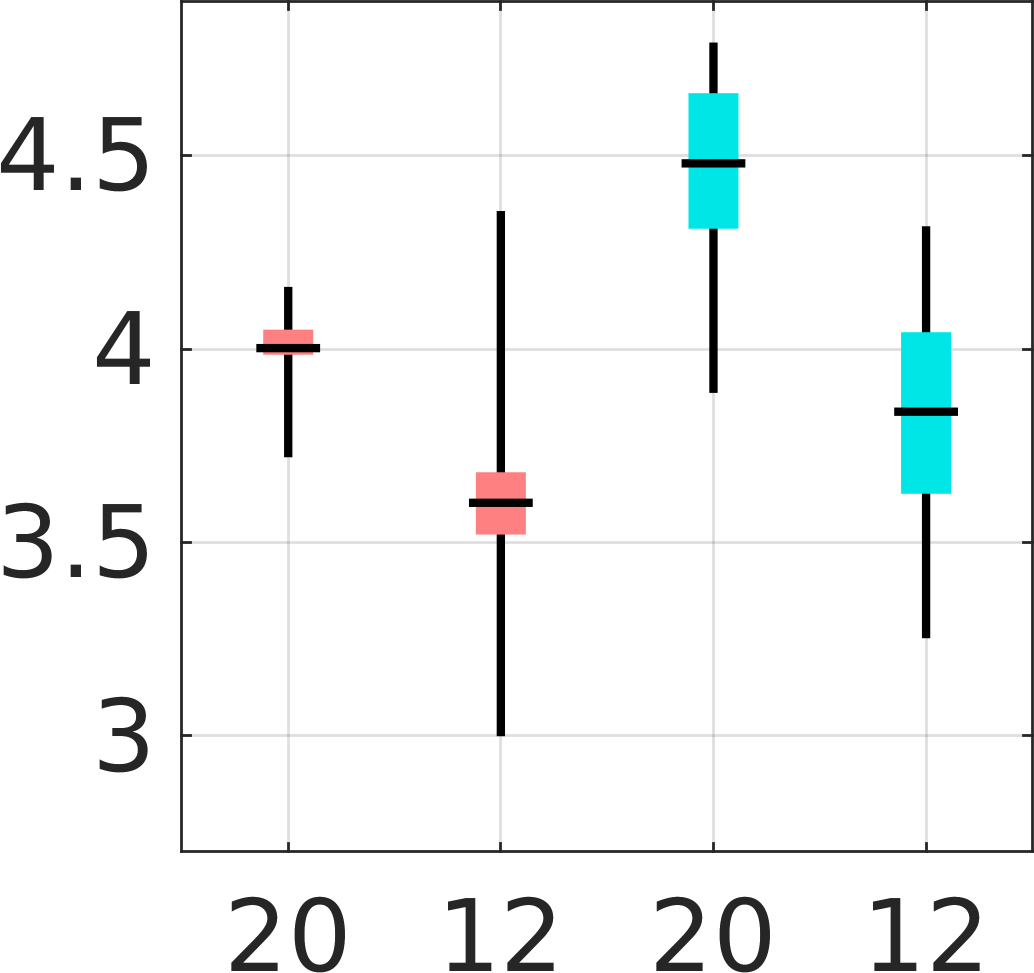}  \end{minipage} \begin{minipage}{1.9cm}  \centering
           \includegraphics[width=1.8cm]{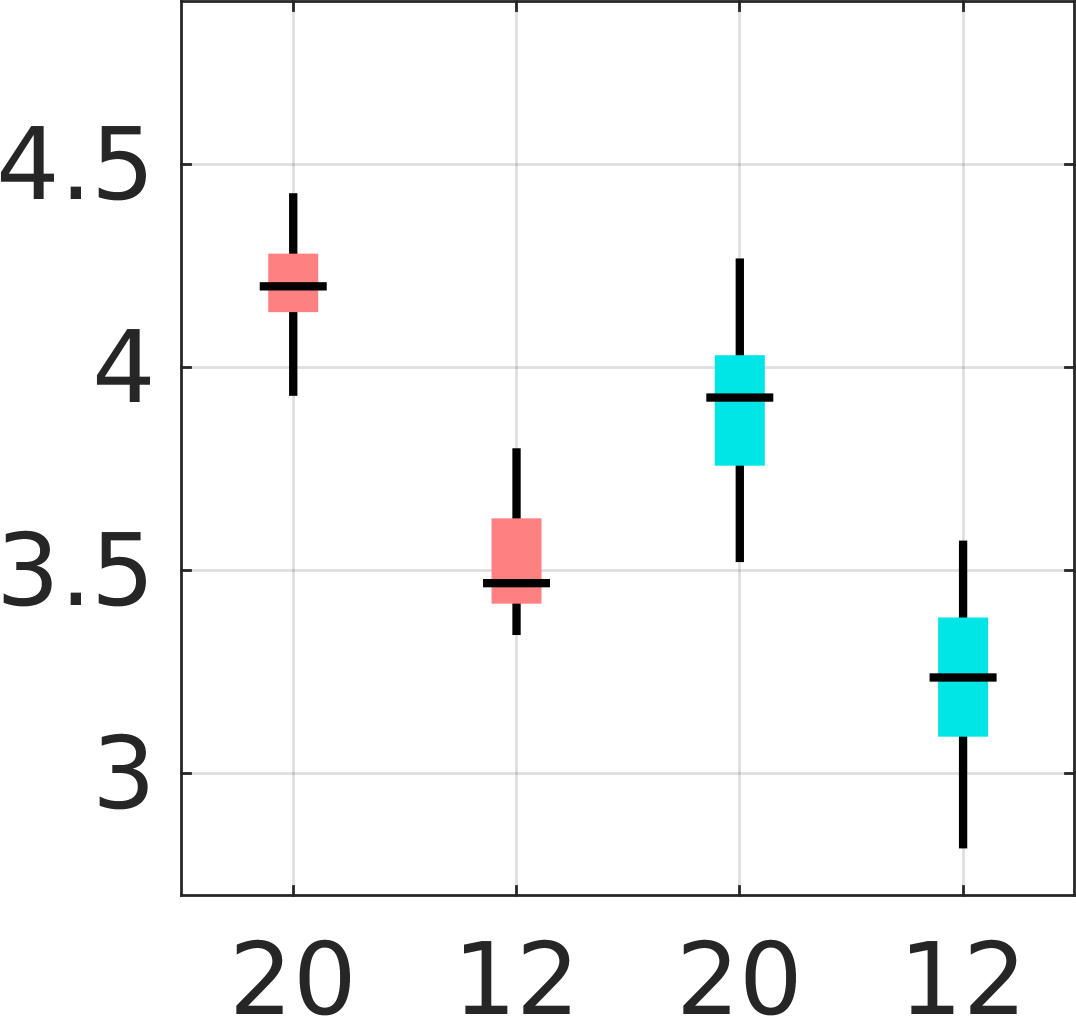}  \end{minipage} \begin{minipage}{1.9cm}
         \includegraphics[width=1.8cm]{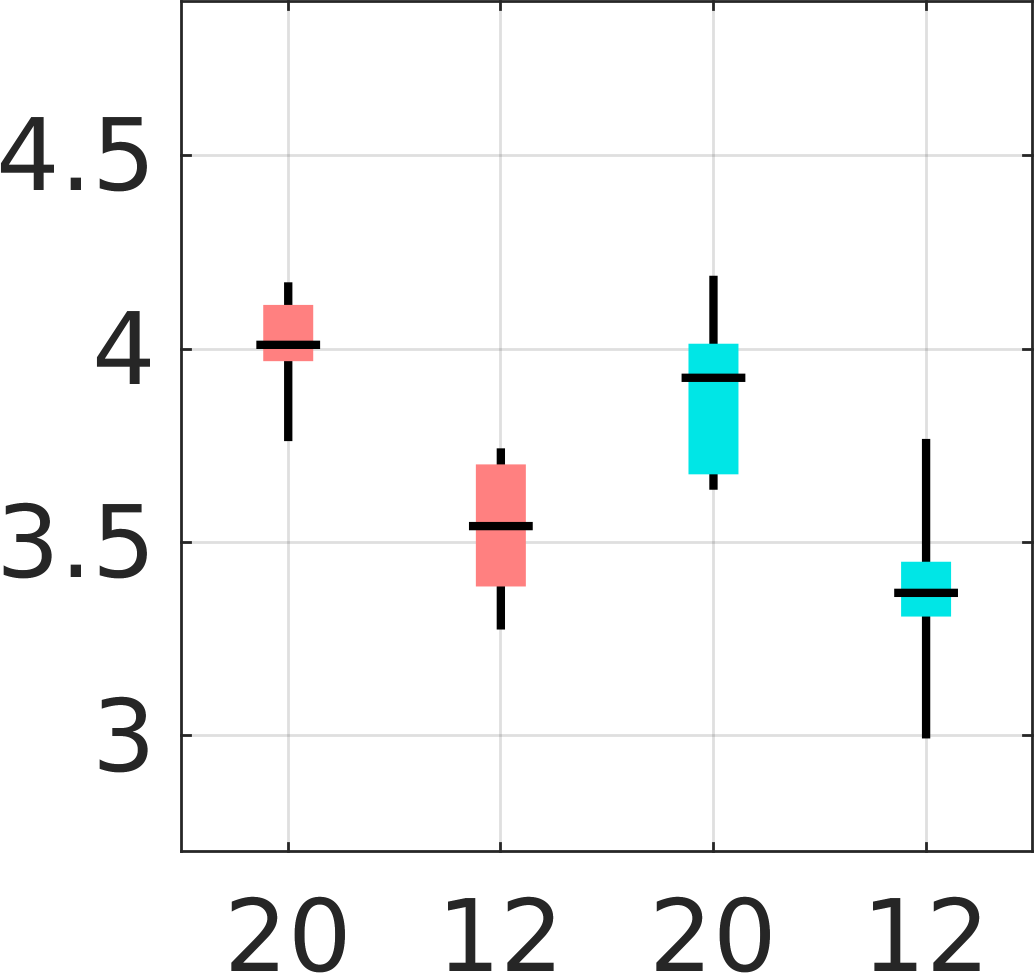} 
          \end{minipage} 
         \\ \vskip0.2cm  Void RMSE  \\ \vskip0.1cm 
          \begin{minipage}{2.0cm}
           \includegraphics[width=2.0cm]{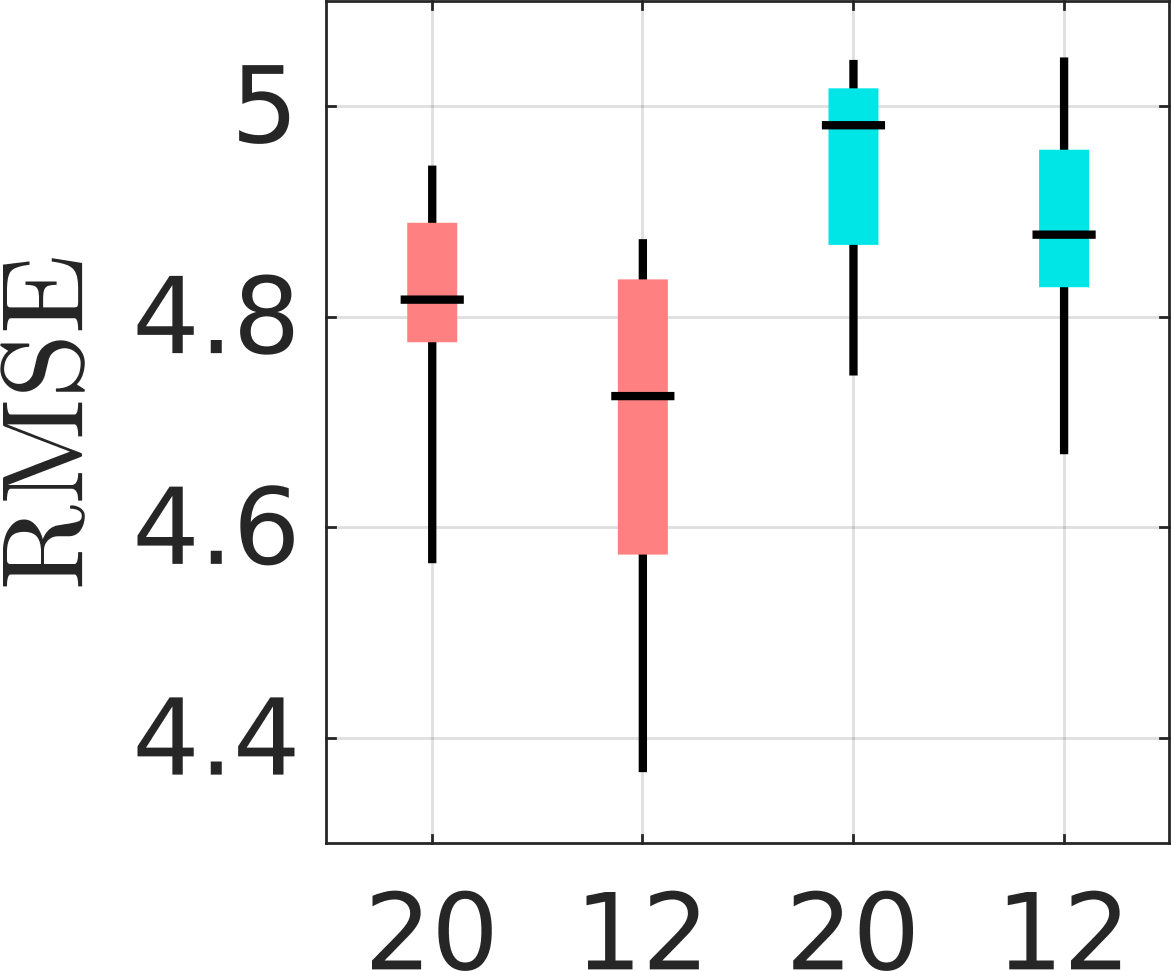}  \end{minipage} \begin{minipage}{1.9cm}
        \includegraphics[width=1.8cm]{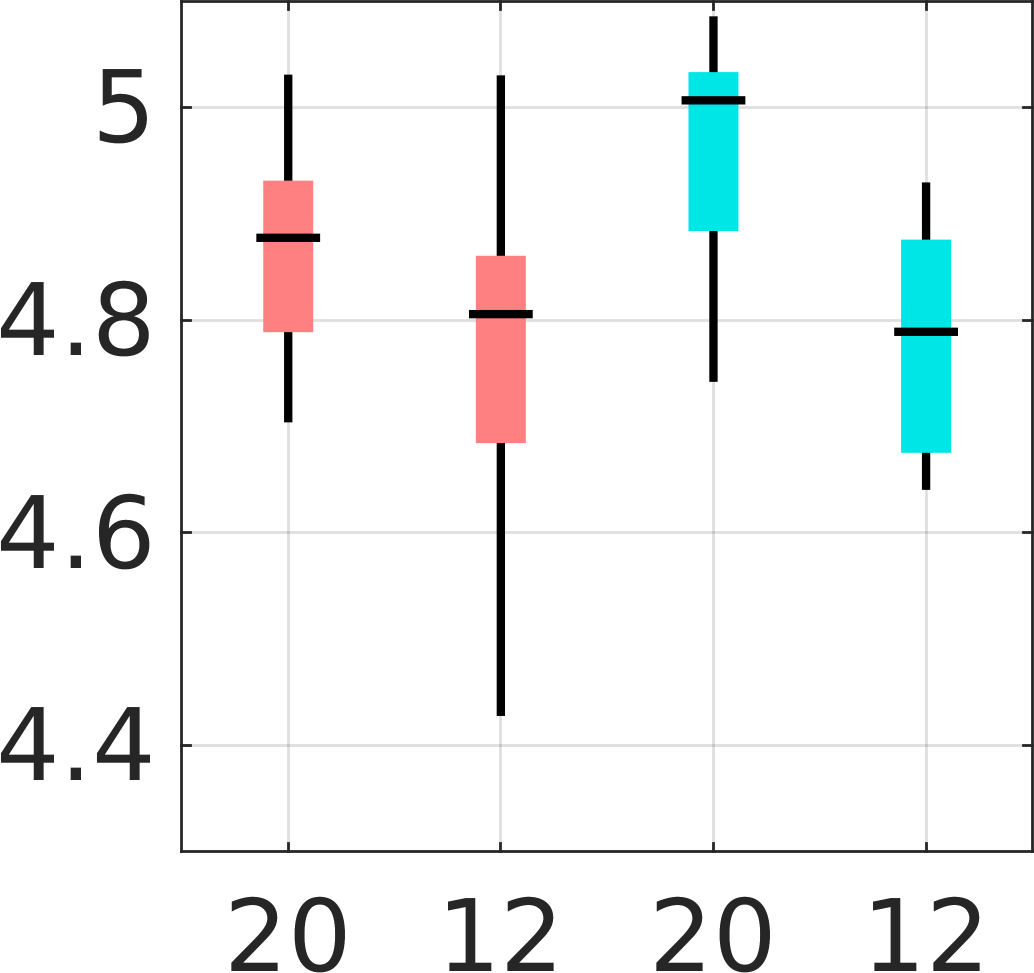}  \end{minipage} \begin{minipage}{1.9cm}
           \includegraphics[width=1.8cm]{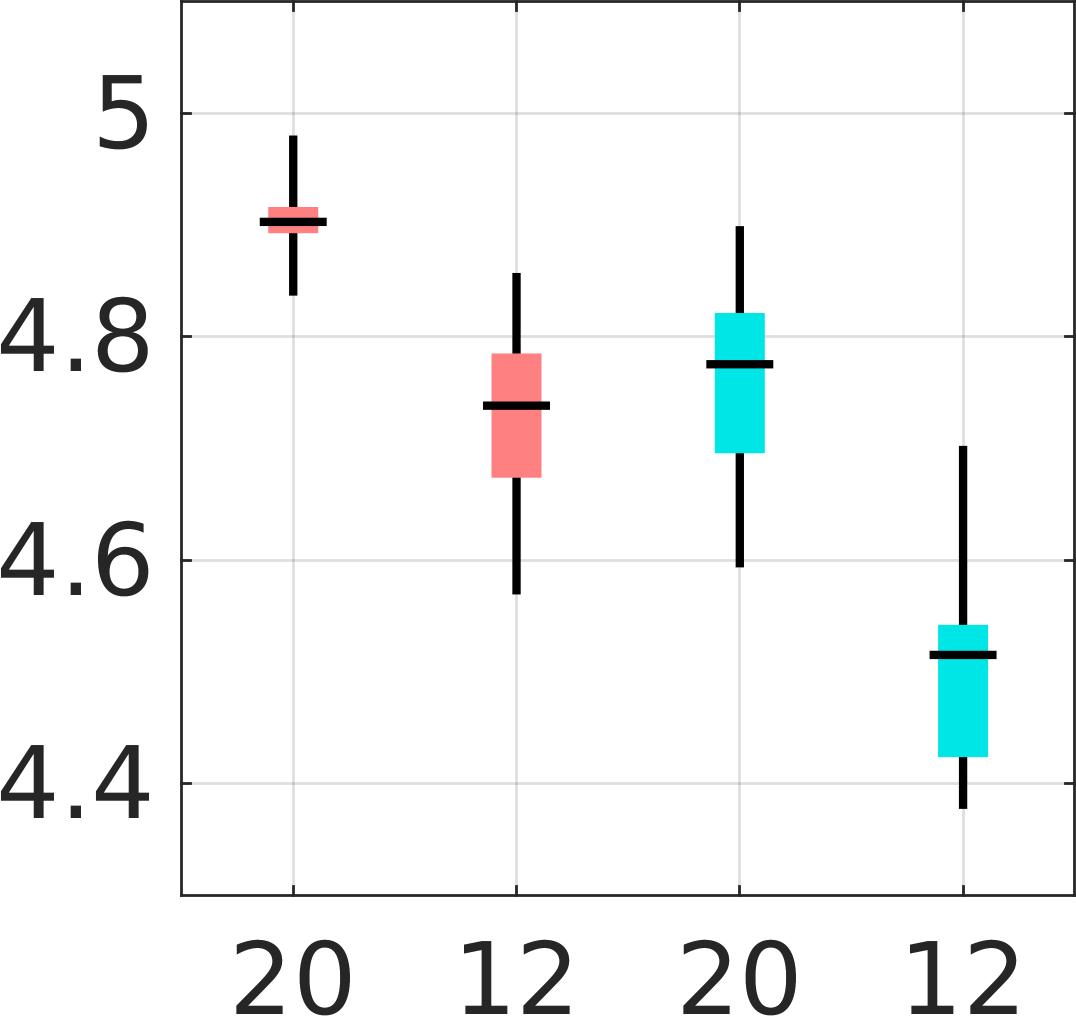}  \end{minipage} \begin{minipage}{1.9cm}
         \includegraphics[width=1.8cm]{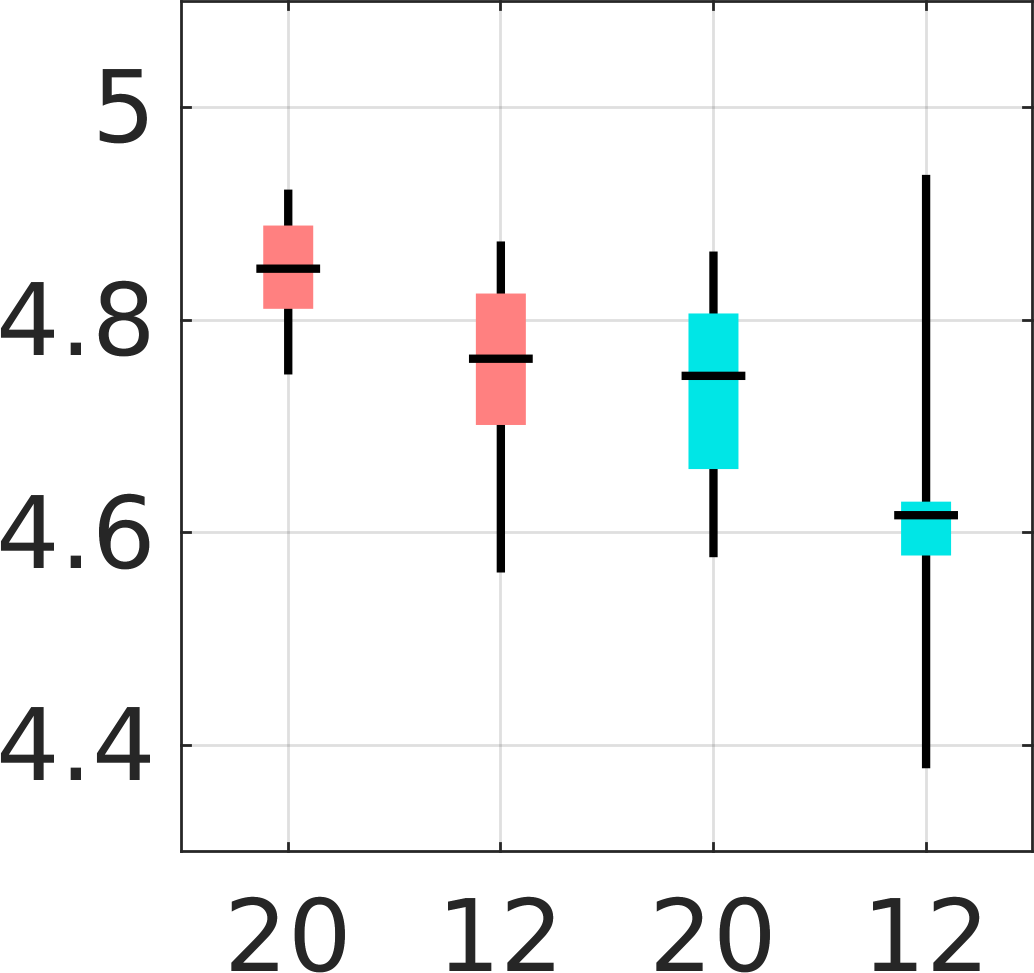} 
          \end{minipage}
         \\ \vskip0.2cm  Surface OE \\  \vskip0.1cm  
         \begin{minipage}{2.0cm}
    \includegraphics[width=2.0cm]{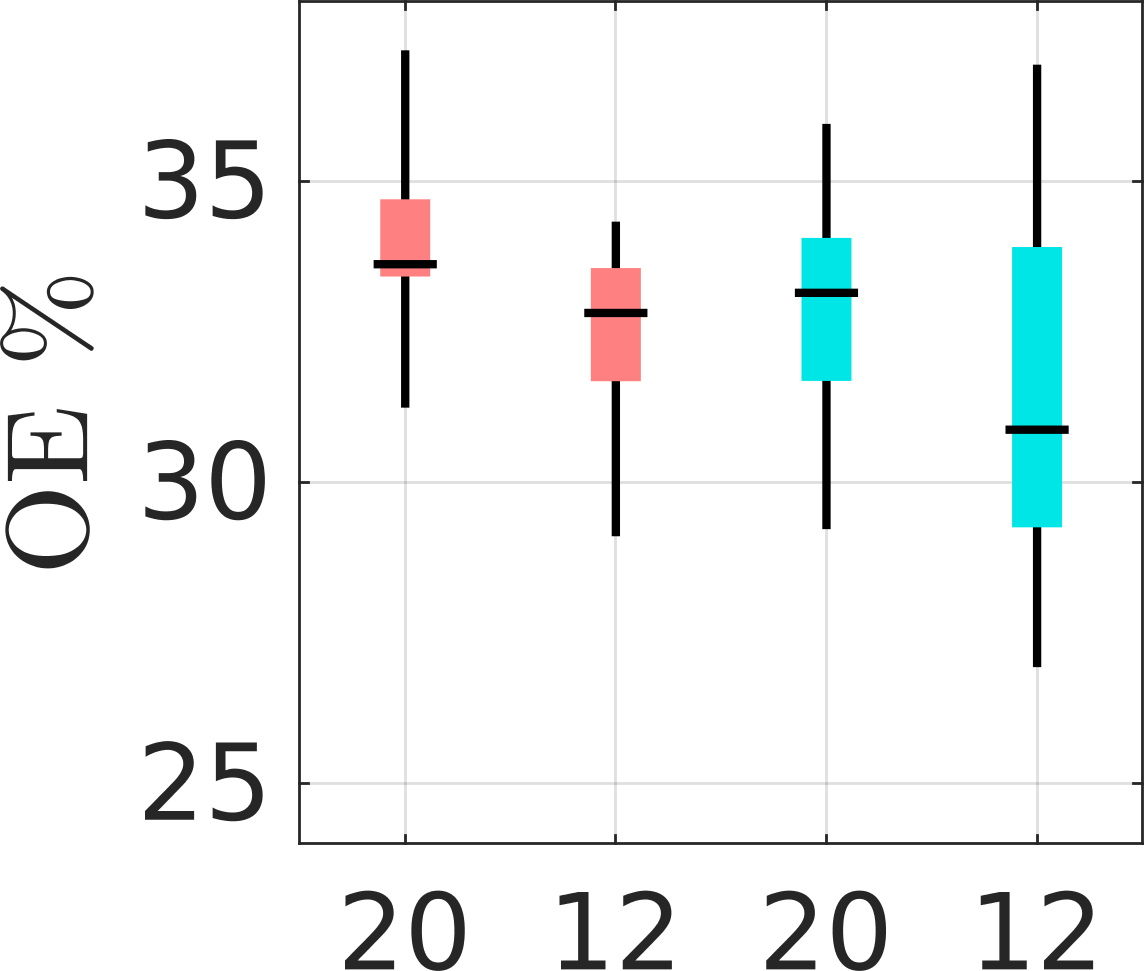}  \end{minipage} \begin{minipage}{1.9cm}
        \includegraphics[width=1.8cm]{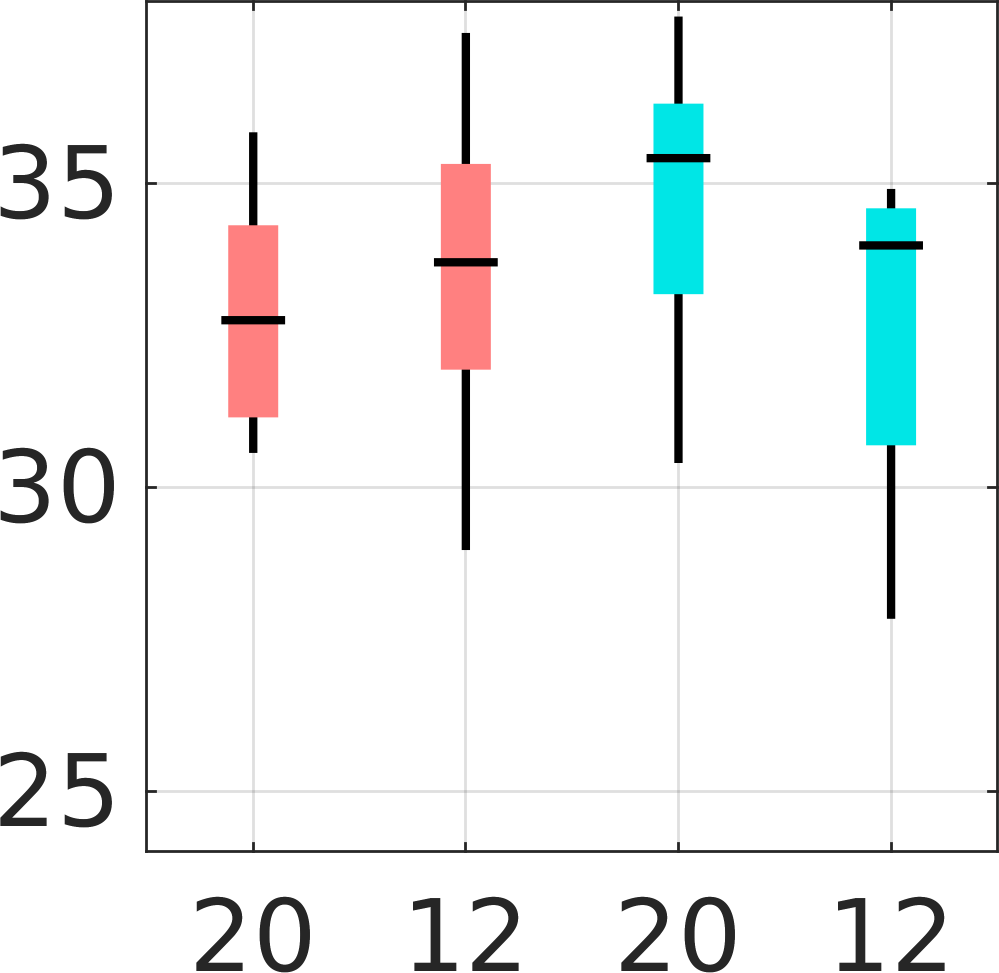}  \end{minipage} \begin{minipage}{1.9cm}
           \includegraphics[width=1.8cm]{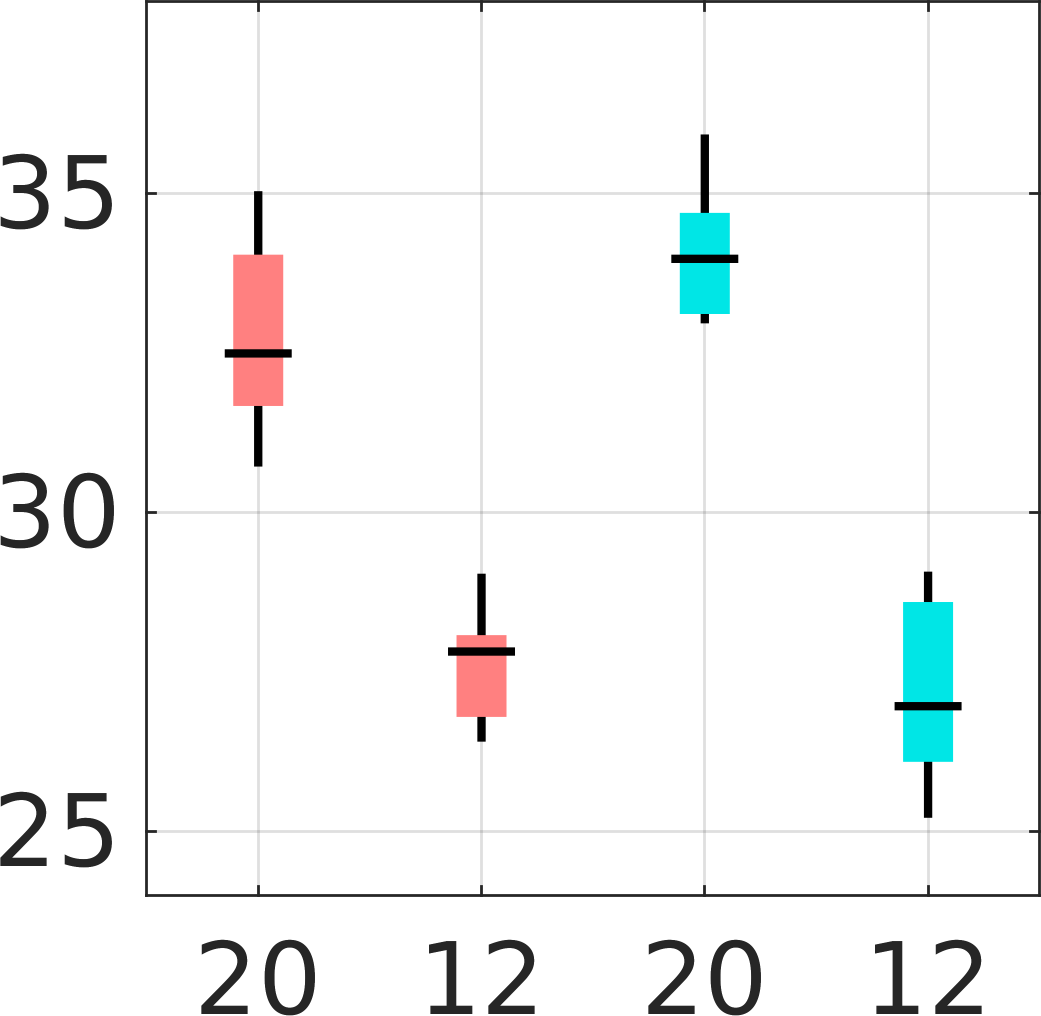} \end{minipage} \begin{minipage}{1.9cm}
         \includegraphics[width=1.8cm]{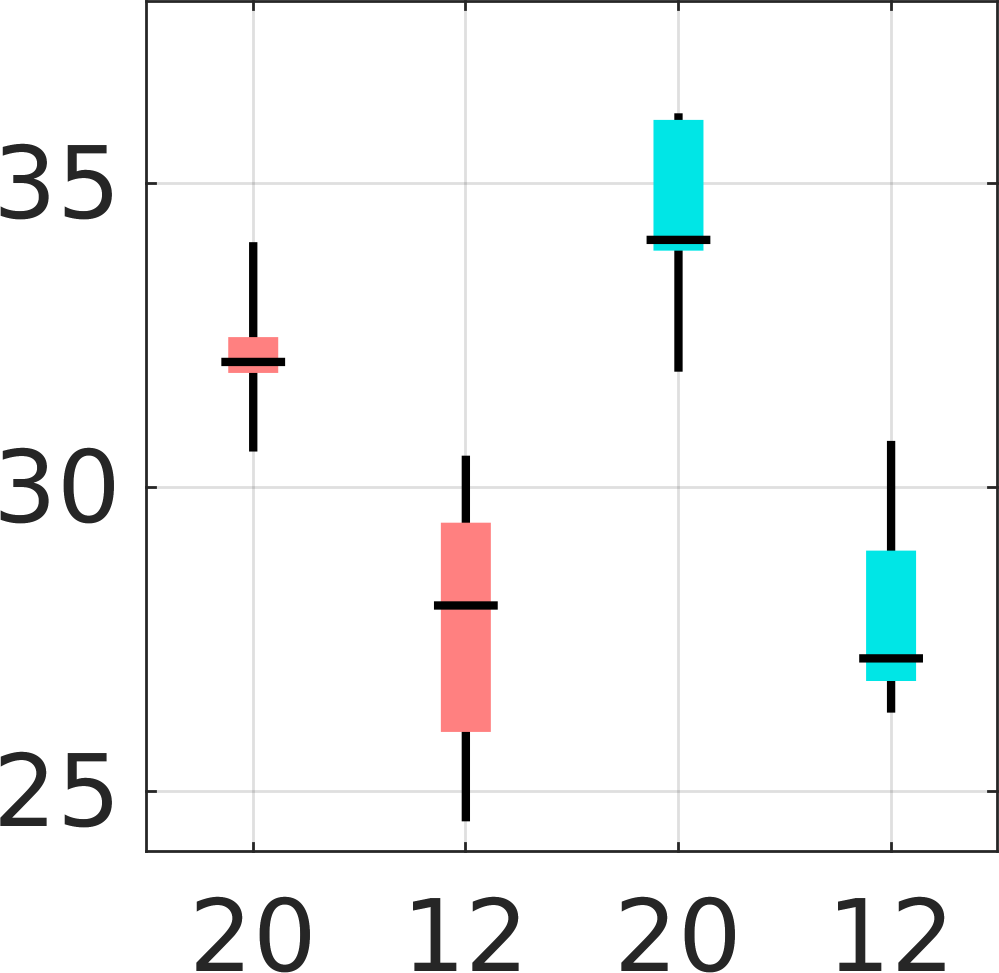} \end{minipage} \\ \vskip0.2cm  Void OE \\ \vskip0.1cm 
            \begin{minipage}{2.0cm}
           \includegraphics[width=2.0cm]{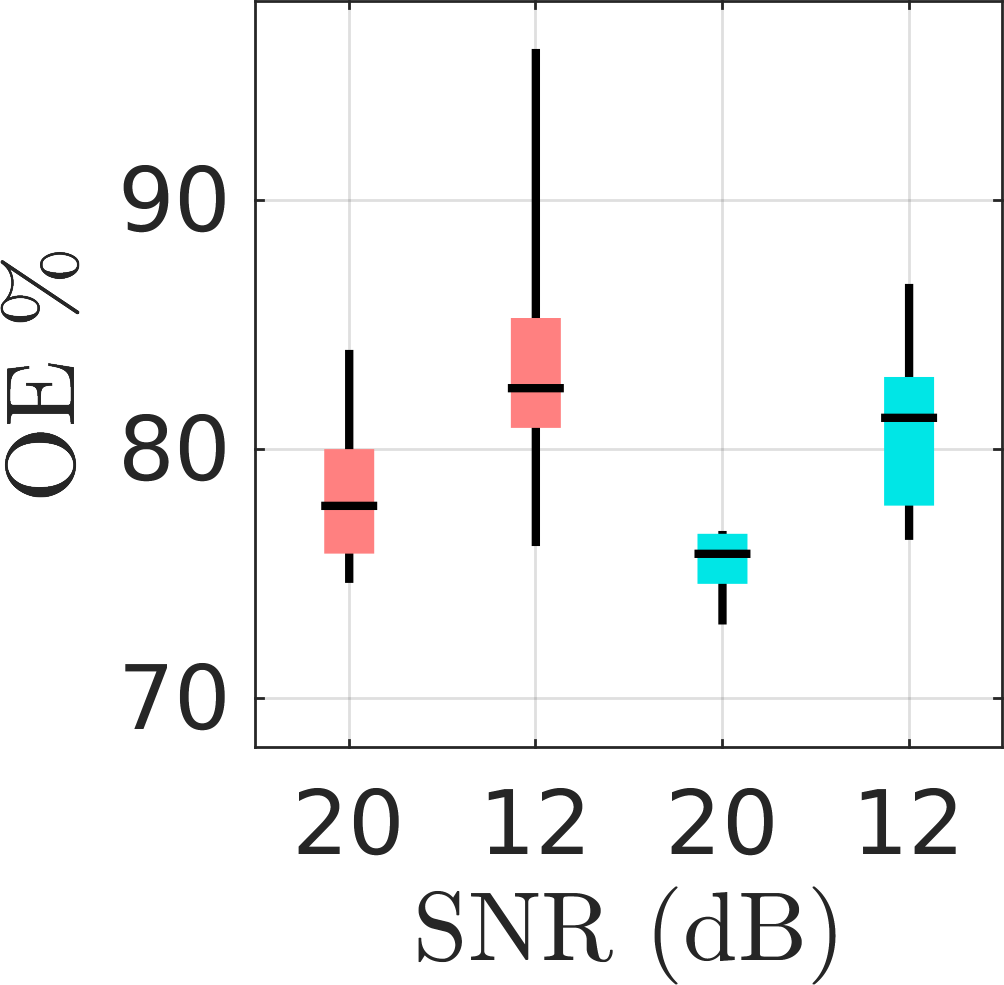}  \end{minipage} \begin{minipage}{1.9cm}
        \includegraphics[width=1.8cm]{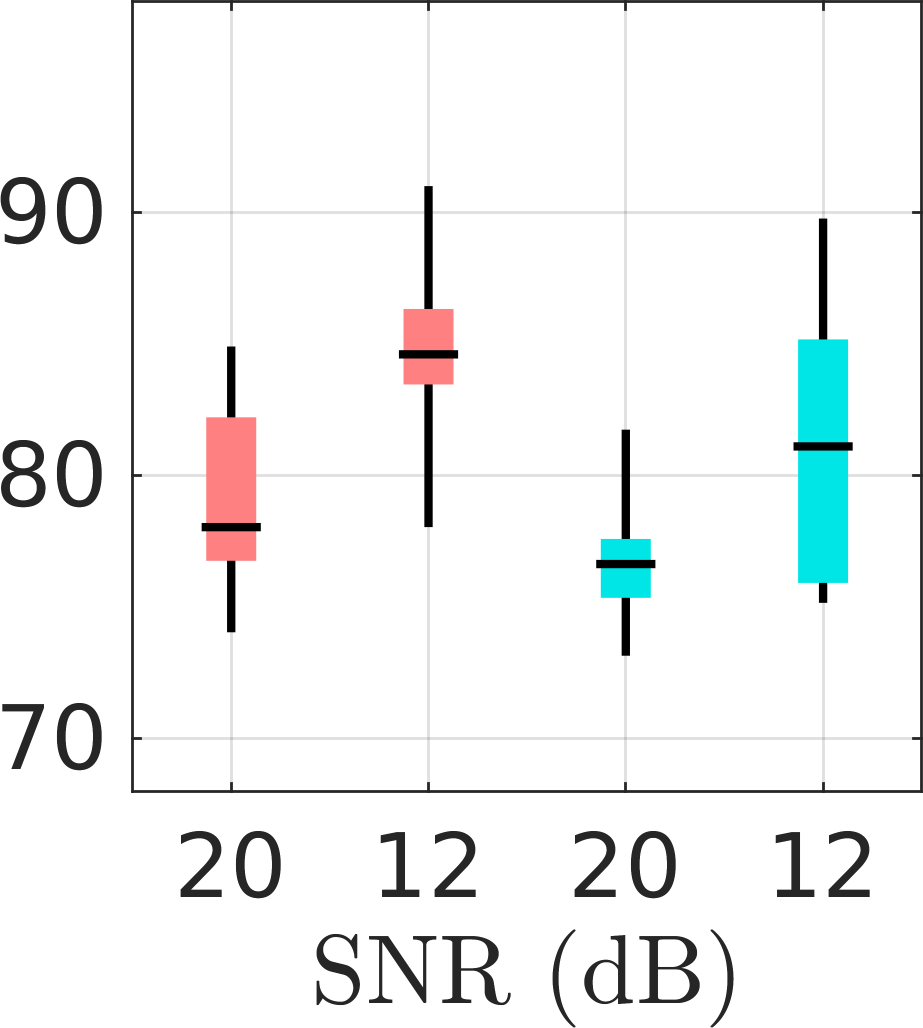}   \end{minipage} \begin{minipage}{1.9cm}
           \includegraphics[width=1.8cm]{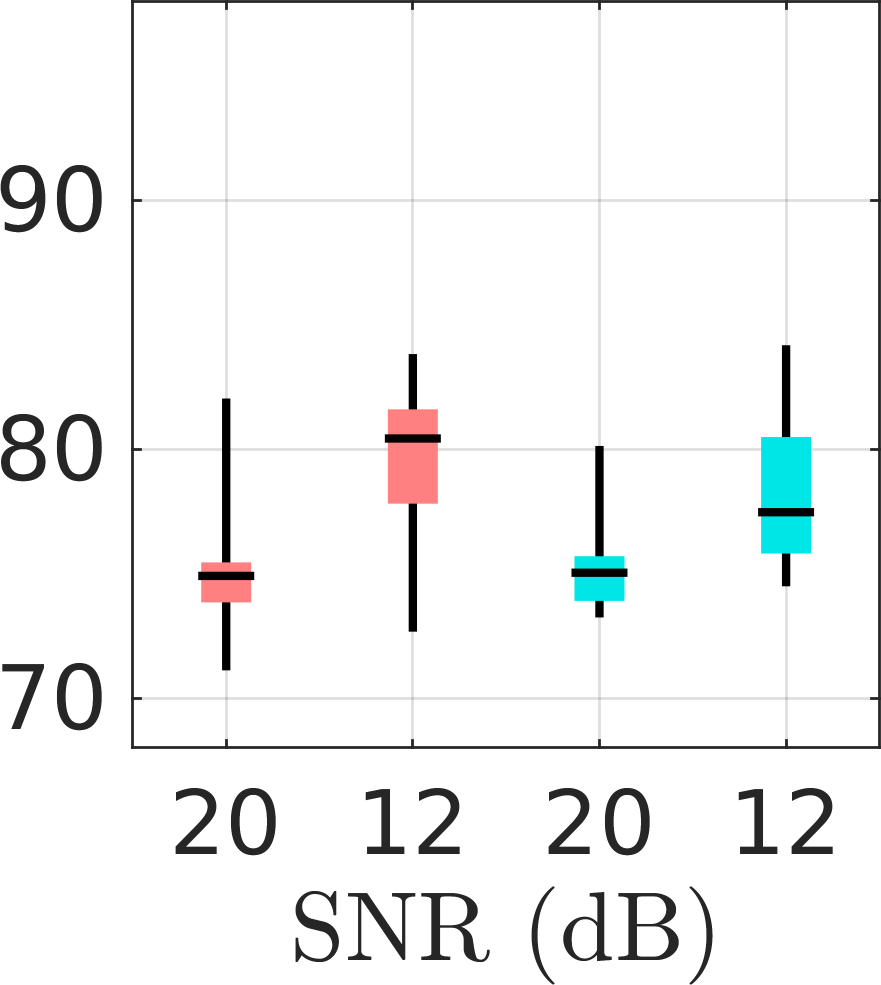}  \end{minipage} \begin{minipage}{1.9cm}
         \includegraphics[width=1.8cm]{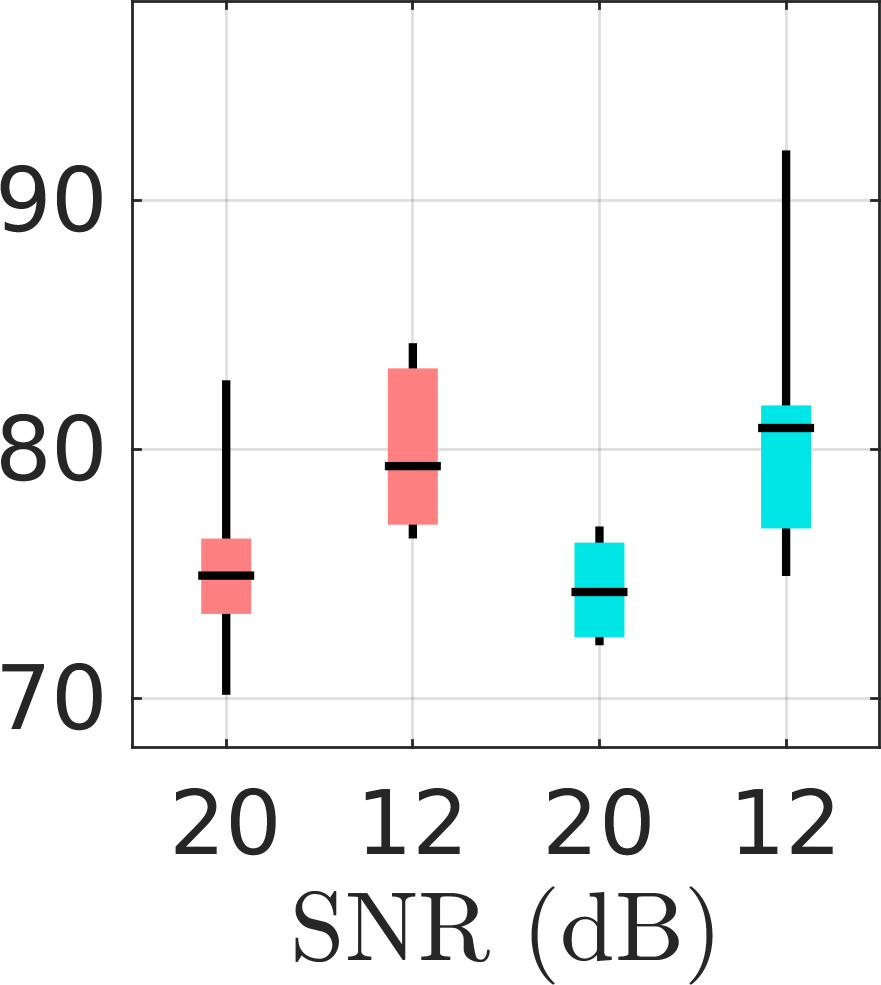}  \end{minipage}   \\ \vskip0.2cm 
          \begin{minipage}{1.9cm} \centering        
        ({\bf I}): Unfiltered 
             \end{minipage}
    \begin{minipage}{1.9cm} \centering 
   ({\bf II}): TSVD 
            \end{minipage}
    \begin{minipage}{1.9cm} \centering 
   ({\bf III}): Unfiltered \&  randomised
               \end{minipage}
    \begin{minipage}{1.9cm} \centering 
   ({\bf IV}): TSVD \&  randomised
             \end{minipage} 
         \end{scriptsize}
    \caption{As in Figure \ref{fig:A_boxplot} obtained for the 60 MHz signal pulse ({\bf B}) with SNR of 20 and 12 dB. The regular configuration has higher SSIM, RMSE \& OE of the surface for the sparse (cyan) point density as compared to the dense (red). This is opposite for the random configuration which has lower SSIM, RMSE of the surface and void for the sparse (green) point density as compared to the dense (red). The OE of the surface is lower for both random and regular configuration for the spare density points. Here the most ideal reconstruction are those obtained from the sparse randomised configurations. }
    \label{fig:B_boxplot}
\end{figure}

The results from the numerical experiments of section \ref{sec:exp} are presented in Figures \ref{fig:B_low_freq}--\ref{fig:sample_summary}. Of these, Figures \ref{fig:B_low_freq} and \ref{fig:B_high_freq} show the amplitude of the reconstructed real relative permittivity perturbation visualised on a linear scale between 0 and 1 after setting an upper threshold with the 98\% quantile of the data as our normalisation. These also show the relative Overlap Error (OE) between the reconstructed details and those of  the actual permittivity distribution.  In Table \ref{tab:compare_results}, Figures \ref{fig:A_boxplot} and \ref{fig:B_boxplot}, the robustness of the reconstructions are analysed in a tabular form and via boxplots obtained with 10 different noise vector realisations. The  norm of the reconstructed permittivity distribution, ${\bf x}$, for the different sampling steps is shown for the randomised configuration and full data in Figure \ref{fig:sample_history}. The results show that the aim of reconstructing the permittivity perturbation distribution of the complex domain incorporating phase errors formulation,  was achieved by a monostatic full-wave simulation and inversion with the spatial averaging, and discrepancy-based frequency averaging methods as stated in Sections \ref{sec:materials}, \ref{sec:grid} and \ref{sec:phase}. The results show that the point selection, spatial density, and  noise level used are highly consequential to how  robust reconstruction can be obtained. The ideal reconstructions obtained for the low frequency case is the Low-noise-Sparse-Random-TSVD configuration with surface overlap error of 36.16\% and void overlap error of 52.77\%, see Figure \ref{fig:B_low_freq} column (IV). The ideal reconstruction obtained for the high frequency case is the High-noise-Spare-Random-Unfiltered configuration with surface overlap error of 27.15\% and void overlap error of 78.21\%, see Figure \ref{fig:B_high_freq} column (III).

%****************************************************

\begin{table*}[!ht]
    \centering 
        \caption{Similarity and error estimates of the reconstructions for different numerical configurations. The 1st column presents the two signal pulses ({\bf A}) 20 and ({\bf B}) 60 MHz, described in Table \ref{tab:signal_frequency}, while the 2nd column shows several configurations over which these frequencies have been experimented. The corresponding levels of the configurations are also highlighted in the 3rd column. The 4th to 8th columns gives the mean values with respect to the different centre frequencies, configurations, and levels for SSIM (scales between 0 \& 1 and 1 shows the best result), RMSE of the void and surface (lower limit of 0, which shows the best result), OE of the void and surface (in \%, scales between 0 \& 100 and 0 shows the best result)). The WRS test values are also highlighted, with $0$ implying no significant differences and $1$ implying a significant difference in the median values of the corresponding  distributions.}
        \resizebox{1.9\columnwidth}{!}{%
    \begin{tabular}{@{}llllc|lc|lc|lc|lc@{}}
        Signal  & Configu-  & Levels  & SSIM & & \multicolumn{4}{c}{RMSE}  & \multicolumn{4}{c}{OE \%} \\
         pulse & ration & && & Void & & Surface   & & Void && Surface& \\
          &  & && WRS &  &WRS &    & WRS &  & WRS & & WRS\\
         
       \toprule
      ({\bf A}) 20 MHz & & & 0.530  && 2.074 && 1.602 & & 58.387 & &38.868 & \\ 
     & \\
           & Spatial  & Dense & 0.527 & \multirow{2}{0.7em}{1} & 2.286 & \multirow{2}{1em}{1} & 1.725 & \multirow{2}{0.7em}{1} & 60.476 & \multirow{2}{1em}{1} & 40.741 & \multirow{2}{0.7em}{1} \\ 
                & averaging& Sparse & 0.533  && 1.863 && 1.480 && 56.298 && 36.995& \\ 
                & \\
            &Point  & Regular &  0.521 &\multirow{2}{0.7em}{1} & 2.224 &\multirow{2}{1em}{1}& 1.751 &\multirow{2}{0.7em}{1}& 60.671 &\multirow{2}{1em}{1}& 43.176 &\multirow{2}{0.7em}{1}\\ 
            & selection & Random & 0.539 && 1.925 && 1.453 && 56.103 && 34.559 &\\ 
                            & \\
           &Filtering & Unfiltered &  0.530 &\multirow{2}{1em}{0}& 2.060 &\multirow{2}{1em}{0}& 1.603 &\multirow{2}{1em}{0}& 58.389 &\multirow{2}{1em}{0}& 38.435&\multirow{2}{1em}{0} \\ 
             && TSVD &  0.530 && 2.088 && 1.602 && 58.384 && 39.301& \\ 
            & \\
           &Noise & Low &   0.530 &\multirow{2}{1em}{0}& 2.085 &\multirow{2}{1em}{0}& 1.657 &\multirow{2}{1em}{1}& 55.933 &\multirow{2}{1em}{1}& 39.795&\multirow{2}{1em}{0} \\ 
             && High & 0.530 && 2.064 && 1.548 && 60.840 && 37.940& \\ 
             &\\
           \toprule
      ({\bf B}) 60 MHz &&& 0.673  && 4.787 && 3.833 && 78.668 && 31.785& \\ 
      & \\
          &Spatial  & Dense &0.672& \multirow{2}{1em}{0}  & 4.797 & \multirow{2}{1em}{0} & 3.787 & \multirow{2}{1em}{0} & 79.527 & \multirow{2}{1em}{1} & 31.612& \multirow{2}{1em}{0} \\ 
            &averaging& Sparse &  0.674  && 4.777 && 3.880 && 77.810 && 31.957&\\ 
                & \\
            &Point  & Regular & 0.680  & \multirow{2}{1em}{0}& 4.840 & \multirow{2}{1em}{1}& 3.957 & \multirow{2}{1em}{1}& 80.045 & \multirow{2}{1em}{1}& 33.058& \multirow{2}{1em}{1} \\ 
               & selection & Random & 0.667  && 4.735 && 3.708 && 77.292 && 30.511& \\ 
            & \\
           &Filtering & Unfiltered &  0.672  & \multirow{2}{1em}{0}& 4.780 & \multirow{2}{1em}{0}& 3.827 & \multirow{2}{1em}{0}& 78.339 & \multirow{2}{1em}{0}& 31.560& \multirow{2}{1em}{0} \\  
           && TSVD &  0.673  && 4.794 && 3.839 && 79.000 && 32.009&\\ 
            & \\
           & Noise & Low &   0.710  & \multirow{2}{1em}{1}& 4.855 & \multirow{2}{1em}{1}& 4.094 & \multirow{2}{1em}{1}& 76.195 & \multirow{2}{1em}{1}& 33.483& \multirow{2}{1em}{1} \\ 
            && High & 0.636  && 4.720 && 3.572 && 81.142 && 30.086& \\ 
           \bottomrule
    \end{tabular}%
    }
    \label{tab:compare_results}
\end{table*}

\subsection{Low vs.\ high centre frequency}

The global structure of the two-dimensional test domain, including the surface layer and voids, was reconstructed with an average surface overlap error of 38.87\% and 31.7\% and void overlap error of 58.39\% and 78.67\% for both 20 and 60 MHz signals, respectively, see Table \ref{tab:compare_results}. The former gave a more ideal reconstruction outcome, see Figure \ref{fig:B_low_freq} columns (I-IV), in terms of RMSE and OE of the void, while the latter with higher SSIM, shows faint details and hot spots on the surface, especially in the concave part of the target and parts close to the receiver, see Figure \ref{fig:B_high_freq} columns (I-IV). This is expected in the high frequency case as there is more reflection at shallow depth and fewer wavefronts are received from the deeper part of the target. This anomaly limits the visibility of the voids which is obvious based on their increased  RMSE and OE of the void, as there is more contrast at the surface compared to the interior part. However, the surface layer has a lesser OE in comparison to the lower frequency case, see Figure \ref{fig:B_low_freq} rows 2 \& 4 vs.\ Figure \ref{fig:B_high_freq} rows 2 \& 4. 

\subsection{Signal configuration}

 For the sparse configurations, the global structure, surface layer, and voids are generally more pronounced and uniform, with an enhanced contrast compared to the dense ones, which are of deteriorated reconstruction quality reflected by, for example, fluctuating artefacts and inconsistent intensity of the voids, see Figure \ref{fig:B_low_freq} row 1 vs.\ 3. Due to these anomalies, the average RMSEs and OEs of the dense configurations are higher than those obtained for the sparse ones, see Table \ref{tab:compare_results}. The difference between the results obtained with dense and sparse point density is less pronounced when the randomisation of the points and frequencies is applied. The randomised and frequency-perturbed configurations have a smooth outcome compared to the cases of regular turn configurations, which result in a weaker visibility of the interior details and structure, as well as fractures within the voids and the surface layer, see Figure \ref{fig:B_low_freq} columns (III-IV). For the distinguishability of the details, the randomised configuration with sparse density is a preferable set-up, especially in the lower frequency case as shown in Figure \ref{fig:B_low_freq} columns (III-IV), which is the benchmark reconstruction in this study. We observe that in the high frequency results, based on the SSIM, RMSE, and OE in Table \ref{tab:compare_results}, the global structure of the reconstructed domain is more conspicuous for the regular turn case, but not necessarily with respect to the void and surface layer.
The choice of signal configuration is more significant with higher frequency which is revealed by the comparison between results from dense regular turn and sparse randomised configurations in Figure \ref{fig:B_high_freq} columns (I-II) vs.\  (III-IV) and Figure \ref{fig:B_boxplot} columns (I-II) vs.\  (III-IV). This is also evident in Table \ref{tab:compare_results} where the average OE of the void and surface are 80.04\% and 33.06\% for the regular turn and 77.29\% and 30.51\% for the randomised configuration.

\subsection{Low vs.\ high noise}

As expected, the reconstruction quality of the void was observed to be maximised with low noise (SNR 20 dB), while the higher noise level (SNR 12 dB) was observed to cause artefacts in the void. However, the latter allowed finding the interior details with a bias towards the surface, see Figure \ref{fig:B_high_freq} (SNR 20dB) vs.\ (SNR 12dB). The differences between the low frequency and the high frequency configurations are maintained regardless of the noise level. The structure obtained with the latter remains smooth while the visibility of the void details weakens with higher noise. This is also revealed in the results presented in Table \ref{tab:compare_results}, where the low noise has a lower overlap error of the void compared to high noise for both 20 and 60 MHz systems. Similarly, the global structure with respect to SSIM suggests that the difference between the noise levels seems comparable to the difference between the configurations in terms of their significance. The high noise case has lower RMSE and OE of the surface as presented in Table \ref{tab:compare_results}, indicating that the surface layer is better reconstructed at higher noise level. This is more pronounced in the high frequency case, see Table \ref{tab:compare_results}, and expected since the noise leaves more signature on the surface layer which is coupled with the intense surface reflection thus reducing the OE of the surface. The void RMSE in the 20 MHz system has a small difference between their respective noise levels as compared to the 60 MHz system, see Table \ref{tab:compare_results}.  The noise levels are insignificant in the 20 MHz system in terms of the SSIM measure, and largely significant in the 60 MHz system, where the low noise level performs better with a similarity of 0.71 compared to the high noise at 0.64. The WRS test shows that all the error measures have significant difference between the two noise levels in the high frequency case while only the void RMSE and surface OE have significant differences between the noise levels for the low frequency case.    

\subsection{Unfiltered vs.\ TSVD }

The TSVD filter can be observed to preserve the main details of the reconstruction, i.e., the global structure, while filtering out minor fluctuations. {The effect of the TSVD filter is pronounced in the low frequency case as it smooths out random background noise from the global structure, see Figure \ref{fig:B_low_freq}}. Similarly, the ideal reconstruction in low frequency case is from the set where the TSVD has been applied, hence suggesting this has an effect in reducing background noise from the data. The reconstructions found with TSVD are biased towards the surface layer in the high frequency case, which can be considered as a predominating structure with respect to the scattered energy, and, therefore, its contribution to the large singular values and vectors constituting the TSVD is enhanced. The unfiltered set, however, has a superior effect on the average measures in Table \ref{tab:compare_results} especially for the high frequency case, hence limiting the importance of the TSVD filtering. The unfiltered configuration has its RMSE and OE of the surface and void lower than that of the TSVD for the high frequency case, while only the surface RMSE is higher than that of the TSVD configuration in the low frequency case. The WRS test indicates that there is no significant difference in the Unfiltered and TSVD median values with respect to the error measures for both 20 and 60 MHz systems.

\subsection{SSIM, RMSE, and  OE}
The RMSE and OE are two complementary measures with the objective of evaluating the quality of the reconstructed domain. The RMSE shows the exact difference between two different distributions, while OE gives an idea of the localisation and contrast of the reconstructed details. The effect of the {signal configurations} on the reconstruction can be observed based on the values of the metrics used in evaluating reconstruction quality, i.e., RMSE and OE. The results show that the RMSE and OE give a similar pattern in evaluating the voids and the surface layer, {see Table \ref{tab:compare_results}}. The SSIM was further used as the evaluating criteria for the global structure of the reconstructed domain, complementing the noise effect in the interior structure, {hence a lower SSIM in the low frequency case as compared to the high frequency case in Table \ref{tab:compare_results}.} 

The WRS test shows that the configuration levels, i.e., dense vs.\ sparse and regular vs.\ random are significantly different for the 20 MHz case as shown in Table \ref{tab:compare_results}. The point selection is significant for all the measures in the low frequency case and  significant for 4 out of 5 measures in the high frequency system.  Similarly, when evaluating the noise level configurations, the RMSE for the surface and OE for the void also have significantly different medians for the low and high noise configurations. The TSVD vs.\ unfiltered levels are insignificant for all their corresponding measures. For the 60 MHz case, only the low vs.\ high noise configurations have significantly different medians for all their corresponding measures.

\begin{figure}[!ht]
\centering
\begin{scriptsize}
  \begin{minipage}{4.5cm} ({\bf A}) 20  MHz  \centering \\ \vskip0.1cm 
    \includegraphics[width=4.5cm]{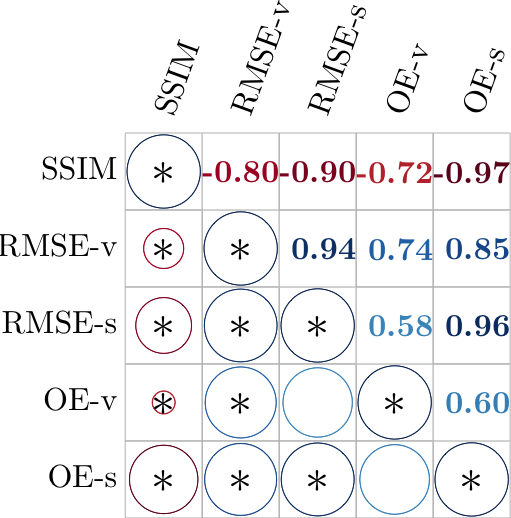}
    \end{minipage}  
   \begin{minipage}{3.5cm} ({\bf B}) 60 MHz \centering \\ \vskip0.1cm
    \includegraphics[width=3.4cm]{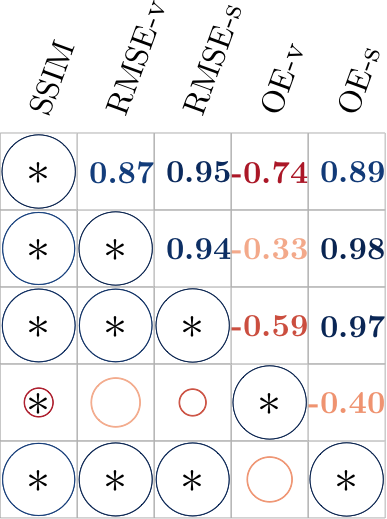}
    \end{minipage}\\ \vskip0.4cm 
     \begin{minipage}{4.9cm} \centering 
             \includegraphics[height= 0.6cm]{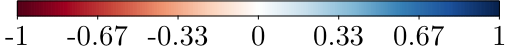}
             \end{minipage} 
    \end{scriptsize}
    \caption{Correlation heatmap between structural similarity (SSIM), Root Mean Squared Error of the void (RMSE-v), Root Mean Squared Error of the surface (RMSE-s), Overlap Error of the void (OE-v) and Overlap Error of the surface (OE-s) with respect to the mean values in Table \ref{tab:compare_results}. The significant correlations are shown with the asterisk sign *, the insignificant correlations are depicted with no asterisk at a 5\% significance level. The size of the circle shows the magnitude of significance and insignificance; larger circle implying high significance or insignificance. The colour shows the magnitude of the correlation for both centre frequencies ({\bf A}) 20 and ({\bf B}) 60 MHz, configurations.} 
    \label{fig:stat_heatmap} 
\end{figure}

The correlation heatmap in Figure \ref{fig:stat_heatmap} shows that the SSIM has a strong negative correlation with other measures ranging from $-0.97$ for OE of the surface, $-0.9$ for RMSE of the surface, $-0.82$ for RMSE of the void, and $-0.72$ for OE of the void for the 20 MHz case. This is expected since the best SSIM values are the highest and they correspond to the best values of other measures which are the lowest and vice versa as shown in Table \ref{tab:compare_results}. The negative correlation here does not relate to the reconstructions, but the measures used in assessing the reconstructions. Other measures are positively correlated with the highest being $0.96$ between RMSE of the surface and OE of the surface, and the least being $0.58$ between OE of the void and RMSE of the surface. In the 60 MHz case, OE of the void is also negatively correlated with other measures, the strongest being $-0.74$ for SSIM and the least being $-0.33$ for RMSE of the void. The other measures have strong positive inter-measure correlations ranging from $0.98$ between RMSE of the void and OE of the surface to $0.87$ between RMSE of the void and SSIM. Here, the

\begin{figure}[!ht]
    \centering
    \begin{scriptsize}  Dense configurations \\ \vskip0.1cm
    \begin{minipage}{4.2cm} \centering
    \includegraphics[width=4.2cm]{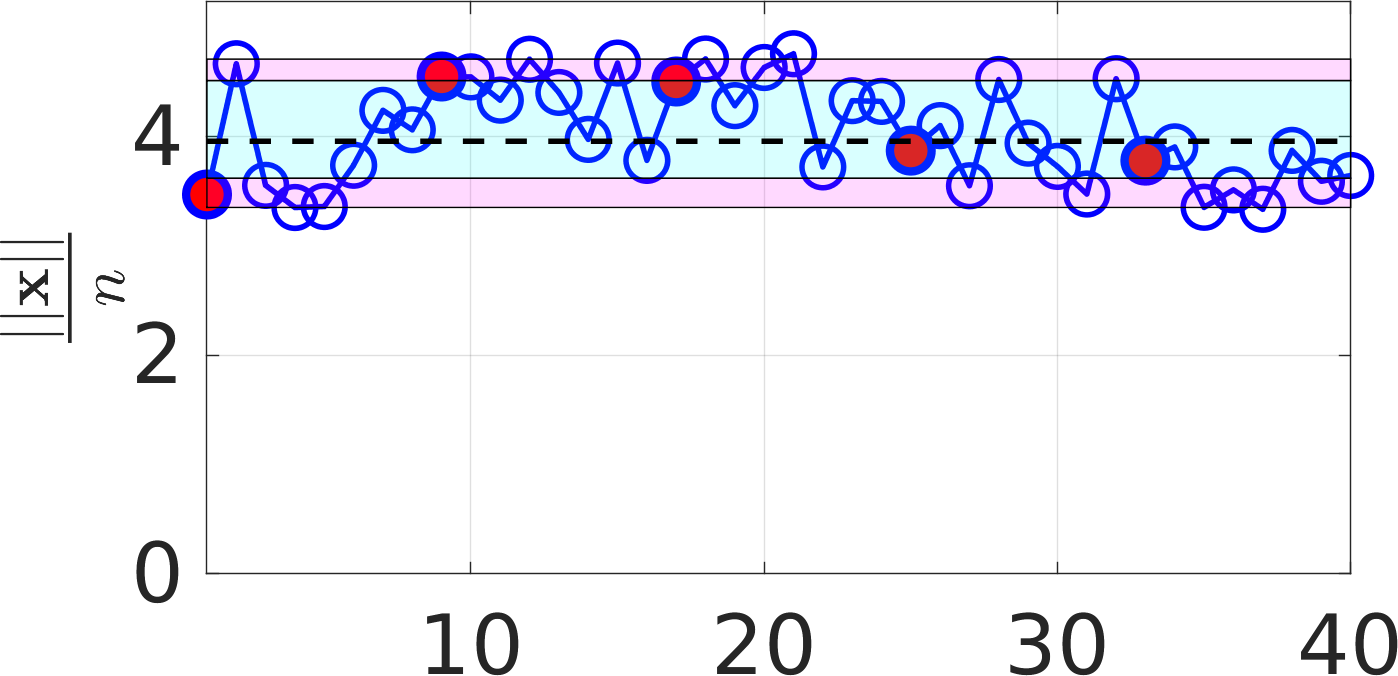} \\ \vskip0.1cm
    ({\bf A}) 20 MHz, SNR 20 dB
    \end{minipage}
     \begin{minipage}{4cm} \centering
    \includegraphics[width=3.8cm]{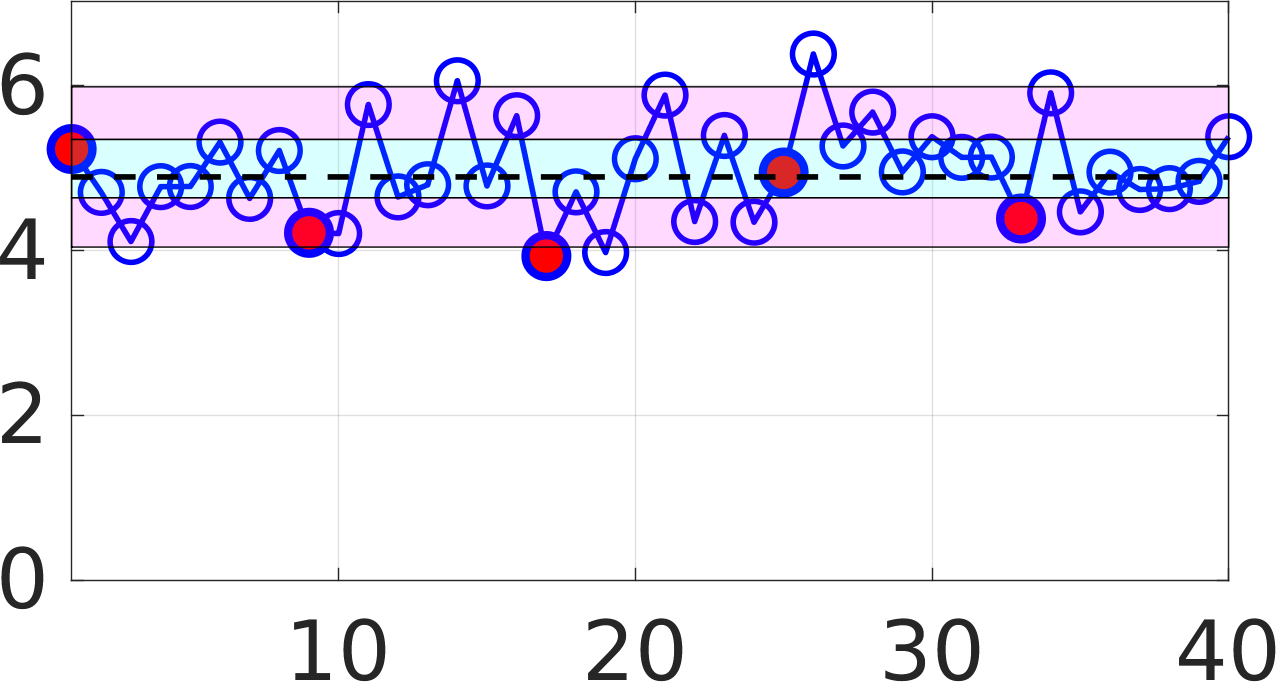} \\ \vskip0.1cm
    ({\bf A}) 20 MHz, SNR 12 dB
    \end{minipage} 
    \begin{minipage}{4cm} \centering
    \includegraphics[width=3.8cm]{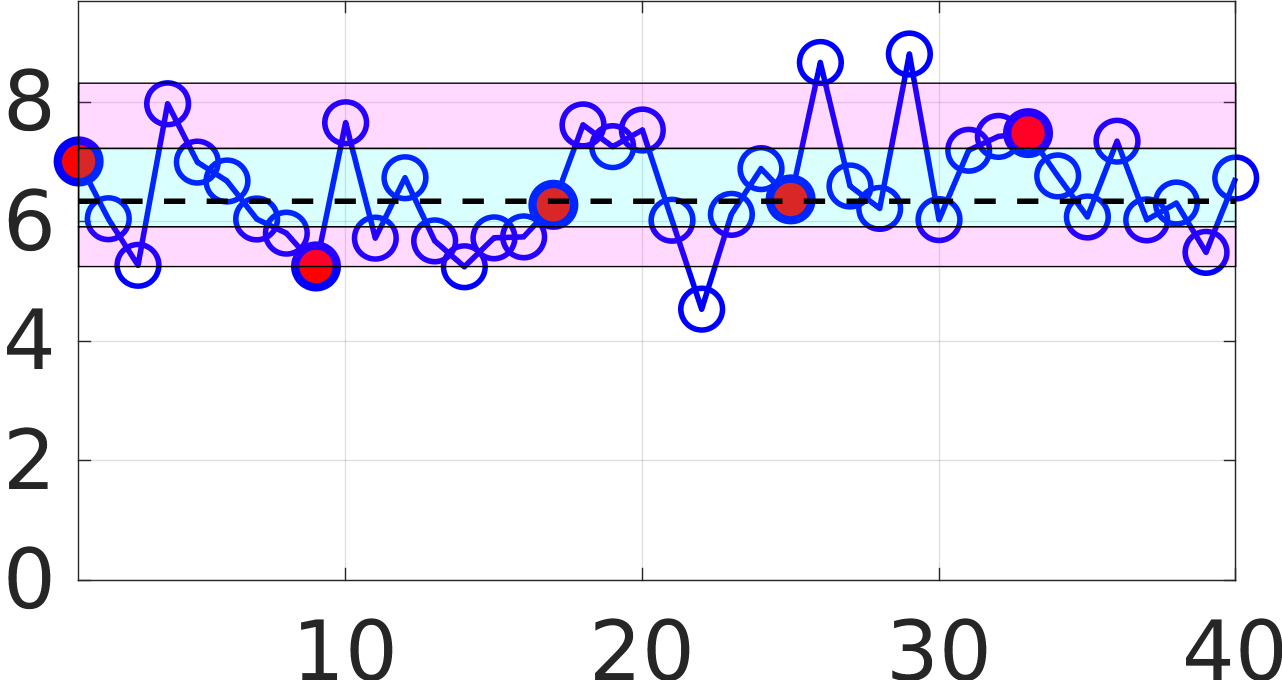} \\ \vskip0.1cm
    ({\bf B}) 60 MHz, SNR 20 dB
    \end{minipage}
     \begin{minipage}{4cm} \centering
    \includegraphics[width=3.8cm]{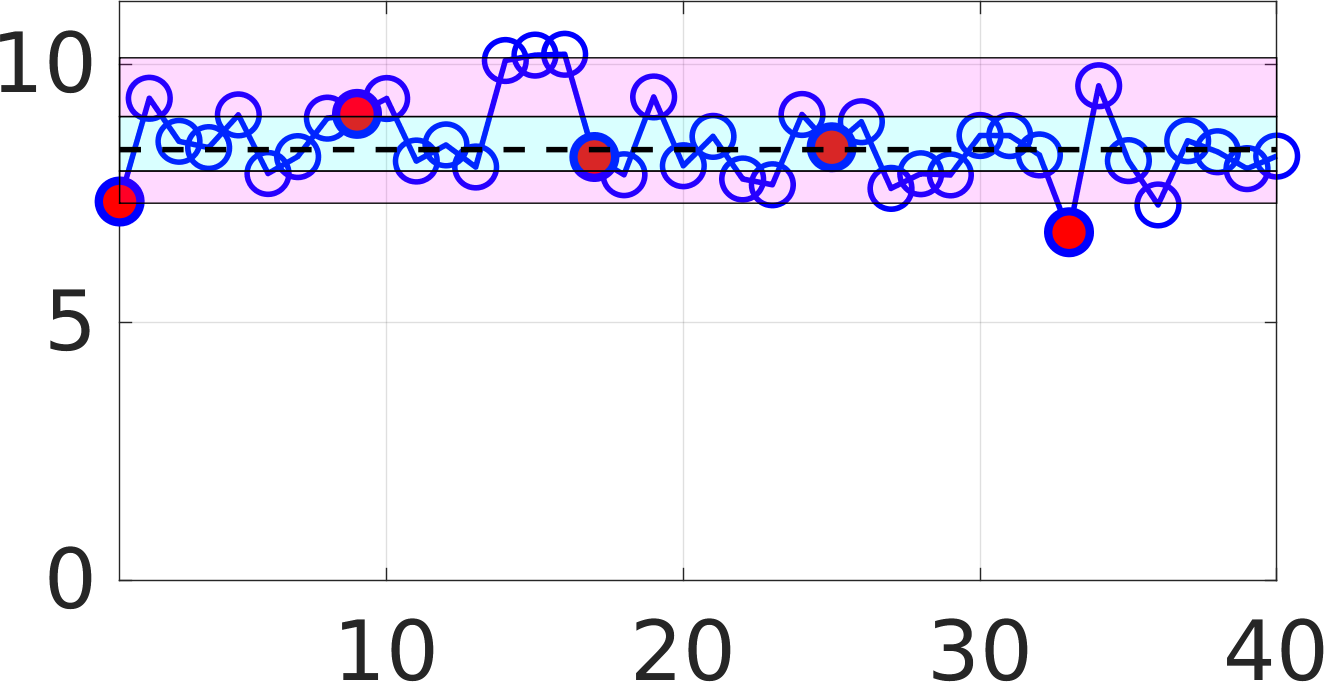} \\ \vskip0.1cm
    ({\bf B}) 60 MHz, SNR 12 dB
    \end{minipage}  \\ \vskip0.2cm  Sparse configurations  \\ \vskip0.1cm
        \begin{minipage}{4.2cm} \centering
    \includegraphics[width=4.2cm]{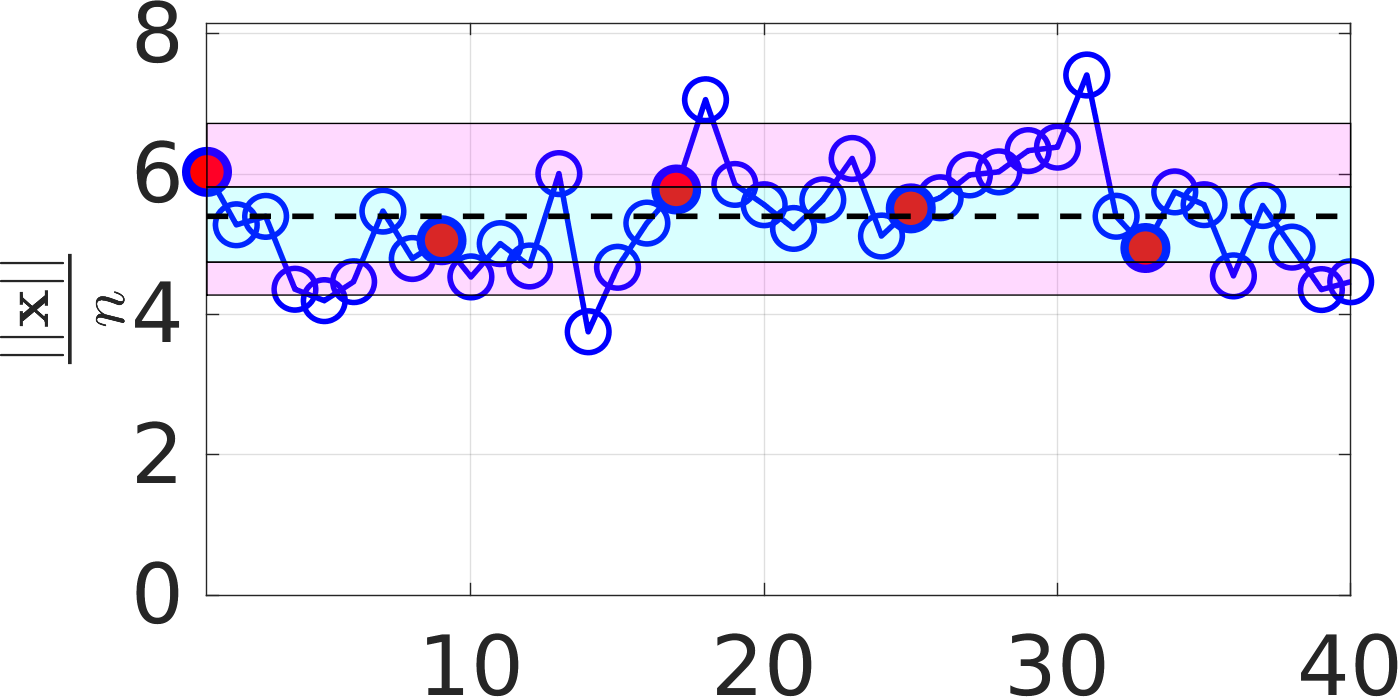} \\ \vskip0.1cm
    ({\bf A}) 20 MHz, SNR 20 dB
    \end{minipage} 
     \begin{minipage}{4cm} \centering
    \includegraphics[width=3.8cm]{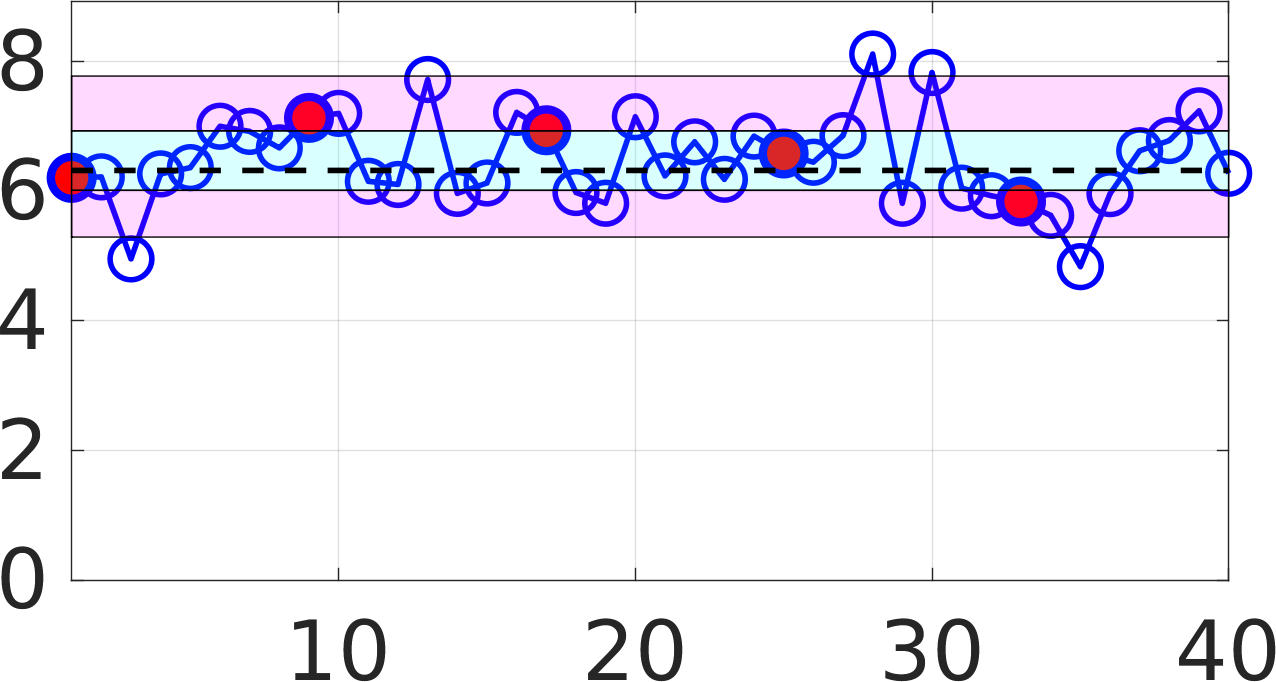} \\ \vskip0.1cm
    ({\bf A}) 20 MHz, SNR 12 dB
    \end{minipage} 
    \begin{minipage}{4cm} \centering
    \includegraphics[width=3.8cm]{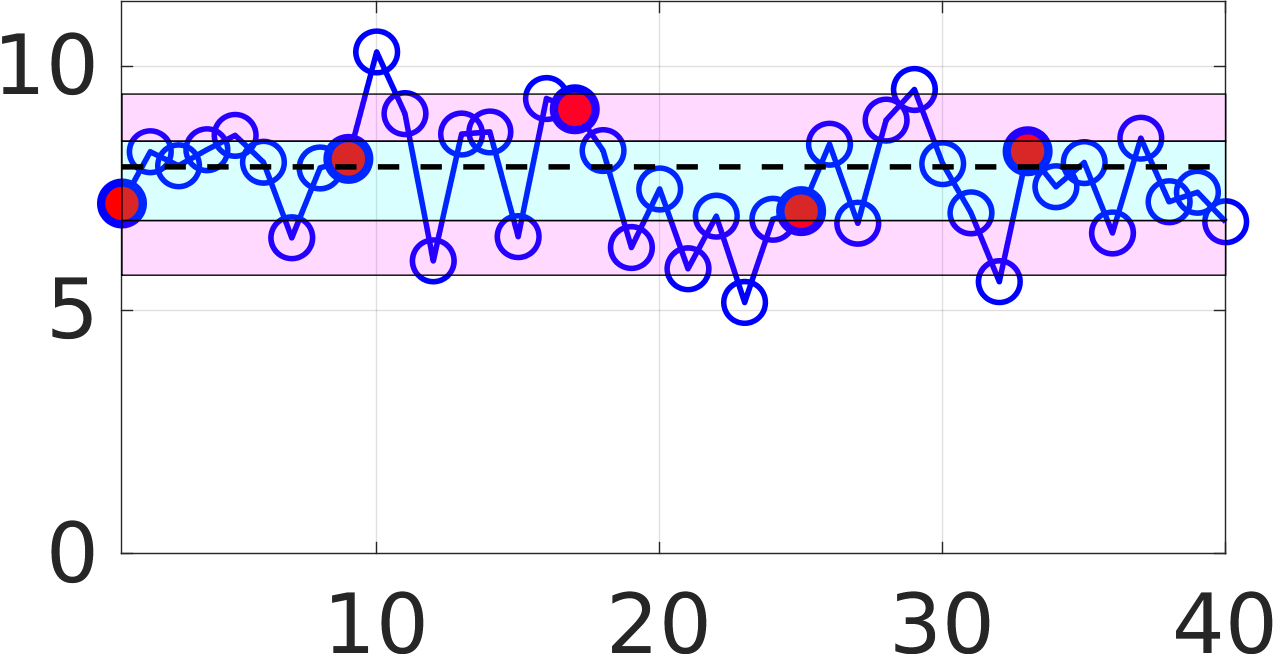} \\ \vskip0.1cm
    ({\bf B}) 60 MHz, SNR 20 dB
    \end{minipage}
     \begin{minipage}{4cm} \centering
    \includegraphics[width=3.8cm]{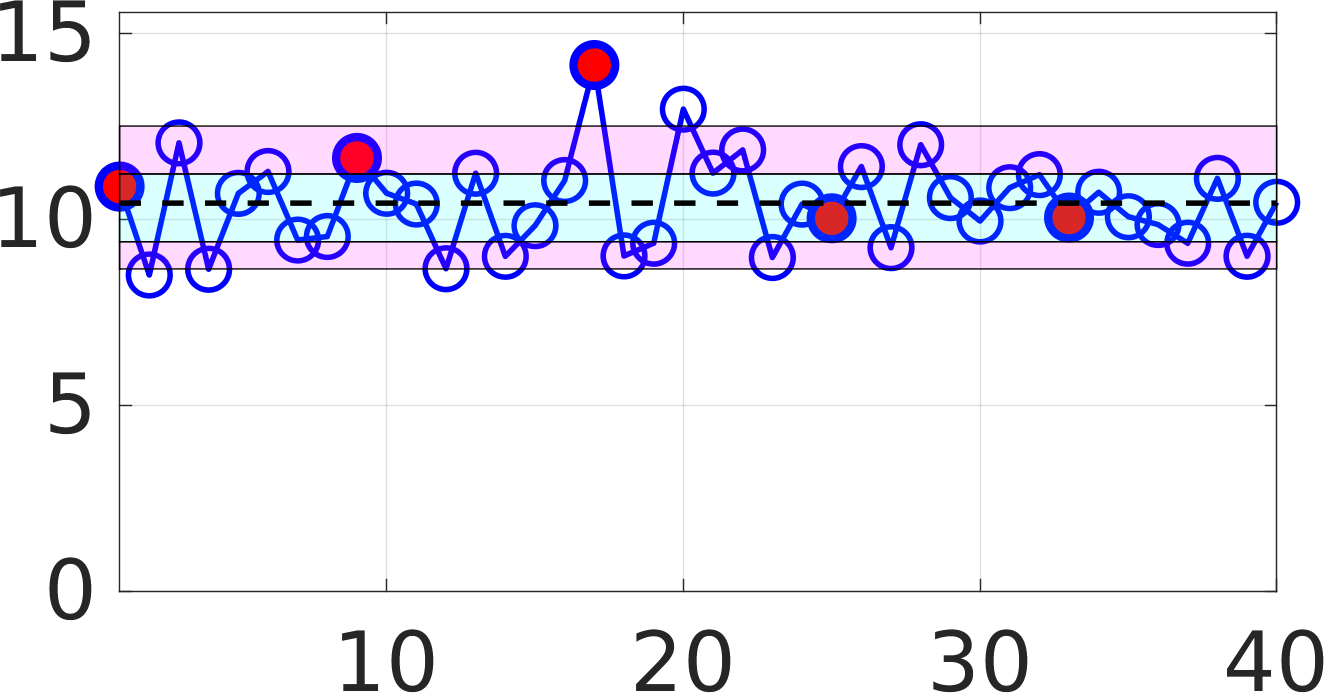} \\ \vskip0.1cm
    ({\bf B}) 60 MHz, SNR 12 dB 
    \end{minipage} \\ \vskip0.2cm
    \begin{minipage}{5cm} \centering   
   \includegraphics[width=5cm]{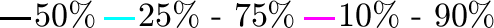}\\
          \end{minipage}\\\vskip0.2cm
    \end{scriptsize}
    \caption{The average absolute value for the entries of ${\bf x}$ as a function of the different sampling steps in the case of the unfiltered \&  randomised data. The 1st row shows the results for the dense configuration, while the 2nd row presents the results for the sparse configuration. The red marker indicates an update in the centre frequency applied in the signal demodulation process. The magenta area shows the range from 5 to 95 \% quantile of the sample variation, the cyan area shows 25 to 75 \% quantile (interquartile range), and the dashed line shows the median of the samples. The magnitude of ${\bf x}$ can be observed to grow as the noise,  signal frequency, and sparsity of the measurement point configuration grows.}
    \label{fig:sample_history}
\end{figure}

\subsection{Convergence}
\begin{figure}[!ht]
\centering
\begin{scriptsize}
  \begin{minipage}{7.5cm} ({\bf A}) 20  MHz  \centering \\ \vskip0.1cm 
    \includegraphics[width=7.5cm]{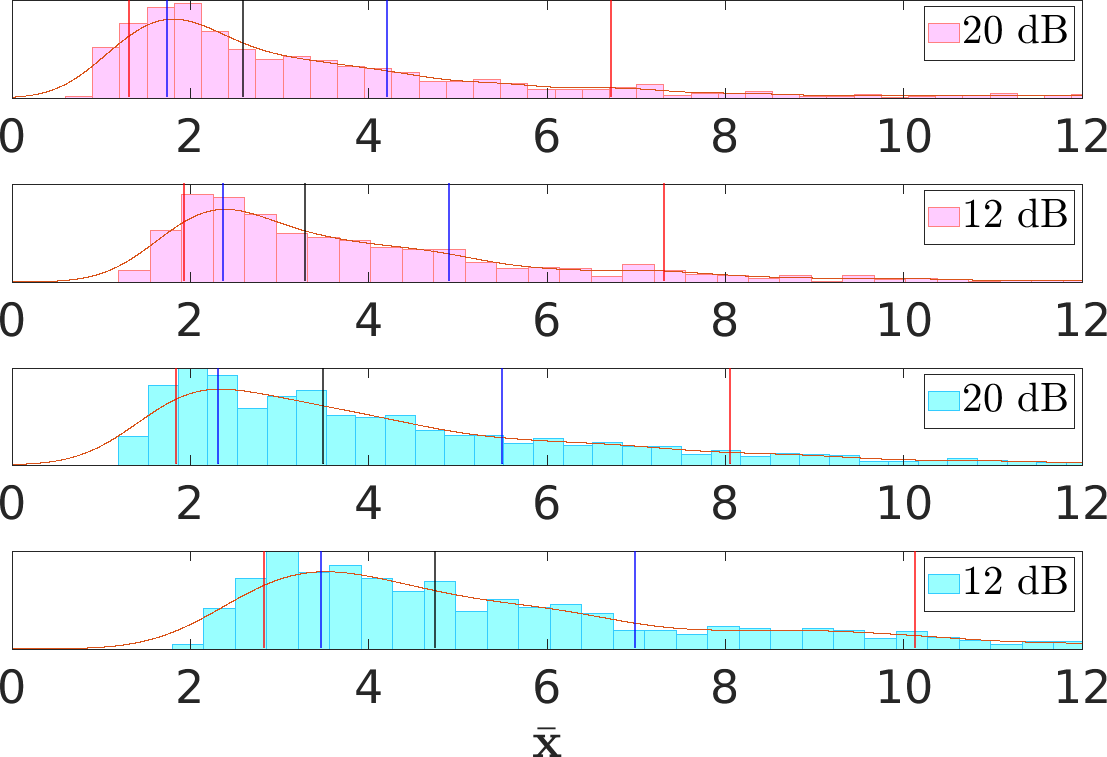}
    \end{minipage}\hspace{0.3cm}
    \begin{minipage}{7.5cm} ({\bf B}) 60 MHz \centering \\ \vskip0.1cm 
    \includegraphics[width=7.5cm]{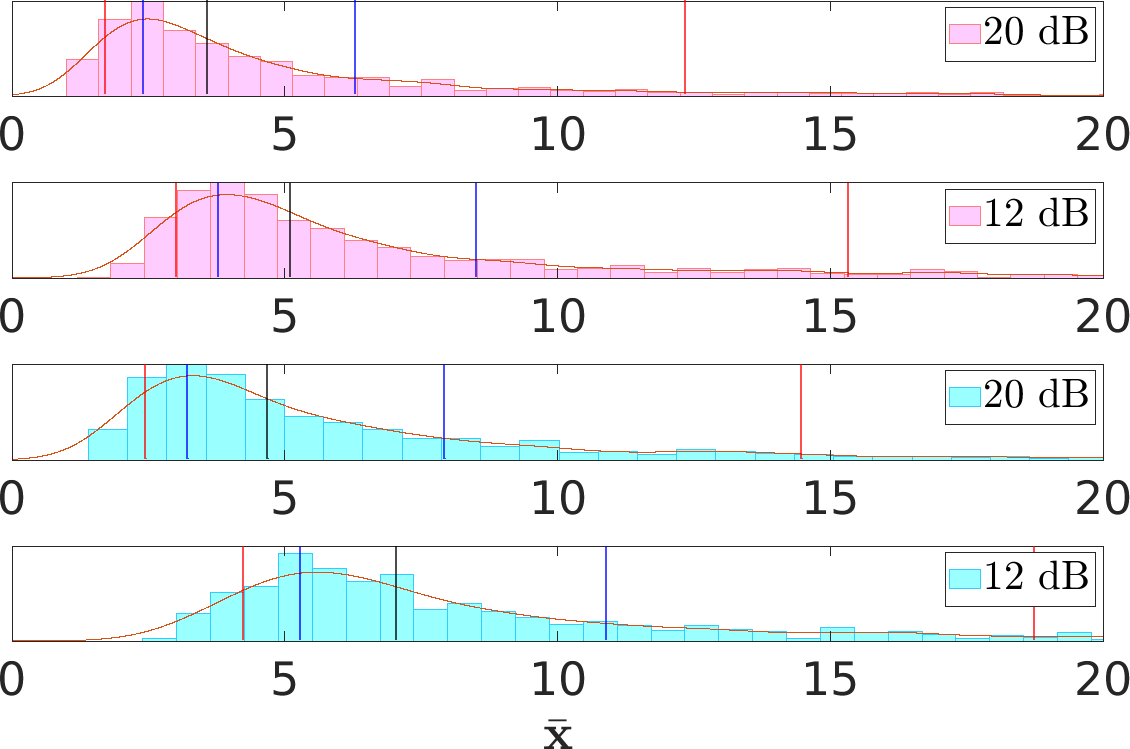}
    \end{minipage}\\\vskip0.2cm
    \begin{minipage}{5cm} \centering   
   \includegraphics[width=5cm]{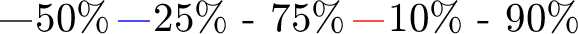}\\
          \end{minipage}
    \end{scriptsize}
    \caption{The average permittivity perturbation distribution for the({\bf A}) 20 and ({\bf B}) 60 MHz systems. The magenta histograms correspond to the dense configuration and the cyan histograms correspond to the sparse configuration, both having their noise levels indicated in their respective legends. The y-axis is the probability density function (PDF) of the estimated permittivity perturbation. This figure indicates that there is a larger median value in the higher frequency estimates and wider spread in the sparse configurations. The lower noise has smaller spread for both density configurations and systems.}
    \label{fig:sample_summary} 
\end{figure}

An investigation of the convergence and performance of several configurations on the full data is presented in Figures \ref{fig:sample_history} and \ref{fig:sample_summary}. The plots in Figure \ref{fig:sample_history} all start from zero on a norm scale to indicate the variation of the magnitude for each configuration. The upper limits are however retained for clear visibility of the spread in the respective distribution. The uncertainty in the reconstructed estimates {\bf x} spreads as the noise increases for both 20 and 60 MHz frequencies.

 The magnitude of the norm in the high frequency case is greater and its spread extends frequently outside the interquartile range. Figure \ref{fig:sample_history} also indicates the effectiveness of the sampling process. Since the sample realisations are computationally demanding to obtain, the discrepancy conditions and averaging are selected such that the randomness in the process is present and the IID assumption holds. The plot in Figure \ref{fig:sample_summary} gives a comparison of the perturbation distribution between the dense (magenta histogram) and sparse configurations (cyan histogram). From this plot, it was evident that the sparse configurations have higher spread in the perturbation distribution for both 20 and 60 MHz frequencies. The dense configuration is of lower magnitude and spread compared to the sparse case, which also shows spurious outliers beyond the 10\% to 90 \% quantile range. The effect of the randomised frequency perturbation is less evident in the dense configuration, especially in the low frequency case. The plot also shows that the outliers are smaller in the low noise case as compared to the high noise.

\subsection{Beyond a piece-wise constant domain}

\begin{figure}[!ht]
    \centering \begin{scriptsize}\begin{minipage}{7.7cm} \centering
 \rotatebox{90}{\hskip-1.9cm 60 MHz \hskip 0.7cm 20 MHz \hskip .8cm Target}   \begin{minipage}{1.8cm} \centering
 \includegraphics[width=1.8cm]{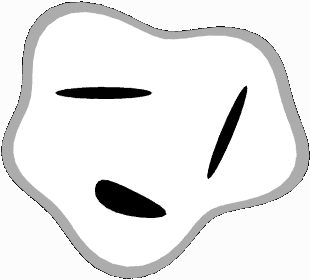} \\
  \includegraphics[width=1.8cm]{images/low_linear/fig_1_2_4_rec.png} \\
  \includegraphics[width=1.8cm]{images/high_linear/fig_2_2_3_rec.png} 
    \end{minipage}
    \begin{minipage}{1.8cm} \centering   
   \includegraphics[width=1.8cm]{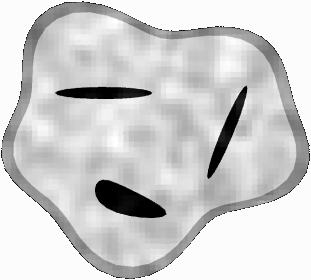}\\
   \includegraphics[width=1.8cm]{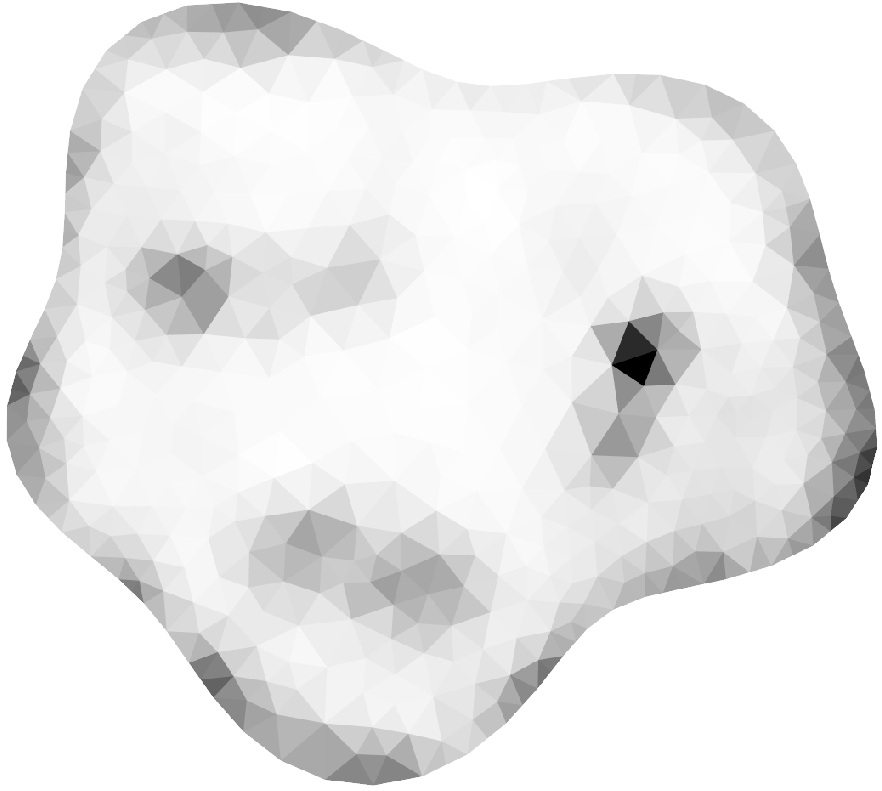}\\ 
   \includegraphics[width=1.8cm]{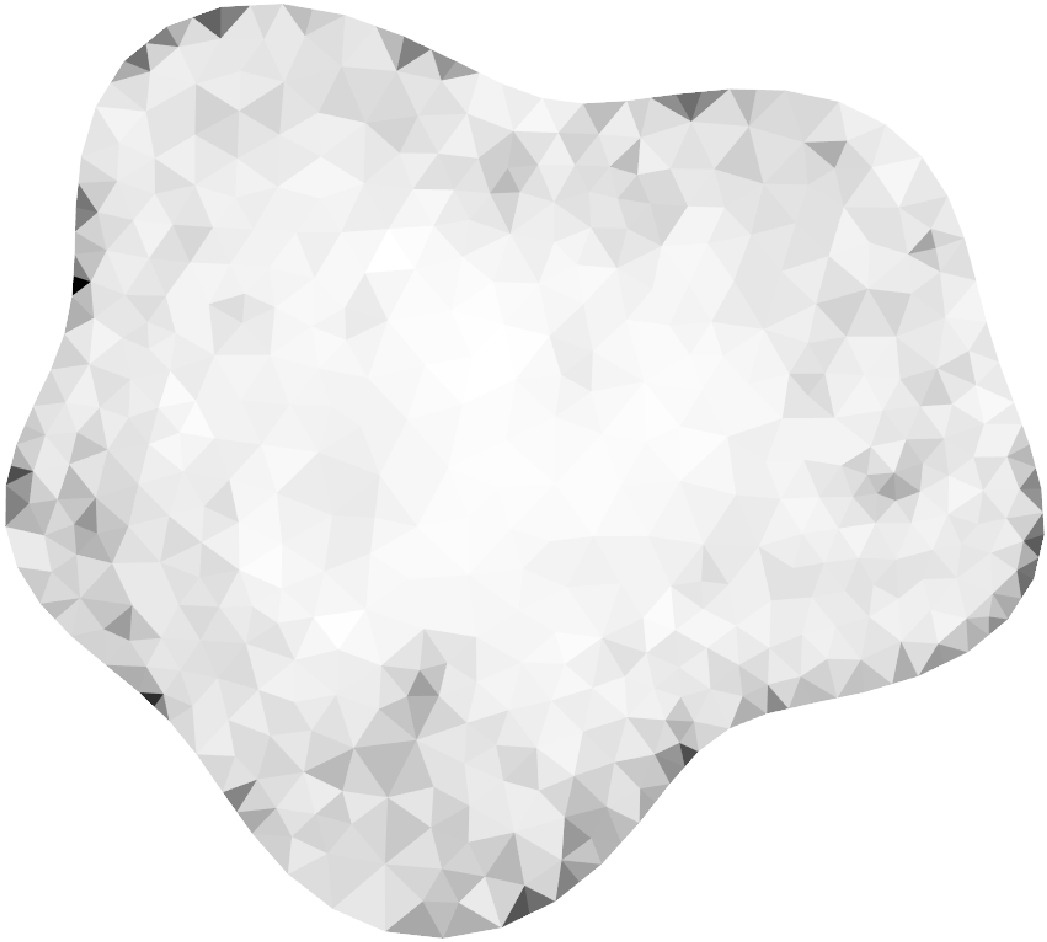} 
          \end{minipage}
    \begin{minipage}{1.8cm} \centering 
           \includegraphics[width=1.8cm]{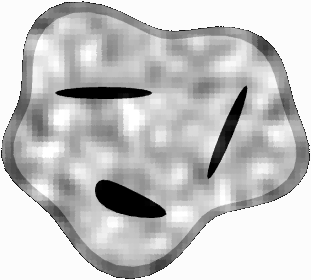}\\ 
           \includegraphics[width=1.8cm]{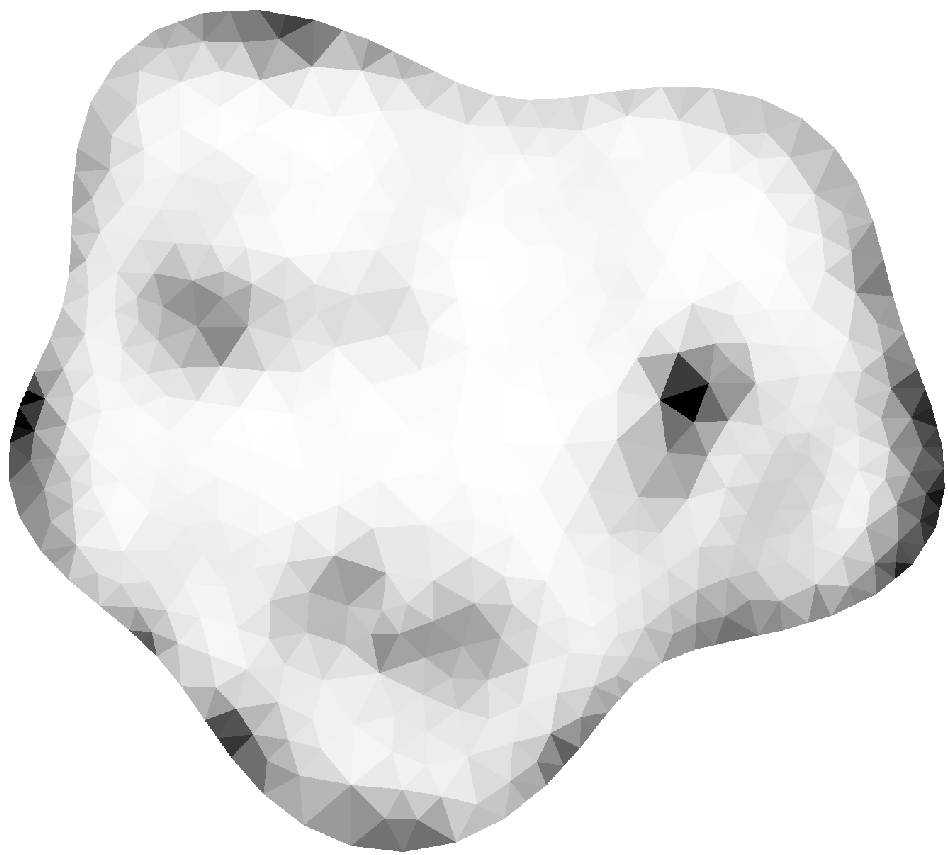}\\
           \includegraphics[width=1.8cm]{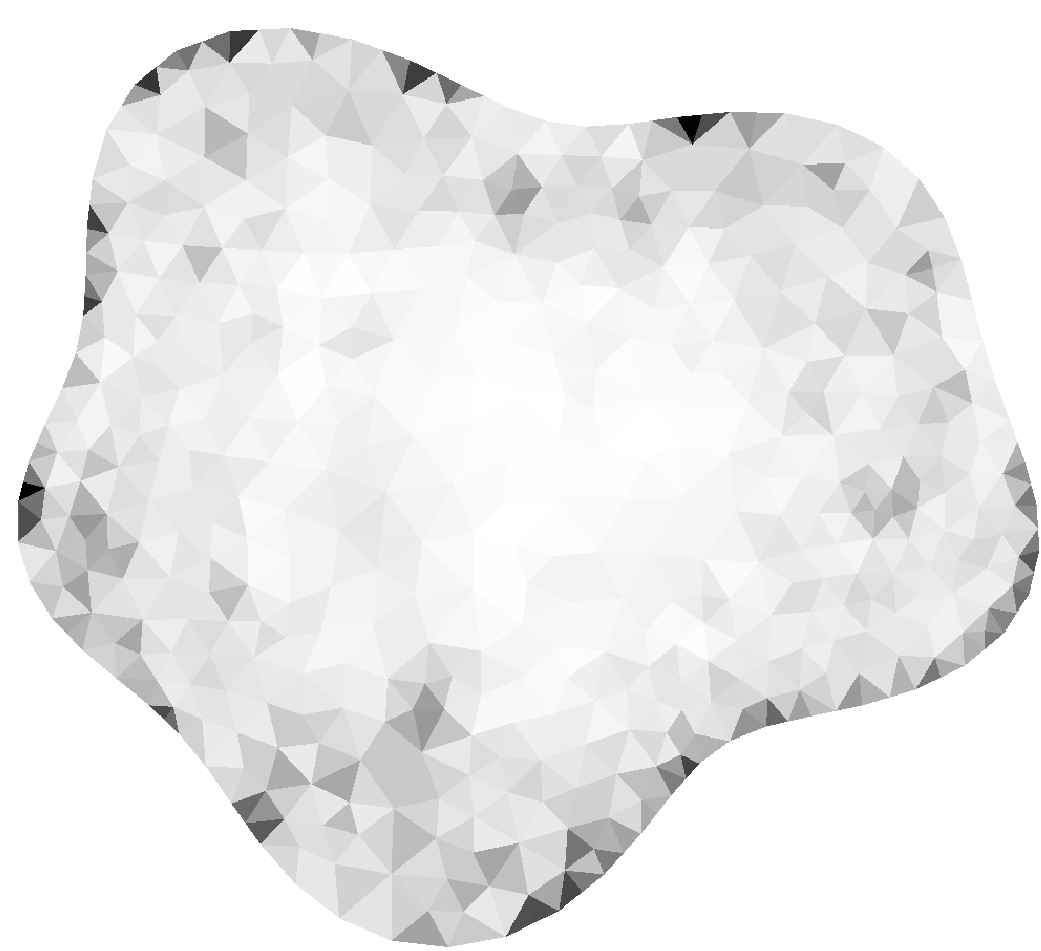} 
               \end{minipage}  \begin{minipage}{0.6cm} \centering
             \includegraphics[height= 5cm]{images/color_bar_lin_1.png}
             \end{minipage}\\\vskip0.2cm
              \begin{minipage}{1.8cm} \centering        
        ({\bf I}): Piece-wise constant
             \end{minipage}
    \begin{minipage}{1.8cm} \centering 
   ({\bf II}): Low  background contrast random field  
            \end{minipage}
    \begin{minipage}{1.8cm} \centering 
   ({\bf III}): High background contrast random field
               \end{minipage} \end{minipage} \\ \vskip0.2cm 
               \begin{minipage}{8.5cm} \centering
     \begin{minipage}{5cm} \centering   
   \includegraphics[width=5cm]{images/point_cloud/outside1.png}\\
          \end{minipage}\\\vskip0.2cm
   \begin{minipage}{4.1cm} \centering 
 \includegraphics[width=4.1cm]{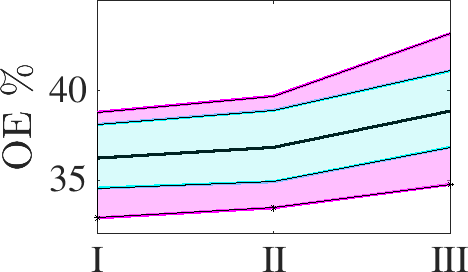} 
 \\ \vskip0.1cm
    ({\bf A}) 20 MHz, OE of the surface
    \end{minipage}
    \begin{minipage}{4.1cm} \centering
 \includegraphics[width=4.1cm]{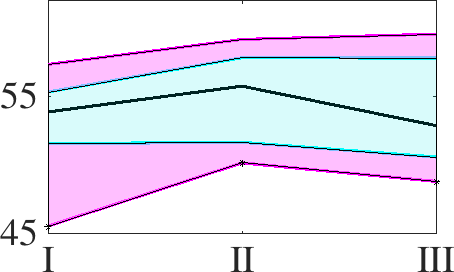}
 \\ \vskip0.1cm
    ({\bf A}) 20 MHz, OE of the void
  \end{minipage}\\ \vskip0.2cm 
  \begin{minipage}{4.1cm} \centering \vskip0.2cm 
      \includegraphics[width=4.1cm]{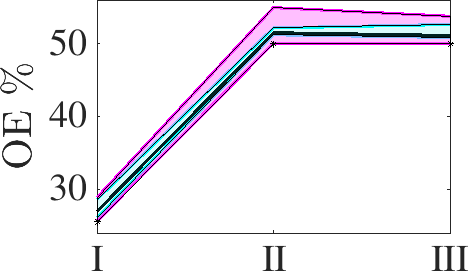}
      \\ \vskip0.1cm
    ({\bf B}) 60 MHz, OE of the surface
      \end{minipage}
      \begin{minipage}{4.1cm} \centering \vskip0.2cm 
     \includegraphics[width=4.1cm]{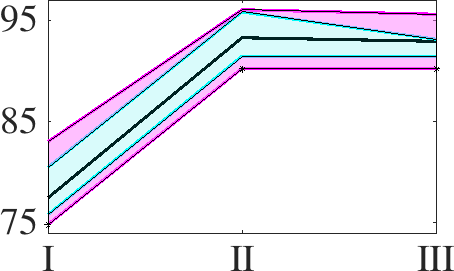}
      \\ \vskip0.1cm
    ({\bf B}) 60 MHz, OE of the void
    \end{minipage}\\ \vskip0.2cm 
          \end{minipage} 
             \\ \vskip0.2cm 
             \end{scriptsize}
    \caption{Benchmark reconstructions compared to the random field model reconstructions for the 20 and 60 MHz systems. The 1st row shows the variation in the background permittivity contrasts for the respective models. The 2nd and 3rd rows present the reconstruction of the models for the 20 and 60 MHz systems, respectively. The surface and void OE for the different background models is shown in the 4th and 5th rows with their interquartile and interdecile range coloured cyan and magenta, respectively. This figure shows that the benchmark reconstruction outperforms the random field models which can be observed by visual inspection and the OE curves.}
    \label{fig:C_point_cloud}
\end{figure}

We carried out a numerical comparison between the benchmark reconstructions for the 20 and 60 MHz systems and two random field models shown in Figure \ref{fig:C_point_cloud}. The first having background amplitude fluctuation of 1 (i.e., surface, and interior part of the domain excluding the void have a permittivity distribution between 3.5 and 4.5) and the second having a background amplitude fluctuation of 4 (i.e., surface, and interior part of the domain  has a permittivity distribution between 1 and 5). The reconstruction obtained for the 20 MHz system for the random field models show clear visibility of the voids and surface layer. However, when compared to the benchmark model as shown in Figure \ref{fig:C_point_cloud}, we see that the benchmark reconstruction has more visibility of the voids and surface layer. This is expected as the background permittivity contrast variation increases the discrepancy formulated in Equation \eqref{discrepancy_inequality}. 
The  benchmark reconstruction of the 60 MHz system  outperforms the reconstruction from the two random field models, in the sense that we can see the voids and surface layer of the benchmark model by visual inspection of the reconstruction. The surface layer and a faint strand of the bottom void is visible in the other reconstruction. The OE of the surface and void are also presented for this investigation in  Figure \ref{fig:C_point_cloud}. The Figure shows that the OE grows almost linearly from the benchmark to the high background contrast variation case for the 20 MHz system. In the 60 MHz system, the OE reaches a plateau at low background contrast variation case which either increases or decreases as it progresses to the high background contrast case depending on the quantile considered.   The interquartile and interdecile range of the samples are shown in the cyan and magenta respectively. The spread in the void OE is wider for both the 20  and 60 MHz systems as compared to the surface OE, indicating that the uncertainty in reconstructing the void is higher than that of the surface.

\section{Discussion}
\label{sec:discuss}

This article describes and evaluates numerically a filtering approach, which allows marginalising wavelength-induced random errors from a large full-waveform dataset. This approach enables obtaining a reconstruction of the surface layer and voids of a 160 m diameter non-convex two-dimensional asteroid model, via a monostatic measurement of a full-wavefield signal, having centre frequencies of 20 and 60 MHz and bandwidths of 10 and 20 MHz, respectively. Finding a reconstruction was feasible with 20- and 12-dB SNR, of which the former has been estimated as the lower bound of the CONSERT measurement of \cite{Kofman2015} and the latter is slightly (2 dB) above the 10 dB limit which has been suggested as the lower bound for finding a reasonable reconstruction for the interior structure of an asteroid model in the recent numerical simulations and laboratory experiments  \cite{takala2018far,Sorsa2019,Eyraud2020analog,sorsa2021analysis,deng2021ei+}.

We demonstrate through numerical experiments on a two-dimensional domain that, when a centre frequency applicable in a space investigation is used in the numerical simulation of the full-wavefield together with quadrature amplitude modulation and demodulation of the simulated signal \cite{sorsa2021analysis}, the reconstruction of the domain is improved i.e., lessening wavelength-induced uncertainty, when it is found via averaging several candidate reconstructions or decomposing the signals into meaningful and noisy parts. The present results show that the expected effect of phase errors is of significantly higher magnitude in comparison to the noise level, and that their effect on the final reconstruction can be reduced via averaging over a sample of candidate reconstructions obtained by a randomising signal configuration in the spatial and spectral domains. Comparing the results obtained with regular and randomised signal configurations, the latter ones demonstrate an overall improved reconstruction quality which we interpret as a consequence of a successfully created larger sample of candidates. Namely, the convergence of the sample means as stated by the generalised central limit theorem \cite{liu2008monte} depends only on the sample size $K$ as $\mathcal{O}(K^{-1/2})$, if it is appropriate otherwise. Here the frequency randomisation constituted an important way to extend the sample size, accounting as a surrogate for the uncertainties due to wave propagation and domain structure.

As for the dense and sparse configurations, the wavelength-induced uncertainties can be identified as the cause of the difference in the reconstructions obtained. The sparse configuration yields a higher contrast as compared to a dense one, since phase fluctuations lead to larger fluctuations in the reconstruction the greater the number of data point in the constellation set. Hence, aggregating small collections of phase fluctuations over a large number of constellations preserves the phase information as compared to aggregating large phase fluctuations over a small number of constellations. We can infer from Figure \ref{fig:sample_summary} that obtaining the exact permittivity values from the inversion process is challenging task if the full-wave data is not augmented with some additional information such as the signal travel-time. The permittivity perturbation norm spreads in the sparse configuration case with the low noise level cases having shorter-tailed distribution compared to the high noise cases. The estimated permittivity perturbations are large, since the formalism in equations \eqref{perturbation}, \eqref{margin}, and \eqref{tv_solution} show that we can only get an estimate ${\bf x}$ for the permittivity perturbation ($\tilde{{\bf x}}$), and this estimate is largely dependent on the model and wavelength-induced uncertainties that cannot be totally marginalised. We have shown distribution of the estimated permittivity perturbation in Figure \ref{fig:sample_summary}. The challenge of estimating exact permittivity distributions and perturbation distribution i.e., minimising the errors in the estimate ${\bf x}$ obtained in this current study is to be considered as a future research goal. Since we assume that the errors in the perturbation estimates obtained are not flat and cannot be fitted to the exact perturbation distribution as that would be an inverse crime, a more robust marginalisation method or supporting data would be needed to solve this task.

In this study, we assume that the averaged total variation regularised estimates in the marginalisation process are independent (enough) and identically distributed (IID) so that the convergence conditions of the central limit theorem or its generalisation to weakly correlated samples hold \cite{liu2008monte}. This assumption  sets a challenge for choosing the measurement point configuration and phase discrepancy appropriately.  We covered two possible strategies for spatial point selection, otherwise known as statistical sparsity-based learning with a suitable prior model \cite{mirbeik2021statistical}.  The sample size of the present randomised configuration was found to reduce the phase-dependent fluctuations by 2\% - 26\% relative to the regular configuration in terms of the OE and RMSE measures, see Table \ref{tab:compare_results}. Further improvement of the current estimates might be obtained by applying a larger sample, relying on the asymptotic central limit theorem, which states that the accuracy of a sample-based estimate is inversely proportional to the square root of the sample size. Alternatively, as the IID condition is only assumed due to the missing {\em a priori} information on the wavelength-induced uncertainties the formation of the sample can be supervised by filtering out the possible outliers. In such a procedure, the principal component analysis \cite{lee2008application}, or machine learning and deep neural network \cite{geng2021deep,li2019survey,ji2021deep} might be utilised. In particular, a deeper investigation on the dependence of the wavelength-induced uncertainties on the frequency and receiver positions might be conducted to gain more {\em a priori} understanding of how the receiver positions could be distributed around the orbit, hence determining if the point selection approach can be improved.

From a potential mission design viewpoint, the results obtained suggest that setting the bandwidth of the modelled signal pulse based on the discrepancy principle and averaging the measurement point density following the Nyquist criterion is crucial w.r.t.\ the stability of the reconstruction process. If the number of averaging points is limited further from the current setup, the size of the reconstructed details grows, or they tend to get blurred as a natural consequence of the lower measurement point density. Due to the importance of selecting the measurement point configuration suggested by the WRS test, it is also obvious that a sparse configuration alone will not allow obtaining an optimal reconstruction quality. However, sampling can be applied without having a point density matching with the Nyquist criterion w.r.t.\ the centre frequency, which can be the case of {\em in-situ} measurements. This is important if the global internal structure of an asteroid is to be explored during a space mission, to provide answers to the scientific questions on the internal composition asteroids and/or comets \cite{carry2012density,jutzi2017formation,Herique2018}.
  
The TSVD filter is based on excluding the less significant signal components out of the inversion process, which is known to improve the outcome of Ground-penetrating radar (GPR) reconstructions \cite{ludeno2020comparison}. The TSVD was found to have a regularising effect on the reconstruction, while it did not affect the major fluctuations due to the wavelength-induced uncertainties but rather increased the error values. As our concentration is on the monostatic data, i.e., the  points of transmission and measurement coincide, similar to the future JuRa investigation, TSVD was considered as a way to decompose the signal. If more complex signal patterns e.g., bistatic or multistatic measurements will need to be processed as suggested in several previous studies \cite{Sorsa2019}, then more advanced decomposition approaches will be necessary. For example, the decomposition of the time reversal operator DORT (Décomposition
de l’opérateur de retournement temporel) \cite{prada1996decomposition,fink2000time} finds the most 'reflective' multistatic patterns by finding the eigenvalues and eigenvectors of a symmetric (reciprocal) transfer matrix between the transmission and measurement points. 

While our concentration is on reconstructing the real part of the relative permittivity structure, motivating real-valued discretization of the forward and inverse problem \cite{sorsa2020time}, a complex formulation follows from the complex-valued QAM signal, where the in-phase and quadrature component compose the real and imaginary part, respectively. Inverting a complex-valued system was found to be advantageous to reduce any possible fluctuations due to wavelength-induced uncertainties. Namely, the wavelength-induced uncertainty might affect only one of the QAM signal components, in which case, a real-value Jacobian matrix might contain two opposite fluctuations, potentially cancelling each other, while a complex Jacobian treats both perturbations appropriately as two complex components with a positive absolute value, thereby, preserving the visibility of the scatterer causing the deviation.

The inversion process applied relies on the linearisation of the full wavefield w.r.t.\ the permittivity perturbation in a given position \cite{virieux2009overview}. Hence, while the wave propagation is modelled as a nonlinear process, taking into account indirect signal paths including reflection and refraction and higher-order or multi-path scattering effects between the scattering obstacles and the structure of the target, the mutual coupling effects between the  obstacles, are omitted in this study. Based on  earlier results of the same model, the improvement obtained with  a higher-order polynomial or Born approximation via a recursive use of the present linearisation  \cite{sorsa2020time} is assumed to be a minor one. In particular, as the higher-order approximation was shown to be sensitive to noise which cannot be avoided in the present application due to the wavelength-induced uncertainties. Notably, the numerical inaccuracies in the modelled material distribution, which gives rise to the requirement for the discrepancy condition, are always present regardless of the order of Born approximation incorporated in the model. However, another modelling approach for the nonlinear higher-order scattering effects might yield an improvement. Therefore, comparisons between different methods and solvers, e.g., ray-tracing techniques \cite{gassot2020sprats,ciarletti2015} and frequency-based methods \cite{Eyraud2019,Eyraud2020analog} will be important.

We have chosen a two-dimensional reference domain for this study, as it allows examining the effects of wavelength-induced uncertainties on  numerical inversion efficiently without the need of high-performance cluster computations. In addition, our domain has a few features that makes it relevant from the  mission design viewpoint, in particular, considering asteroids as potential targets. Firstly, our model includes a surface layer which is a potential structure in asteroid interiors as shown by impact simulation studies \cite{jutzi2017formation}. As a surface layer is known to significantly hinder the detection of buried objects \cite{dogan2017detection}, it can be considered as important from the practical applicability aspect. Many studies, however, omit this aspect, e.g.,  \cite{deng2021ei+}, thus, potentially somewhat exaggerating the detectability of the deep interior structures. In addition to the surface layer, another important modelling feature ensuring the relevance of our model is the realistic value of the real background relative permittivity which, based on the assumed mineral composition of the target \cite{herique2002dielectric,Herique2018}, was set to be 4. A significantly lower value 1.6  has been used in \cite{haynes2021small}, hence, reducing the complexity of the wave propagation inside the target as compared to our present study. Furthermore, the extension of the piece-wise constant domain to a random field domain gives an indication of how the complexity of the wavelength-induced uncertainties grows. In essence, the high frequency (60 MHz) is more sensitive to all kinds of uncertainty which in turn, reduces the reconstruction quality of the target domain. The experimental relevance of the present   structural model composition can be motivated by the recent successful laboratory experiments with 3D-printed analogue targets \cite{sorsa2021analogue,sorsa2021analysis,Eyraud2020analog} with a comparable structure compared to the present case.

The next step will be to investigate the present methodology with a three-dimensional analogue object and with experimental scattering data similar to  \cite{sorsa2021analysis,sorsa2021analogue,Eyraud2020analog}. In such a process, the outcome of this study to understand combining point cloud data into a single reconstruction is important. The current two-dimensional implementation can, in principle, be directly extended to  the three-dimensional laboratory coordinates as done in the previous analogue studies. This study also allows for further numerical investigation of the uncertainties relating to the  wavelength-induced uncertainties, which is essential due to the significance of their magnitude based on the current results.

 \bibliographystyle{elsarticle-num-names} 
 \bibliography{Referencias}

\section*{Acknowledgement}
This study was carried out as a part of the Academy of Finland project, ICT 2023 (FETD-Based Tomographic Full-Wave Radar Imaging of Small Solar System Body Interiors; project number 336151) and was supported by the Academy of Finland Centre of Excellence in Inverse Modelling and Imaging, 2018-2025. Y.O. Yusuf was also supported by the Magnus Ehrnrooth Foundation through the graduate student scholarship.
\end{document}